\documentclass{article}
\usepackage[utf8]{inputenc}
\usepackage{xcolor}
\usepackage{hyperref}
\usepackage{amsmath}
\usepackage{amsfonts}
\usepackage{amssymb}
\usepackage{stmaryrd} 
\usepackage{physics}
\usepackage{graphicx}
\usepackage[margin=1in]{geometry}
\usepackage{adjustbox}
\usepackage{appendix}
\usepackage{bm}
\usepackage{braket}
\usepackage{cite}

\usepackage[normalem]{ulem}

\newcommand{\UU}{$U_q(\mathfrak{sl}_n)$}

\graphicspath{ {./diagrams/} }

\newtheorem{remark}{Remark}

\begin{document}

\thispagestyle{empty}

\begin{center}

\Large{$U_{\mathfrak{q}}(\mathfrak{sl}_3)$ web models: Locality, phase diagram and geometrical defects}

\vspace{1.0cm}

\large{Augustin Lafay$^{1}$, Azat M.\ Gainutdinov$^{2,3}$, and Jesper Lykke Jacobsen$^{1,4,5,6}$}

\vspace{1.0cm}

{\sl\small $^1$ Laboratoire de Physique de l'\'Ecole Normale Sup\'erieure, ENS, Universit\'e PSL, \\
CNRS, Sorbonne Universit\'e, Universit\'e de Paris, F-75005 Paris, France\\}

{\sl\small $^2$
Institut Denis Poisson, CNRS, Universit\'e de Tours, Universit\'e d'Orl\'eans, \\Parc de Grandmont, F-37200 Tours, France\\}

 {\sl\small $^3$  
National Research University Higher School of Economics,\\ Usacheva str., 6, Moscow, Russia \\}

{\sl\small $^4$ Sorbonne Universit\'e, \'Ecole Normale Sup\'erieure, CNRS, \\
Laboratoire de Physique (LPENS), F-75005 Paris, France\\}

{\sl\small $^5$ Universit\'e Paris Saclay, CNRS, CEA, Institut de Physique Th\'eorique, \\ F-91191 Gif-sur-Yvette, France\\}

{\sl\small $^6$ Institut des Hautes \'Etudes Scientifiques, Universit\'e Paris Saclay, CNRS, \\
Le Bois-Marie, 35 route de Chartres, F-91440 Bures-sur-Yvette, France\\}
 
\end{center}

\begin{abstract}

We continue investigating the generalisations of geometrical statistical models introduced in \cite{LGJ21}, in the form of
models of webs on the hexagonal lattice
 $\mathbb{H}$ having a $U_{\mathfrak{q}}(\mathfrak{sl}_n)$ quantum group symmetry. 
We focus here on the $n=3$ case of cubic webs, based on the Kuperberg $A_2$ spider,
and illustrate its properties by comparisons with the well-known dilute loop model (the $n=2$ case)
throughout. A local vertex-model reformulation is exhibited, analogous to the correspondence
between the loop model and a three-state vertex model. The $n=3$ representation uses seven states
per link of $\mathbb{H}$, displays explicitly the geometrical content of the webs and their $U_{\mathfrak{q}}(\mathfrak{sl}_3)$
symmetry, and permits us to study the model on a cylinder via a local transfer matrix. A numerical study of
the central charge reveals that for each $\mathfrak{q} \in \mathbb{C}$ in the critical regime, $|\mathfrak{q}| = 1$, the web model possesses
a dense and a dilute critical point, just like its loop model counterpart. In the dense $\mathfrak{q}=-{\rm e}^{i \pi/4}$ case,
the $n=3$ webs can be identified with spin interfaces of the critical three-state Potts model defined on the
triangular lattice dual to $\mathbb{H}$. We also provide another mapping to a $\mathbb{Z}_3$ spin model
on $\mathbb{H}$ itself, using a high-temperature expansion. We then discuss the sector
structure of the transfer matrix, for generic $\mathfrak{q}$, and its relation to defect configurations in both the strip and the cylinder
geometries. These defects define the finite-size precursors of electromagnetic operators. This discussion paves the road
for a Coulomb gas description of the conformal properties of defect webs, which will form the object of a
subsequent paper. Finally, we identify the fractal dimension of critical webs in the $\mathfrak{q}=-{\rm e}^{i \pi/3}$ case,
which is the $n=3$ analogue of the polymer limit in the loop model.

\end{abstract}

\newpage 

\section{Introduction}

Two-dimensional lattice models of loops have been widely studied for many years and have proved to be a focal
point of a diverse array of methods, including quantum integrability \cite{Nienhuis,Baxter86,WNS92,IntReview},
algebra \cite{AlgReview,GL1,GS16}, conformal field theory (CFT) \cite{DS-NPB87,RS01,LoopReview}, and
probabilistic approaches \cite{SLE,CLE}. An important feature of the loop models that we have in mind---a defining ingredient for
some of the methods mentioned, and better hidden but still implicit in others---is the presence of an underlying
quantum group symmetry. For the most fundamental loop models---the ones covered by the given set of references---this
symmetry is $U_{-q}(\mathfrak{sl}_n)$ with $n=2$.

In a recent paper \cite{LGJ21} we have defined a series of statistical models on the hexagonal lattice $\mathbb{H}$
that extend this symmetry to any $n \ge 2$. For the cases $n > 2$, these models define geometrical configurations
of cubic (and bipartite, for $n=3$) graphs, called webs, on $\mathbb{H}$. The configurations reduce to the usual loops when $n=2$, in which case bifurcations
are suppressed. The present paper is the second in a series, in which we intend to lay the foundations for the study of
such web models. The algebra underlying the description of the loop model is the Temperley-Lieb algebra~\cite{TL71},
while the $n=3$ webs are built on the Kuperberg spider~\cite{Kuperberg_1996}, and more precisely on its $A_2$ variant.

The most interesting feature of loop and web models is that the partition sum carries over configurations of a set of extended,
geometrical objects, whose statistical weight contains a {\em non-local} part. For the loop model ($n=2$) this non-locality simply amounts
to replacing each loop by a real number, while for the web model the weight results from a quite non-trivial reduction
of each connected component to a set of loops which are then replaced by their corresponding weights \cite{Kuperberg_1996,LGJ21}.

The transfer matrix is a powerful tool to study statistical models, especially critical models in two dimensions, where fundamental results
relate the finite-size scaling of the transfer matrix eigenvalues to the central charge \cite{Cardy_c,Affleck_c} and conformal weights \cite{Cardy_x}
of the corresponding CFT.
It is of course not immediately clear whether non-local weights can be accommodated by the transfer matrix formalism. More precisely,
one may ask, for the model defined on a cylinder of circumference $L$ (or a strip of width $L$), whether there exists a finite-dimensional Hilbert space ${\cal H}_L$,
defined on a time slice in the usual radial quantisation, and a suitable representation of the transfer matrix within that space, which will allow
one to compute the non-local weights ``on the fly'' in the transfer process.

The answer to that question is positive for the loop model \cite{BN89}: one
uses for ${\cal H}_L$ the space of link patterns, which are the pairwise connections between loop strands within the time slice, the connections being defined
by the evolution prior to that time. This Hilbert space thus contains non-local information that allows one to compute the non-local
weights of loops. But for the $n=3$ web model it is not at all obvious how to achieve a similar goal.

An alternative for setting up such a non-local transfer matrix is to search for a local reformulation of the model, in which the non-local part of the weight
is rewritten locally in terms of other degrees of freedom than the original ones. For the loop model this can be done \cite{BKW76}, at the expense
of introducing complex Boltzmann weights (which is not a problem for the transfer matrix formalism). The result is a vertex model,
where each link of $\mathbb{H}$ can be in three different states \cite{WBN92}, for which a standard, local transfer matrix can readily be written down.

We show in Section~\ref{sec:kuploc} that a local reformulation can be obtained for the $n=3$ web model as well, now in the form of a coloured vertex
model, in which each link of $\mathbb{H}$ can be in seven different states. This number comes from the three colours and two orientations possible for states of links covered by webs, in addition to a vacuum state carried by an empty link. An example configuration of this seven-state vertex model on $\mathbb{H}$ is given in the following picture, where we show the cylinder geometry (periodic boundary conditions identify the left and right boundaries): 
\begin{center}
    \includegraphics[scale=0.3]{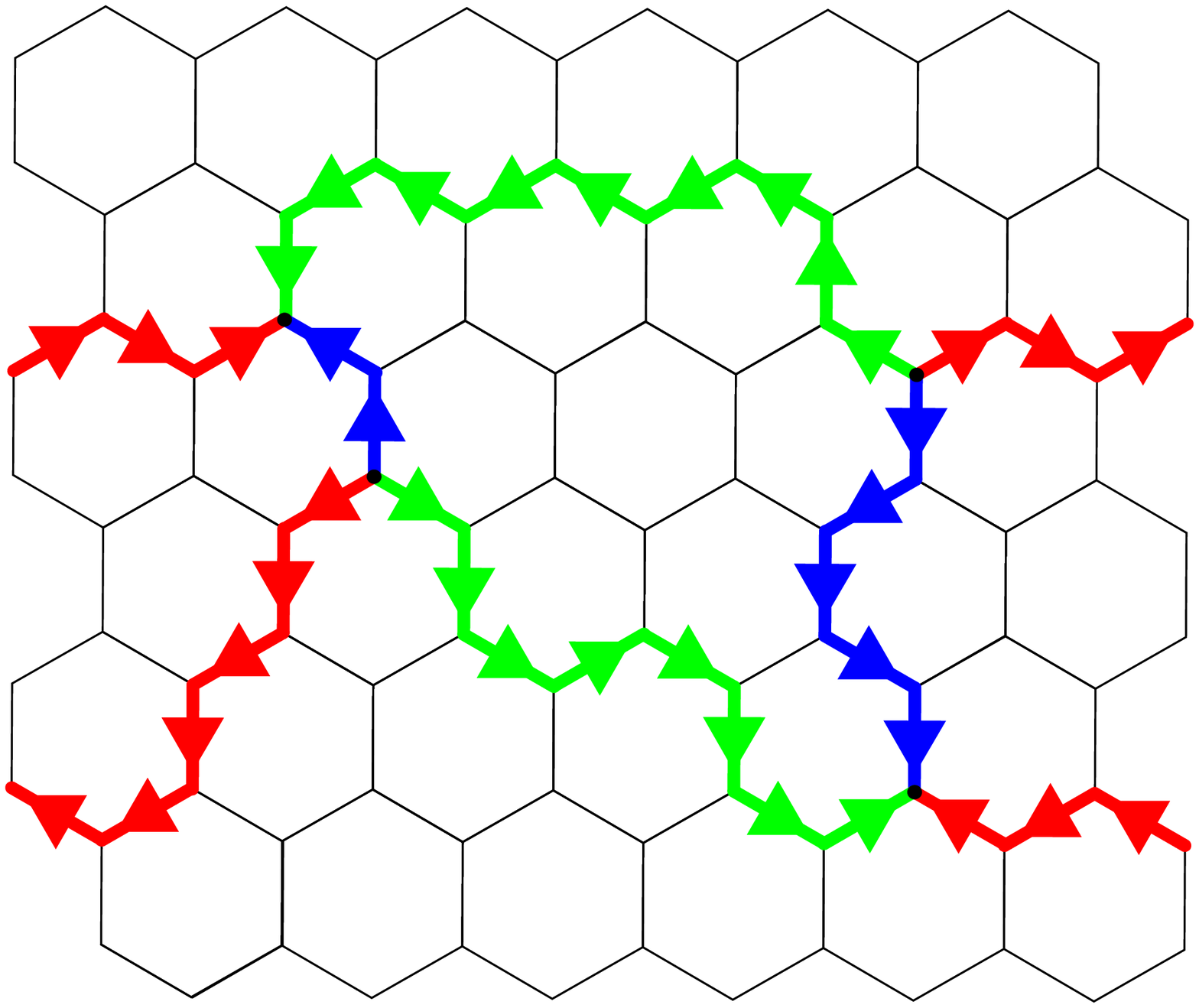}
\end{center}
This local rewriting has a twofold advantage. On one hand, it sheds
more light on the model's underlying 
$U_{-q}(\mathfrak{sl}_3)$ symmetry, as we shall explain in details in Section~\ref{sec:kuploc}. On the other hand, it enables us to carry out a numerical
investigation of the web model's phase diagram. This is done in Section~\ref{sec:pd}, and we conclude that for each $q$ in the critical regime,
\begin{equation}
\label{crit-reg-loop}
 q = {\rm e}^{i \gamma} \,, \mbox{ with } \gamma \in [0,\pi] \,,
\end{equation}
the web model possesses a dense and a dilute critical point, just like its loop model counterpart \cite{Nienhuis}.

One of our principal goals is to identify the CFT of the web model and compute its critical exponents by the Coulomb gas method.
Although this will be deferred to a subsequent paper \cite{LGJ-CG}, we shall find it convenient to prepare the ground here, by tackling
some of the issues that are most conveniently discussed in the lattice model setting. In particular, in Section~\ref{sec:em}
we discuss the conservation laws and hence the sector decomposition of the transfer matrix, both in the cylinder and strip geometries.
Each sector is related to a certain defect configuration, which can be imposed by the boundary conditions and an appropriate modification of the transfer matrix,
and it provides a finite-size precursor of a pair of electromagnetic operators within the field theory.

The main combinatorial objects in Section~\ref{sec:em} are open Kuperberg webs, embedded in a rectangle or a cylinder, and their three-colourings. We describe the transfer-matrix sectors in terms of the (coloured) open webs subject to conditions which are analogous to the Temperley-Lieb standard modules in the loop models case, i.e.\ no contractions of through lines. However, it is worth noticing that, contrary to the loop models case, the classification of irreducible open webs in the cylinder geometry is rather non-trivial, the most technical problem we solve at the end of Section~\ref{sec:em}.

\medskip

We also consider applications in a few models.
It follows from Section~\ref{sec:pd} and \cite{LGJ21} that for $q={\rm e}^{i \pi/4}$ the critical point in the dilute phase of
the $n=3$ webs can be identified with spin interfaces of the critical three-state Potts model defined on the
triangular lattice $\mathbb{T}$, dual to $\mathbb{H}$. This equivalence is analogous to the well-known identification
of domain walls of the critical Ising model within the $n=2$ loop model.
In Section~\ref{sec:defect1} we  provide another mapping between the $n=3$ webs and a
$\mathbb{Z}_3$ spin model defined on $\mathbb{H}$ itself, by means of a high-temperature expansion. We discuss in
particular defects within this formulation.

Another interesting special case of the $n=3$ web model occurs for $q={\rm e}^{i \pi/3}$, where the model is trivial because every non trivial web is weighted by $0$. However a renormalization of the partition function defines an interesting web model where only   one-connected-component webs contribute. This case is analogous to the
polymer limit of the $n=2$ loop model. We discuss this special case and conjecture a relation between the fractal dimension of critical webs and electromagnetic conformal weights in Section~\ref{sec:defect2}.

Finally we give our conclusions and some perspectives for further developments in Section~\ref{sec:disc}. Appendix~\ref{sec:quantumgroupconventions} contains our
conventions and notations for quantum groups and the pivotal structure, while other technical details are relegated to Appendix~\ref{App:MagWebs}.

\section{Vertex-model representation of Kuperberg $A_2$ web models} \label{sec:kuploc}
\subsection{Geometrical definition}
\begin{figure}
\begin{center}
    \includegraphics[scale=0.3]{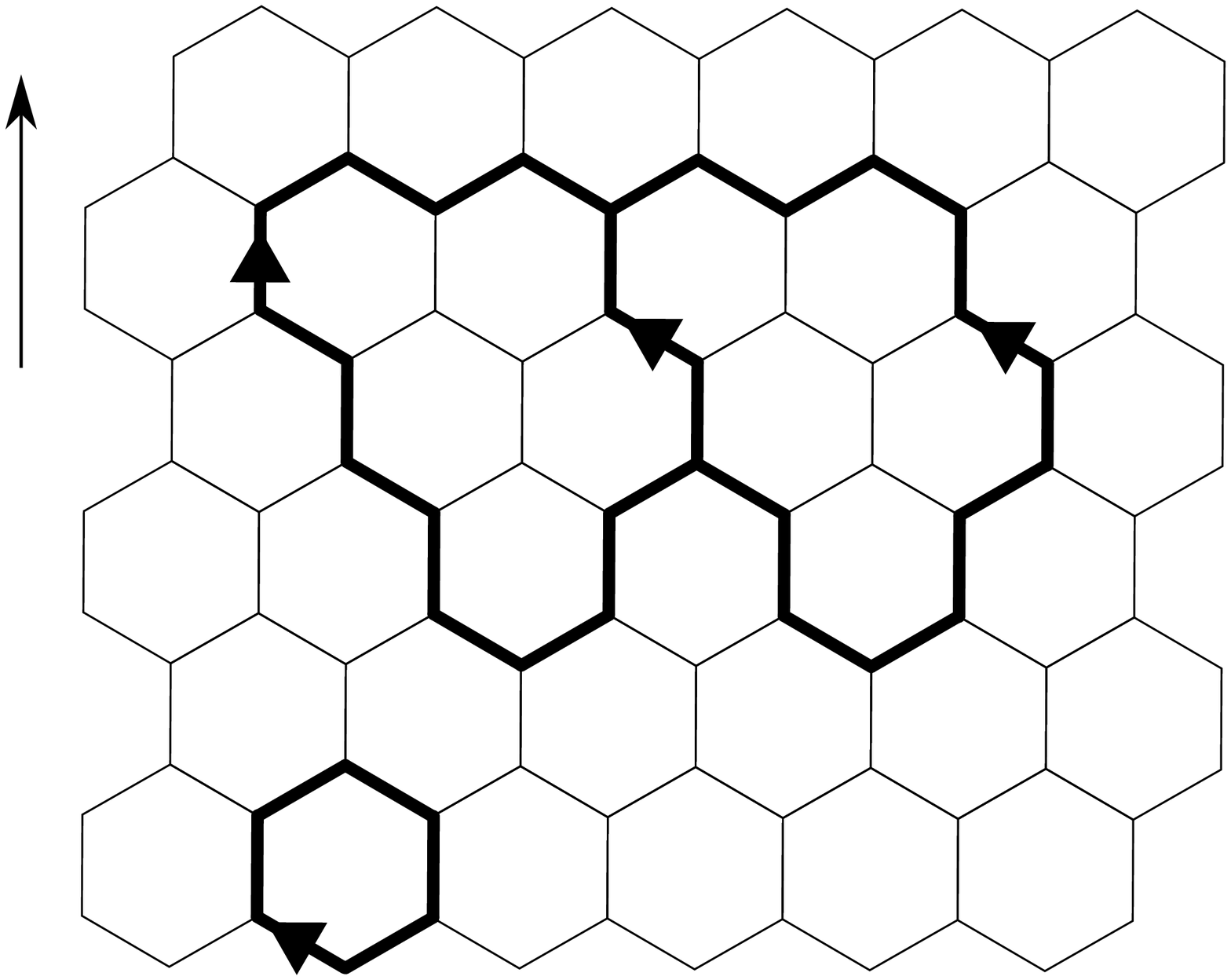} \qquad \qquad \includegraphics[scale=0.3]{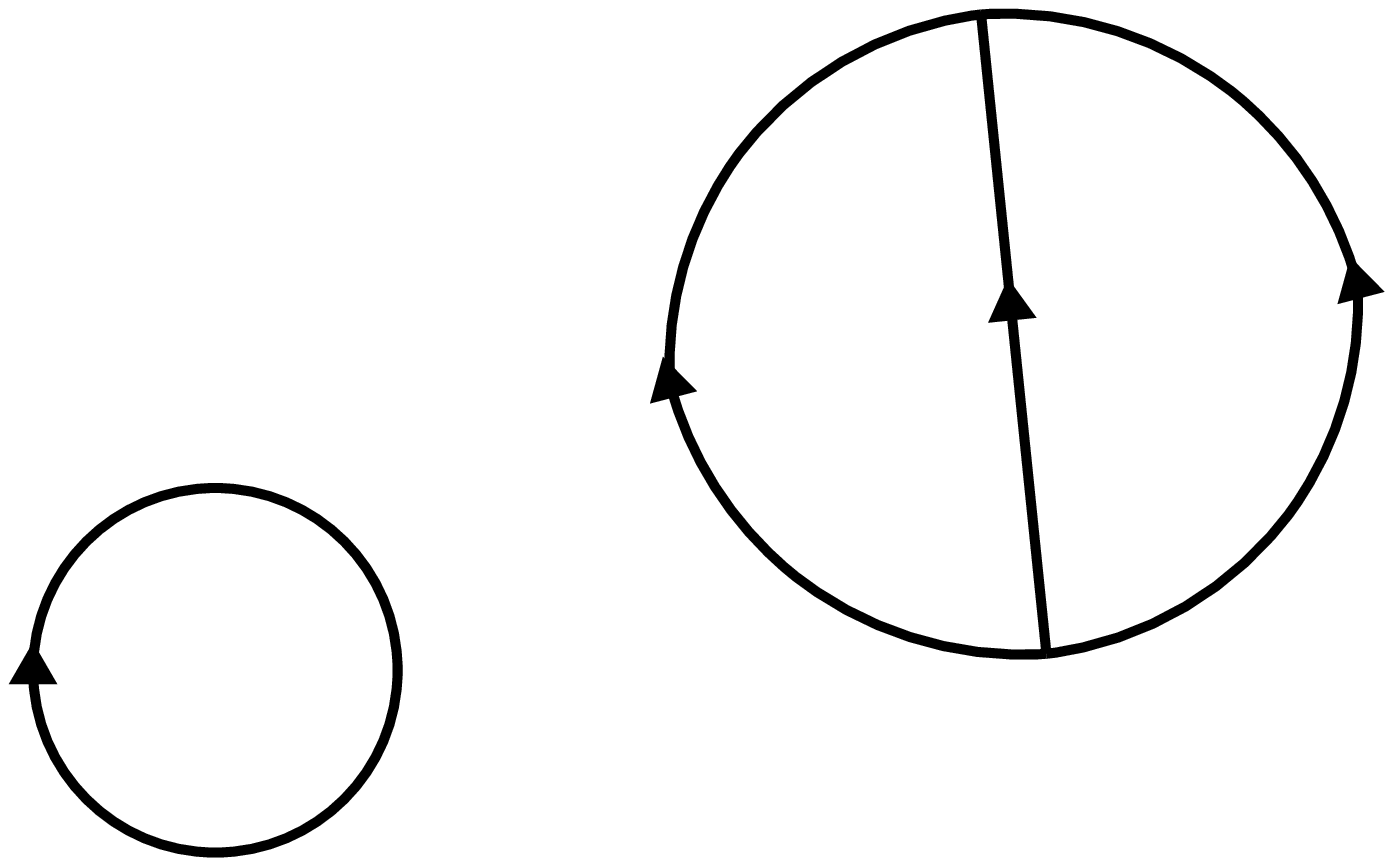}
\end{center}
    \caption{Left panel: A configuration on $\mathbb{H}$ with $2M=6$ rows and $2L=12$ columns. The arrow is parallel to the axis of the cylinder. The left and right sides of the drawing are identified by periodic boundary conditions. Right panel: The same configuration drawn as a web.}
    \label{fig:config}
\end{figure}
We first recall the definition of the Kuperberg web model, as given in our first paper \cite{LGJ21}. The model is defined on an hexagonal lattice $\mathbb{H}$ made of $2M$ rows and $2L$ columns embedded in a vertical strip or a vertical cylinder. What is meant by row and column can be read from the example in Figure \ref{fig:config}. To fix vocabulary, $\mathbb{H}$ is comprised of \textit{nodes} and \textit{links}. It is oriented such that one third of its links are parallel to the vertical axis of the strip, or cylinder. Configurations are \textit{closed} Kuperberg webs embedded in the lattice (see Figure \ref{fig:config}). We refer to such webs by the label K. Kuperberg webs are planar oriented trivalent bipartite graphs. They are comprised of \textit{vertices} and \textit{edges}. The two types of vertices are either \textit{sources} or \textit{sinks} with respects to the arrows located on edges. A \textit{bond} is a link of the hexagonal lattice covered by an edge of a web. Each bond inherits the orientation of its corresponding edge. Thus, when a path of several links is covered by one edge, the corresponding bonds must be consistently oriented.

The weight of a configuration $c$ is the product of a local part and a non-local part. A fugacity $x_1$ (respectively $x_2$) is given to a bond covered by an edge flowing upward (respectively downward). Remark that upward and downward are well defined as no link of $\mathbb{H}$ is drawn horizontally. In addition, a fugacity $y$ is given to each sources, and a fugacity $z$ is given to each sink. The fugacities $x_1$, $x_2$, $y$, and $z$ define the local part of the weight of $c$. The non-local part is given by a number $w_{\rm K}(c)$ assigned to each closed web; it is computed by reducing $c$ to the empty web by means of the relations
\begin{subequations}
\label{3rules}
\begin{align}
    \vcenter{\hbox{\includegraphics[scale=0.2]{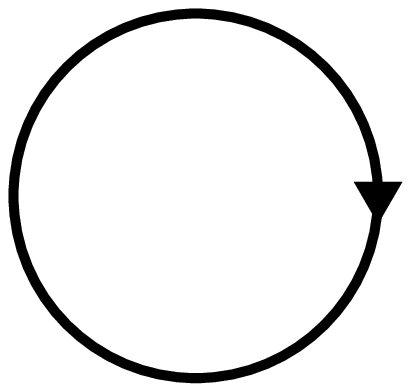}}}&\;=\;[3]_q \label{3rulesa} \\
    \vcenter{\hbox{\includegraphics[scale=0.2]{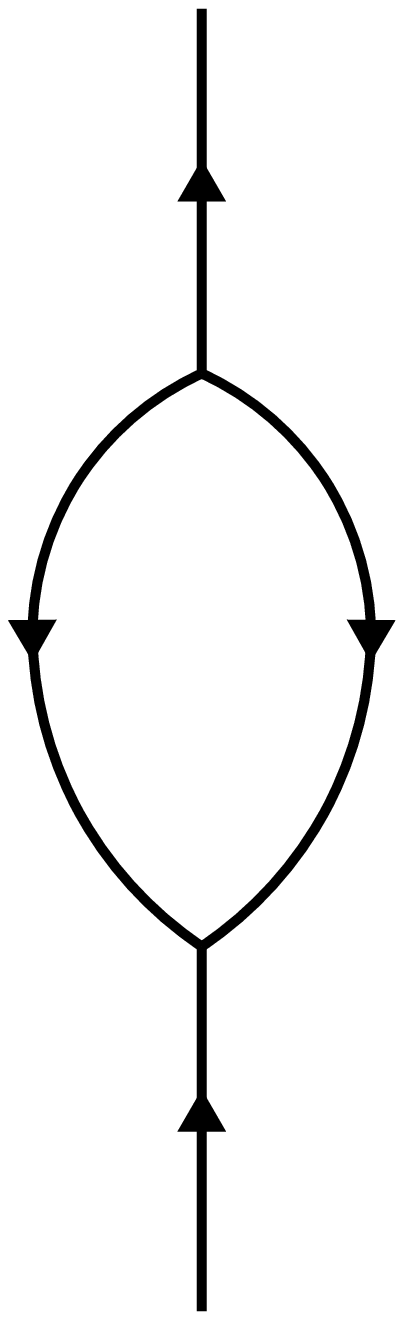}}}&\;=\;[2]_q\;\vcenter{\hbox{\includegraphics[scale=0.2]{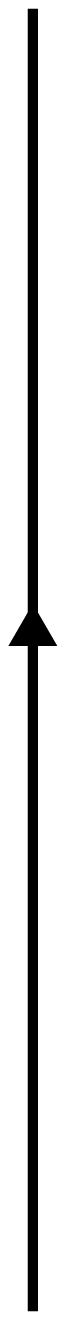}}} \label{3rulesb} \\
    \vcenter{\hbox{\includegraphics[scale=0.2]{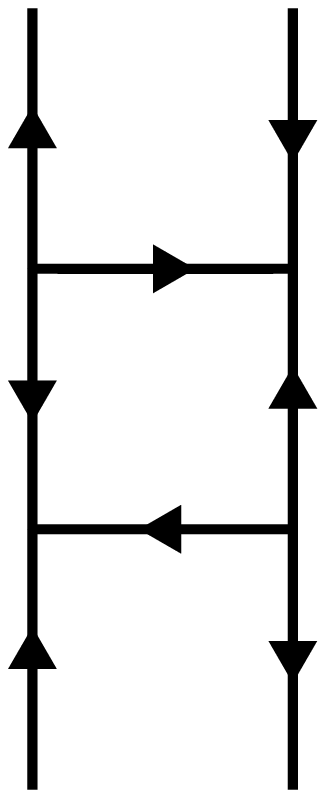}}}&\;=\;\vcenter{\hbox{\includegraphics[scale=0.2]{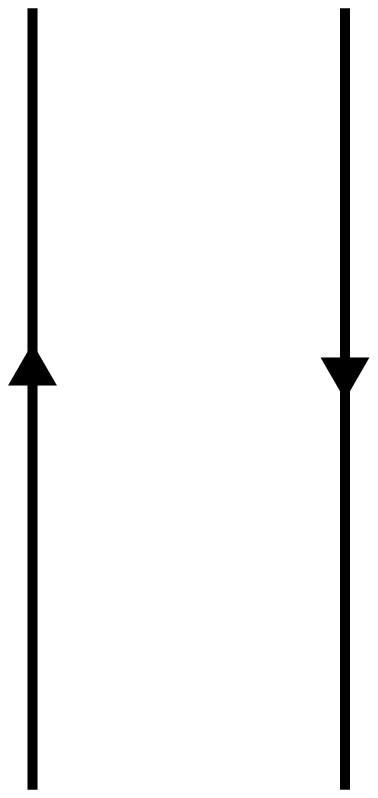}}}\;+\;\vcenter{\hbox{\includegraphics[scale=0.2]{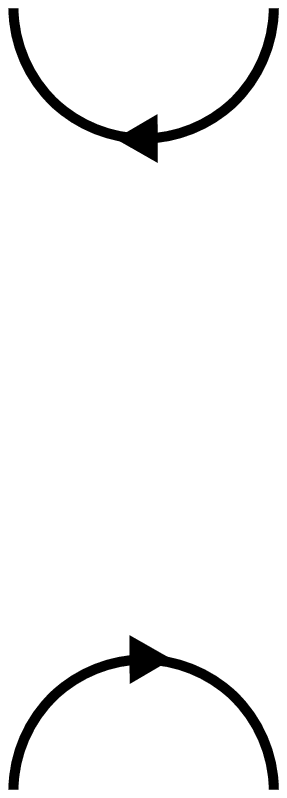}}}\label{kupsquare}
\end{align}
\end{subequations}
The non-local weight is well-defined, in the sense that any planar closed Kuperberg web $c$ can indeed be reduced to the empty web by means of the rules \eqref{3rules}, and moreover $w_{\rm K}(c)$ does not depend on the order in which the three rules are applied \cite{Kuperberg_1996}. Note that the embedding of graphs in either the strip or the cylinder ensures planarity.

The partition function then reads
\begin{align}
    Z_{\rm K} =\sum_{c\in \mathcal{K}} x_1^{N_1}x_2^{N_2}(yz)^{N_V}w_{\rm K}(c) \,,
 \label{Z_K}
\end{align}
where $N_1$ (respectively $N_2$) is the number of upward (respectively downward) bonds, $N_V$ is the number of sink/source pairs of vertices, 
and $\mathcal{K}$ denotes the set of embedded Kuperberg webs.
The model is discretely rotationally invariant when $x_1=x_2$. The vertex fugacities $y$, $z$ and the Kuperberg weight $w_{\rm K}(c)$ do not depend on how a given web is embedded in $\mathbb{H}$ but only on the abstract graph. Thus, we dub the product of these two parts, $(yz)^{N_V}w_{\rm K}(c)$, the {\em topological weight} of the configuration $c$.

Note that the partition function is invariant under the transformation
\begin{subequations}
\begin{align}
    yz&\mapsto -yz \,, \\
    q&\mapsto -q
\end{align}
\end{subequations}
and under the transformation
\begin{align}
    q\mapsto q^{-1} \,.
\end{align}
In this paper, we will focus on the following subspace of the parameter space:
\begin{eqnarray}
    & & x_1,\ x_2,\ yz\geq 0 \,, \nonumber \\
    & & q={\rm e}^{i\gamma} \,, \mbox{ with } \gamma\in [0,\pi] \,. \label{crit-reg}
\end{eqnarray}

\subsection{Combinatorial vertex-model formulation}
\label{sec:kupvertex}
We shall now describe a combinatorial vertex-model formulation of the above Kuperberg web model. It is similar in spirit to the localisation of the loop weight in the $O(N)$ loop model in terms of a corresponding oriented loop model \cite{BKW76,LoopReview}. We first give a quick reminder of the latter construction.

The configurations of the $O(N)$ loop model are collections of self-avoiding and mutually avoiding loops embedded in $\mathbb{H}$. The weight of a configuration is the product of local fugacity $x$ assigned to each bond (i.e., a link covered by a loop) and a non-local factor $N$ for each loop. In order to localise the latter, one needs a way to relate $N$ to local degrees of freedom. A convenient trick is to first assign orientations to each loop. One then gives a weight $q$ (respectively $q^{-1}$) to a loop oriented clockwise\footnote{Our convention on loop orientations is opposite to what one may find in part of the literature. It follows from our convention for the coproduct of the quantum group (see Appendix~\ref{sec:quantumgroupconventions}).}  (respectively anticlockwise) such that, with $N=q+q^{-1}=[2]_q$, the original loop weight is retrieved. Next, one can localise the weight of an oriented loop by requiring that a piece of it carry a weight $q^{-\frac{\theta}{2\pi}}$ when it bends an angle $\theta$. The angle $\theta$ of an oriented edge bending will always be counted positive in the anti-clockwise direction:
\begin{center}
    \includegraphics[scale=0.2]{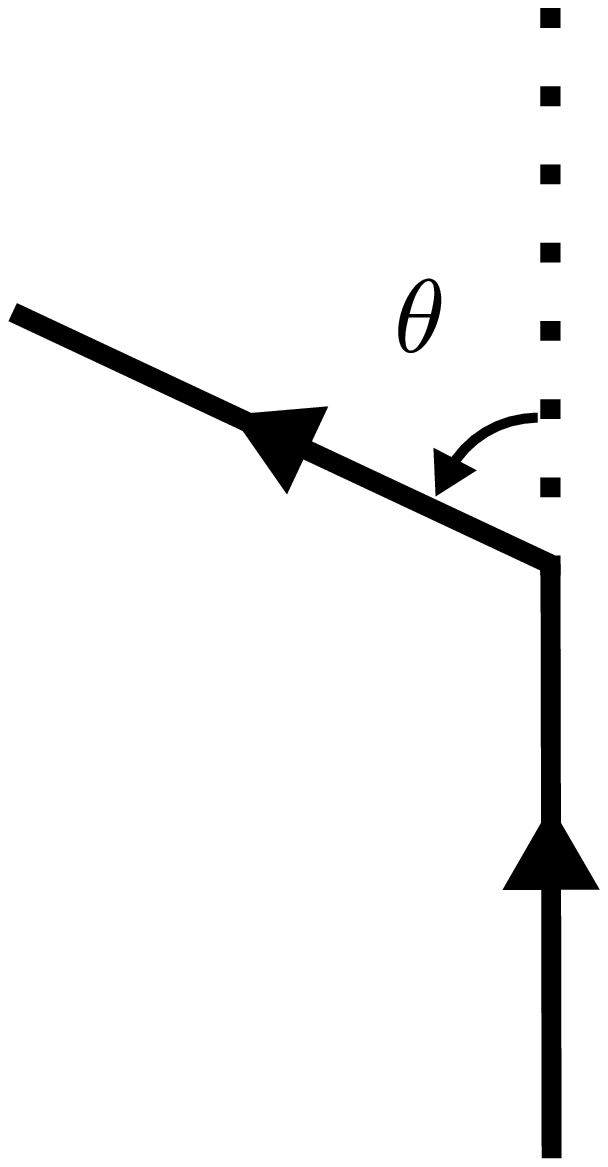}
\end{center}
We will use the same definition regarding the bending of (coloured) oriented edges in the Kuperberg web model.
On the hexagonal lattice $\mathbb{H}$ embedded in the strip, the weight of an oriented loop can be accounted for by the followings local weights (where the bond fugacity is taken care of as well):
\begin{align}
\label{loopvertexweights}
\begin{split}
    &\vcenter{\hbox{\includegraphics[scale=0.2]{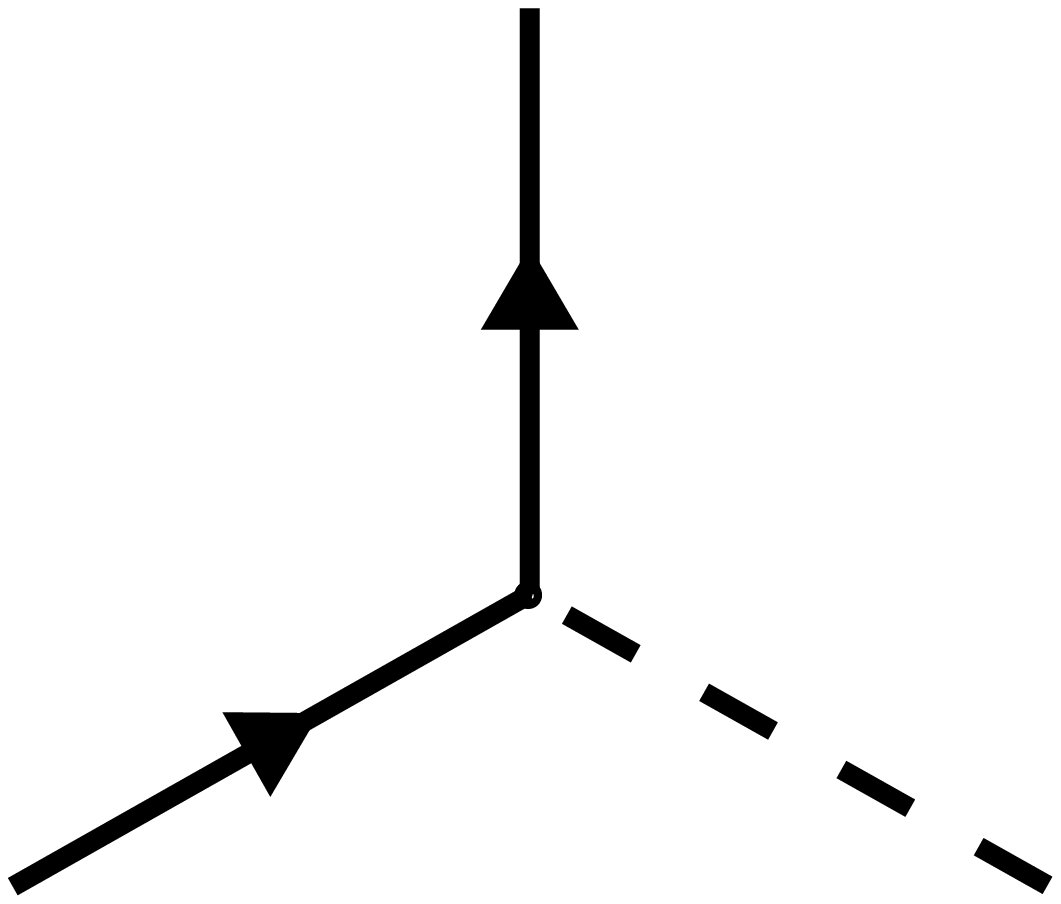}}}=xq^{-\frac{1}{6}} \,, \qquad
    \vcenter{\hbox{\includegraphics[scale=0.2]{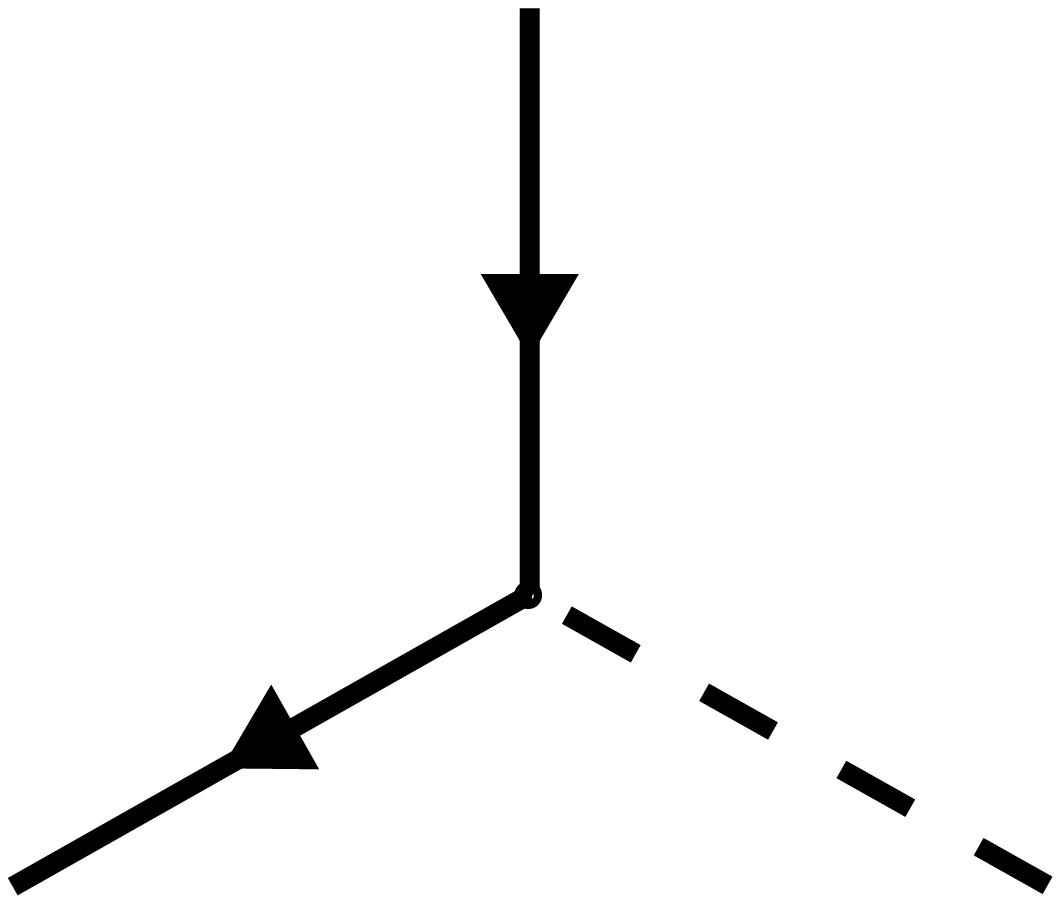}}}=xq^{\frac{1}{6}} \,,  \qquad
    \vcenter{\hbox{\includegraphics[scale=0.2]{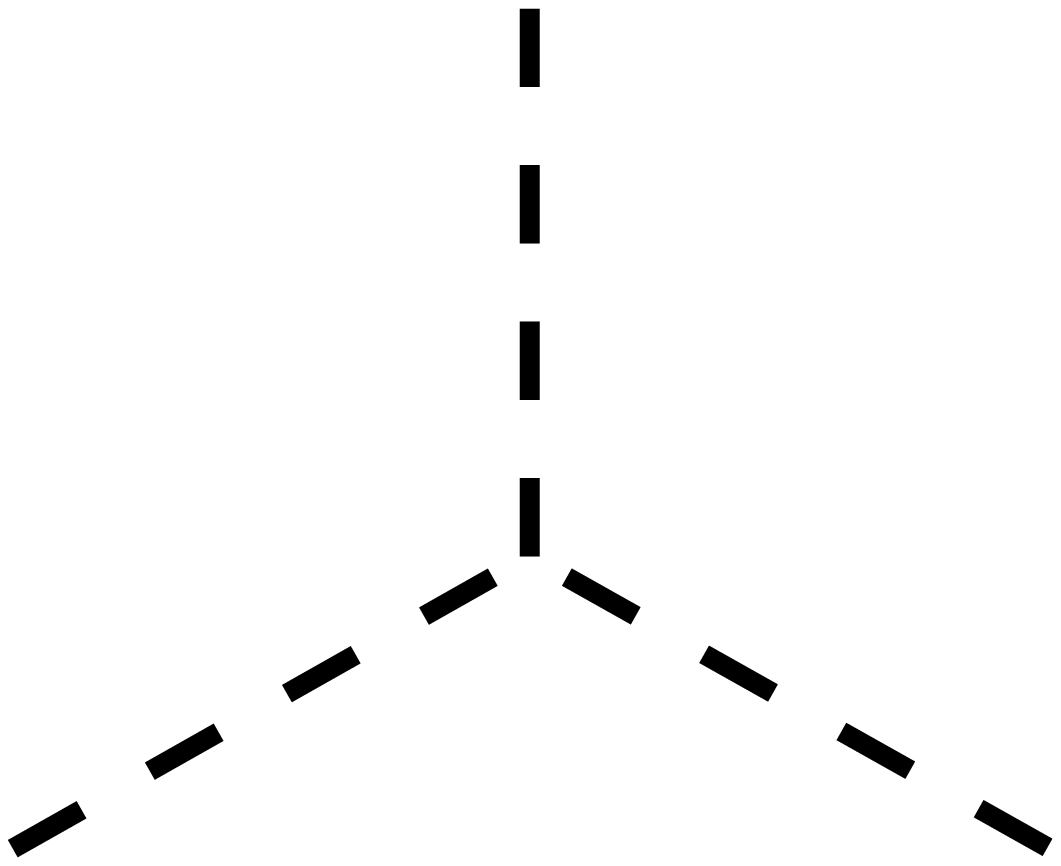}}}=1 \,,
\end{split}
\end{align}
where the dashed line represent a link unoccupied by a loop. Here we only drew some of the possible node configurations, omitting those related to the above ones by a rotation. All node configurations related by a rotation are given the same weight. Hence the model possess the discrete rotation symmetry of $\mathbb{H}$.

When $\mathbb{H}$ is embedded in the cylinder, the above vertex weights give an uncorrect weight $2$ to non-contractible loops. This situation can be remedied by introducing an oriented seam line running along the cylinder and avoiding nodes, such that additional weights are given to links crossing the seam line as follows:
\begin{align}
\begin{split}
    &\vcenter{\hbox{\includegraphics[scale=0.2]{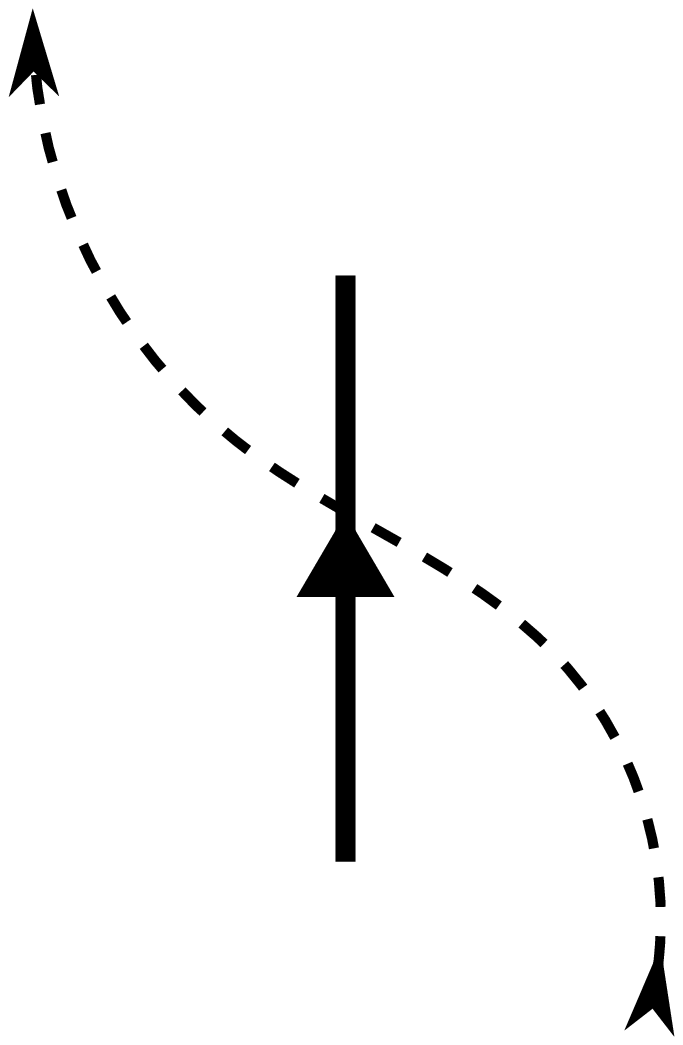}}}=q \,, \qquad
    \vcenter{\hbox{\includegraphics[scale=0.2]{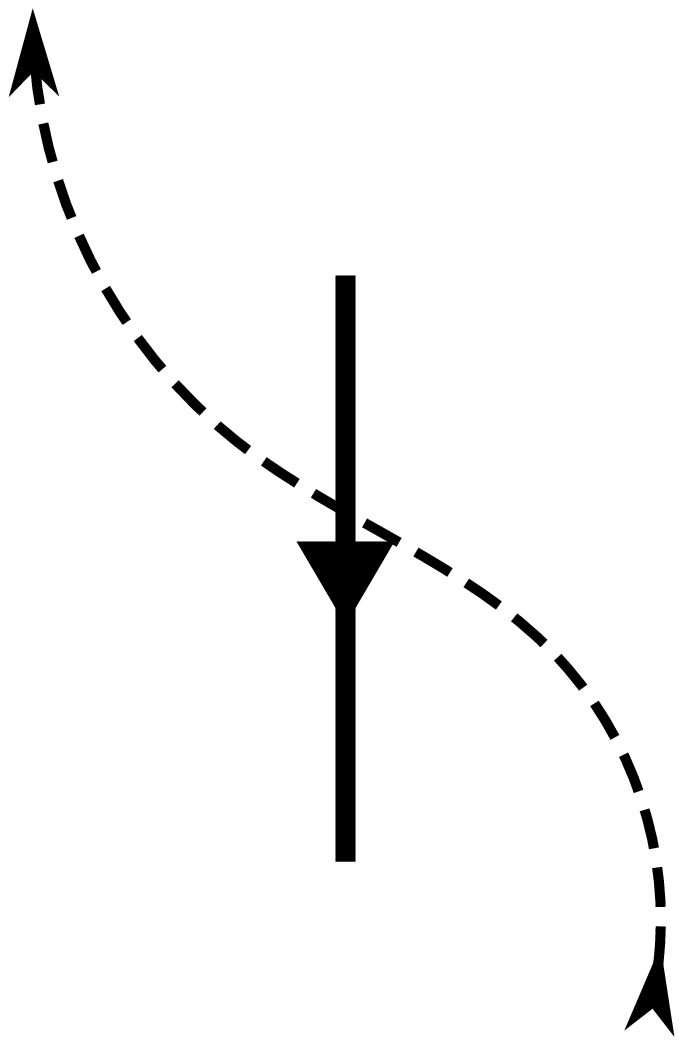}}}=q^{-1} \,, \qquad
    \vcenter{\hbox{\includegraphics[scale=0.2]{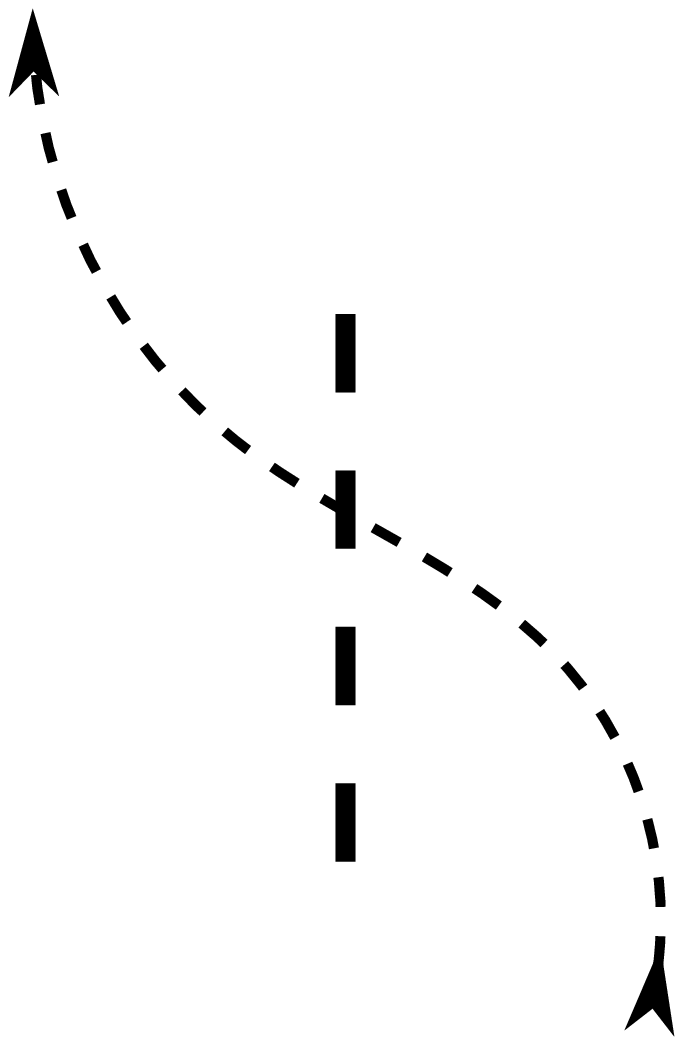}}}=1 \,.
\end{split}
\end{align}
Indeed, these weights just compensate the lack of bending for configurations that wrap around the periodic direction.
\begin{remark}
\label{rem:1}
From this point in the discussion we shall refer to oriented loops as having instead one of two possible {\rm colours}. We introduce this non-standard
usage in order to parallel the terminology of Kuperberg webs to be discussed below.
Indeed, the cases $n=2$ (loop model) and $n=3$ (Kuperberg webs) both use $n$ distinct colours. In addition, the Kuperberg webs are endowed with orientations
(arrows), but these are {\rm not} analogous to what we have hitherto called the `orientation' of a loop. In fact, it will be made clear below that the
orientation (in the sense of Kuperberg webs) is a redundant information for the loop model, motivating our change of terminology.
The weight of a coloured configuration $d$ given by the above local weights will henceforth be denoted as $w_{\rm col} (d)$.
\end{remark}

Let us now get back to the Kuperberg web model. We again begin by decorating the webs, as a first step in making the weights local. We call a {\em three-colouring} of a Kuperberg web $c$, a map from the set of edges of~$c$ into the set $\{\rm red, blue, green\}$, subject to the constraint that each vertex be incident on three edges with different colours. As usual, a bond inherits the colour of the edge it belongs to. By the above Remark~\ref{rem:1}, these colours are the analogue of what we called `orientations' in the above discussion of the loop model (but which we shall now refer to as colours as well). A precise algebraic interpretation of the concepts of colours (for loops and webs) and orientations (for webs only) will be given in Section \ref{sec:localtmKup}.

Each coloured web $d$ is assigned a weight $w_{\rm col} (d)$, to be defined shortly, such that the sum of these weights over all possible three-colourings $d$ of a given Kuperberg web $c$ will give back the non-local weight $w_{\rm K}(c)$. Note that we use the same notation for the weight of coloured configurations in both the loop and Kuperberg case, since the model being considered should always be clear from the context. We will describe the weight given to a coloured web directly in terms of its local pieces.

Consider first the strip geometry. We now restrict to $x_1=x_2=x$ and will come back to the general case later. The local weights of the model are given by
factors when a node is incident on three bonds
\begin{subequations}
\label{kupvertex}
\begin{align}
    &\vcenter{\hbox{\includegraphics[scale=0.2]{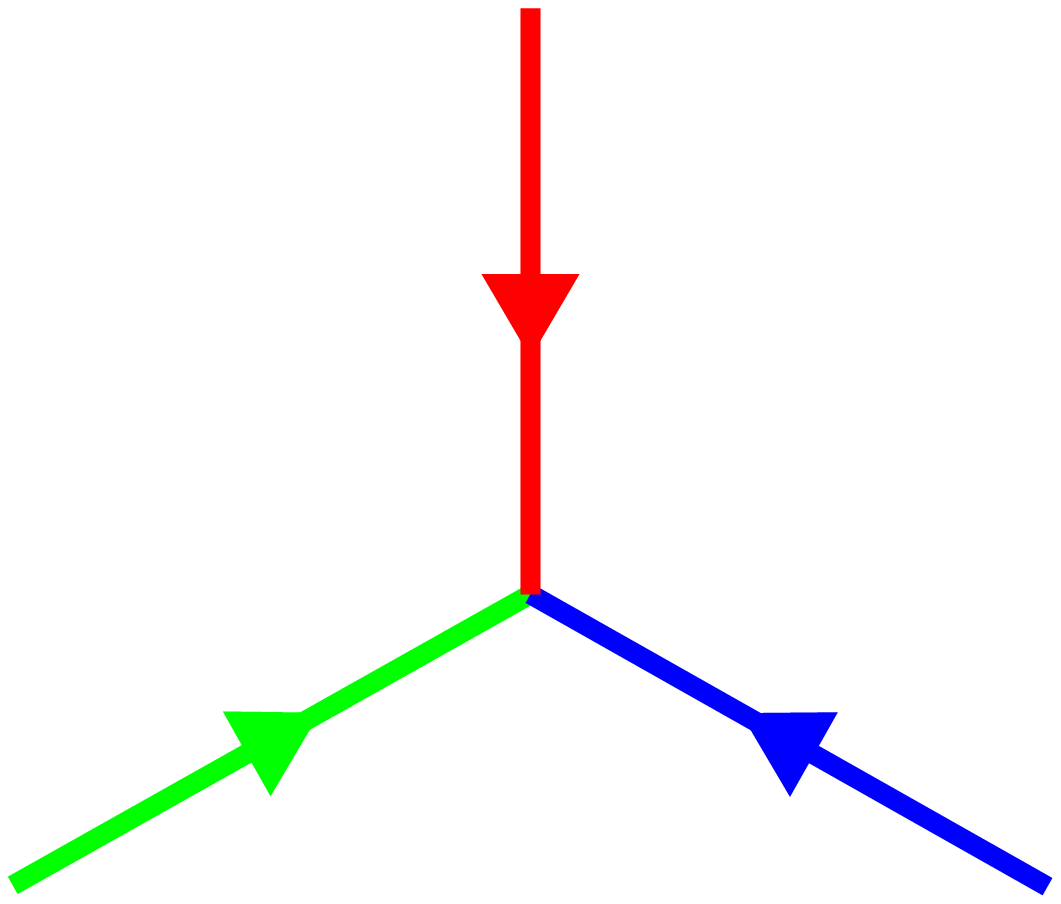}}}=zx^{\frac{3}{2}}q^{-\frac{1}{6}} \,, \qquad
    \vcenter{\hbox{\includegraphics[scale=0.2]{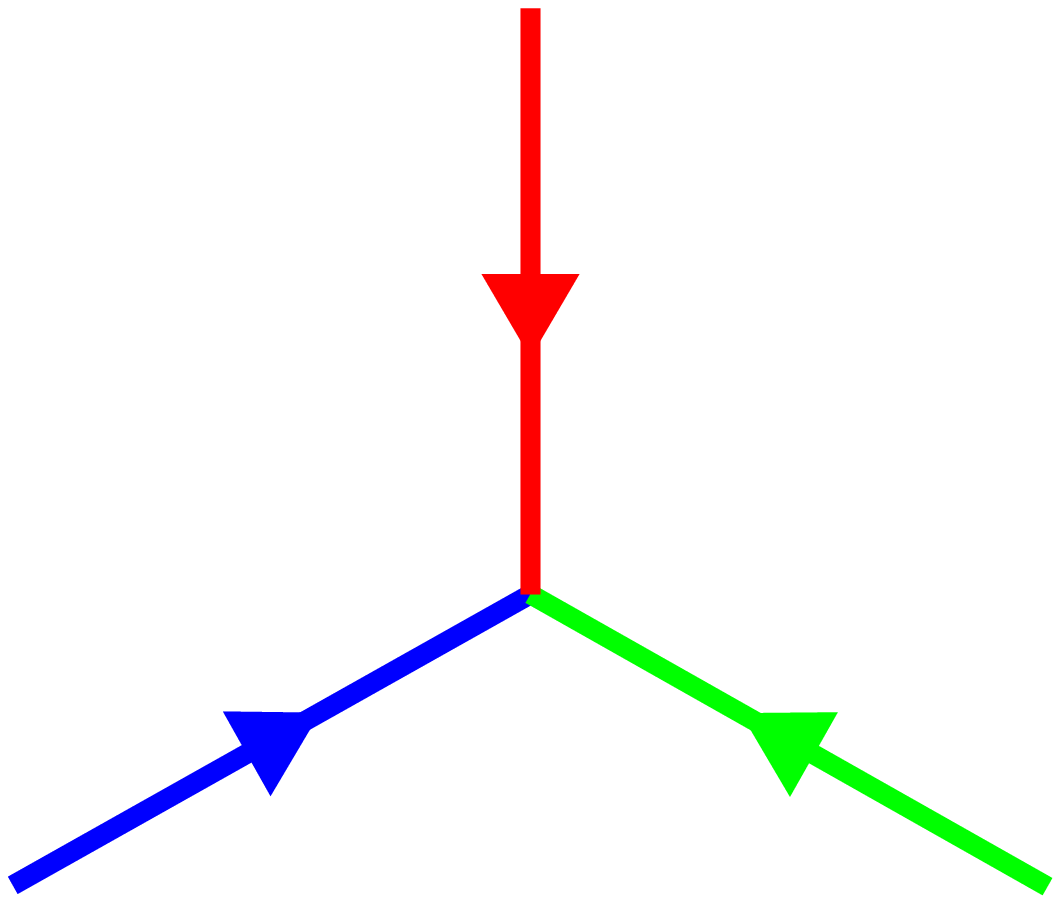}}}=zx^{\frac{3}{2}}q^{\frac{1}{6}} \,, \nonumber\\
    &\vcenter{\hbox{\includegraphics[scale=0.2]{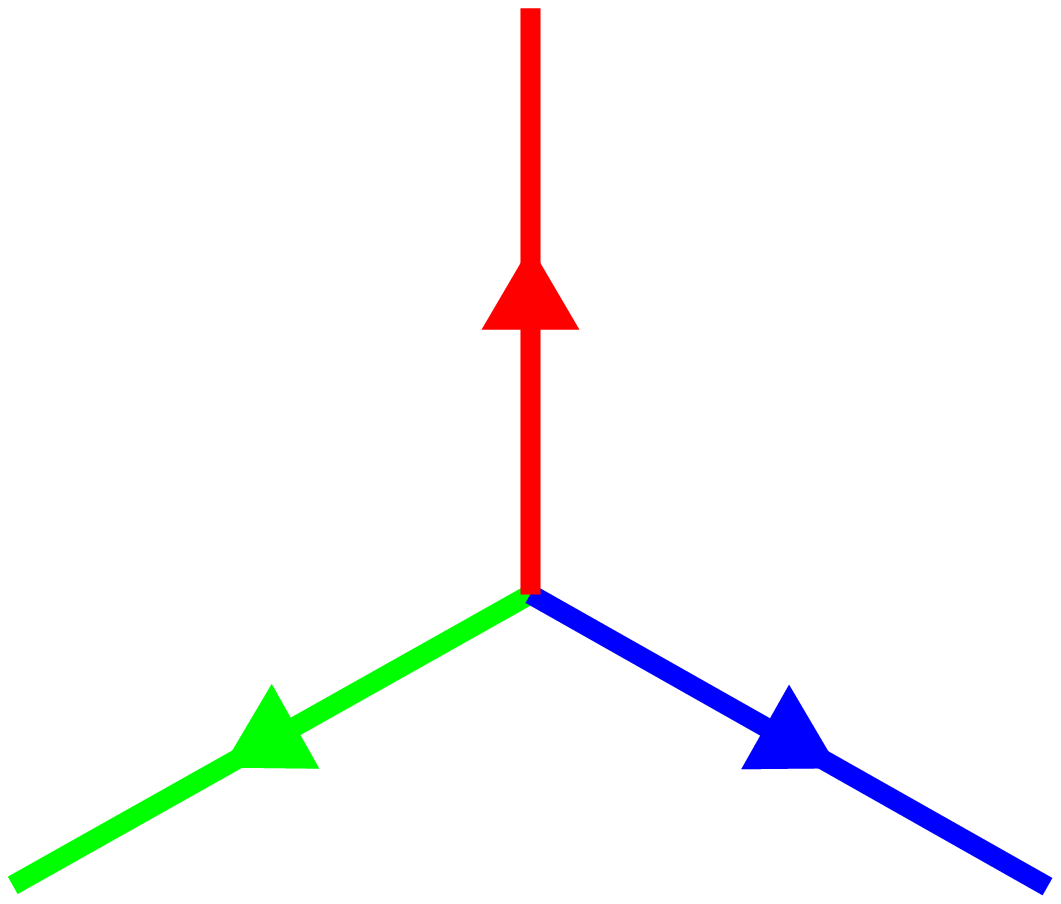}}}=yx^{\frac{3}{2}}q^{\frac{1}{6}} \,, \ \, \, \qquad
    \vcenter{\hbox{\includegraphics[scale=0.2]{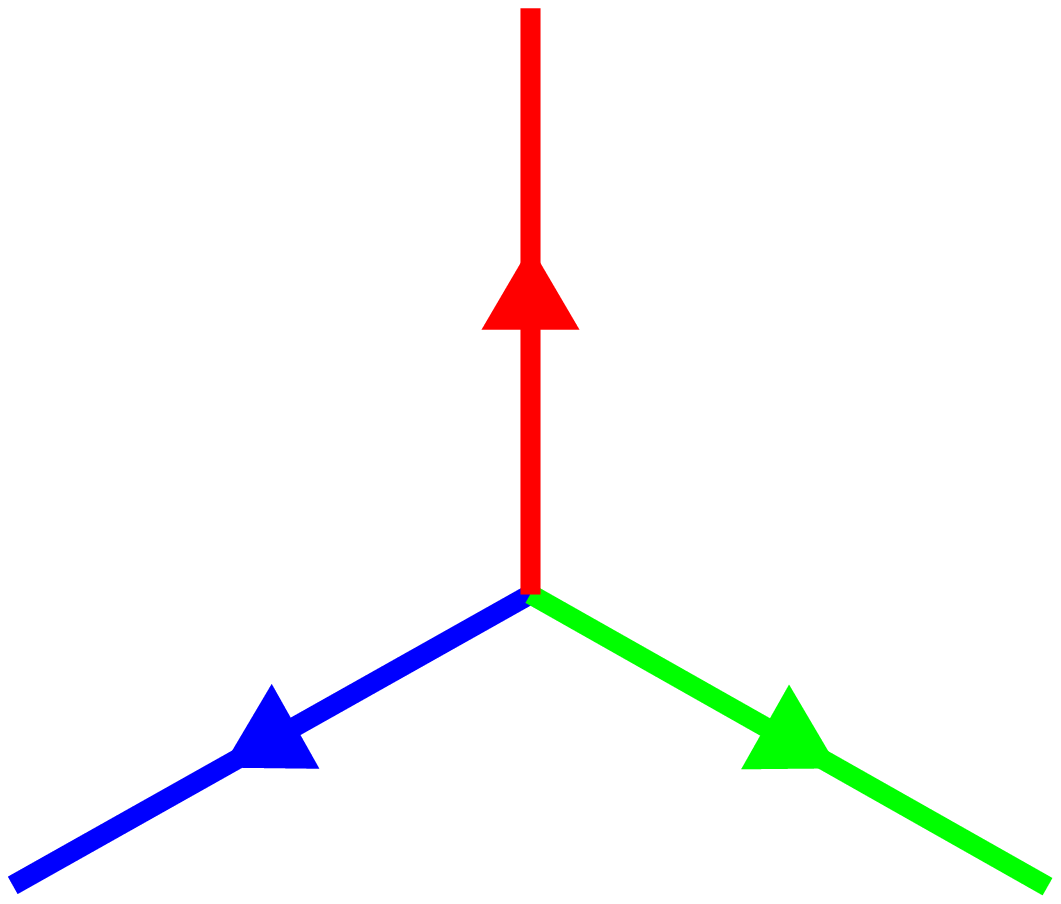}}}=yx^{\frac{3}{2}}q^{-\frac{1}{6}} \,,
\end{align}
or on only two bonds (which are then parts of the same embedded edge, hence having consistent orientations and colourings)
\begin{align}
    &\vcenter{\hbox{\includegraphics[scale=0.2]{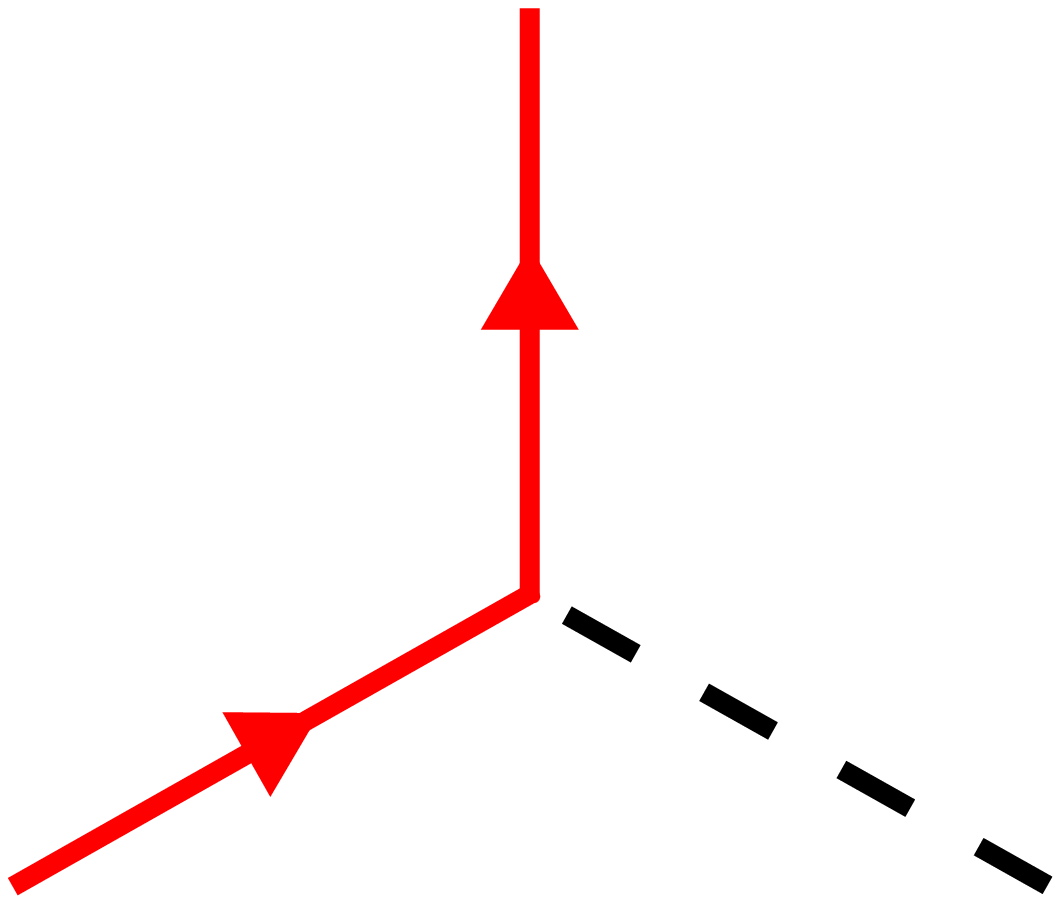}}}=xq^{-\frac{1}{3}} \,, \qquad
    \vcenter{\hbox{\includegraphics[scale=0.2]{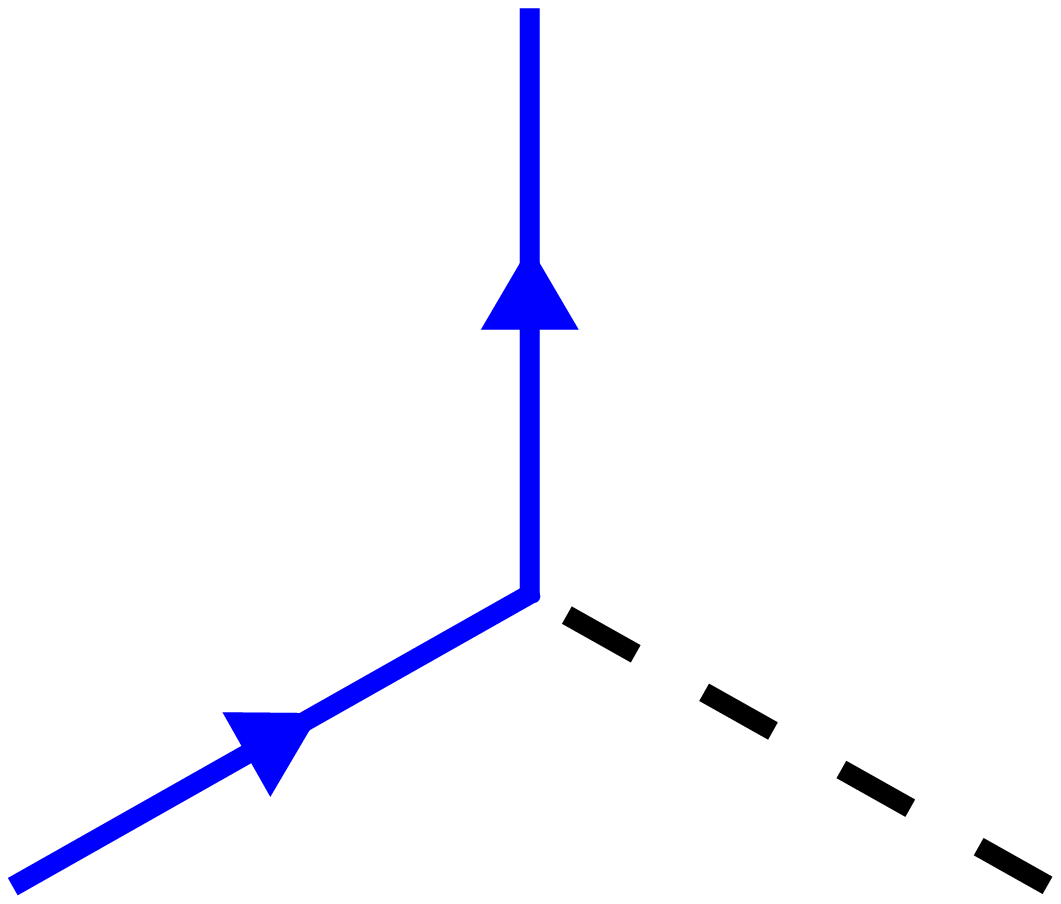}}}=x \,, \qquad
    \vcenter{\hbox{\includegraphics[scale=0.2]{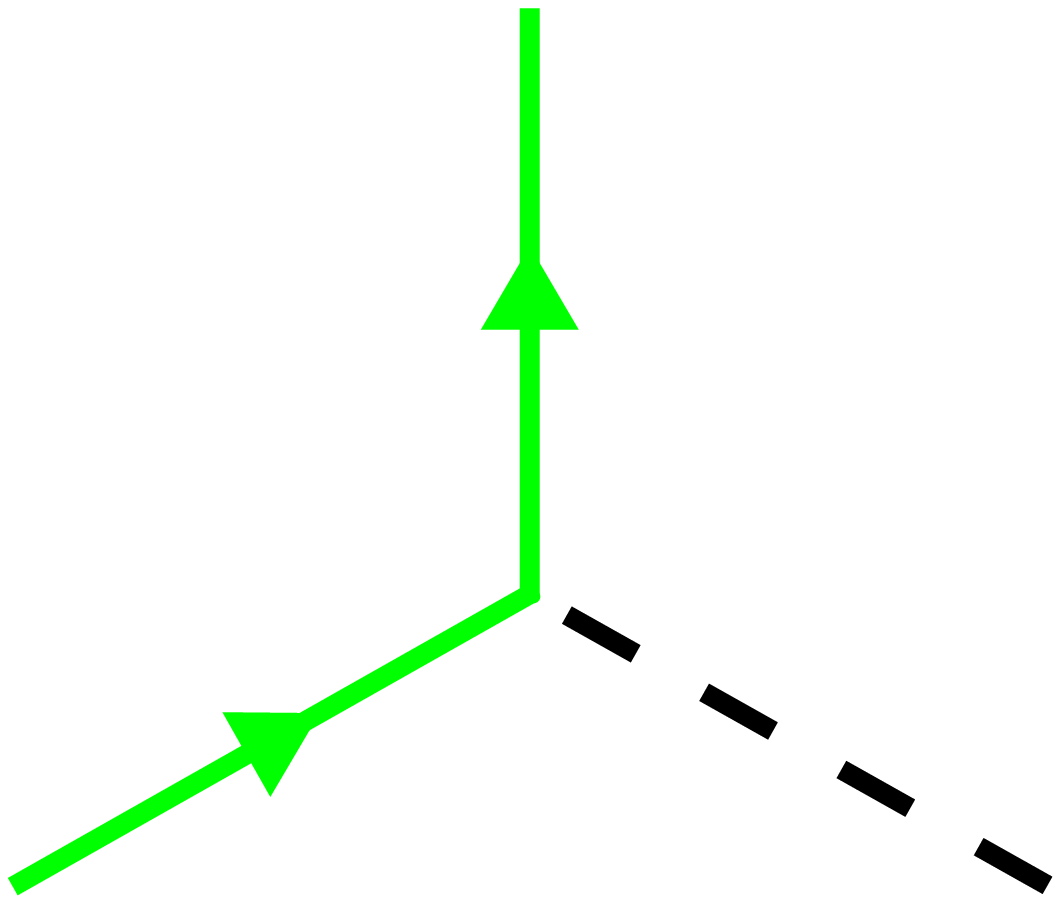}}}=xq^{\frac{1}{3}} \,, \nonumber \\
    &\vcenter{\hbox{\includegraphics[scale=0.2]{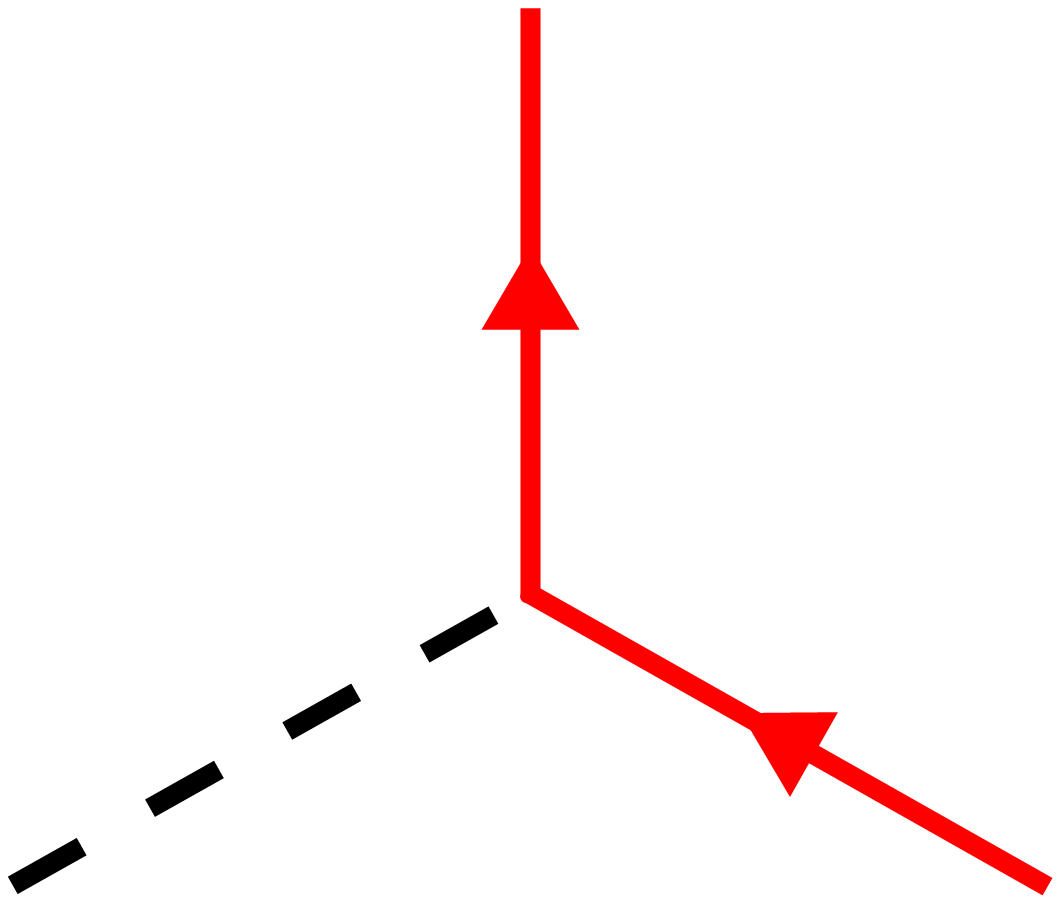}}}=xq^{\frac{1}{3}} \,, \ \, \, \qquad
    \vcenter{\hbox{\includegraphics[scale=0.2]{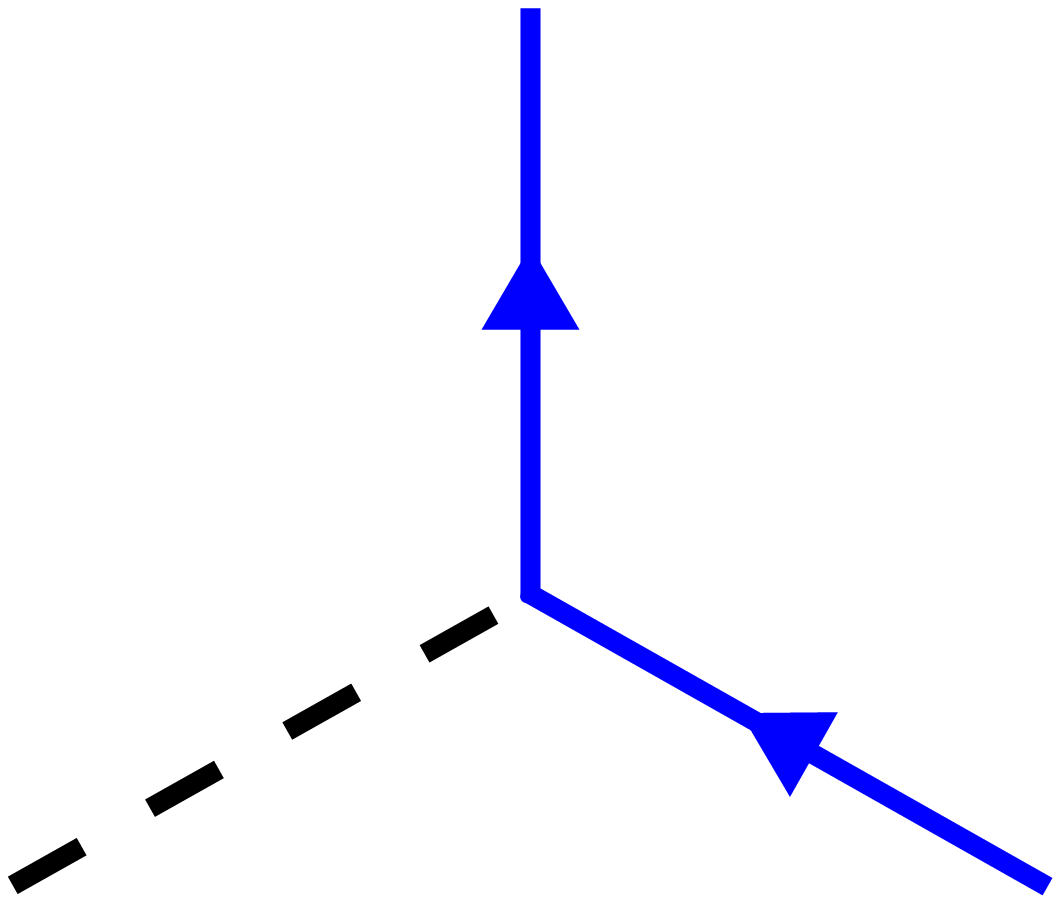}}}=x \,, \qquad
    \vcenter{\hbox{\includegraphics[scale=0.2]{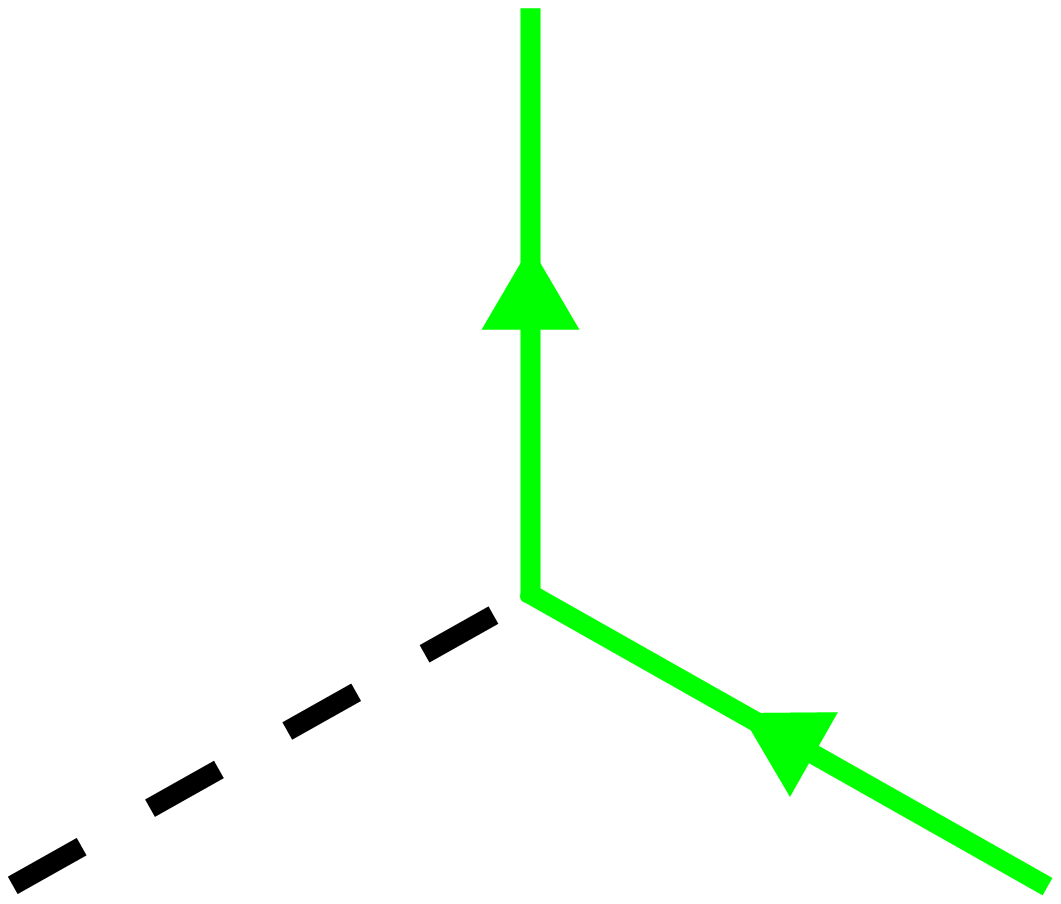}}}=xq^{-\frac{1}{3}} \,,
\end{align}
or finally when the node is empty
\begin{align}
    &\vcenter{\hbox{\includegraphics[scale=0.2]{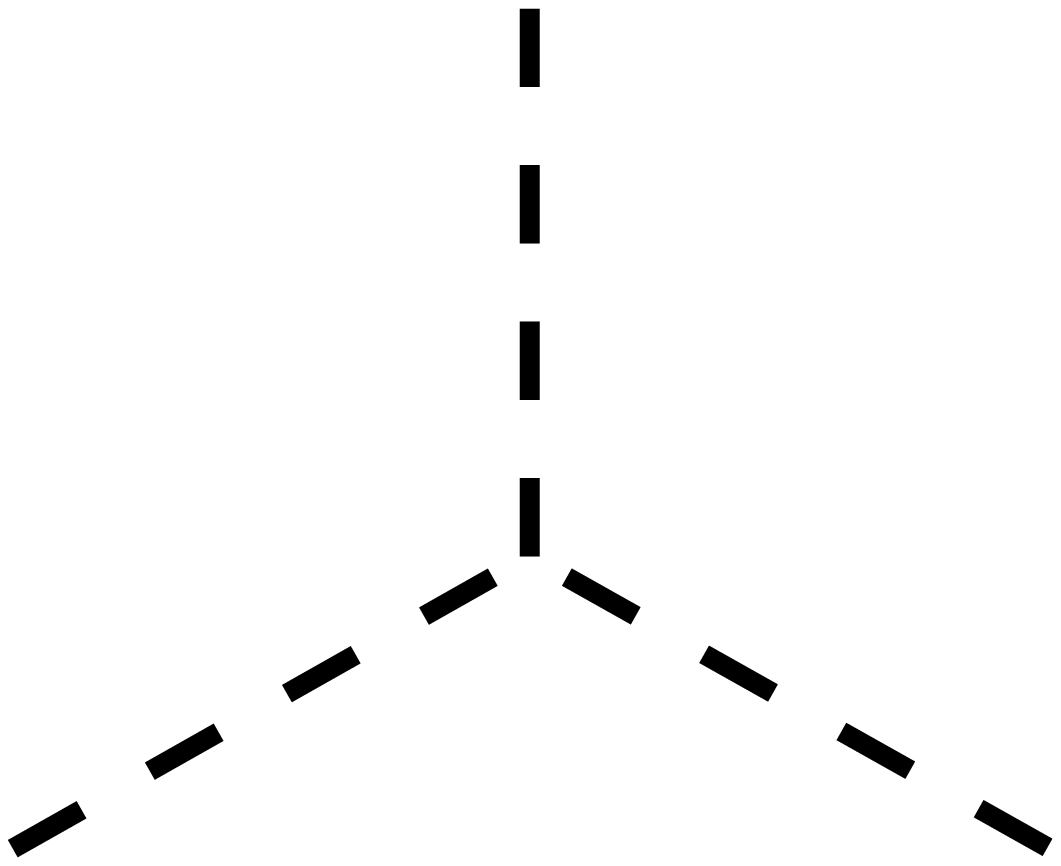}}}=1 \,.
\end{align}
\end{subequations}
In other words, in addition to the bond and vertex fugacities of the original Kuperberg web model, red edges get a weight $q^{-\frac{\theta}{\pi}}$  when they bend an angle $\theta$, green edges get the weight $q^{\frac{\theta}{\pi}}$ when they bend an angle $\theta$, whereas blue edges do not get any weight. There is also a special weight $q^{\pm \frac{1}{6}}$ when three colours meet at a vertex. The sign in the exponent changes when the cyclic order of the colours meeting at a vertex is reversed or when the orientations of the three edges meeting at the vertex are flipped simultaneously. Again, we draw only a subset of the node configurations, omitting those related to the above by a rotation. All node configurations related by a rotation are weighted the same way. Observe that in \eqref{kupvertex}, the three lines adjacent to a given node are understood as half-links of $\mathbb{H}$, hence a half-bond is weighted by $x^{\frac{1}{2}}$.

It is not difficult to deduce from this the local weights in the general case where $x_1$ and $x_2$ are arbitrary. In this case, two node configurations related by a rotation are weighted differently, in general. We will not draw the complete set of node configurations but it should be clear from the following examples how any of them is weighted:
\begin{align}
\label{kupvertex-gen}
    &\vcenter{\hbox{\includegraphics[scale=0.2]{diagrams/coloredvertex1.eps}}}=zx_1x_2^{\frac{1}{2}}q^{-\frac{1}{6}} \,, \qquad
    \vcenter{\hbox{\includegraphics[scale=0.2]{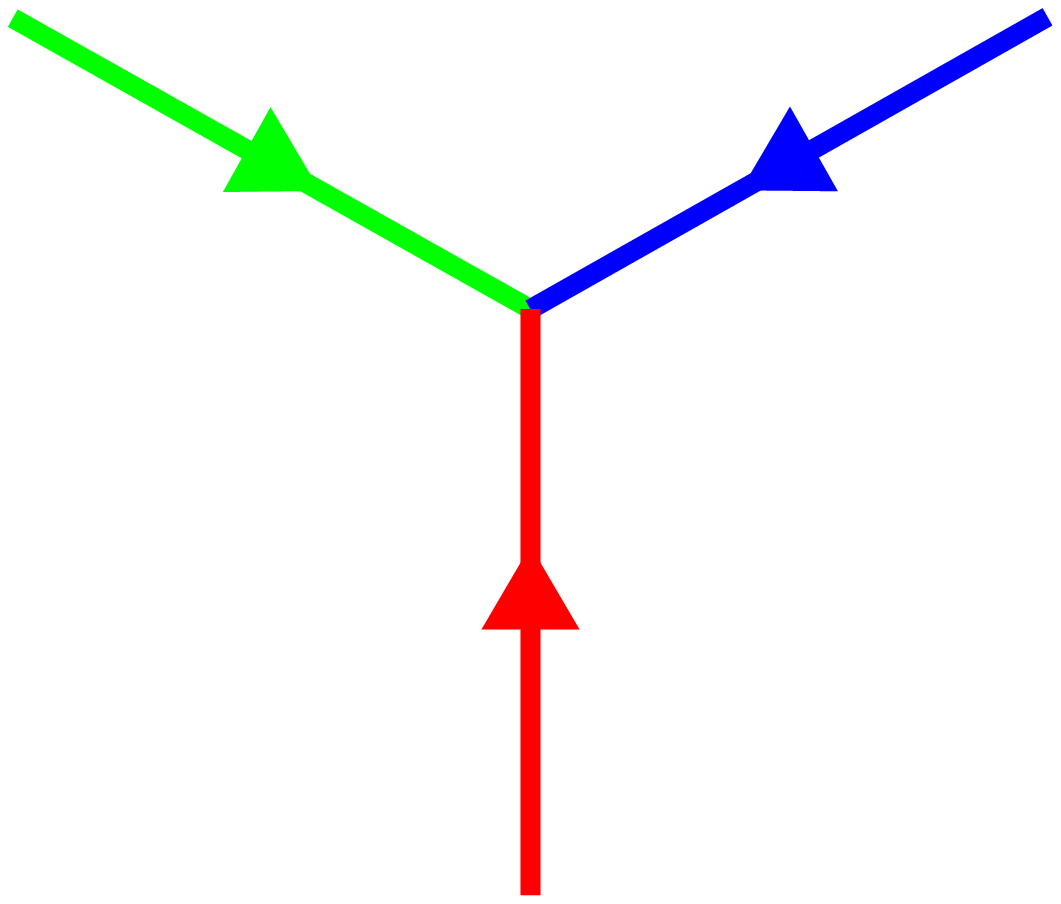}}}=zx_1^{\frac{1}{2}}x_2q^{\frac{1}{6}} \,, \nonumber\\
    &\vcenter{\hbox{\includegraphics[scale=0.2]{diagrams/coloredvertex8.eps}}}=x_1q^{-\frac{1}{3}} \,, \ \ \ \ \ \qquad
    \vcenter{\hbox{\includegraphics[scale=0.2]{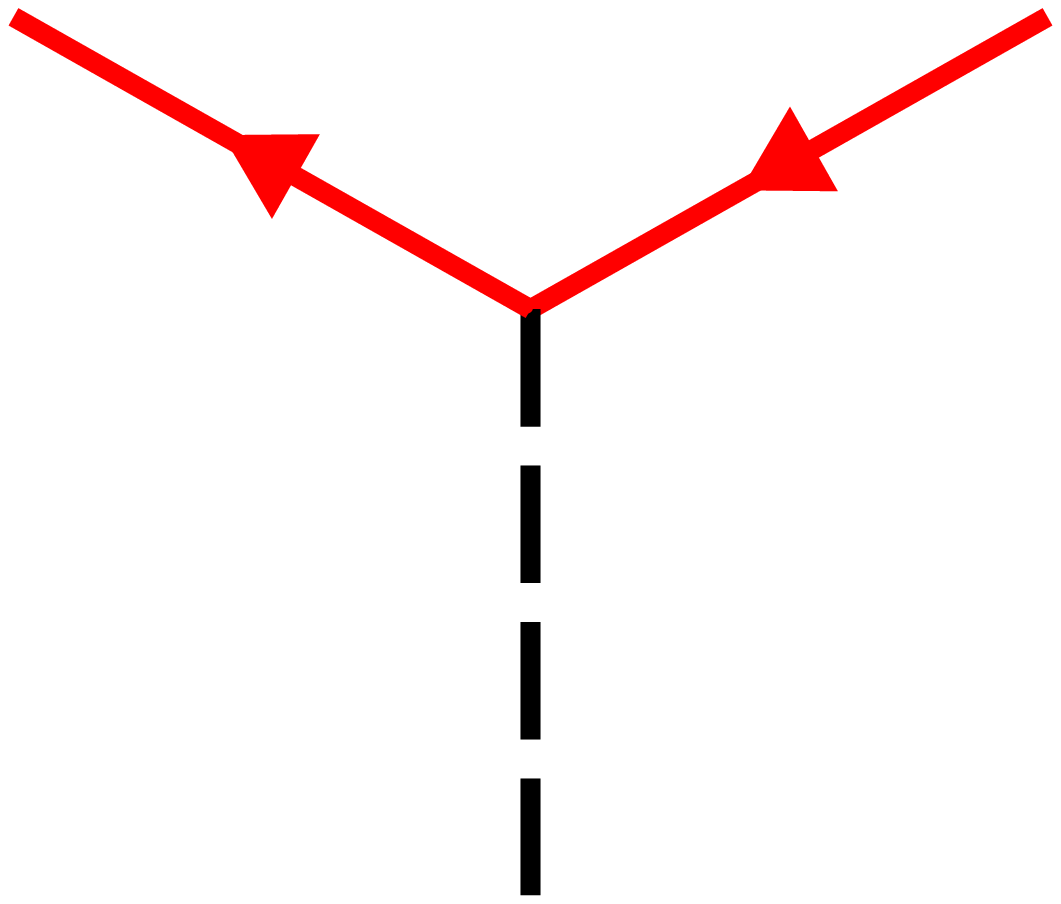}}}=x_1^{\frac{1}{2}}x_2^{\frac{1}{2}}q^{\frac{1}{3}} \,.
\end{align}\\

We now show that the above local weights recover the weight of a web configuration $c$. The local weights define the weight $w_{\rm col}(c_i)$ for a given 
three-colouring 
$c_i$ of $c$. We want to show that the sum of these weights over all three-colourings recovers the Kuperberg web model weight,
\begin{align}
    \sum_{i} w_{\text{col}}(c_i) = x_1^{N_1}x_2^{N_2}(yz)^{N_V}w_{\rm K}(c) \,,
    \label{colorsum}
\end{align}
where $N_1$, $N_2$ and $N_V$ were defined after \eqref{Z_K}.

The weight of a coloured web $w_{\rm col}(c_i)$ is given by a product of bond and vertex fugacities as well as some power of $q$, conveniently denoted $q^{n(c_i)}$. It is clear that the product of bond and vertex fugacities is the same for any three-colouring of $c$ and is equal to $x_1^{N_1}x_2^{N_2}(yz)^{N_V}$, the same factor appearing on the right-hand side of \eqref{colorsum}. Hence it remains to show that 
\begin{align}
    \sum_{i} q^{n(c_i)} = w_{\rm K}(c) \,,
    \label{coloredsum2}
\end{align}
where the sum is over all three-colourings of $c$ and $q^{n(c_i)}$ is the product of local weights given by 
the local factors
\begin{subequations}
\label{coloredkupvertex}
\begin{align}
    &\vcenter{\hbox{\includegraphics[scale=0.2]{diagrams/coloredvertex1.eps}}}=q^{-\frac{1}{6}} \,, \qquad
    \vcenter{\hbox{\includegraphics[scale=0.2]{diagrams/coloredvertex2.eps}}}=q^{\frac{1}{6}} \,, \nonumber\\
    &\vcenter{\hbox{\includegraphics[scale=0.2]{diagrams/coloredvertex3.eps}}}=q^{\frac{1}{6}} \,, \ \ \qquad
    \vcenter{\hbox{\includegraphics[scale=0.2]{diagrams/coloredvertex4.eps}}}=q^{-\frac{1}{6}} \,,
\end{align}
together with
\begin{align}
    &\vcenter{\hbox{\includegraphics[scale=0.2]{diagrams/coloredvertex8.eps}}}=q^{-\frac{1}{3}} \,, \qquad
    \vcenter{\hbox{\includegraphics[scale=0.2]{diagrams/coloredvertex10.eps}}}=1 \,, \qquad
    \vcenter{\hbox{\includegraphics[scale=0.2]{diagrams/coloredvertex9.eps}}}=q^{\frac{1}{3}} \,, \nonumber \\
    &\vcenter{\hbox{\includegraphics[scale=0.2]{diagrams/coloredvertex5.eps}}}=q^{\frac{1}{3}} \,, \ \ \qquad
    \vcenter{\hbox{\includegraphics[scale=0.2]{diagrams/coloredvertex7.eps}}}=1 \,, \qquad
    \vcenter{\hbox{\includegraphics[scale=0.2]{diagrams/coloredvertex6.eps}}}=q^{-\frac{1}{3}} \,,
\end{align}
and
\begin{align}
    &\vcenter{\hbox{\includegraphics[scale=0.2]{diagrams/coloredvertex11.eps}}}=1 \,,
\end{align}
\end{subequations}
where again any local node configuration related by a rotation to one of the above is weighted accordingly.

It will turn out convenient to generalise the reasoning by considering the coloured web as an abstract web, i.e., as a coloured web embedded in the plane. In other words, we forget about the underlying lattice~$\mathbb{H}$ and allow edges to bend in any possible way, rather than through the discrete angles dictated by $\mathbb{H}$. A coloured abstract web is given a weight which is again a product of powers of $q$. Red edges get a weight $q^{\frac{-\theta}{\pi}}$  when they bend an angle $\theta$, green edges get a weight $q^{\frac{\theta}{\pi}}$ when they bend an angle $\theta$, whereas blue edges do not get any weight. Moreover vertices account for a weight depending on the angle $\alpha$ between the red and green edges, measured from the red edge to the green one as shown here:
\begin{align}
\label{coloredkupvertexgeneralized}
    &\vcenter{\hbox{\includegraphics[scale=0.2]{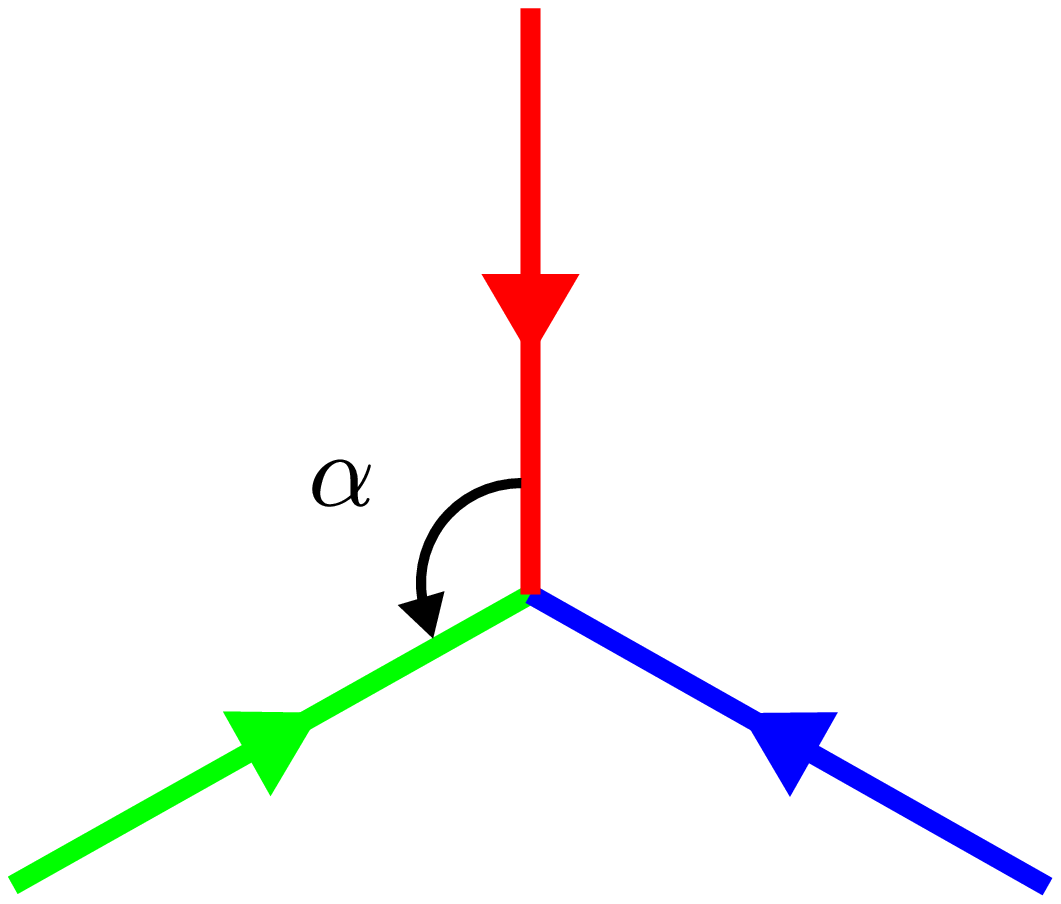}}}=q^{-\frac{\alpha}{\pi}+\frac{1}{2}} \,, \qquad
    \vcenter{\hbox{\includegraphics[scale=0.2]{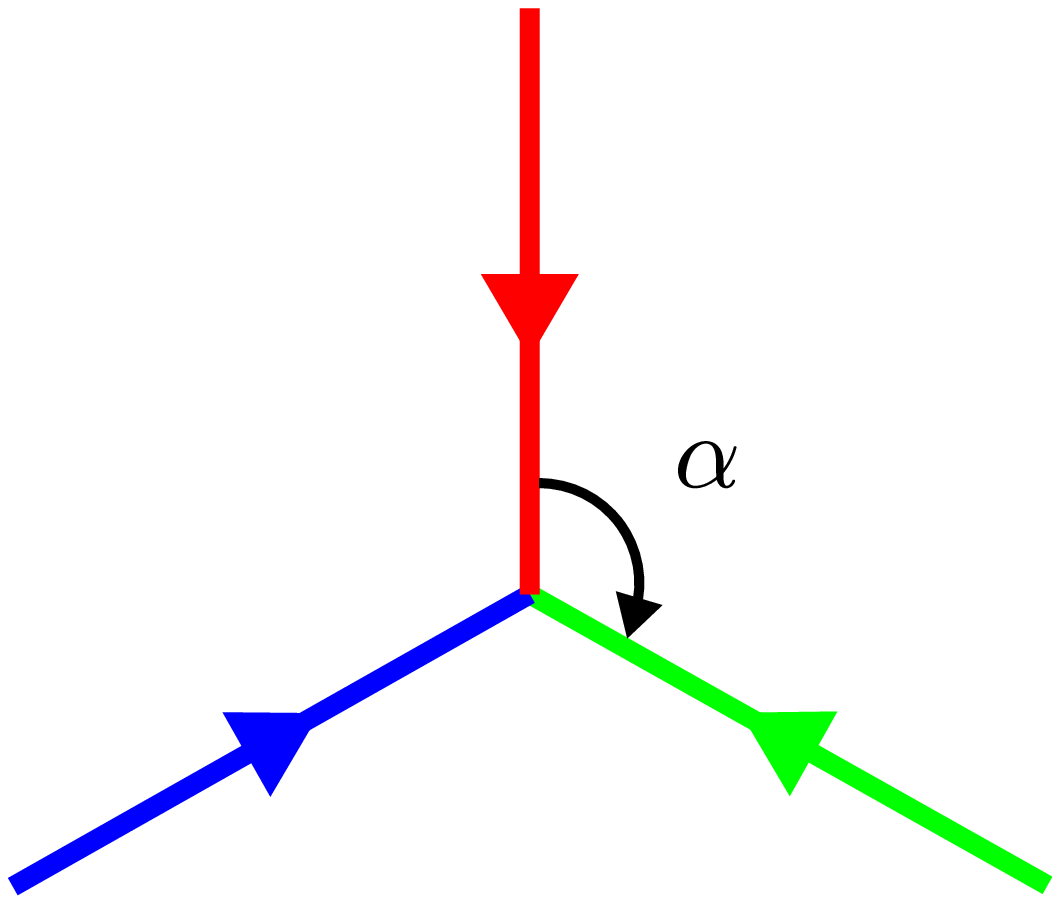}}}=q^{-\frac{\alpha}{\pi}-\frac{1}{2}}\,, \nonumber \\
    &\vcenter{\hbox{\includegraphics[scale=0.2]{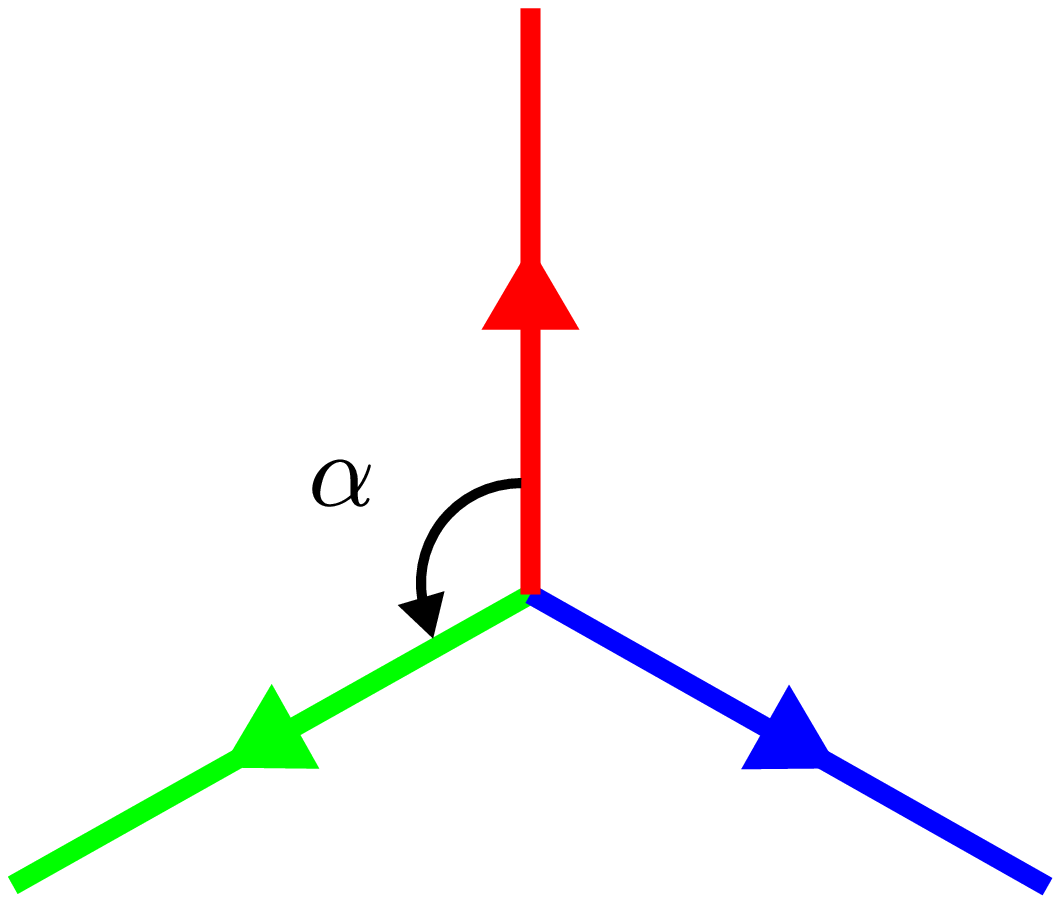}}}=q^{\frac{\alpha}{\pi}-\frac{1}{2}} \,, \ \  \qquad
    \vcenter{\hbox{\includegraphics[scale=0.2]{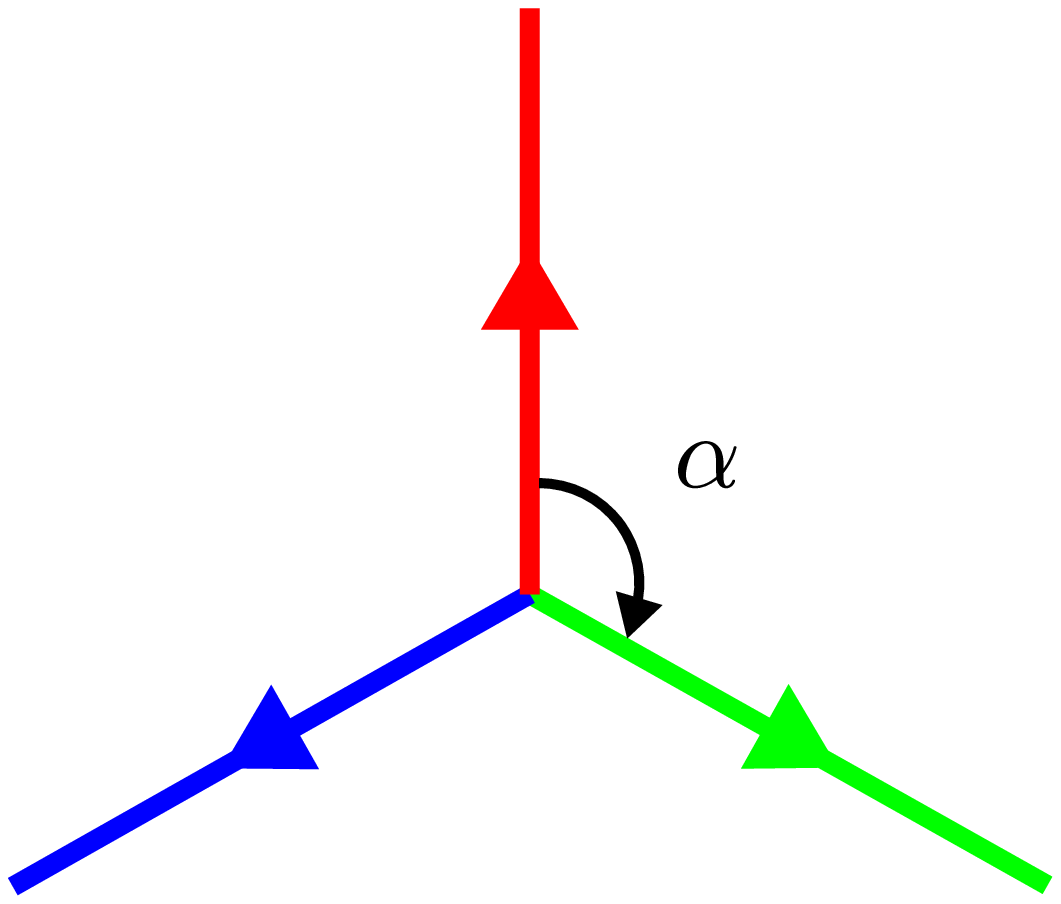}}}=q^{\frac{\alpha}{\pi}+\frac{1}{2}} \,.
\end{align}
For a coloured web embedded in $\mathbb{H}$, this agrees with \eqref{coloredkupvertex}. The total weight of a coloured web defined by the above local weights is invariant under isotopy. Indeed, straightening a coloured edge does not change the total weight of the coloured web. Moreover, bending an edge incident on a vertex, the local weight associated to the bending compensates the change in the local weight of the given vertex. We shall use this freedom in the following.

In order to show \eqref{coloredsum2}, it is sufficient to show that the local relations \eqref{3rules} are satisfied by the local weights. That is, for any relation, if we fix the colours of the external edges and sum over the possible colourings of the internal ones, the two sides must be weighted the same. The loop rule \eqref{3rulesa} is obviously satisfied as a clockwise (respectively anticlockwise) oriented red loop gives a factor $q^2$ (respectively $q^{-2}$), a blue one gives a factor $1$ regardless of its orientation, and a clockwise (respectively anticlockwise) oriented green loop gives a factor $q^{-2}$ (respectively $q^{2}$). The sum over colours indeed produces the required loop weight, $[3]_q=q^2+1+q^{-2}$, for any of the two possible orientations.

Regarding the second rule \eqref{3rulesb}, we have, for the case where the external edges are red (the other cases being similar):
\begin{equation}
 \label{digon-gym}
    \vcenter{\hbox{\includegraphics[scale=0.15]{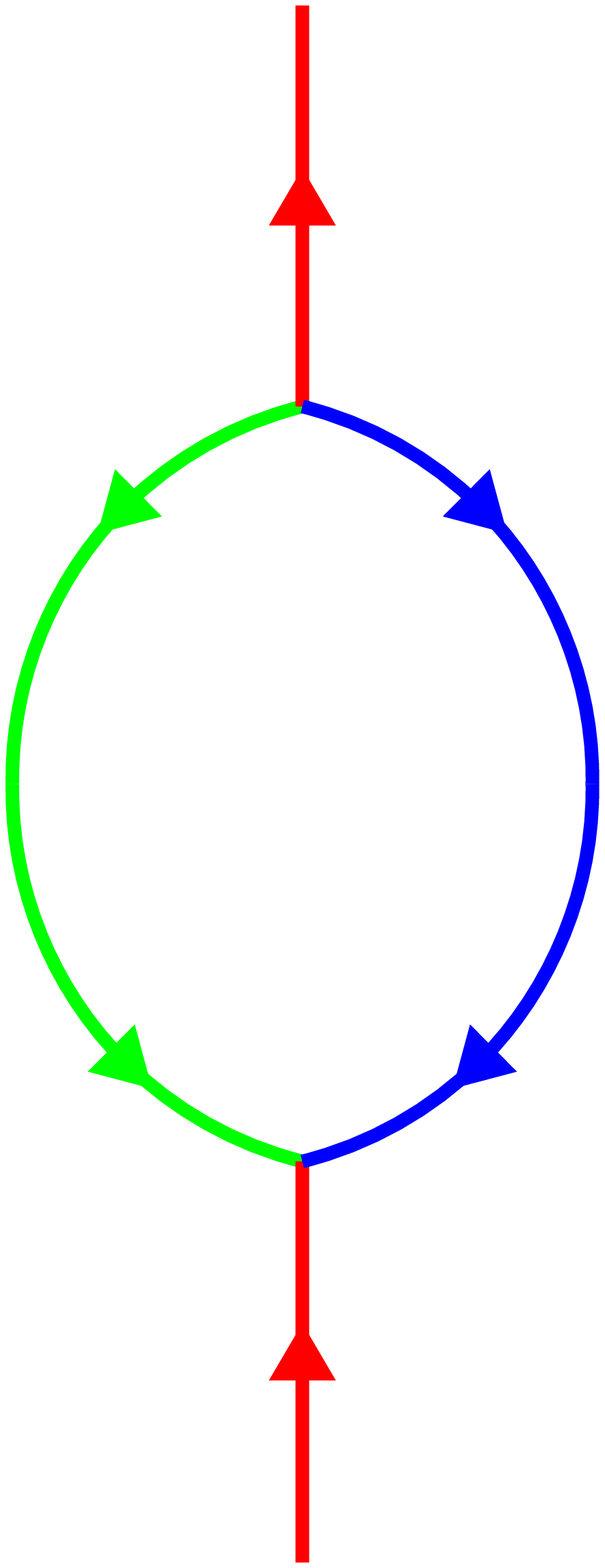}}}\ \;+\;\ \vcenter{\hbox{\includegraphics[scale=0.15]{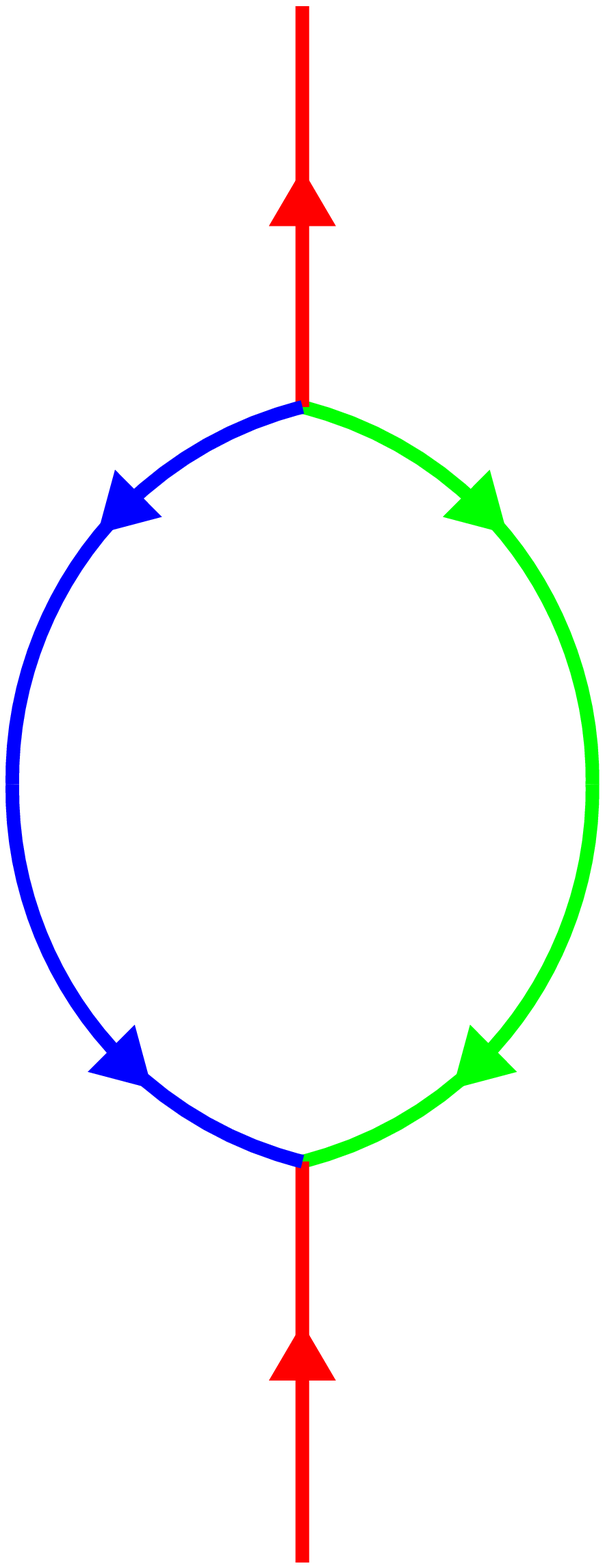}}}\ \;=\;q^{\frac{1}{6}\times2+\frac{2}{3}}\ \vcenter{\hbox{\includegraphics[scale=0.15]{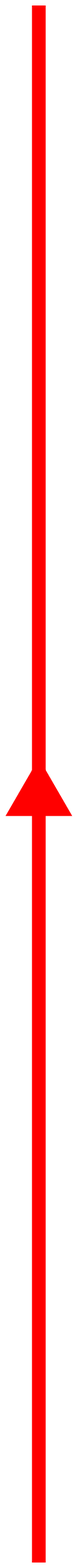}}}\;+\;q^{-\frac{1}{6}\times2-\frac{2}{3}}\ \vcenter{\hbox{\includegraphics[scale=0.15]{diagrams/coloredrule3.eps}}}\;=\;[2]_q\ \vcenter{\hbox{\includegraphics[scale=0.15]{diagrams/coloredrule3.eps}}}
\end{equation}

Regarding the last rule \eqref{kupsquare}, there are two cases for the colourings of the four external edges to be considered.
In the first case, all external edges have the same colour, and we find (for the case of green external edges):
\begin{equation}
\label{square-gym}
    \vcenter{\hbox{\includegraphics[scale=0.15]{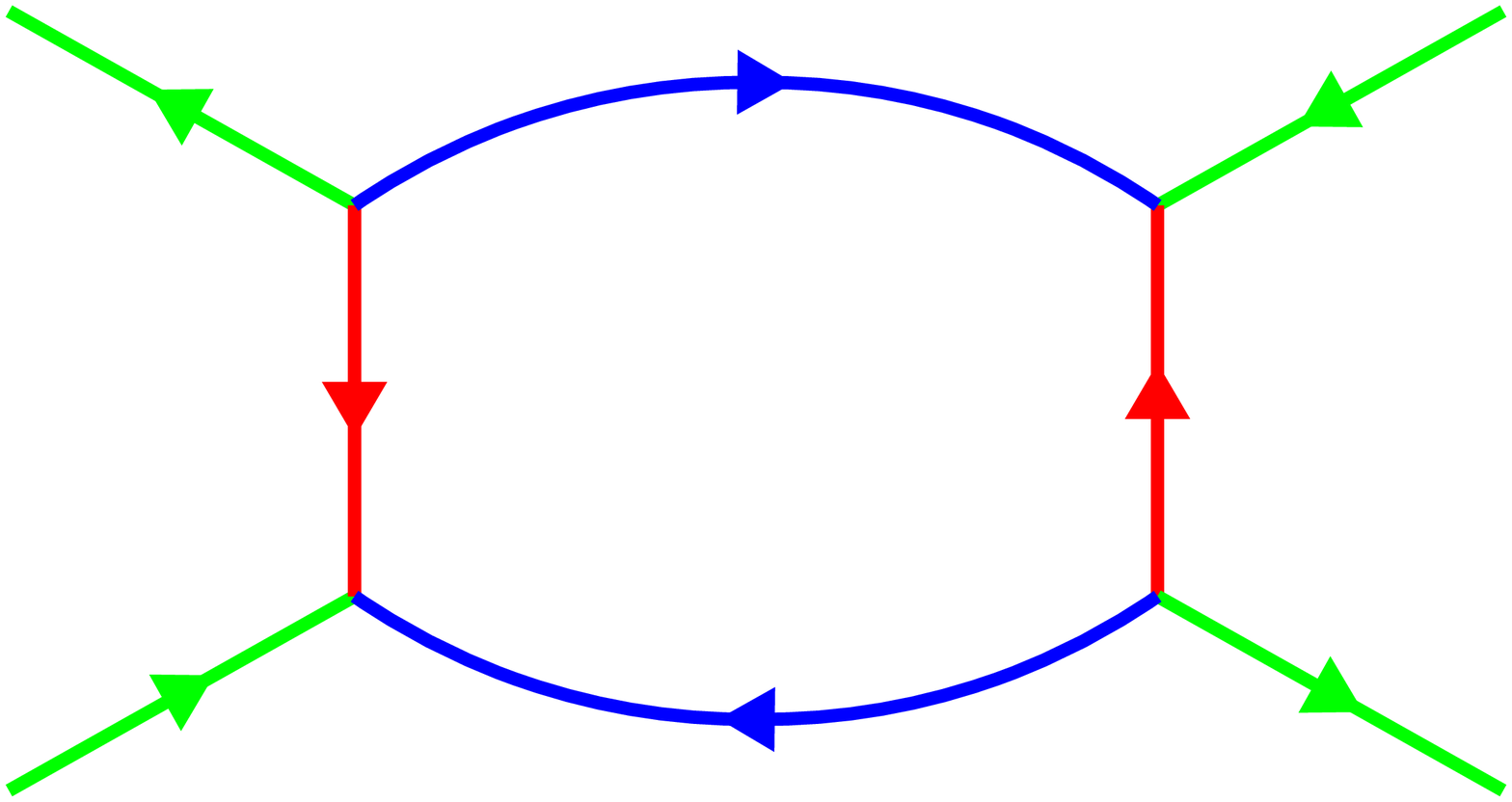}}}\ +\ \vcenter{\hbox{\includegraphics[scale=0.15]{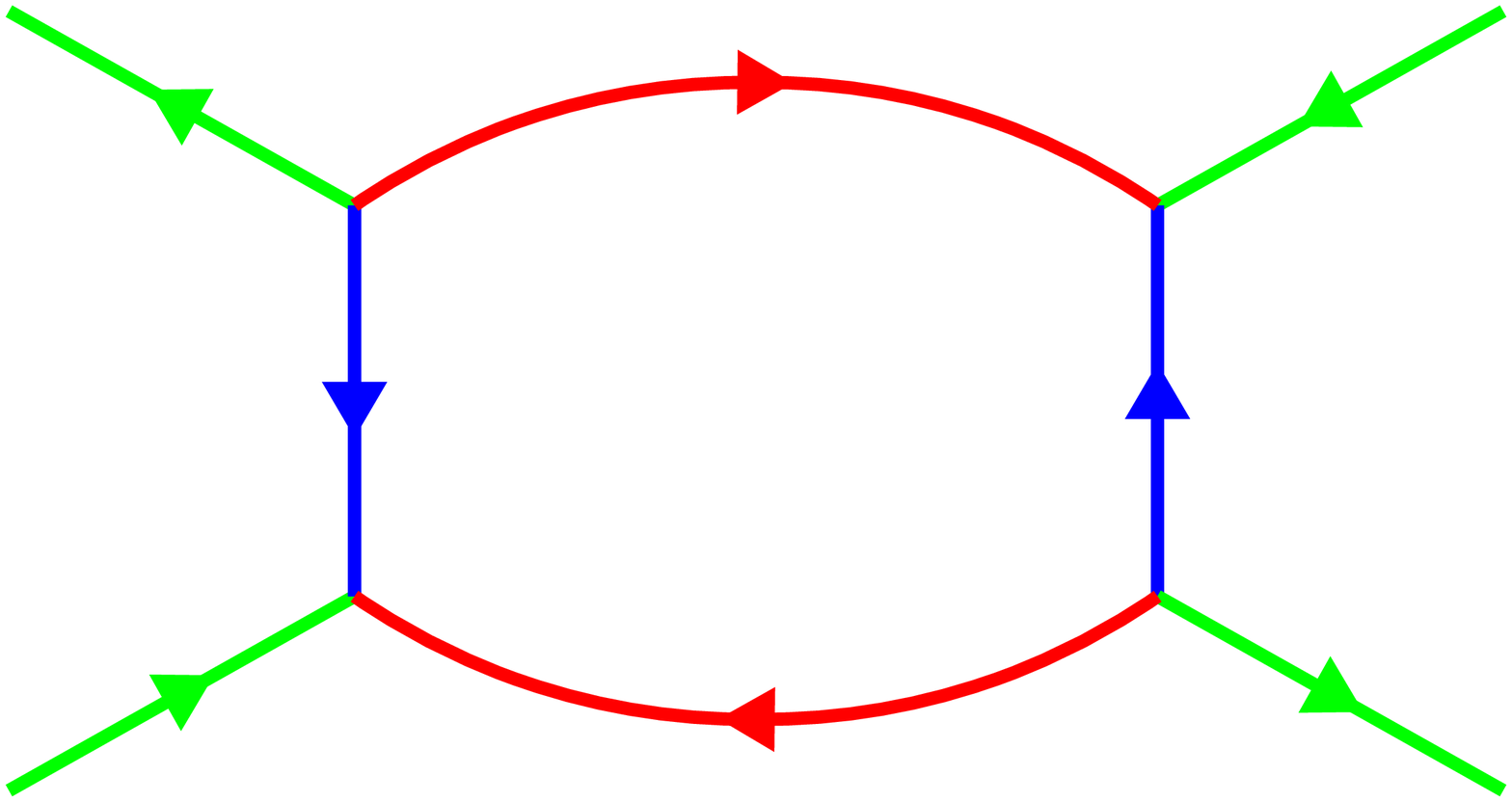}}} =
    \vcenter{\hbox{\includegraphics[scale=0.15]{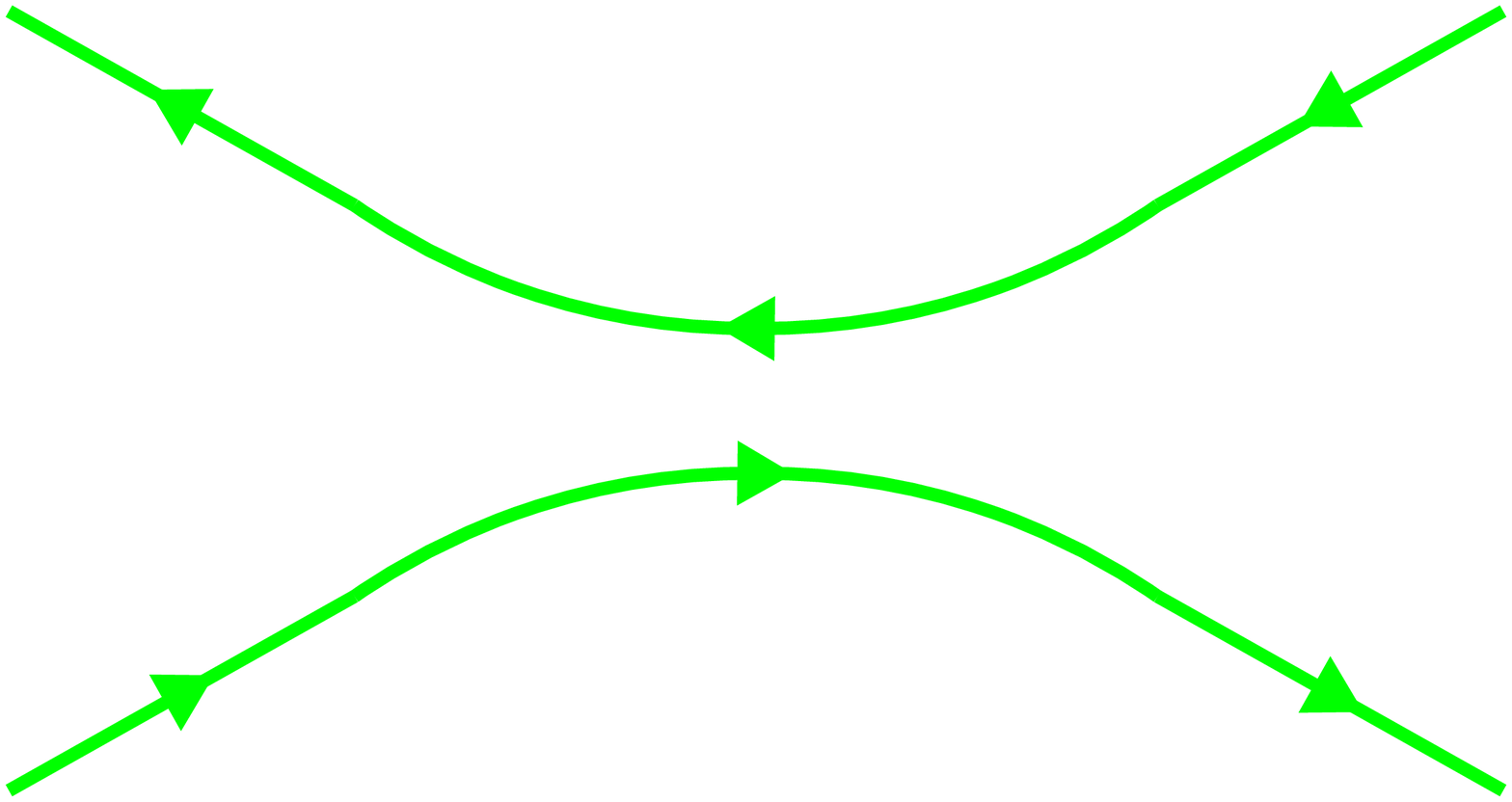}}}\ +\ \vcenter{\hbox{\includegraphics[scale=0.15]{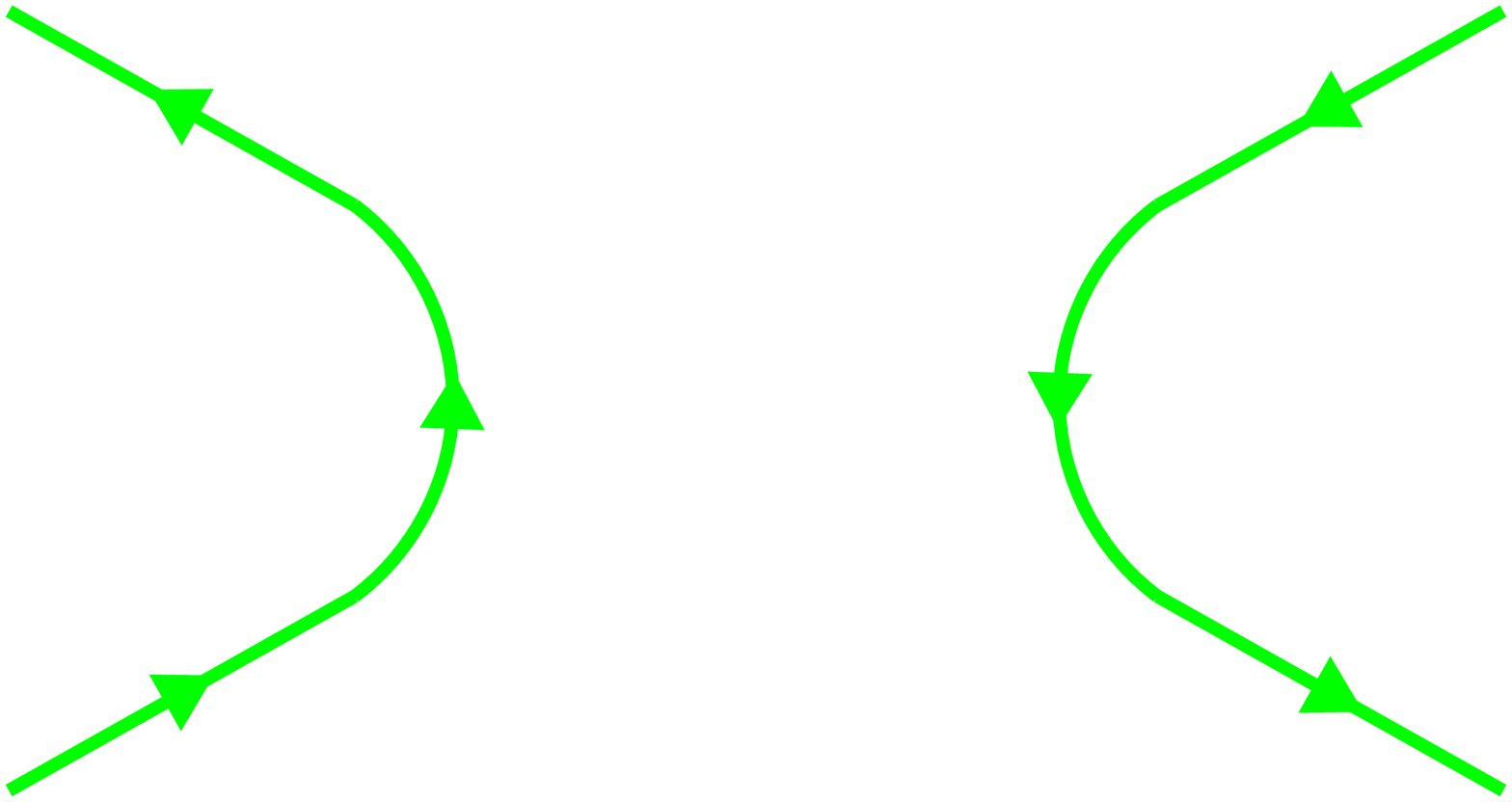}}}
\end{equation}
where indeed summing out the local weights over the possible colourings on the left-hand side gives
$q^{-\frac{1}{6}\times 4}+q^{\frac{1}{6}\times 4+ \frac{1}{3}\times 2} = q^{-\frac{2}{3}}+q^{\frac{4}{3}}$,
while summing out the local weights over the possible contractions on the right-hand side gives the same result.
In the second case, the external edges have two different colours, one on each side of the square. In that case, the left- and right-hand sides of \eqref{kupsquare} are again weighted the same:
\begin{align}
\label{square2colours}
    \vcenter{\hbox{\includegraphics[scale=0.15]{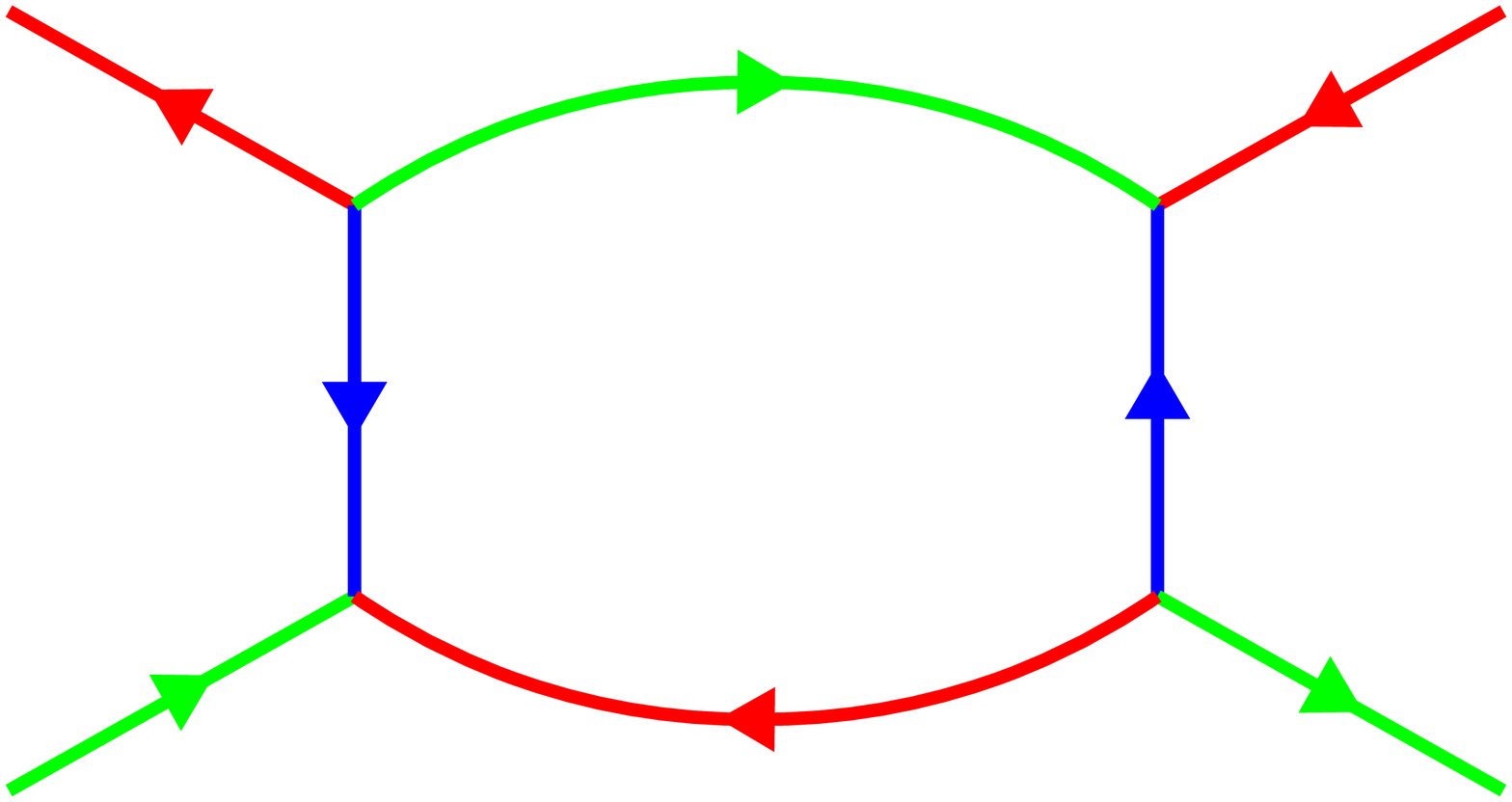}}} = \vcenter{\hbox{\includegraphics[scale=0.15]{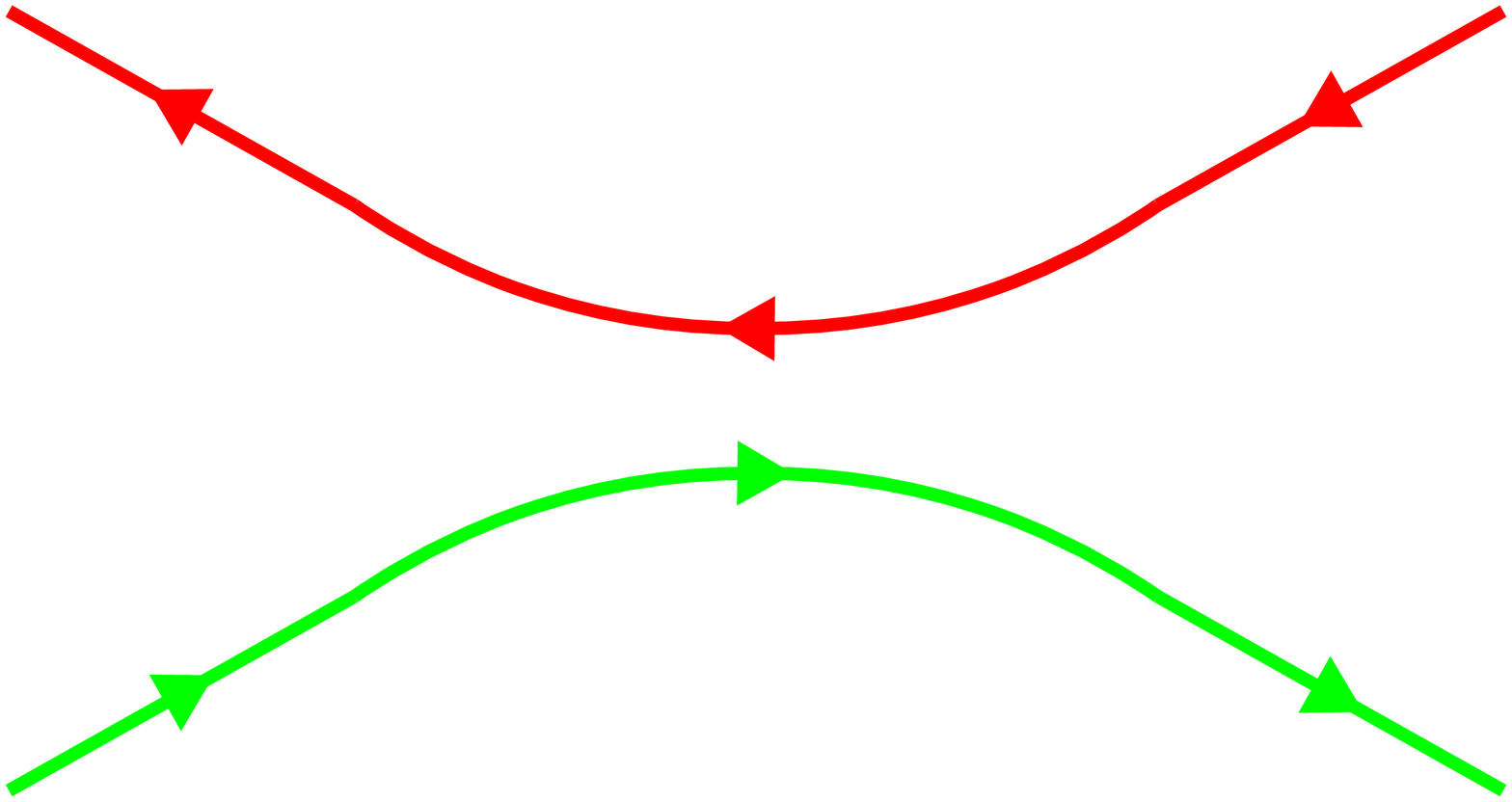}}}
\end{align}
The computations for other colourings of external edges are similar. Thus we have shown \eqref{coloredsum2}.

\medskip

When $\mathbb{H}$ is embedded in the cylinder, we must again introduce an oriented seam line with local weights given by the weight carried by oriented coloured curves when they do a full turn:
\begin{align}
\label{kupseamline}
    &\vcenter{\hbox{\includegraphics[scale=0.2]{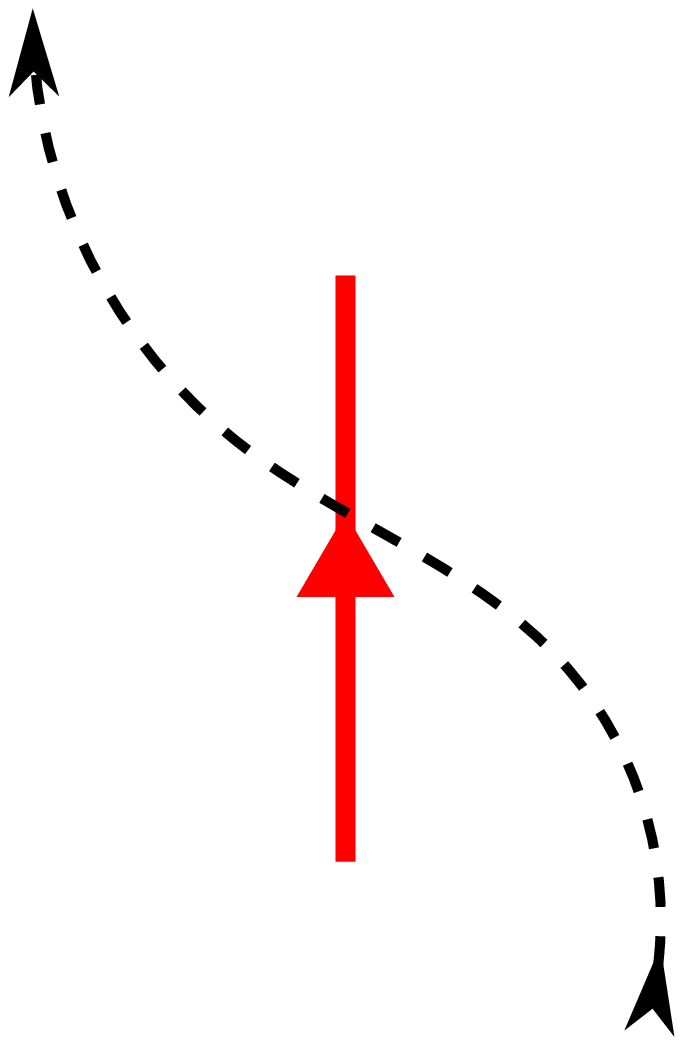}}}=q^2 \,, \ \ \qquad \vcenter{\hbox{\includegraphics[scale=0.2]{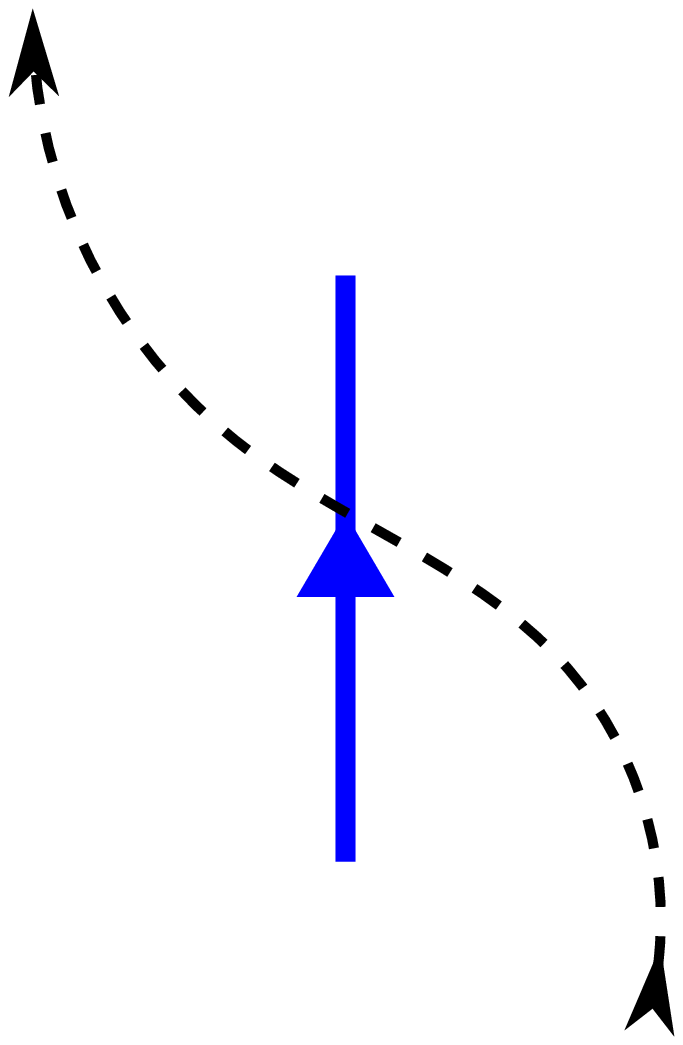}}}=1 \,, \qquad \vcenter{\hbox{\includegraphics[scale=0.2]{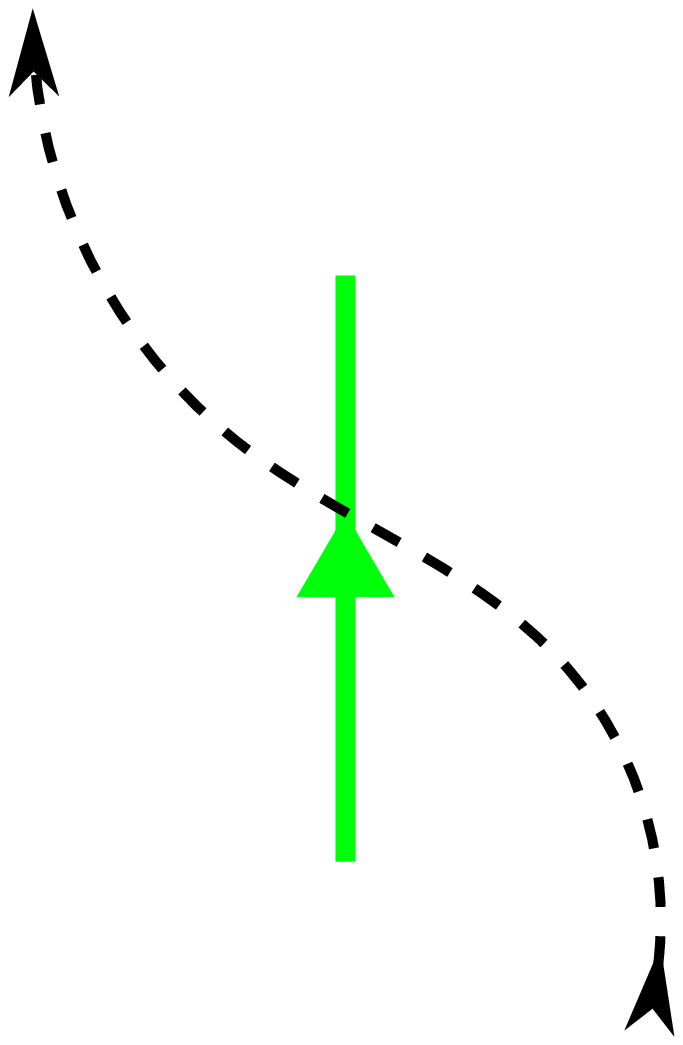}}}=q^{-2} \,, \nonumber\\
    &\vcenter{\hbox{\includegraphics[scale=0.2]{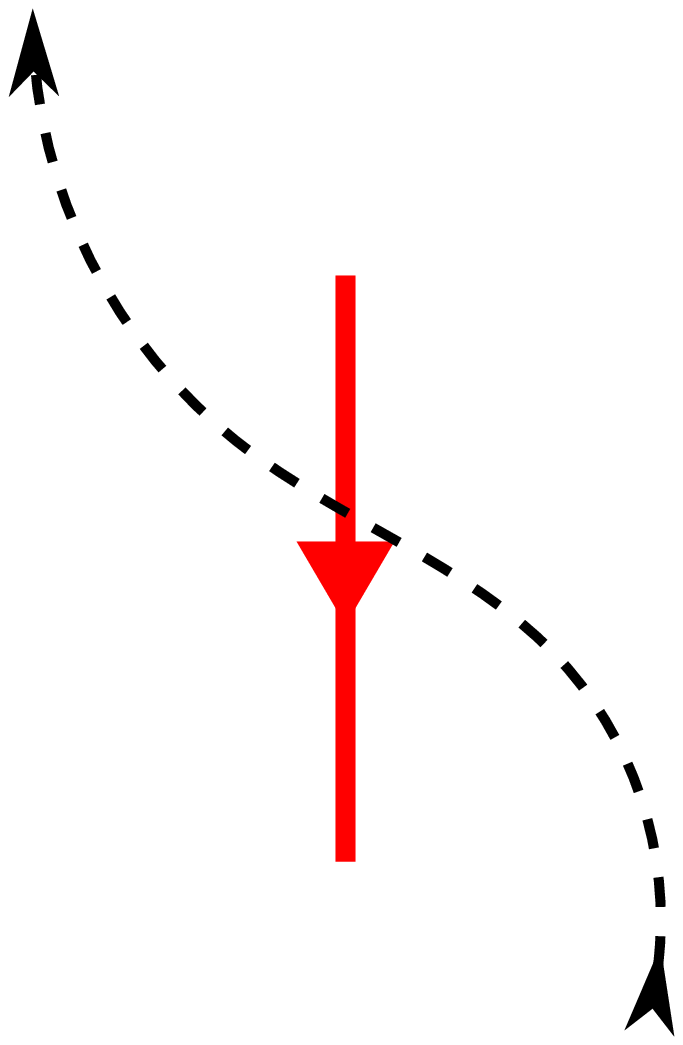}}}=q^{-2} \,, \qquad \vcenter{\hbox{\includegraphics[scale=0.2]{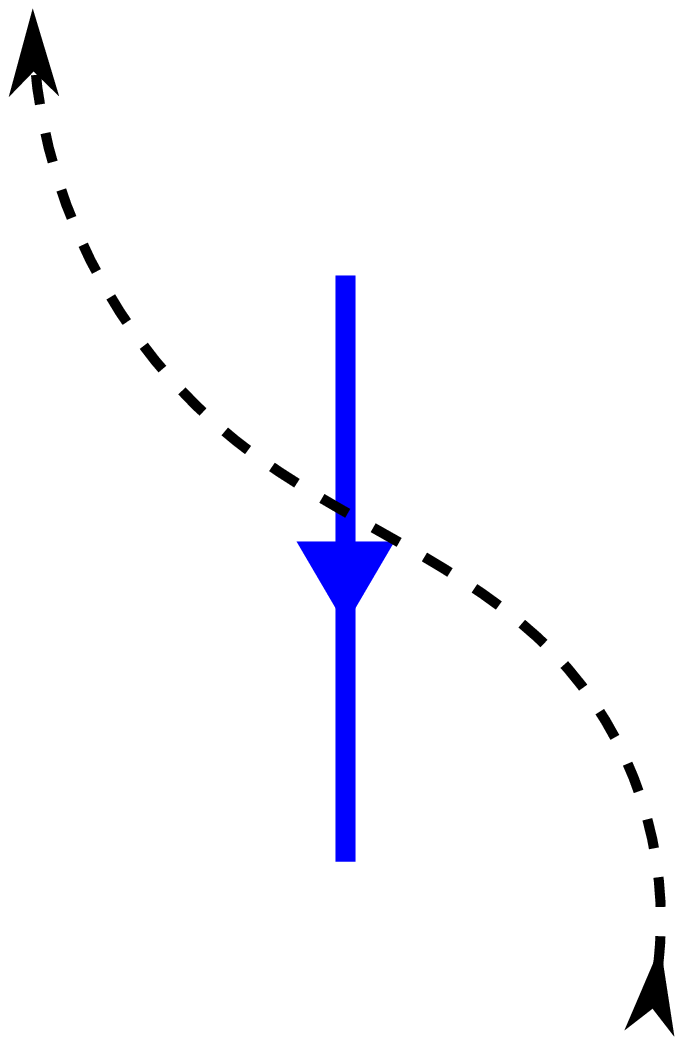}}}=1 \,, \qquad \vcenter{\hbox{\includegraphics[scale=0.2]{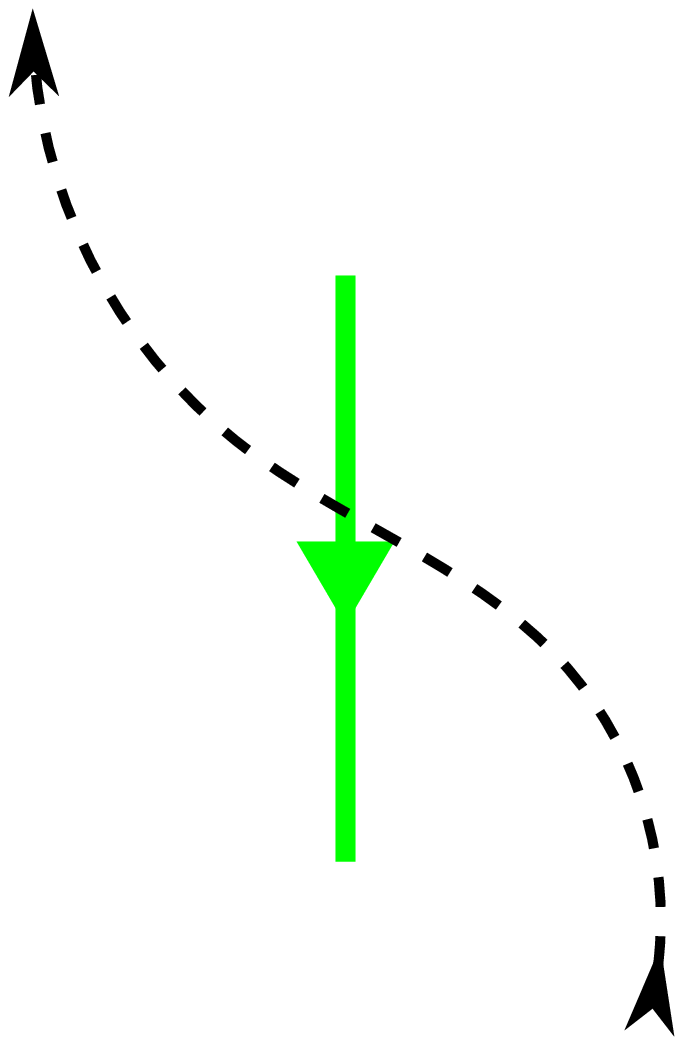}}}=q^2 \,, \\
    &\vcenter{\hbox{\includegraphics[scale=0.2]{diagrams/crossseamleftrightloop3.eps}}}=1 \,. \nonumber
\end{align}
It is obvious that these weights just compensate the lack of bending for configurations that wrap around the periodic direction, so the remainder of the proof
of \eqref{colorsum} can be taken over from the strip case discussed above.

\subsection{Local weights for any trivalent lattice}

As a side-effect of the above proof, we remark that the vertex weights \eqref{kupvertex} can be generalised to account for the local formulation of the
Kuperberg web model defined on any trivalent lattice embedded in the plane. 

This is more easily seen when we restrict to only one bond fugacity $x$. When an edge undergoes bending (by passing through a node incident on two 
bonds), it is given the appropriate weight depending on the colour and the bending angle (as defined above~\eqref{coloredkupvertexgeneralized}) times the  bond fugacity $x$. When three colours meet at a vertex, the weight depends on the angle $\alpha$  between the red and green edges, measured from the red edge to the green one:
\begin{align*}
    &\vcenter{\hbox{\includegraphics[scale=0.2]{diagrams/generalisedcoloredvertex1.eps}}}=zx^{\frac{3}{2}}q^{-\frac{\alpha}{\pi}+\frac{1}{2}} \,, \qquad
    \vcenter{\hbox{\includegraphics[scale=0.2]{diagrams/generalisedcoloredvertex2.eps}}}=zx^{\frac{3}{2}}q^{-\frac{\alpha}{\pi}-\frac{1}{2}} \,, \\
    &\vcenter{\hbox{\includegraphics[scale=0.2]{diagrams/generalisedcoloredvertex3.eps}}}=yx^{\frac{3}{2}}q^{\frac{\alpha}{\pi}-\frac{1}{2}} \,, \ \  \qquad
    \vcenter{\hbox{\includegraphics[scale=0.2]{diagrams/generalisedcoloredvertex4.eps}}}=yx^{\frac{3}{2}}q^{\frac{\alpha}{\pi}+\frac{1}{2}} \,.
\end{align*}
That the Kuperberg web weight is retrieved follows from the fact, that, in the last subsection, we have in fact shown \eqref{coloredsum2} for any coloured web embedded in the plane.

It is also possible to generalise further to account for two types of bond fugacities, $x_1$ and $x_2$, once one chooses an appropriate time foliation of the plane.

\medskip

The study of two-dimensional statistical models defined on arbitrary trivalent lattices---or by duality, on arbitary triangulations of the plane---is relevant
for the discretisation of models of two-dimensional quantum gravity. In such models the partition function is a double sum over the triangulations, with
a certain weighting (the so-called cosmological term) coupling to the area of the corresponding surface, and over the statistical model defined on a given 
triangulation. There are many interesting connections from this approach to random matrix integrals, combinatorics and graph theory.
We refer the reader to the review \cite{QGReview} for further details. It should be noticed in particular that the O($N$) loop model has been solved in this context,
using random matrix techniques \cite{Kostov89}, and we leave for future research to determine whether the Kuperberg web model coupled to quantum
gravity can be treated by similar means.

\subsection{Algebraic transfer matrix formulation}
\label{sec:localtmKup}

Our next goal is to define the transfer matrix corresponding to the vertex models of Section \ref{sec:kupvertex}. To this end, we associate to each link of $\mathbb{H}$ a local space of states whose basis is given by the link degrees of freedom. In the loop model case, this leads to a three-dimensional local states space $\mathcal{H}_{\text{loop}}=\text{span}(\ket{\uparrow},\ket{\downarrow},\ket{\ })$. In the Kuperberg web model, the local state space has dimension seven and is written in terms of colours and orientations:
\begin{equation}
 \mathcal{H}_{\rm K}=\text{span}(\ket{\color{red} \uparrow},\ket{\color{blue} \uparrow},\ket{\color{green} \uparrow},\ket{\color{red} \downarrow},\ket{\color{blue} \downarrow},\ket{\color{green} \downarrow},\ket{\ }) \,.
\end{equation}
The vertex weights are then understood as matrix elements between states, but to define them we need tensor products of several local state spaces. The operators built this way are the local transfer matrices. The weights associated to the seam line are interpreted as matrix elements of twist operators, as they introduce twisted boundary conditions. 

\begin{figure}
\begin{center}
    \includegraphics[scale=0.4]{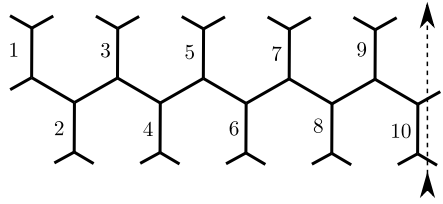}
\end{center}
    \caption{The row to row transfer matrix in the case of periodic boundary conditions with $2L=10$.}
    \label{fig:transfermatrix}
\end{figure}

We shall call node of type $1$ (respectively type $2$) a node situated at the bottom (respectively top) of a vertical link. For example, \eqref{loopvertexweights} and \eqref{kupvertex} show nodes of type 1 for the loop and Kuperberg web models, respectively. As usual, we first discuss the loop model.

We first recall how to build the full transfer matrix from the local transfer matrices and the twist operators. Denote by $t^{\text{loop}}_{(k)}$ the local transfer matrices propagating through a node of type $k \in \{1,2\}$. They are linear maps:
\begin{subequations}
\begin{eqnarray}
  t^{\text{loop}}_{(1)} &:& \mathcal{H}_{\text{loop}}\otimes \mathcal{H}_{\text{loop}} \to \mathcal{H}_{\text{loop}} \,, \\
  t^{\text{loop}}_{(2)} &:& \mathcal{H}_{\text{loop}} \to \mathcal{H}_{\text{loop}}\otimes \mathcal{H}_{\text{loop}} \,,
\end{eqnarray}
\end{subequations}
and we use their pictorial notation $\vcenter{\hbox{\includegraphics[scale=0.05]{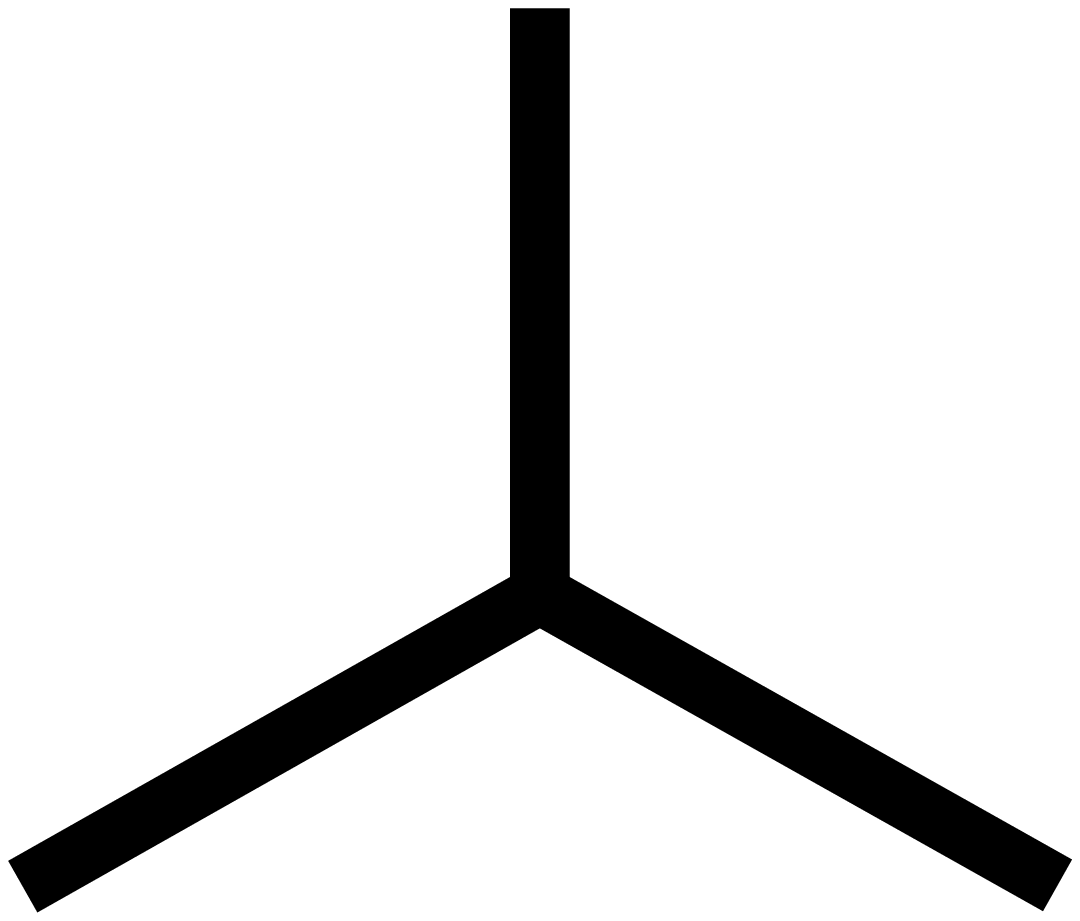}}}$ and $\vcenter{\hbox{\includegraphics[scale=0.05]{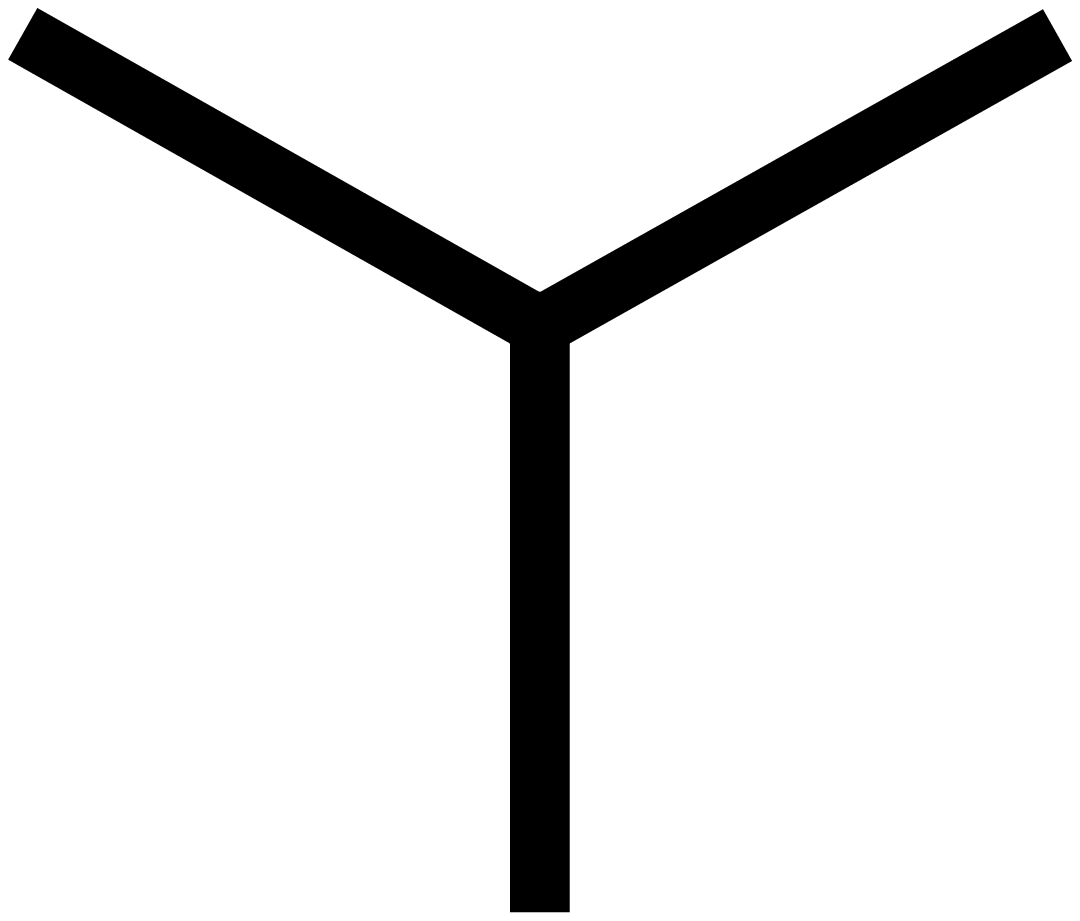}}}$, respectively, in Figure \ref{fig:transfermatrix}. Their matrix elements are given by~\eqref{loopvertexweights} in the case $k=1$ plus rotations, and similarly for $k=2$.
Hence their composition $t^{\text{loop}}=t^{\text{loop}}_{(2)}t^{\text{loop}}_{(1)}$ is a linear map from $\mathcal{H}_{\text{loop}}\otimes \mathcal{H}_{\text{loop}}$ to itself (i.e., an endomorphism of $\mathcal{H}_{\text{loop}}\otimes \mathcal{H}_{\text{loop}}$).%
\footnote{Remark that $t^{\text{loop}}$ corresponds to summing over the state of a vertical link, so that a pair of vertices on $\mathbb{H}$ is effectively transformed into a single vertex on a (tilted) square lattice.}
We index the copies $t^{\text{loop}}_i$ of these operators by their position $i$ in a row as in Figure~\ref{fig:transfermatrix}. 

Denote by $S_{\text{loop}}$ the twist operator associated with crossing the seam line running from right to left, and its inverse $S_{\text{loop}}^{-1}$ associated with the seam line running from left to right. Then the row-to-row\footnote{Note that with our definition, the row-to-row transfer matrix propagates states through {\em two} rows of the lattice.} transfer matrix $T_{\text{loop}}$ in the cylinder geometry reads
\begin{align}
\label{looptransfermatrixcyl}
    T_{\text{loop}}=\left(\prod_{k=0}^{L-1}t^{\text{loop}}_{2k+1}\right)\left(\prod_{k=1}^{L-1}t^{\text{loop}}_{2k}\right)S_{\text{loop}}t^{\text{loop}}_{2L}S_{\text{loop}}^{-1} \,,
\end{align}
where $S_{\text{loop}}$ acts non-trivially on site $1$ only.
In case of open boundary conditions we have instead\footnote{Remark that no non-trivial boundary operator is used in this setup. Generalisations to non-trivial boundary interactions are however possible \cite{DubailSp1,DubailSp2}.}
\begin{align}
\label{looptransfermatrixstrip}
    T_{\text{loop}}=\left(\prod_{k=0}^{L-1}t^{\text{loop}}_{2k+1}\right)\left(\prod_{k=1}^{L-1}t^{\text{loop}}_{2k}\right) \,.
\end{align}
It is an endomorphism of $\mathcal{H}_{\text{loop}}^{\otimes 2L}$. The partition function is then recovered as the vacuum expectation value of powers of the row-to-row transfer matrix:
\begin{align}
\label{loopPF-TM}
    Z_{\text{loop}}=\braket{\ T_{\text{loop}}^M\ } \,.
\end{align}
By the vacuum expectation value, we mean the matrix element from $\ket{\ }^{\otimes 2L}$ to itself. To be precise, the right-hand-side of \eqref{loopPF-TM} expresses the partition function $Z_{\text{loop}}$ on a hexagonal lattice with $2M-2$ rows, because while $T_{\text{loop}}^M$ builds loop configurations on a lattice with $2M$ rows, the degrees of freedom on the first and last row are constrained to be empty due to our choice of vacuum state.

\medskip

Next we discuss the symmetries of the local transfer matrices. Let $V$ be the fundamental representation of $U_{-q}(\mathfrak{sl}_2)$.%
\footnote{The ``$-q$'' in $U_{-q}(\mathfrak{sl}_2)$ may seam unusual but it is actually convenient in order not to introduce additional minus signs in expressions like \eqref{loopvertexweights}.}
Let $(v_1,v_2)$ be the basis of $V$ such that the generators of $U_{-q}(\mathfrak{sl}_2)$ are represented by the matrices
\begin{align}
    (-q)^H=\begin{pmatrix} -q & 0\\
    0 & -q^{-1}
    \end{pmatrix} \,,\qquad 
    E=\begin{pmatrix} 0 & 1\\
    0 & 0
    \end{pmatrix} \,,\qquad 
    F=\begin{pmatrix} 0 & 0\\
    1 & 0
    \end{pmatrix} \,.
\end{align}

Each local state space $\mathcal{H}_{\text{loop}}$ carries an action of $U_{-q}(\mathfrak{sl}_2)$, as $\mathcal{H}_{\text{loop}}\cong V\oplus \mathbb{C}$, where $\mathbb{C}$ denote the trivial representation (corresponding to the empty state). We define the action on $\mathcal{H}_{\text{loop}}$ by relating the basis $\{\ket{\uparrow},\ket{\downarrow},\ket{\ }\}$ with the basis $\{v_1,v_2,1\}$ on each link. We shall here need to distinguish between the three possible spatial orientations of links, that we call {\em inclinations} for convenience. On links of inclination $\diagdown$ we have
\begin{subequations}
\begin{align}
    (\ket{\uparrow},\ket{\downarrow},\ket{\ })=\text{diag}(q^{\frac{1}{6}},q^{-\frac{1}{6}},1)(v_1,v_2,1) \,,
\end{align}
whereas on links of inclination $\diagup$
\begin{align}
    (\ket{\uparrow},\ket{\downarrow},\ket{\ })=\text{diag}(q^{-\frac{1}{6}},q^{\frac{1}{6}},1)(v_1,v_2,1) \,,
\end{align}
and finally on vertical links we have
\begin{align}
    (\ket{\uparrow},\ket{\downarrow},\ket{\ })=(v_1,v_2,1) \,.
\end{align}
\end{subequations}

It can be showed that the local transfer matrices $t$ and $t'$ are intertwiners with respect to the above action of $U_{-q}(\mathfrak{sl}_2)$. Remark also that the seam line operators (also called twist operators) are given by the action of an element belonging to the Cartan subalgebra
\begin{align}
\label{looppivot}
    S_{\text{loop}}=q^{2H_{\bm{\rho}}}=q^H \,,
\end{align}
where $\bm{\rho}$ is the Weyl vector of $\mathfrak{sl}_2$. As local transfer matrices are intertwiners, this means that the seam line can be deformed passing through nodes of $\mathbb{H}$.

There is a convenient way to write the local transfer matrices in terms of diagrams where each diagram represents a particular intertwiner%
\footnote{This comes from the fact that these diagrams are morphisms in the Temperley-Lieb category which is equivalent as a pivotal category to a subcategory of the category of representations of $U_{-q}(\mathfrak{sl}_2)$.}:
\begin{subequations}\begin{align}
    t^{\text{loop}}_{(1)}&=x\vcenter{\hbox{\includegraphics[scale=0.15]{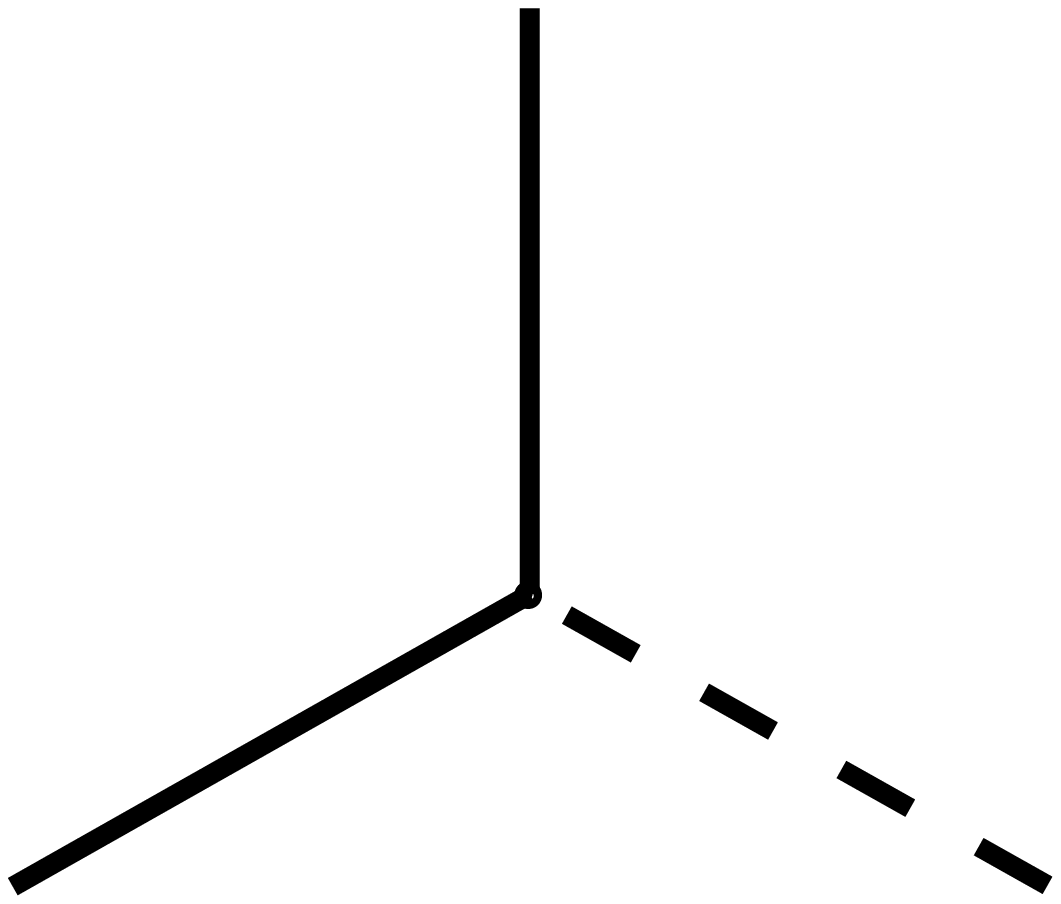}}}+x\vcenter{\hbox{\includegraphics[scale=0.15]{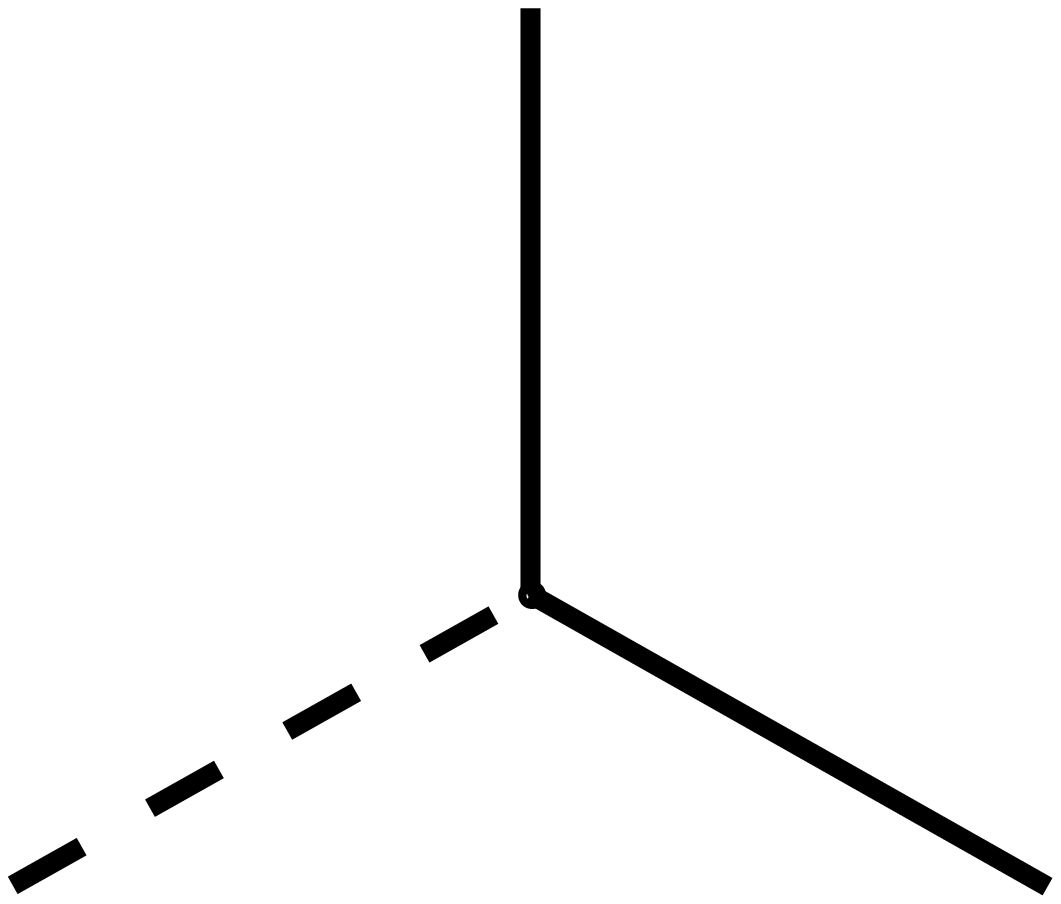}}}+x\vcenter{\hbox{\includegraphics[scale=0.15]{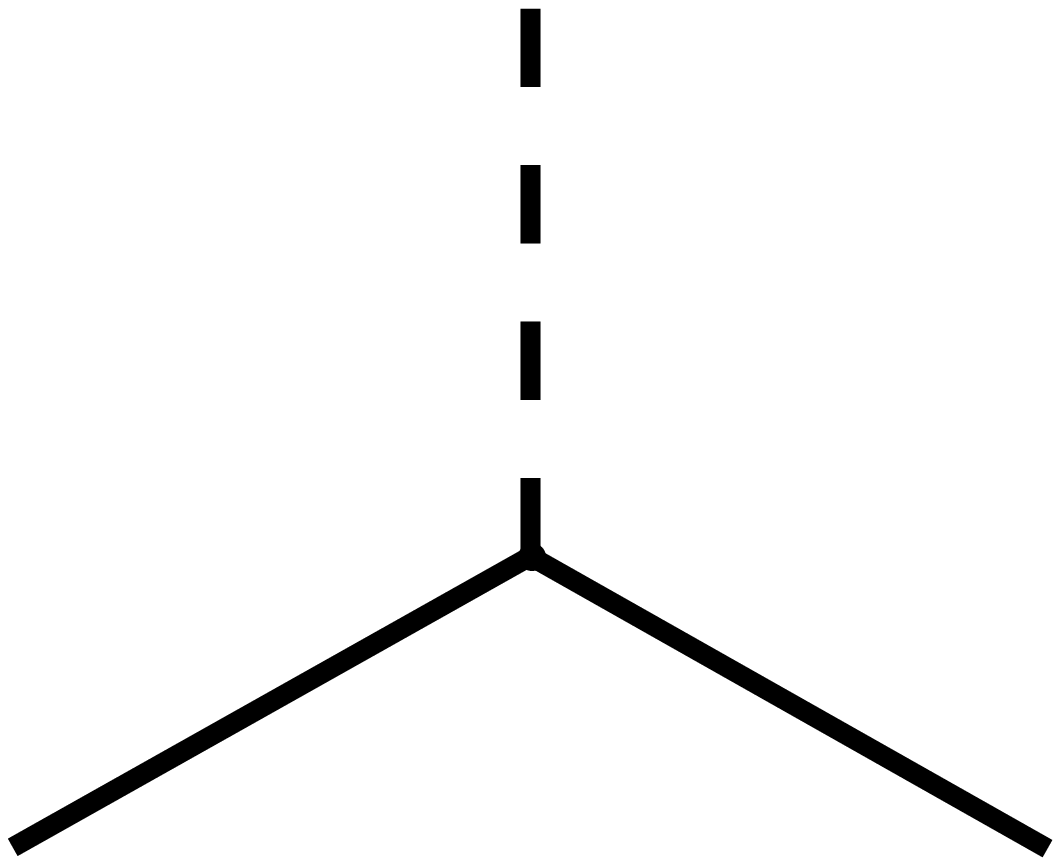}}}+\vcenter{\hbox{\includegraphics[scale=0.15]{diagrams/loopvertex3.eps}}}\label{transfermatrixloop1}\\[10pt]
    t^{\text{loop}}_{(2)}&=x\vcenter{\hbox{\includegraphics[scale=0.15]{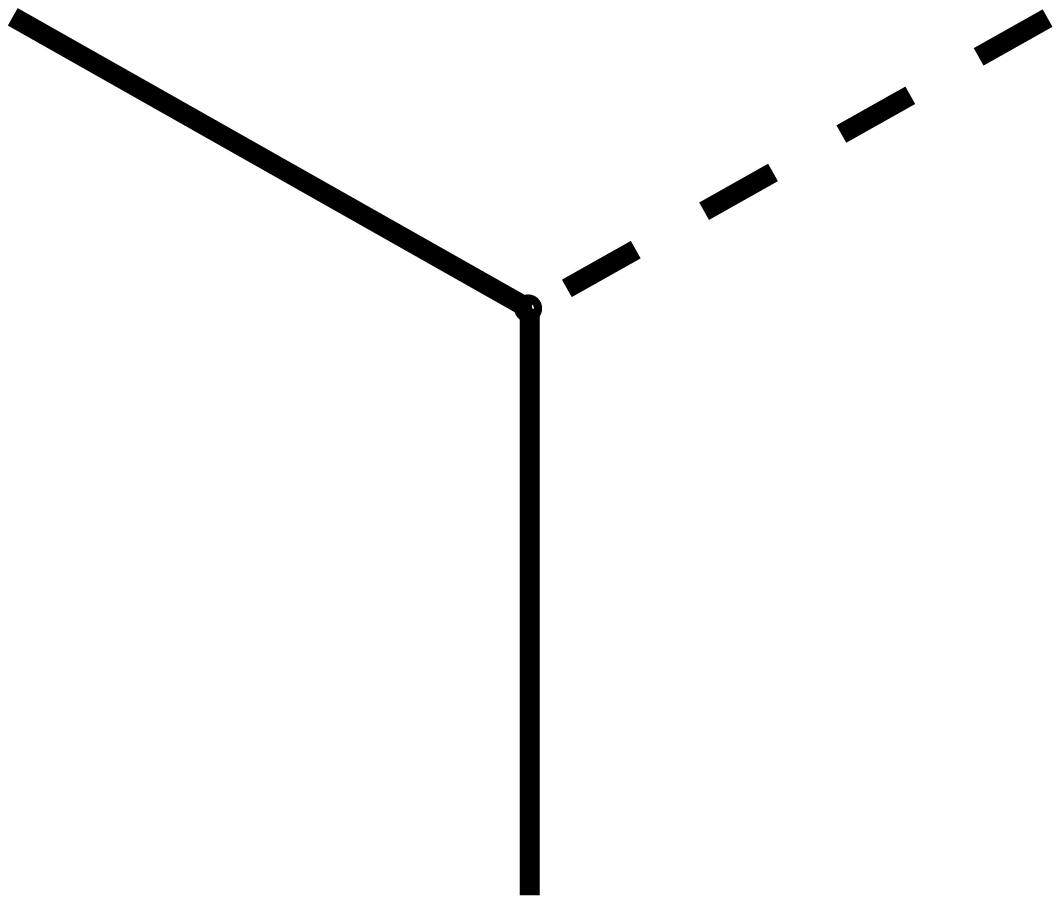}}}+x\vcenter{\hbox{\includegraphics[scale=0.15]{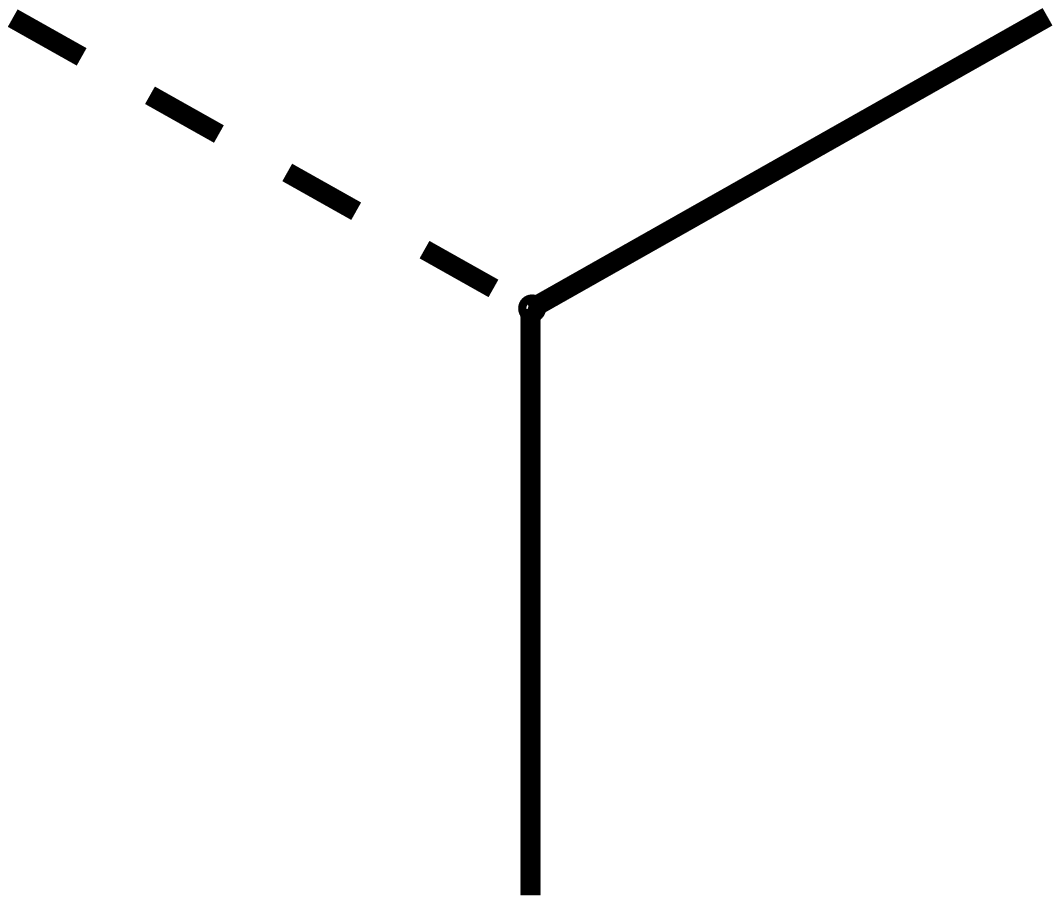}}}+x\vcenter{\hbox{\includegraphics[scale=0.15]{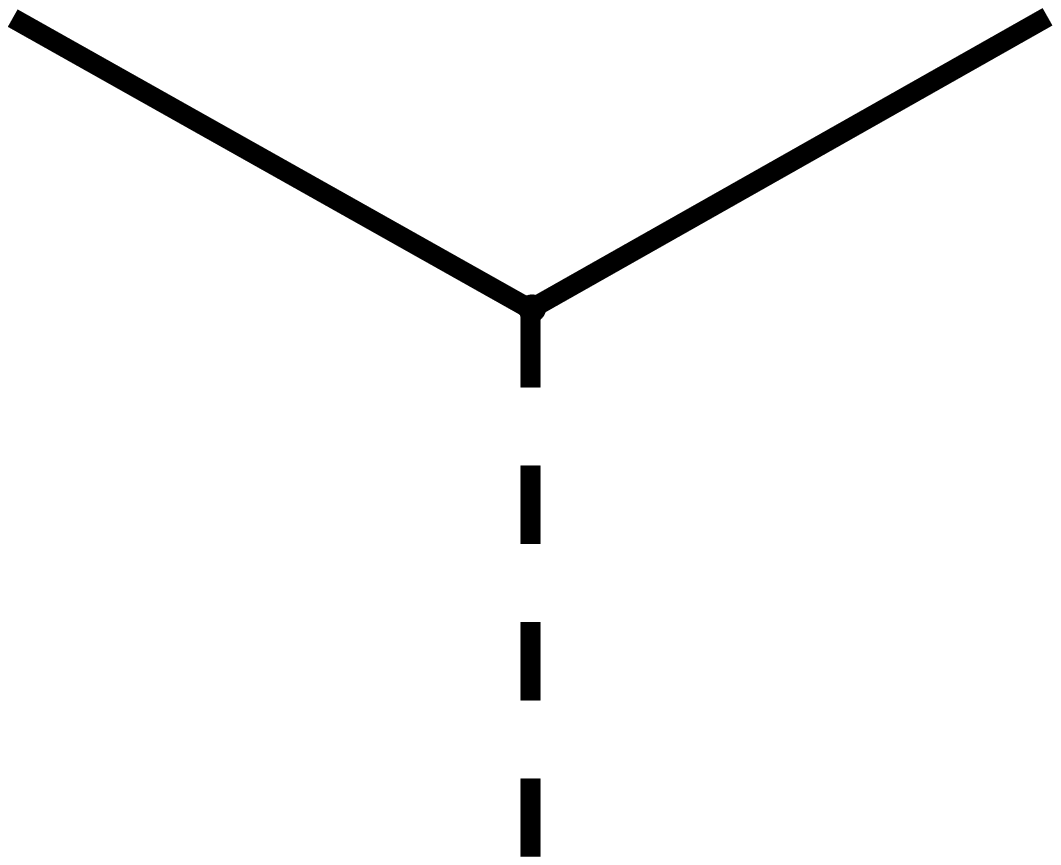}}}+\vcenter{\hbox{\includegraphics[scale=0.15]{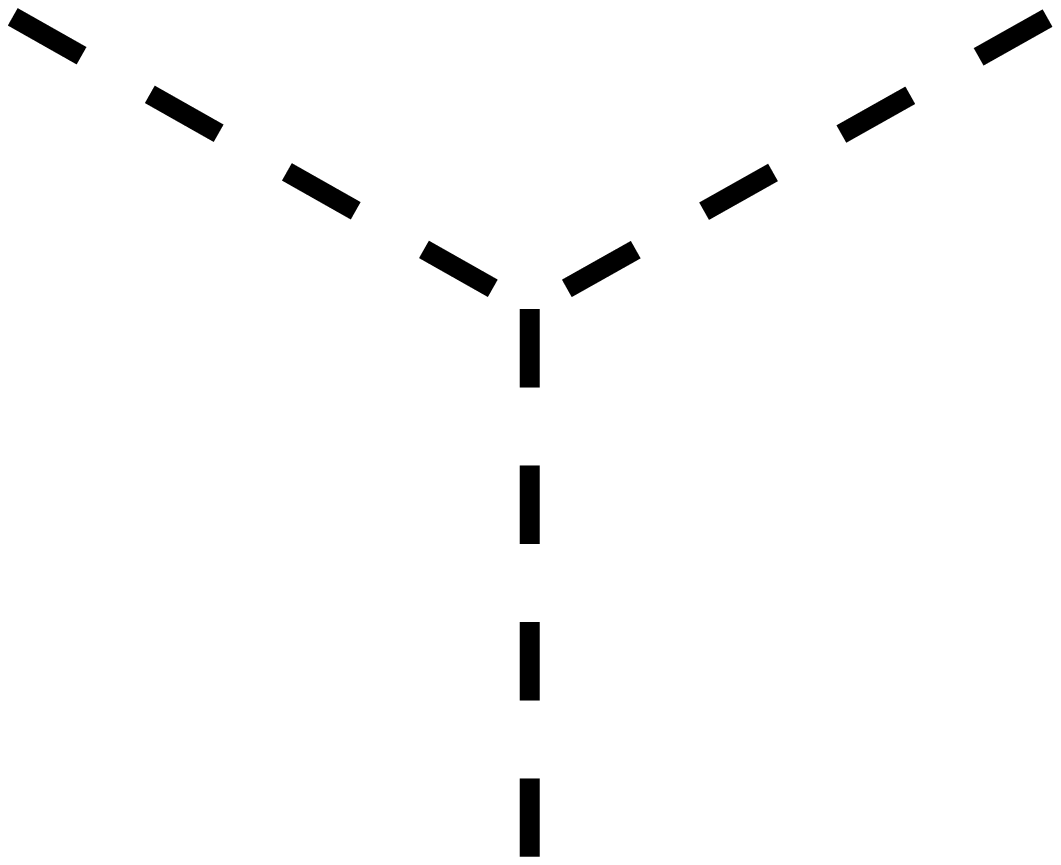}}}\label{transfermatrixloop2}
\end{align}
\label{transfermatrixloop}
\end{subequations}
Here, full lines represent the propagation of states living in $V$ whereas dashed lines represent the vacuum state living in $\mathbb{C}$. Diagrams are to be read from bottom to top. For instance, the first diagram in \eqref{transfermatrixloop1} represent the isomorphism $V\otimes \mathbb{C}\cong V$ as $U_{-q}(\mathfrak{sl}_2)$ representations. The non-zero matrix elements of this isomorphism in the basis $\{v_1,v_2,1\}$ are
\begin{align}
    v_1\otimes 1 \mapsto v_1\,, \qquad v_2\otimes 1 \mapsto v_2 \,,
\end{align}
which in the basis $\{\ket{\uparrow},\ket{\downarrow},\ket{\ }\}$ give
\begin{align}
\label{looparrow1}
   \ket{\uparrow}\otimes \ket{\ } \mapsto q^{-\frac{1}{6}}\ket{\uparrow}\,, \qquad \ket{\downarrow}\otimes \ket{\ } \mapsto q^{\frac{1}{6}}\ket{\downarrow} \,.
\end{align}

As another example, the third diagram in \eqref{transfermatrixloop1} represents the projection onto the trivial representation appearing in the decomposition of the tensor product $V\otimes V$, whose non zero matrix elements in the basis $\{v_1,v_2,1\}$ are
\begin{align}
    v_1\otimes v_2 \mapsto q^{\frac{1}{2}}\,, \qquad v_2\otimes v_1 \mapsto q^{-\frac{1}{2}} \,.
\end{align}
In the basis $\{\ket{\uparrow},\ket{\downarrow},\ket{\ }\}$ we thus have
\begin{align}
\label{looparrow2}
   \ket{\uparrow}\otimes \ket{\downarrow} \mapsto q^{\frac{1}{6}}\ket{\ }\,, \qquad \ket{\downarrow} \otimes \ket{\uparrow} \mapsto q^{-\frac{1}{6}}\ket{\ } \,.
\end{align}

We see from \eqref{looparrow1} and \eqref{looparrow2} that we indeed recover the corresponding matrix elements of $t^{\text{loop}}_{(1)}$ in the basis $\{\ket{\uparrow},\ket{\downarrow},\ket{\ }\}$ given by the vertex weights \eqref{loopvertexweights}.

The diagrams appearing in \eqref{transfermatrixloop} can be concatenated when their boundary edges agree. Such a concatenation represents a composition of the operators associated to the diagrams. For example, in the diagrammatic langage, $t^{\text{loop}}=t^{\text{loop}}_{(2)}t^{\text{loop}}_{(1)}$ reads
\begin{align}
    t^{\text{loop}}=&x^2\vcenter{\hbox{\includegraphics[scale=0.15]{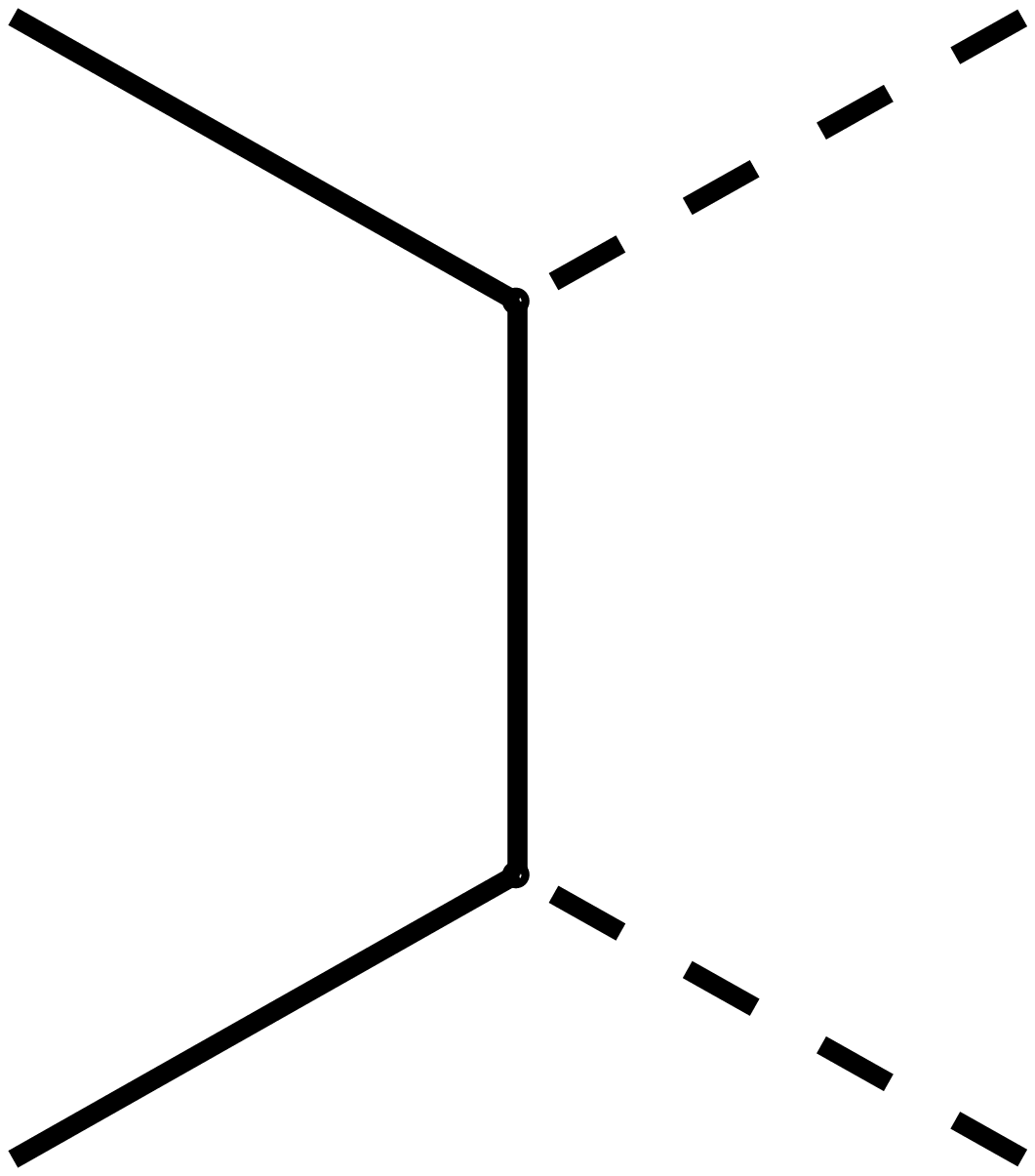}}}+x^2\vcenter{\hbox{\includegraphics[scale=0.15]{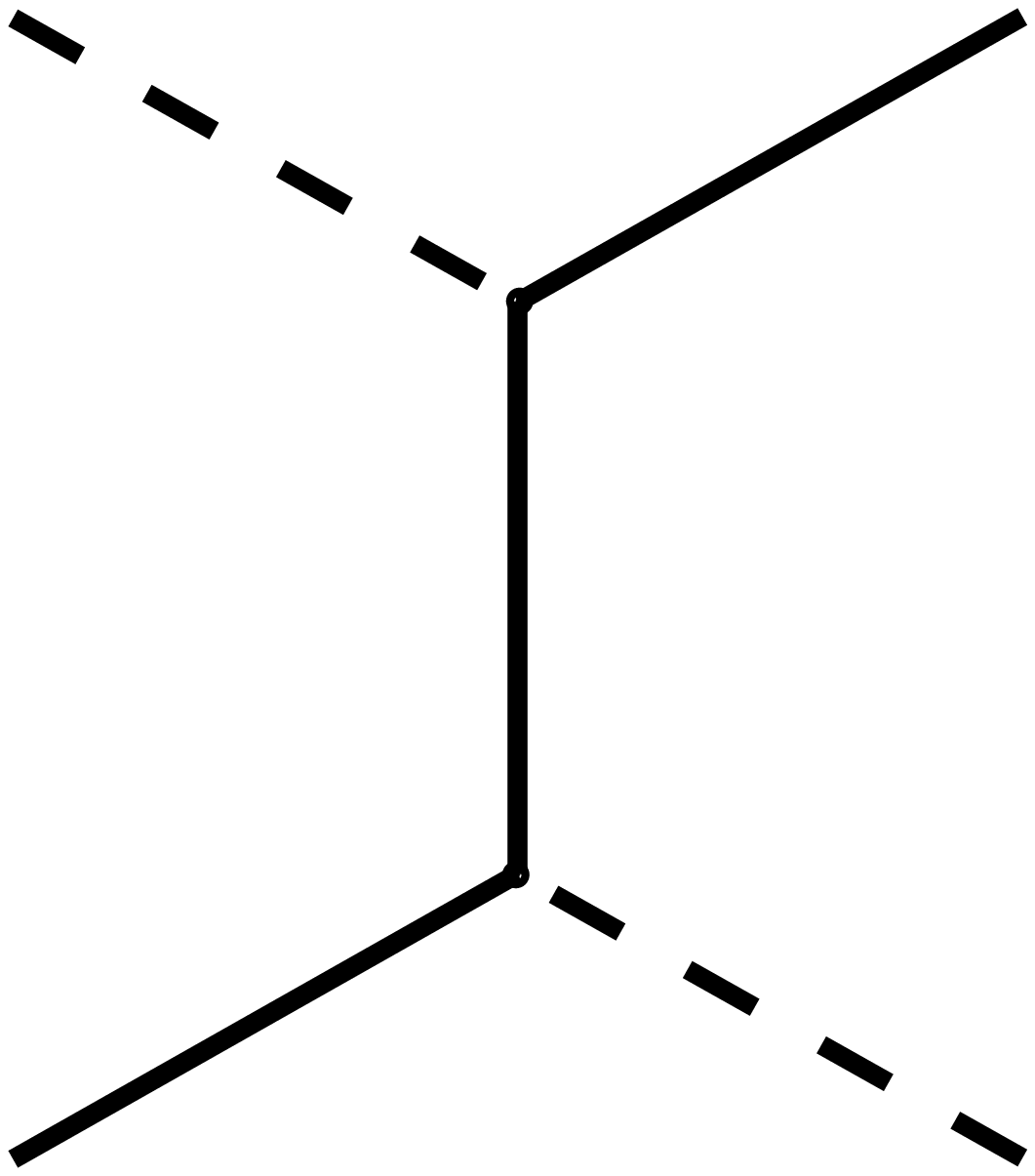}}}+x^2\vcenter{\hbox{\includegraphics[scale=0.15]{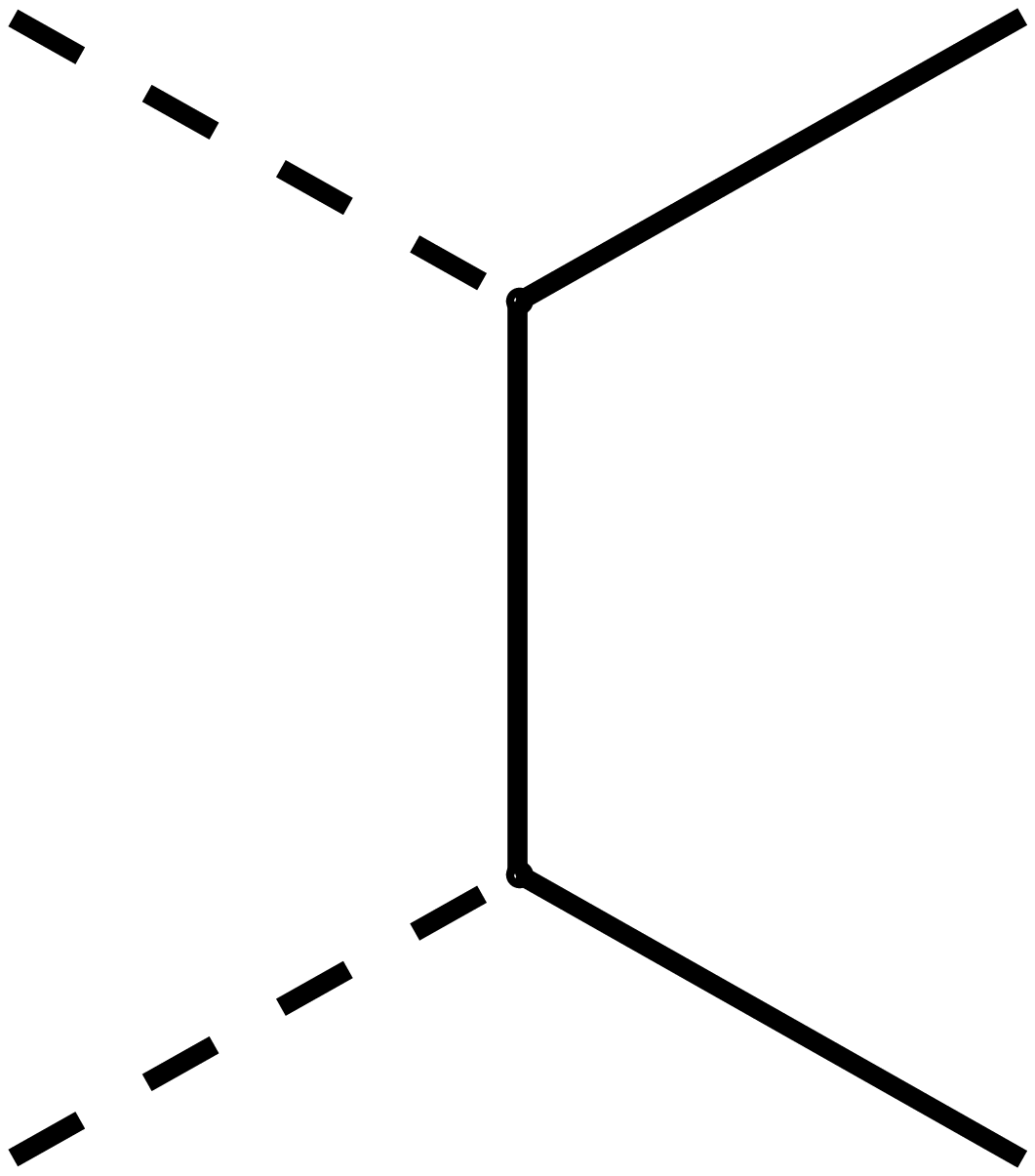}}}+x^2\vcenter{\hbox{\includegraphics[scale=0.15]{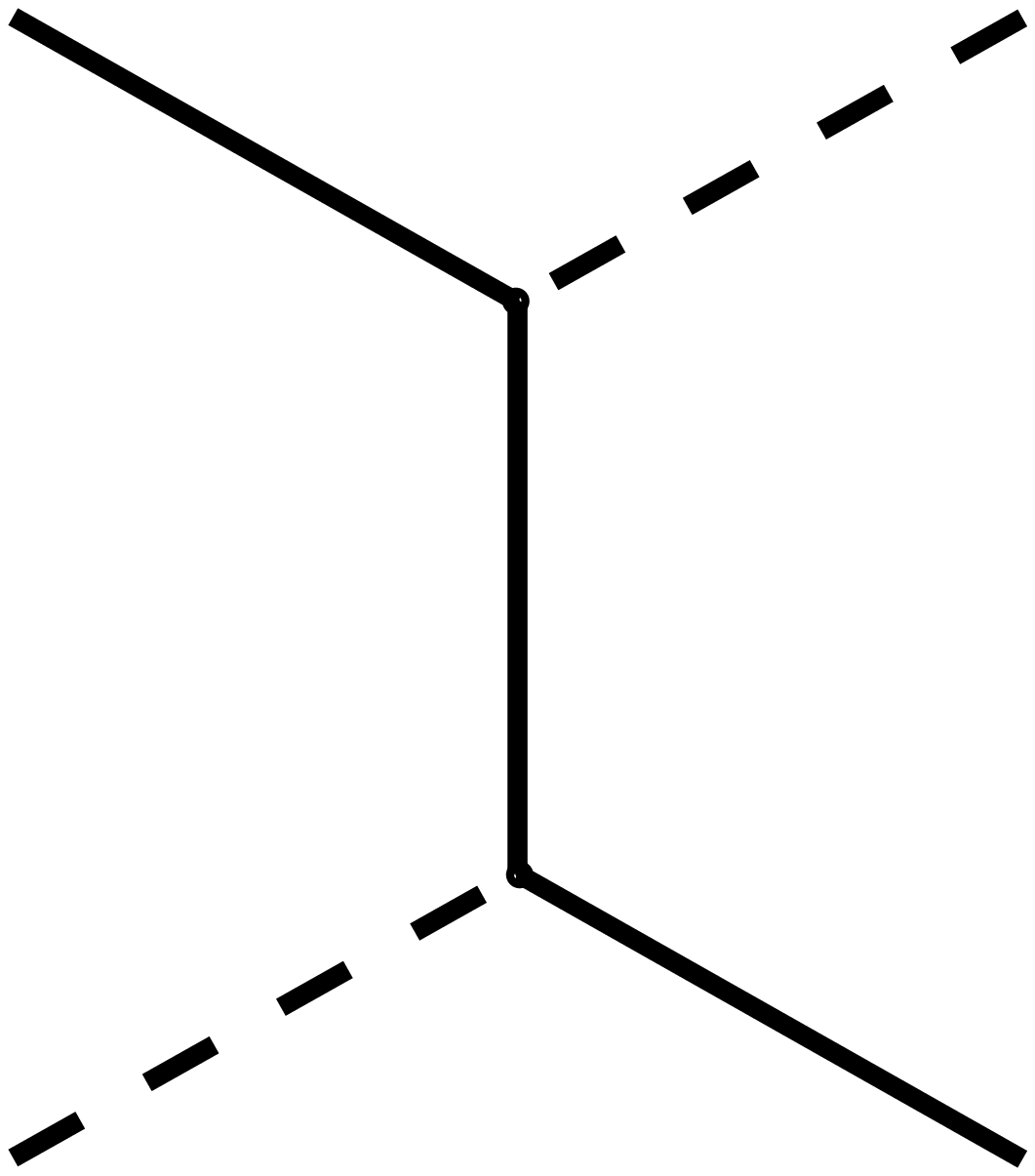}}}\nonumber\\[10pt]
    &+x^2\vcenter{\hbox{\includegraphics[scale=0.15]{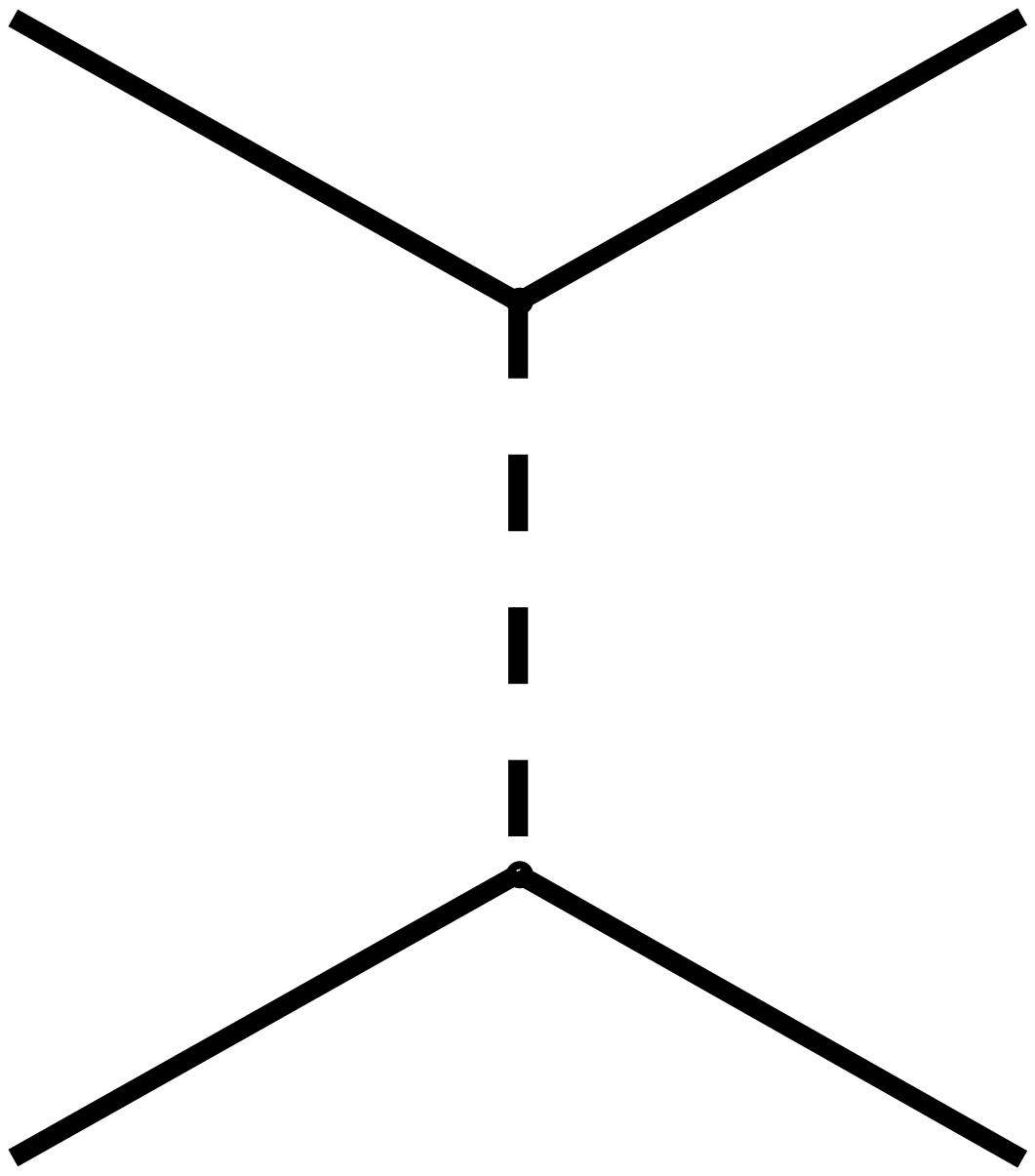}}}+x\vcenter{\hbox{\includegraphics[scale=0.15]{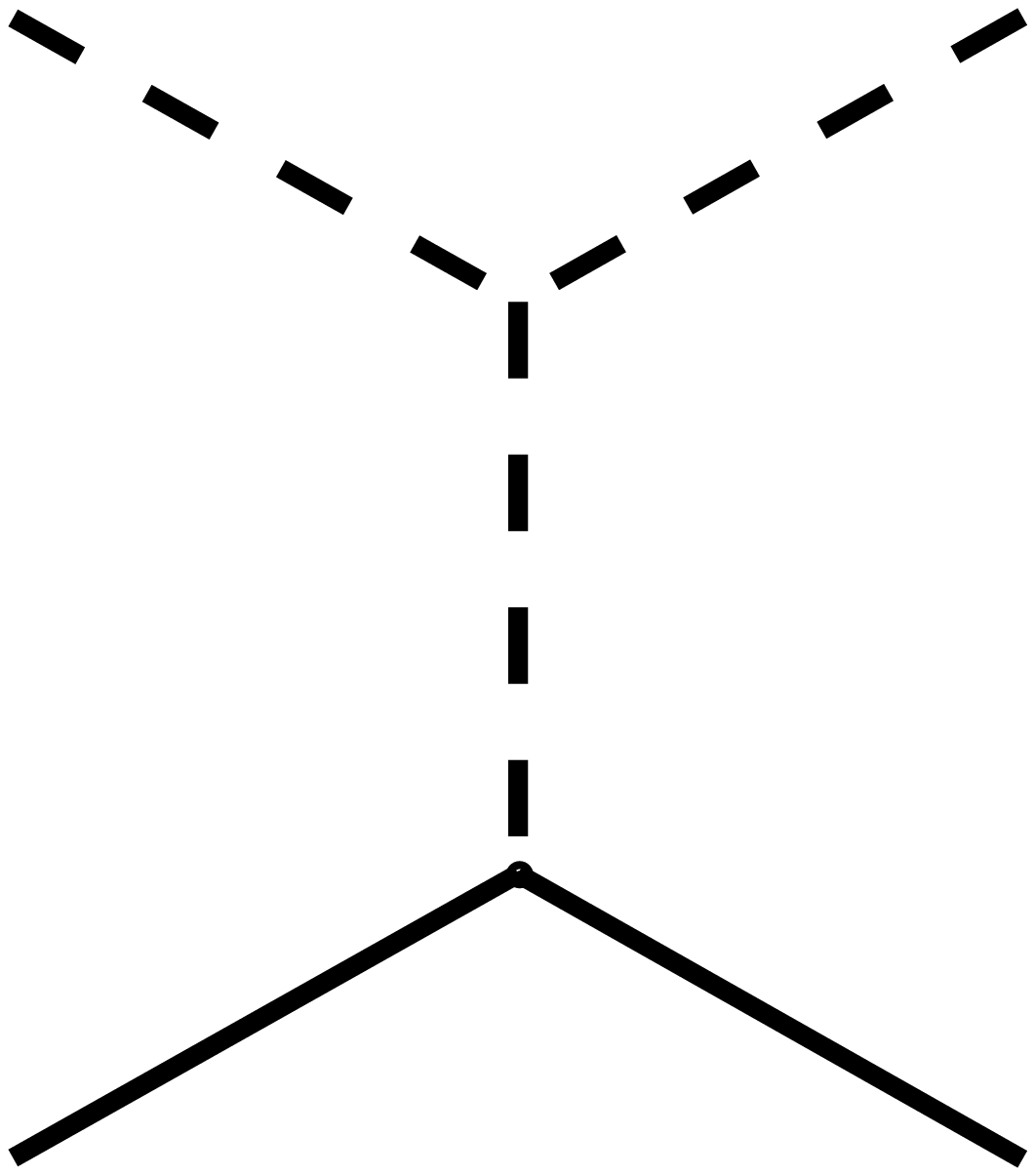}}}+x\vcenter{\hbox{\includegraphics[scale=0.15]{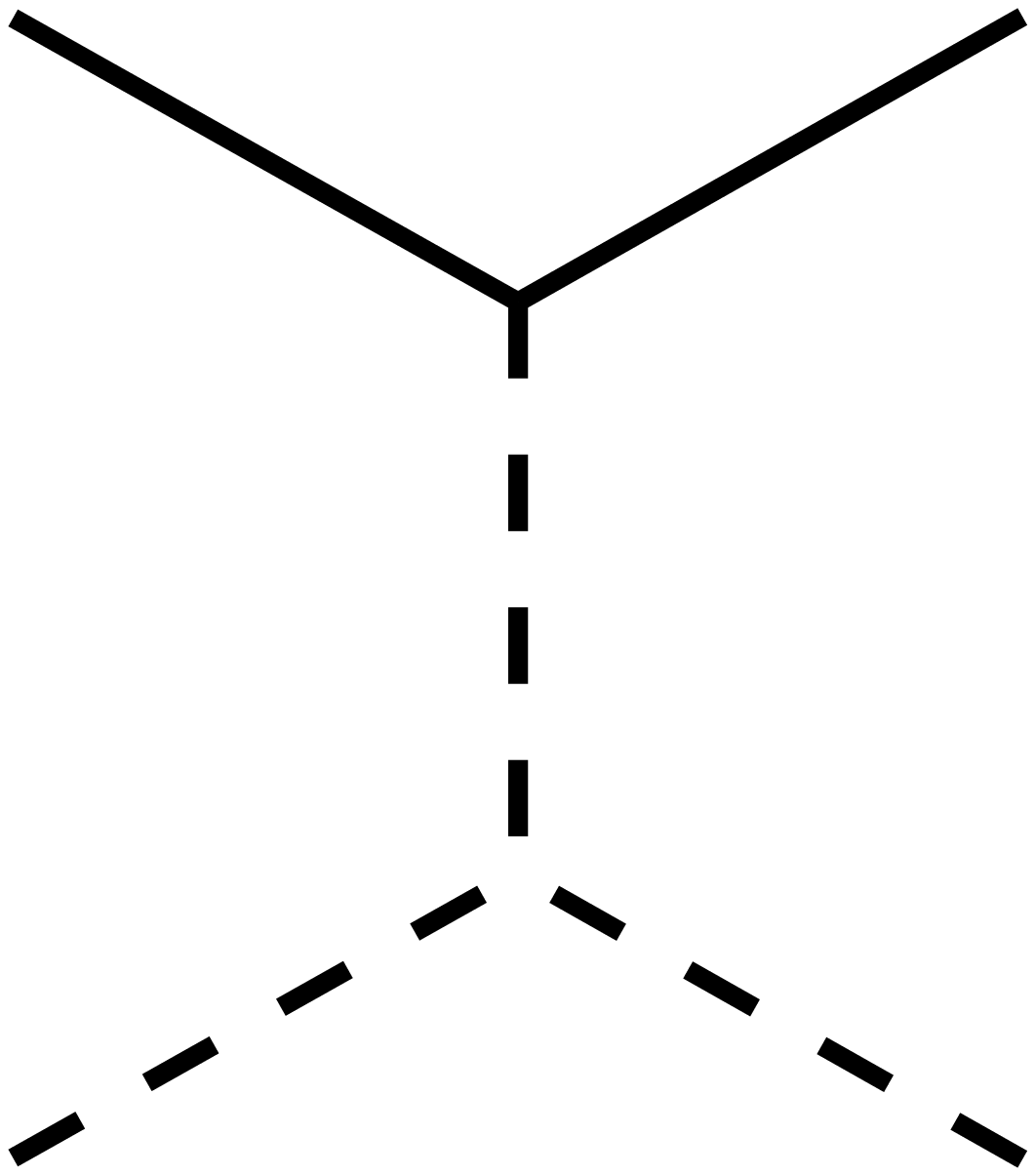}}}+\vcenter{\hbox{\includegraphics[scale=0.15]{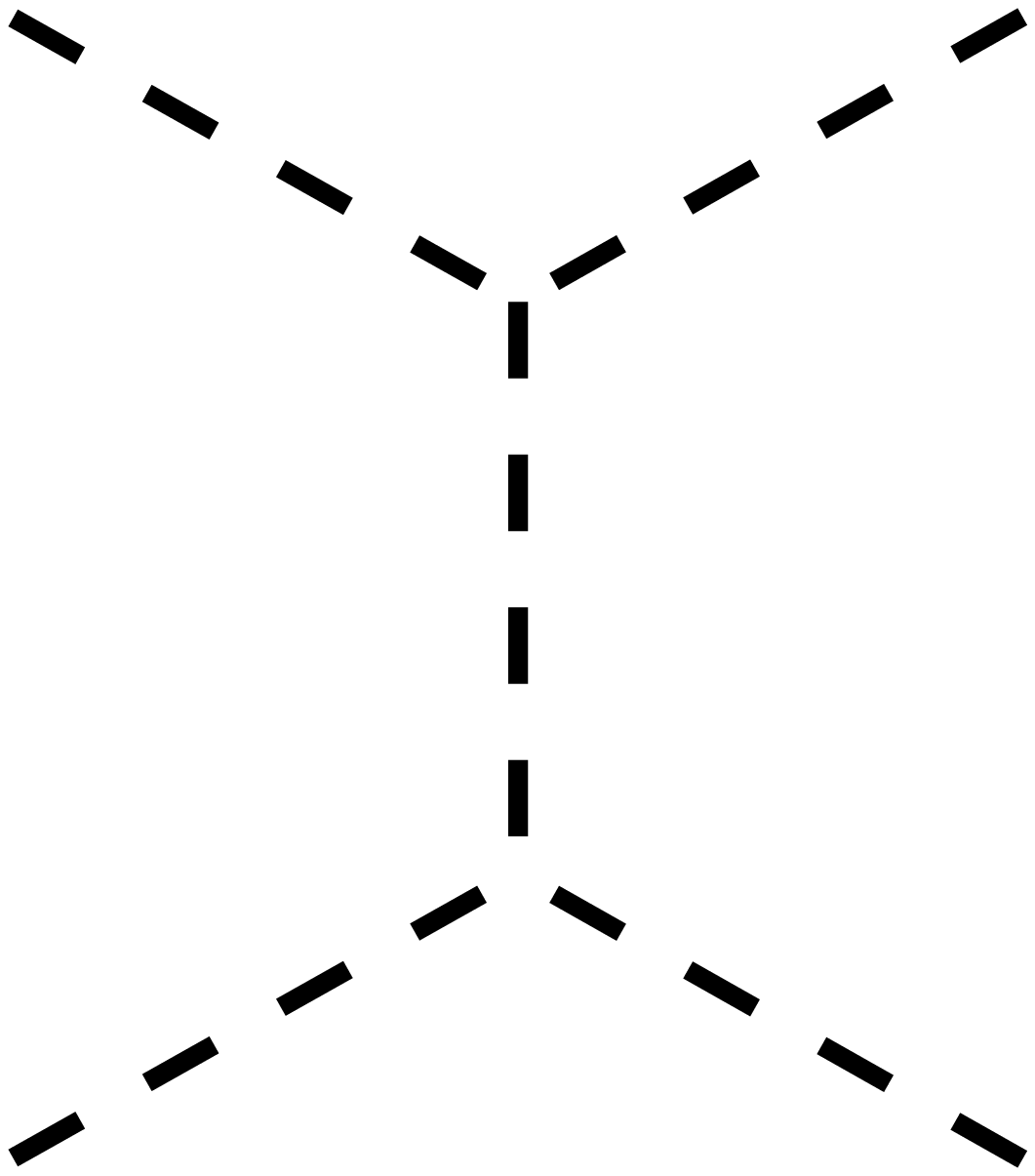}}}
\end{align}
We recognise here elements of the dilute Temperley-Lieb algebra.

Note that, in the case of the strip geometry, the row-to-row transfer matrix is an intertwiner, whereas in the cylinder case this is generally not the case. However in this latter situation the row-to-row transfer matrix is still symmetric with respect to the action of the Cartan subalgebra.

\medskip

The same story goes for the Kuperberg web model. The local transfer matrices are symmetric with respect to an action of $U_{-q}(\mathfrak{sl}_3)$. Let $V_1$ be the first fundamental representation of $U_{-q}(\mathfrak{sl}_3)$ and $\{v_1,v_2,v_3\}$ be its basis such that the action of the generators reads
\begin{align}
\begin{split}
    (-q)^{H_1}&=\begin{pmatrix} -q & 0 & 0\\
    0 & -q^{-1} & 0\\
    0 & 0 & 1
    \end{pmatrix}\,,\qquad 
    E_1=\begin{pmatrix} 0 & 1 & 0\\
    0 & 0 & 0 \\
    0 & 0 & 0
    \end{pmatrix}\,,\qquad 
    F_1=\begin{pmatrix} 0 & 0 & 0\\
    1 & 0 & 0\\
    0 & 0 & 0
    \end{pmatrix}\,,    \\[10pt]
    (-q)^{H_2}&=\begin{pmatrix} 1 & 0 & 0\\
    0 & -q & 0\\
    0 & 0 & -q^{-1}
    \end{pmatrix}\,,\qquad 
    E_2=\begin{pmatrix} 0 & 0 & 0\\
    0 & 0 & 1 \\
    0 & 0 & 0
    \end{pmatrix}\,,\qquad 
    F_2=\begin{pmatrix} 0 & 0 & 0\\
    0 & 0 & 0\\
    0 & 1 & 0
    \end{pmatrix}  \,.
\end{split}
\end{align}
Let $\{w_1,w_2,w_3\}$ be the basis of $V_1^*$ dual to $\{v_1,v_2,v_3\}$, i.e. $w_i(v_j)=\delta_{ij}$. The action of the generators in this basis reads :
\begin{align}
\begin{split}
    (-q)^{H_1}&=\begin{pmatrix} -q^{-1} & 0 & 0\\
    0 & -q & 0\\
    0 & 0 & 1
    \end{pmatrix}\,,\qquad 
    E_1=\begin{pmatrix} 0 & 0 & 0\\
    q & 0 & 0 \\
    0 & 0 & 0
    \end{pmatrix}\,,\qquad 
    F_1=\begin{pmatrix} 0 & q^{-1} & 0\\
    0 & 0 & 0\\
    0 & 0 & 0
    \end{pmatrix}\,,    \\[10pt]
    (-q)^{H_2}&=\begin{pmatrix} 1 & 0 & 0\\
    0 & -q^{-1} & 0\\
    0 & 0 & -q
    \end{pmatrix}\,,\qquad 
    E_2=\begin{pmatrix} 0 & 0 & 0\\
    0 & 0 & 0 \\
    0 & q & 0
    \end{pmatrix}\,,\qquad 
    F_2=\begin{pmatrix} 0 & 0 & 0\\
    0 & 0 & q^{-1}\\
    0 & 0 & 0
    \end{pmatrix} \,.
\end{split}
\end{align}

Each local space of states $\mathcal{H}_{\rm K}$ carries an action of $U_{-q}(\mathfrak{sl}_3)$ as $\mathcal{H}_{\rm K}\cong V_1\oplus V_1^* \oplus \mathbb{C}$ where $\mathbb{C}$ denotes the trivial representation of $U_{-q}(\mathfrak{sl}_3)$. We define this action by relating the basis $\{\ket{\color{red} \uparrow},\ket{\color{blue} \uparrow},\ket{\color{green} \uparrow},\ket{\color{red} \downarrow},\ket{\color{blue} \downarrow},\ket{\color{green} \downarrow},\ket{\ }\}$ with the basis $\{v_1,v_2,v_3,w_1,w_2,w_3,1\}$ on each link. On links of inclination $\diagdown$ we have
\begin{align}
\label{reptocolor1}
    (\ket{\color{red} \uparrow},\ket{\color{blue} \uparrow},\ket{\color{green} \uparrow},\ket{\color{red} \downarrow},\ket{\color{blue} \downarrow},\ket{\color{green} \downarrow},\ket{\ })=\text{diag}(q^{\frac{1}{3}},1,q^{-\frac{1}{3}},q^{-\frac{4}{3}},1,q^{\frac{4}{3}},1)(v_1,v_2,v_3,w_1,w_2,w_3,1) \,,
\end{align}
while on links of inclination $\diagup$
\begin{align}
    (\ket{\color{red} \uparrow},\ket{\color{blue} \uparrow},\ket{\color{green} \uparrow},\ket{\color{red} \downarrow},\ket{\color{blue} \downarrow},\ket{\color{green} \downarrow},\ket{\ })=\text{diag}(q^{-\frac{1}{3}},1,q^{\frac{1}{3}},q^{-\frac{2}{3}},1,q^{\frac{2}{3}},1)(v_1,v_2,v_3,w_1,w_2,w_3,1) \,,
\end{align}
and finally on vertical links we find
\begin{align}
\label{reptocolor3}
    (\ket{\color{red} \uparrow},\ket{\color{blue} \uparrow},\ket{\color{green} \uparrow},\ket{\color{red} \downarrow},\ket{\color{blue} \downarrow},\ket{\color{green} \downarrow},\ket{\ })=\text{diag}(1,1,1,q^{-1},1,q,1)(v_1,v_2,v_3,w_1,w_2,w_3,1) \,.
\end{align}

The local transfer matrices can then be expressed in terms of diagrams, where each diagram represent a particular intertwiner: 
\begin{subequations}
\label{kuptransfermatrix}
\begin{align}
    t^{\rm K}_{(1)}=&zx_1x_2^{\frac{1}{2}}\vcenter{\hbox{\includegraphics[scale=0.2]{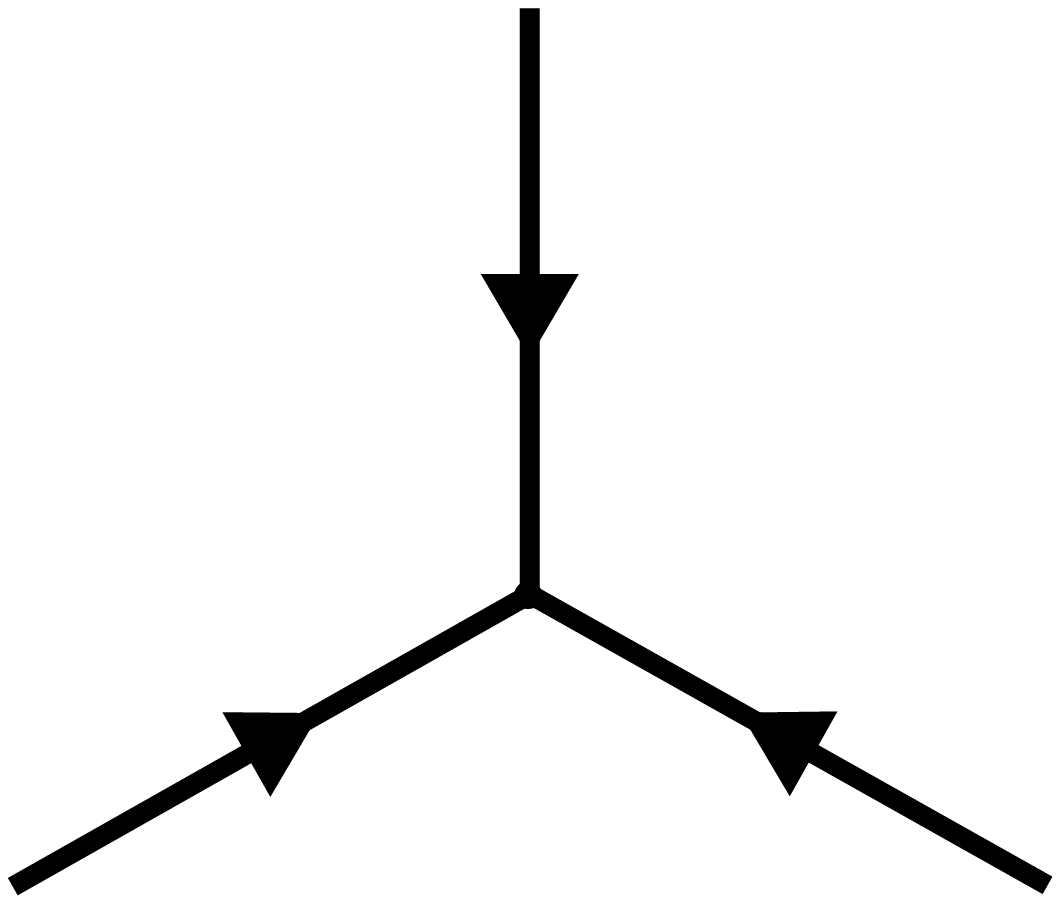}}}+ yx_1^{\frac{1}{2}}x_2 \vcenter{\hbox{\includegraphics[scale=0.2]{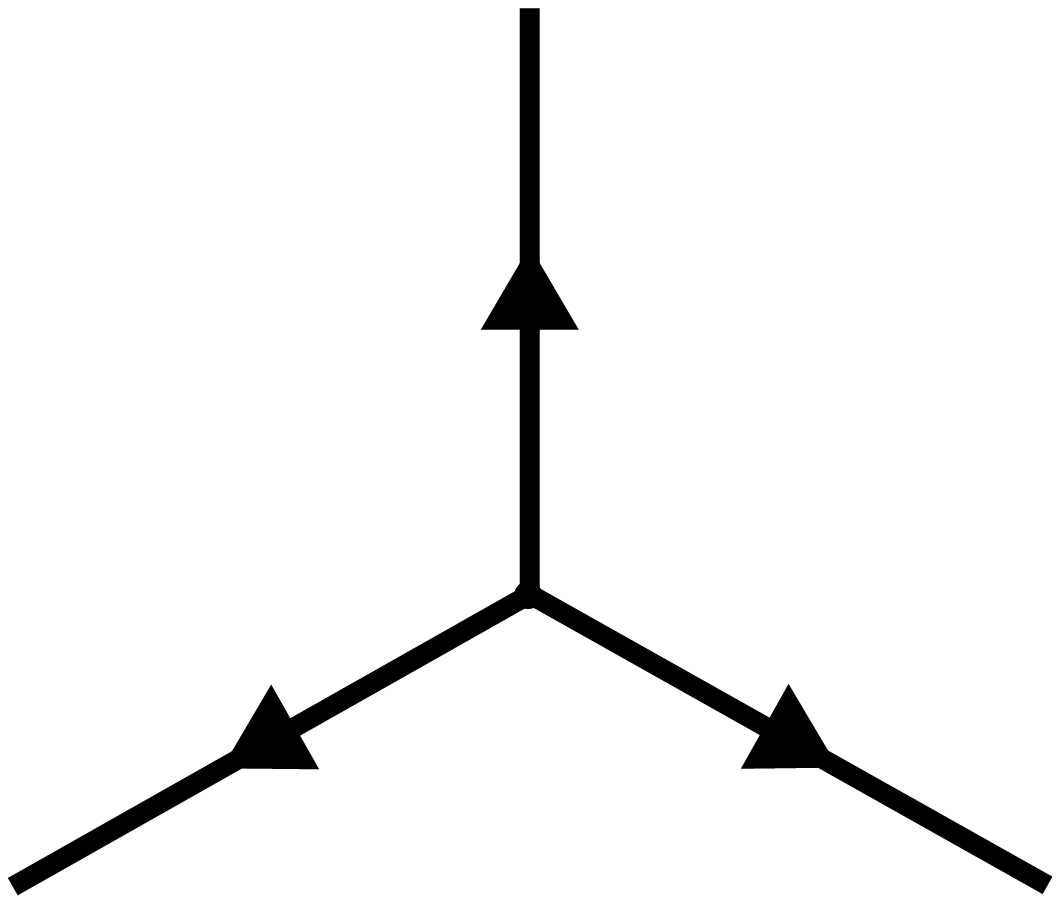}}}+ x_1\vcenter{\hbox{\includegraphics[scale=0.2]{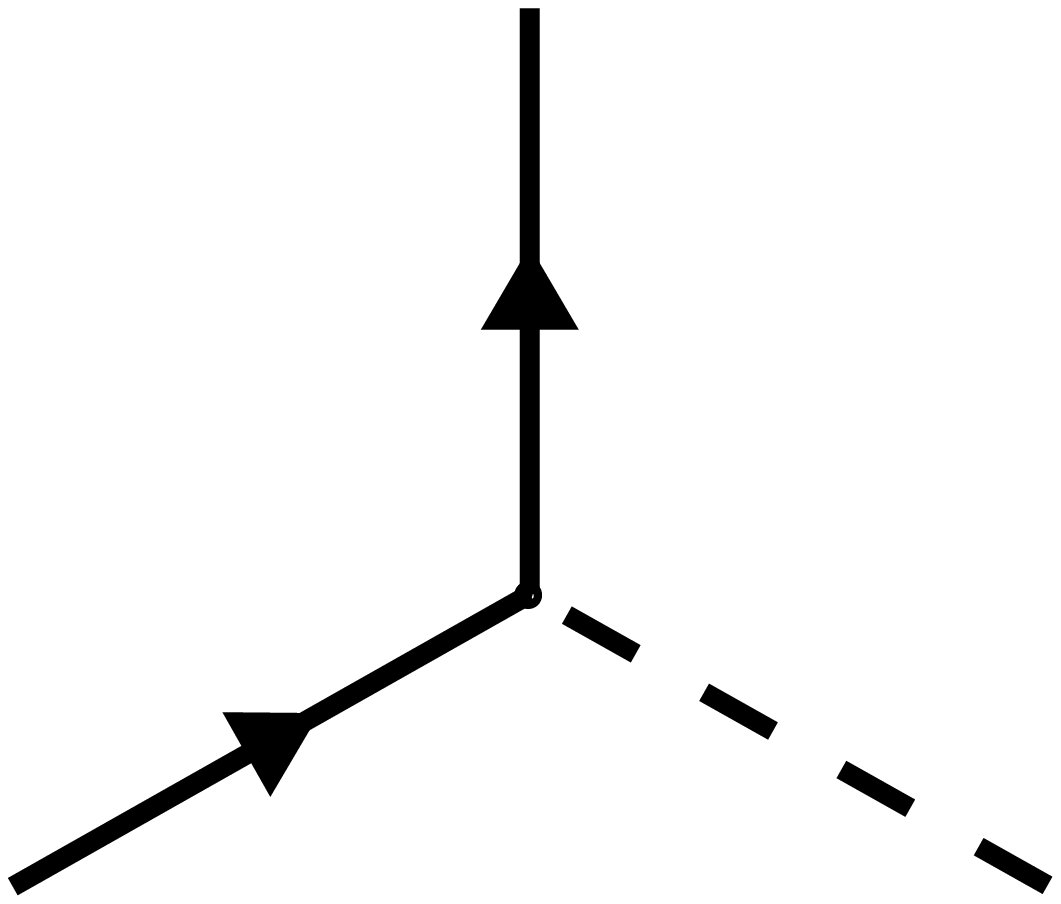}}}+ x_1\vcenter{\hbox{\includegraphics[scale=0.2]{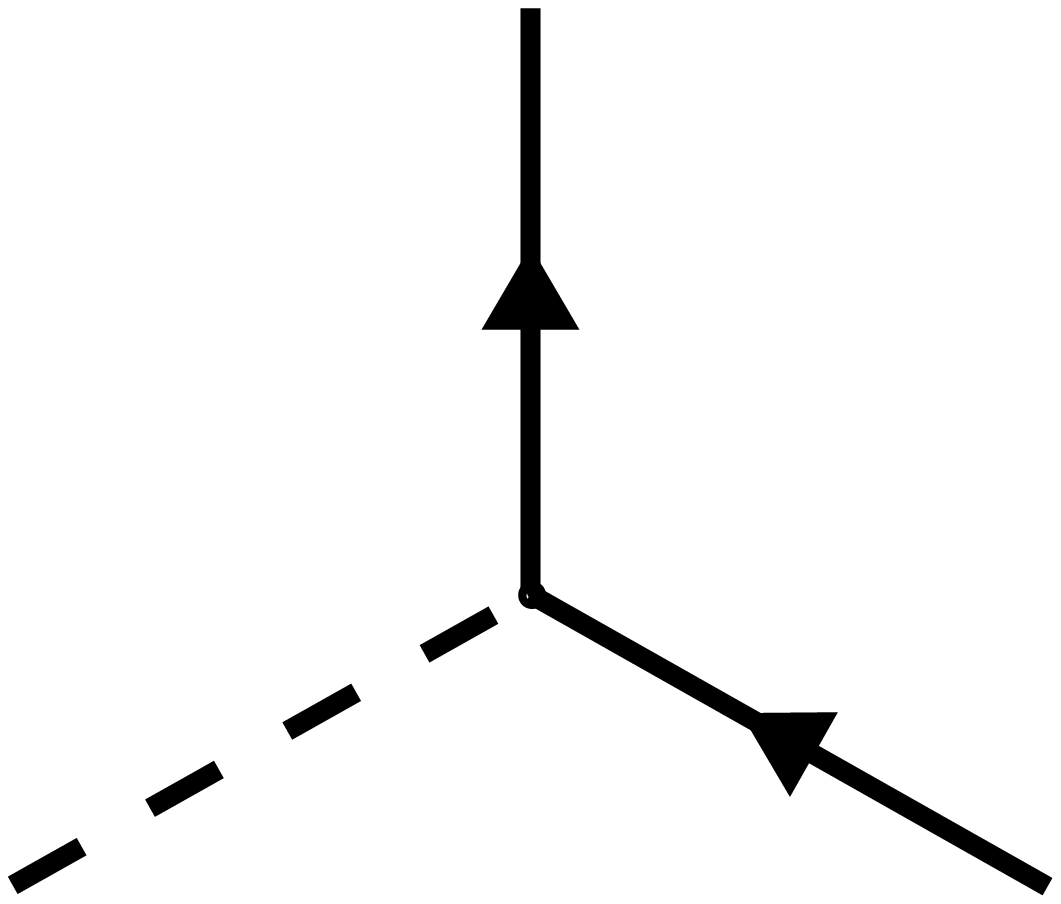}}}+ x_2\vcenter{\hbox{\includegraphics[scale=0.2]{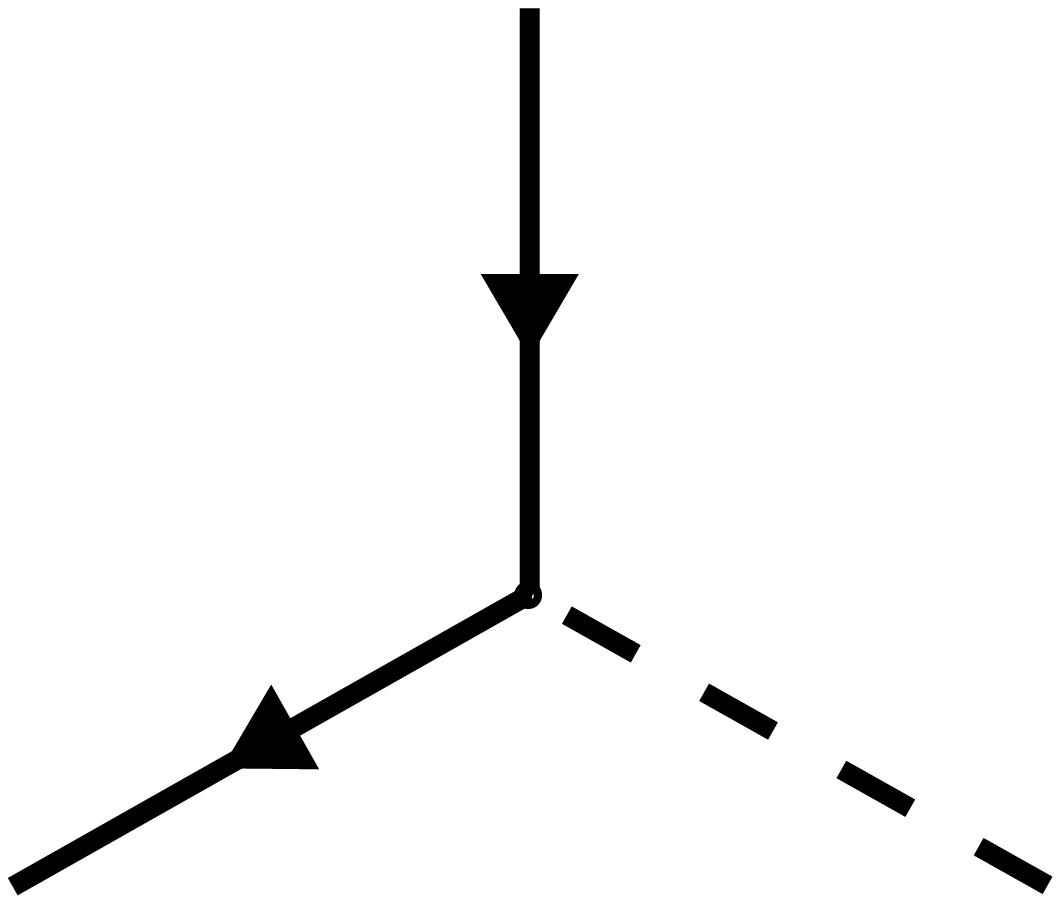}}}\nonumber\\ &+ x_2\vcenter{\hbox{\includegraphics[scale=0.2]{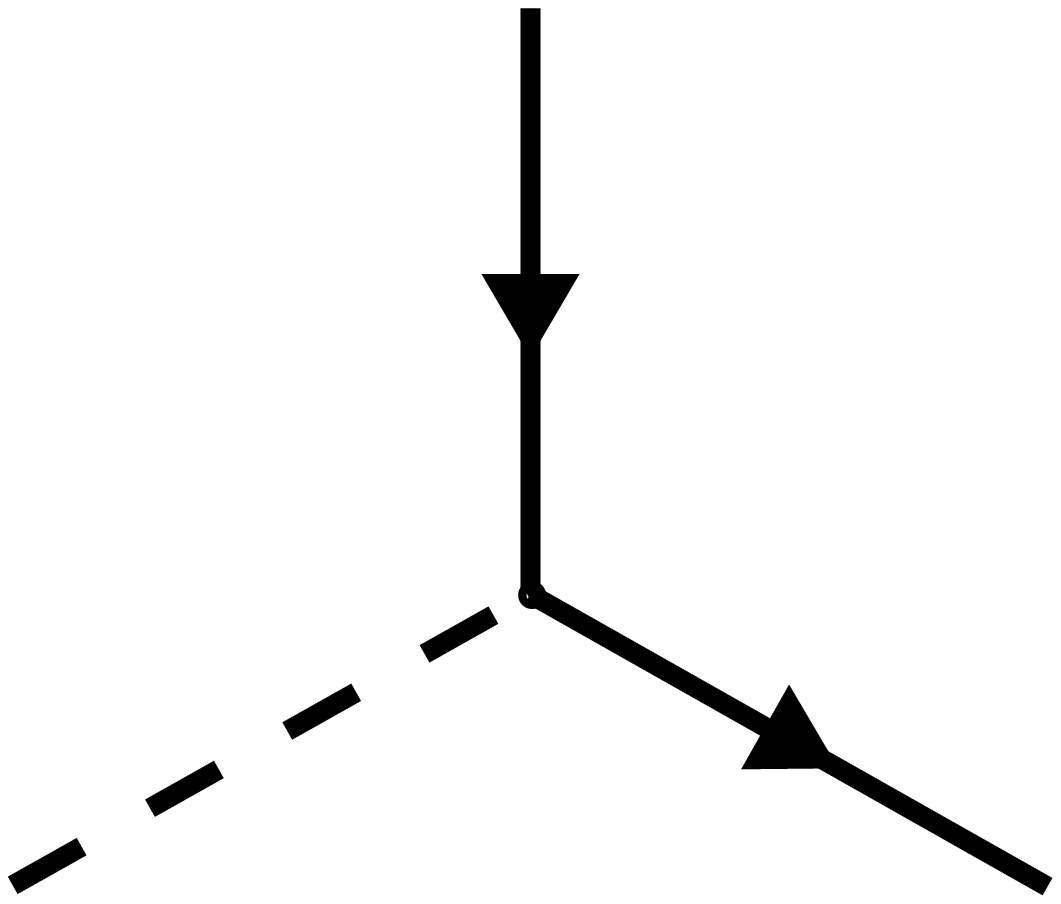}}}+ x_1^{\frac{1}{2}}x_2^{\frac{1}{2}}\vcenter{\hbox{\includegraphics[scale=0.2]{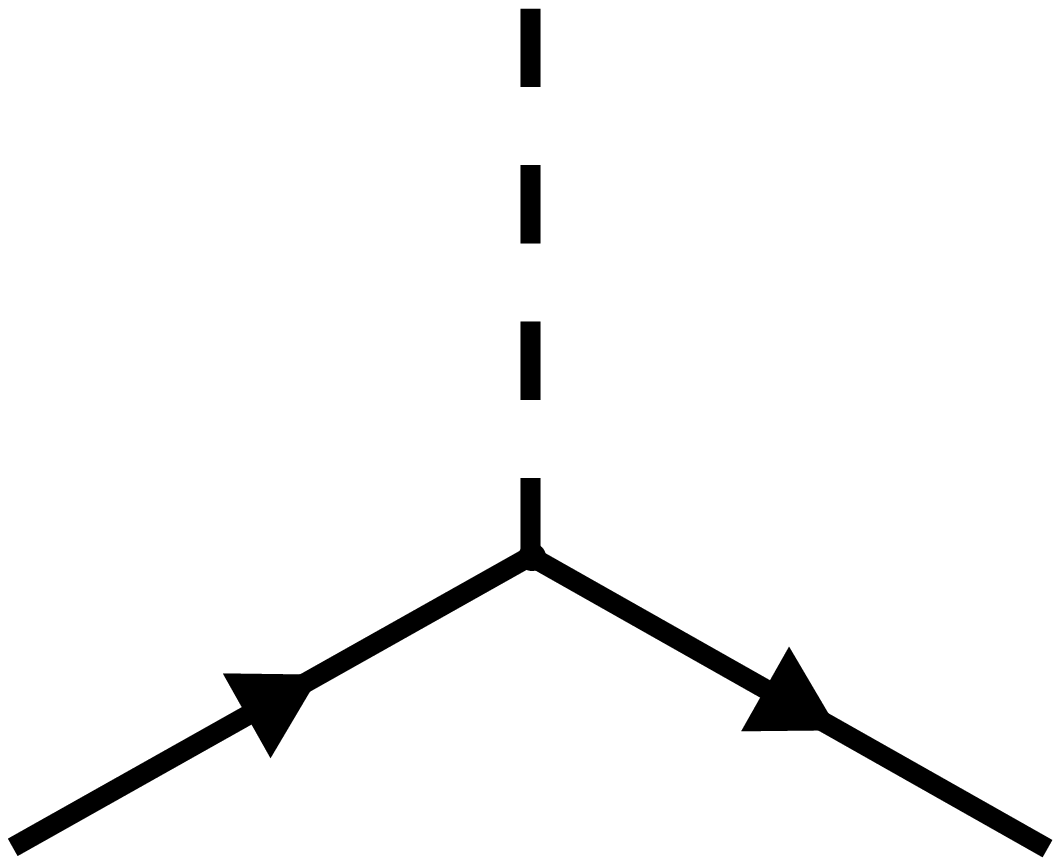}}}+ x_1^{\frac{1}{2}}x_2^{\frac{1}{2}}\vcenter{\hbox{\includegraphics[scale=0.2]{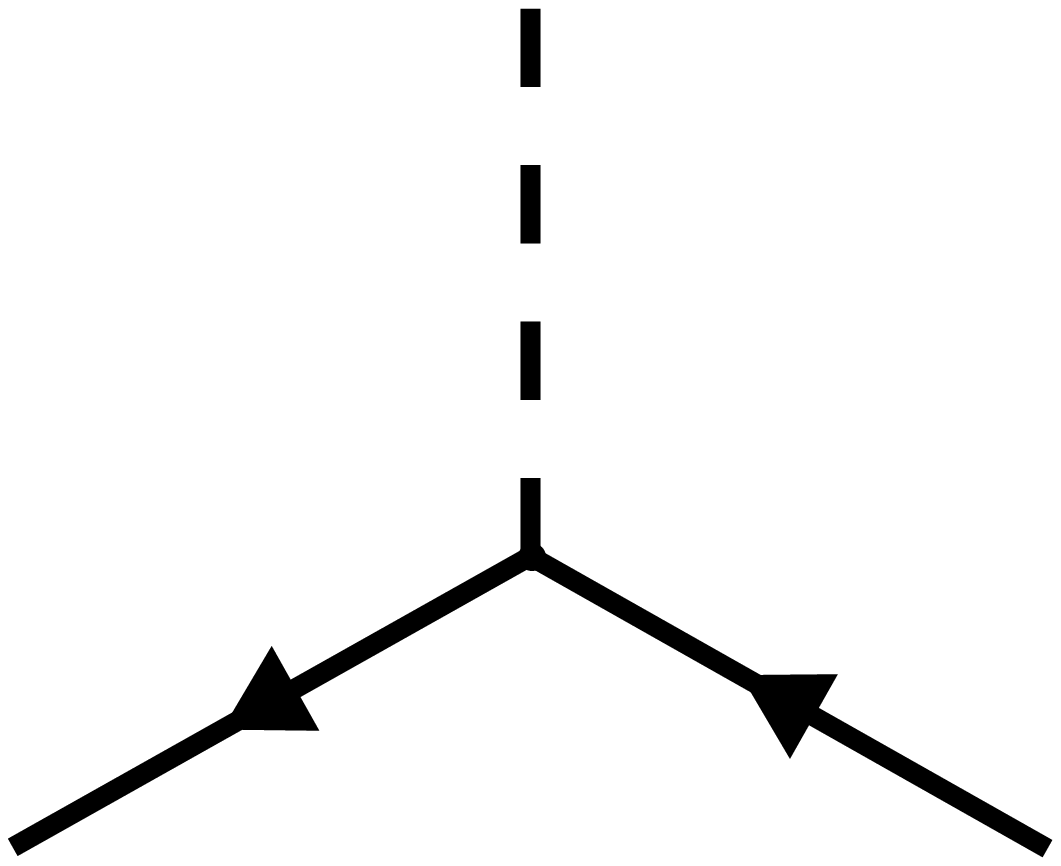}}}+ \vcenter{\hbox{\includegraphics[scale=0.2]{diagrams/loopvertex3.eps}}}\label{kuptransfermatrix1}\\[20pt]
    t^{\rm K}_{(2)}=&zx_1^{\frac{1}{2}}x_2\vcenter{\hbox{\includegraphics[scale=0.2]{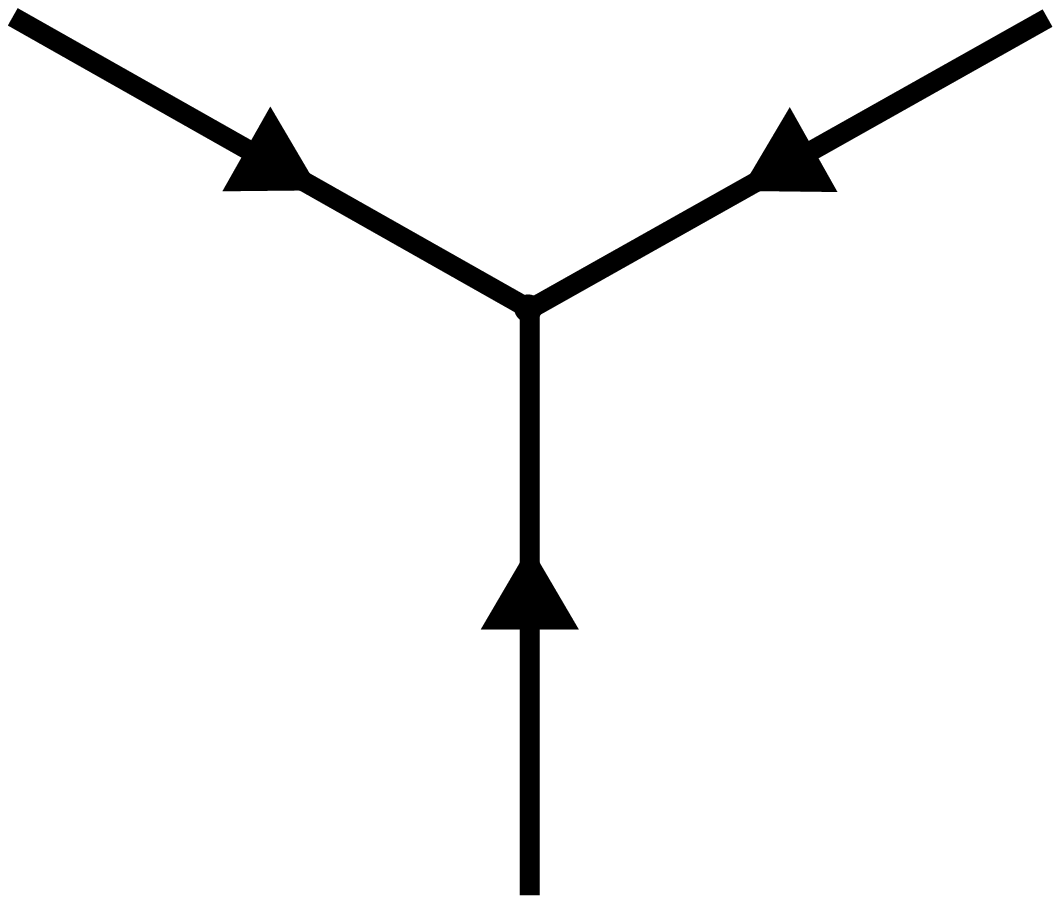}}}+ yx_1x_2^{\frac{1}{2}}\vcenter{\hbox{\includegraphics[scale=0.2]{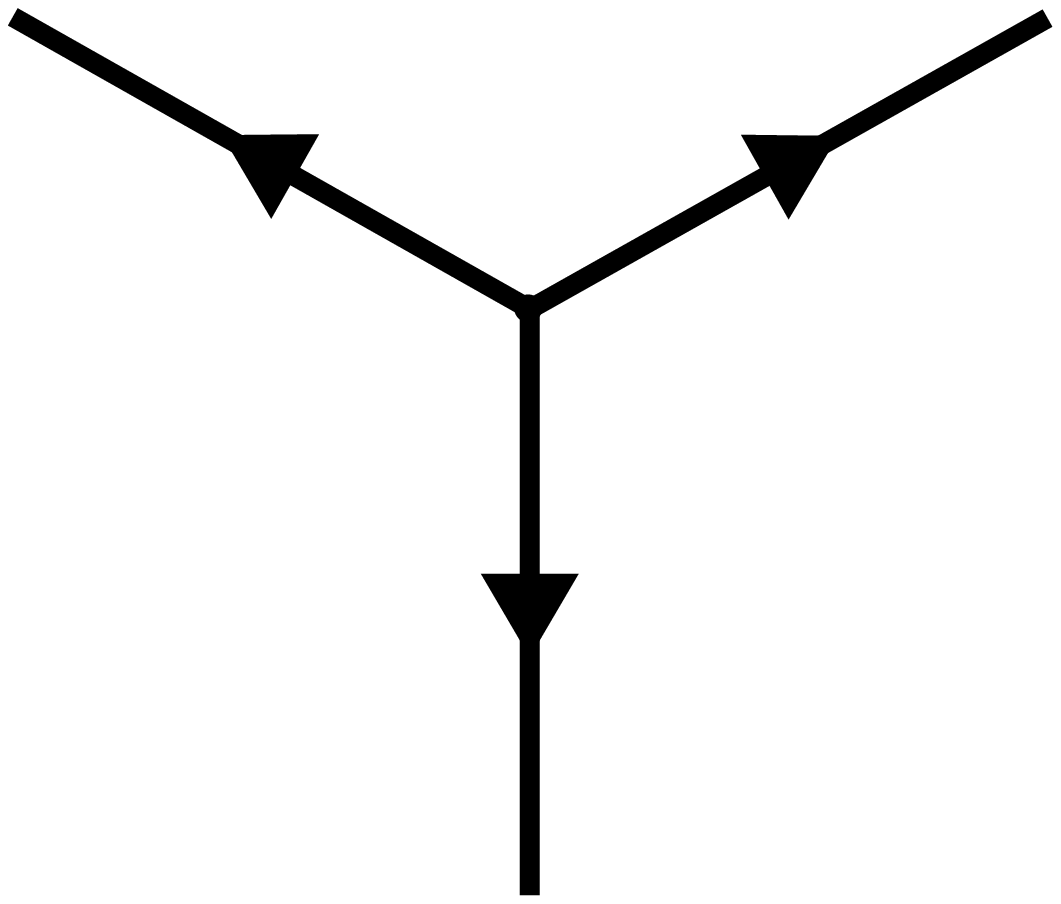}}}+ x_1\vcenter{\hbox{\includegraphics[scale=0.2]{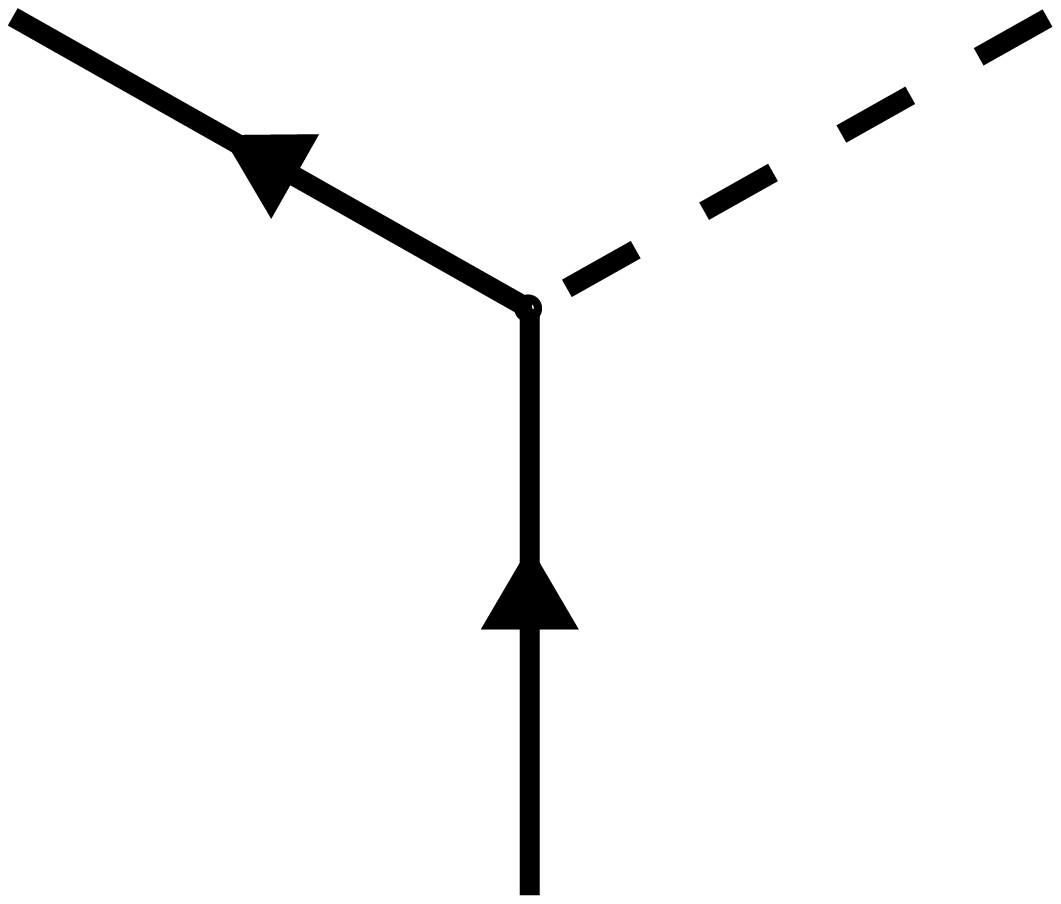}}}+ x_1\vcenter{\hbox{\includegraphics[scale=0.2]{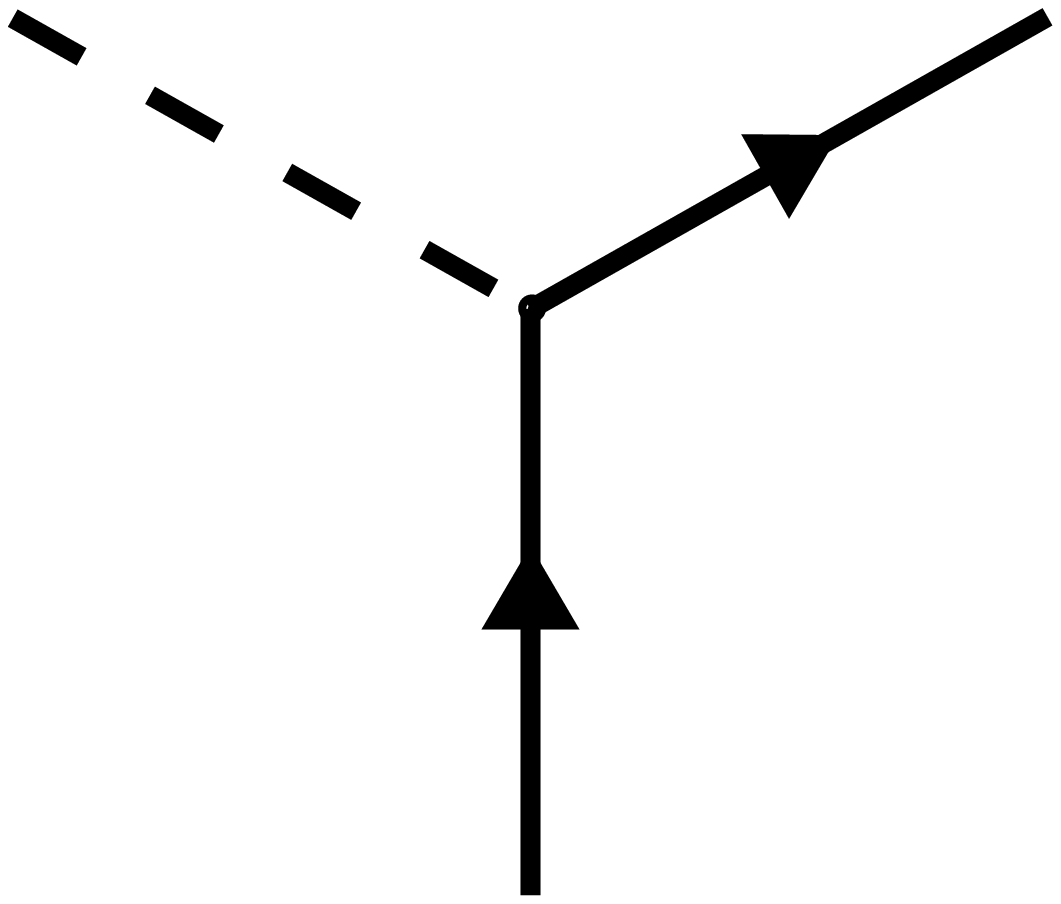}}}+ x_2\vcenter{\hbox{\includegraphics[scale=0.2]{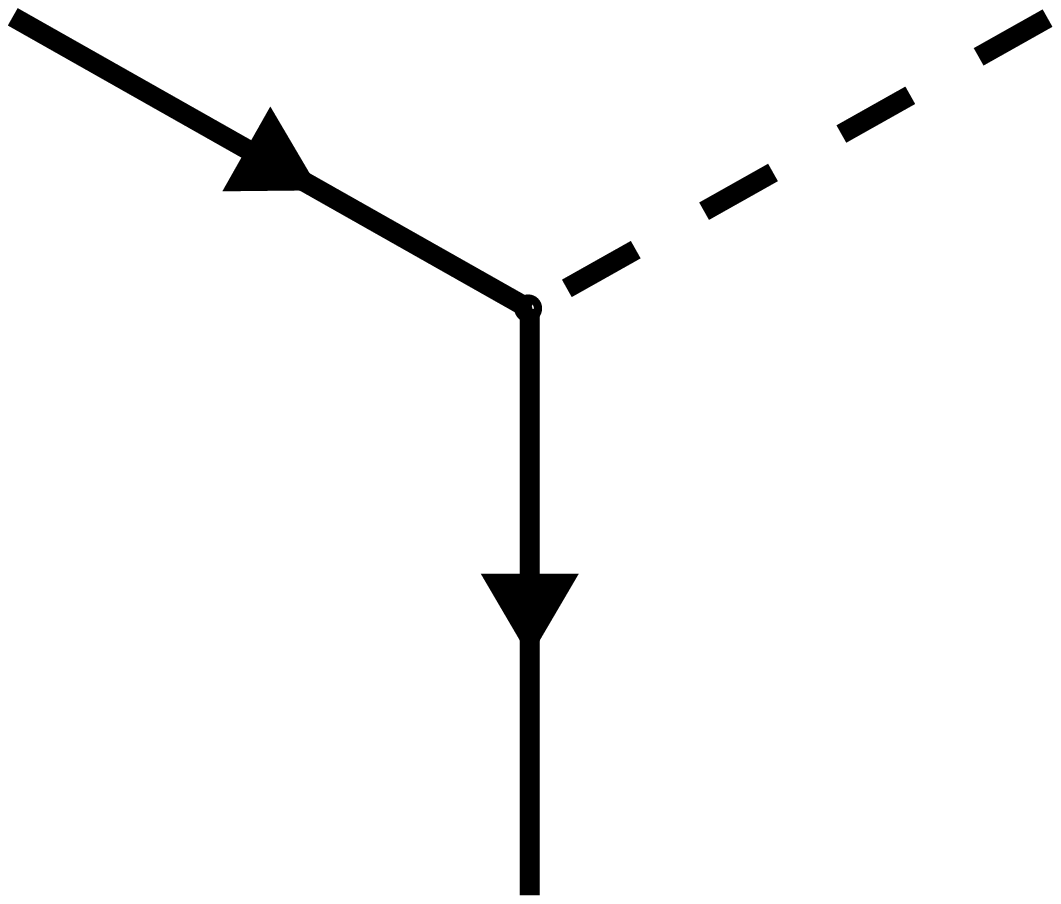}}}\nonumber\\ &+ x_2\vcenter{\hbox{\includegraphics[scale=0.2]{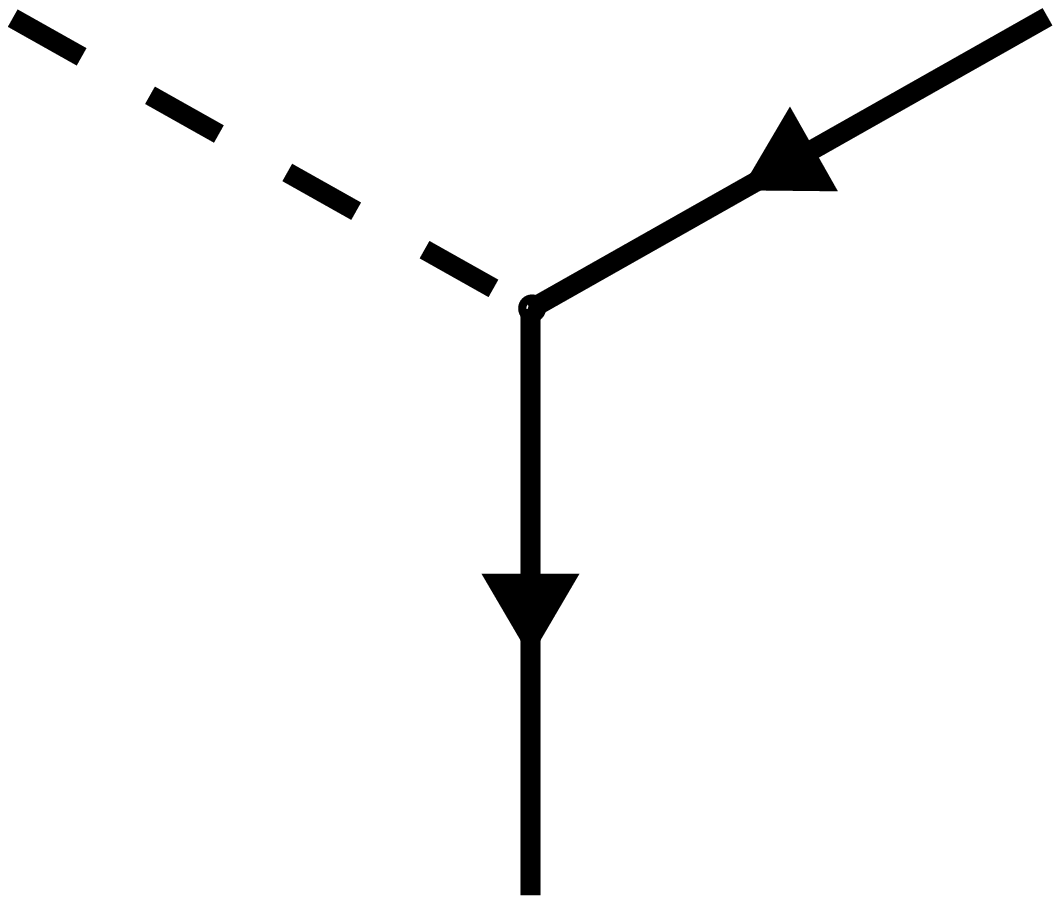}}}+ x_1^{\frac{1}{2}}x_2^{\frac{1}{2}}\vcenter{\hbox{\includegraphics[scale=0.2]{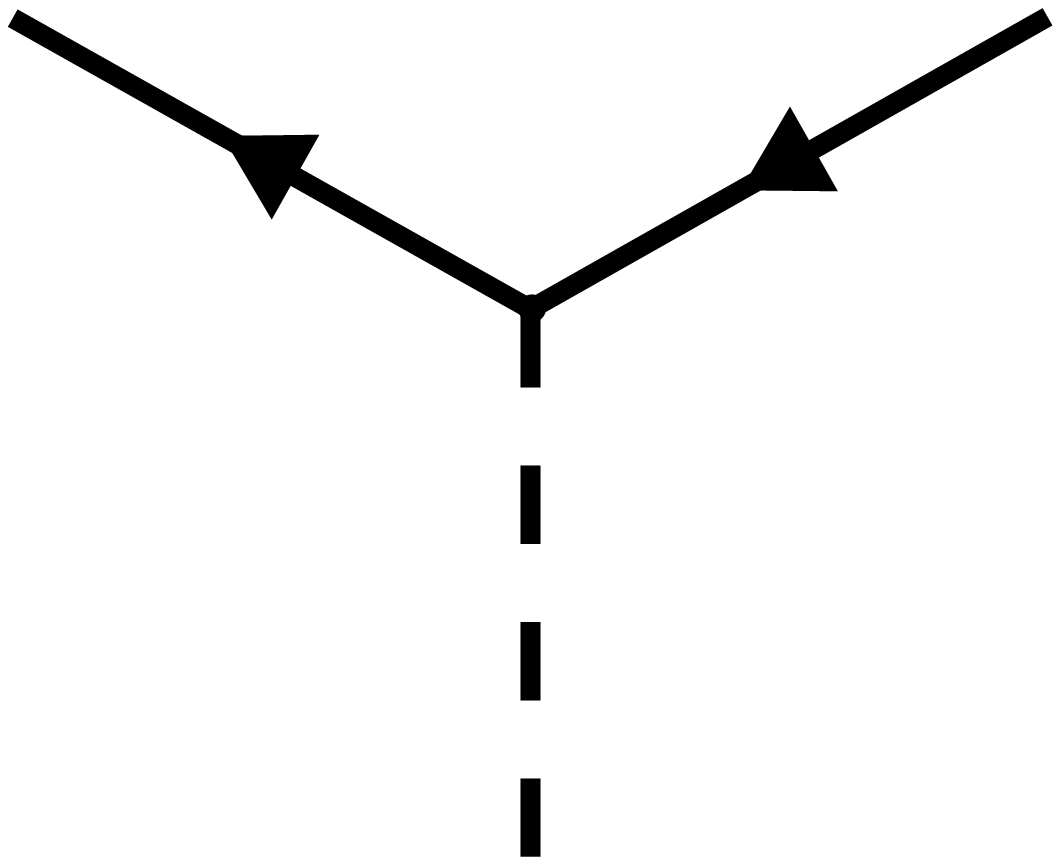}}}+ x_1^{\frac{1}{2}}x_2^{\frac{1}{2}}\vcenter{\hbox{\includegraphics[scale=0.2]{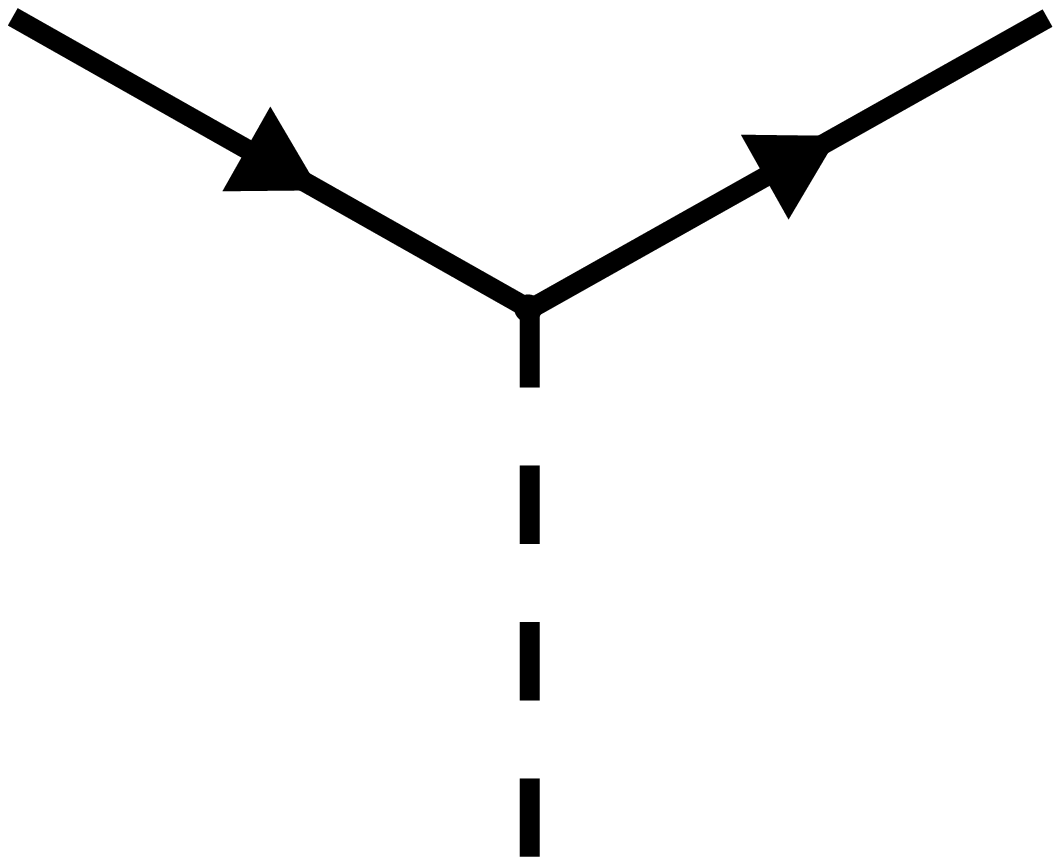}}}+ \vcenter{\hbox{\includegraphics[scale=0.2]{diagrams/loopTLvertex7.eps}}}\label{kuptransfermatrix2}
\end{align}
\label{transfermatrixkup}
\end{subequations}
An oriented full line represents the propagation of states inside $V_1$ (respectively $V_1^*$) if the arrow is pointing up (respectively down) and a dashed line represents the vacuum. In the diagrammatic formulation of the local transfer matrices of the loop model, it was possible to avoid arrows on edges because the fundamental representation of $U_{-q}(\mathfrak{sl}_2)$, $V$, is self-dual. Here this is not the case anymore, as $V_1^*$ is isomorphic to the second fundamental representation of $U_{-q}(\mathfrak{sl}_3)$. 

The diagrams appearing in \eqref{transfermatrixkup} are \textit{open} Kuperberg webs. Let us discuss briefly how these webs are related to intertwiners \cite{Kuperberg_1996}. Any web can be obtained as a combination of horizontal juxtaposition and vertical concatenation of the following elementary blocks
\begin{align}
\label{kupgendiagrams}
    v&\;=\;\vcenter{\hbox{\includegraphics[scale=0.2]{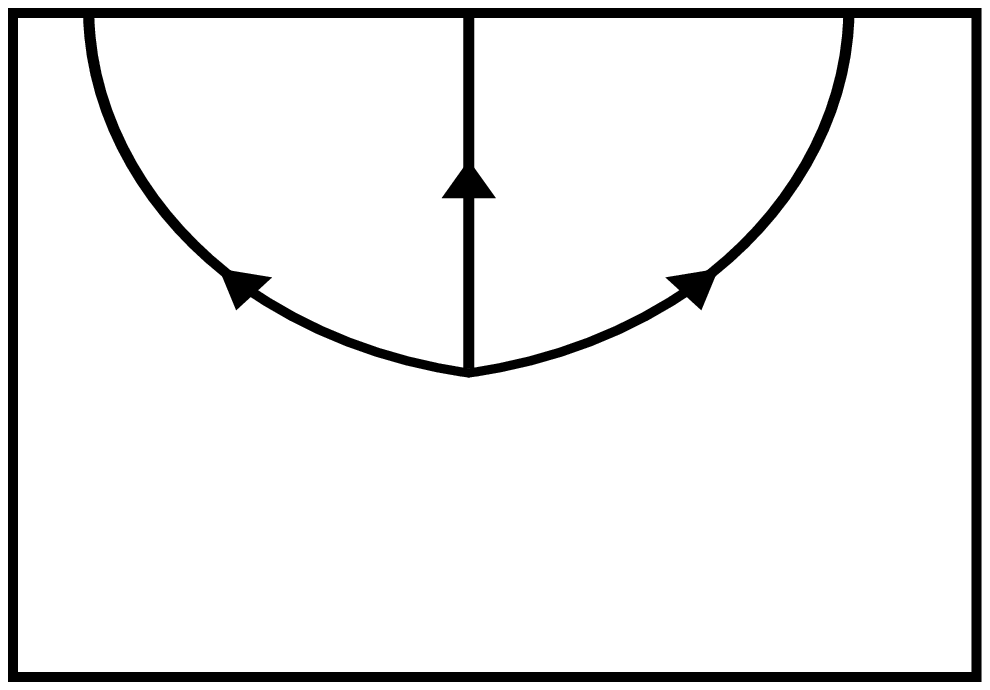}}}\qquad\quad\, w\;=\;\vcenter{\hbox{\includegraphics[scale=0.2]{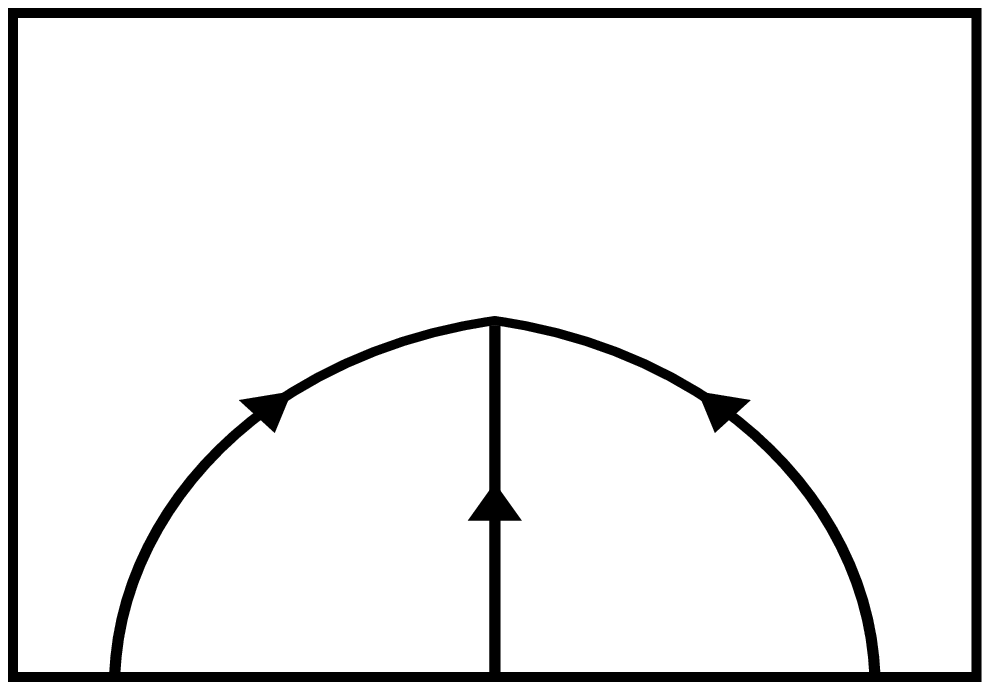}}}\nonumber\\
    \text{ev}&\;=\;\vcenter{\hbox{\includegraphics[scale=0.2]{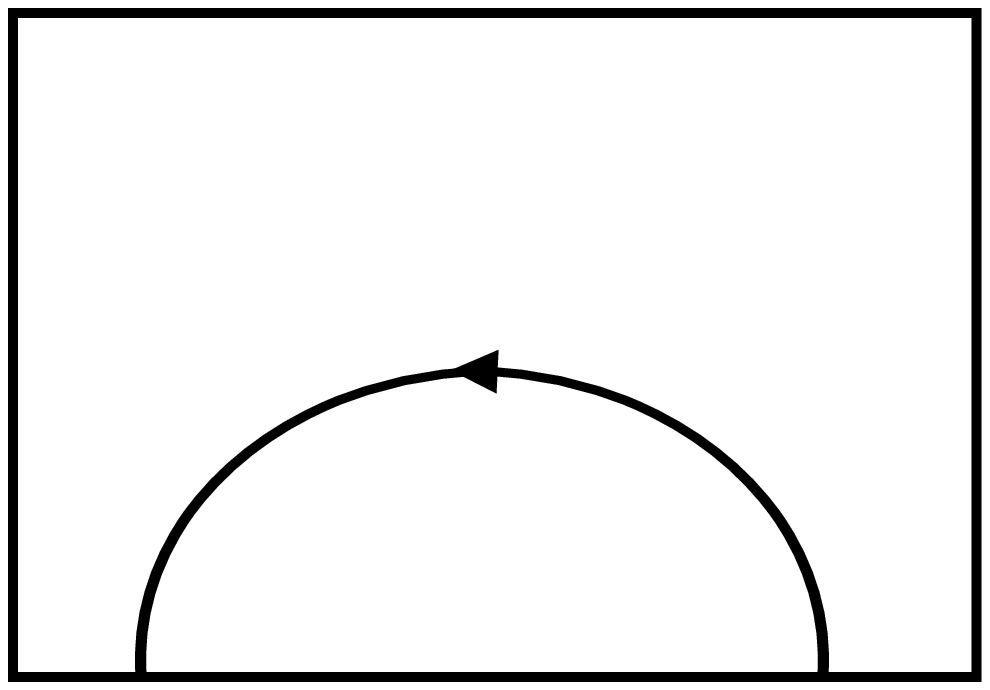}}}\qquad \text{coev}\;=\;\vcenter{\hbox{\includegraphics[scale=0.2]{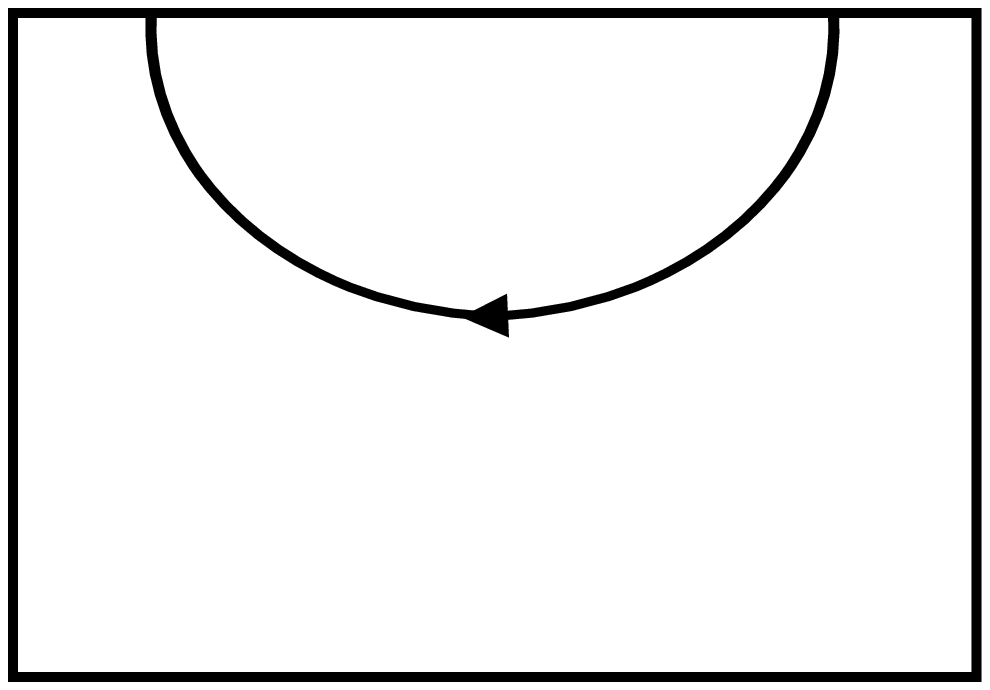}}}\\
    \widetilde{\text{ev}}&\;=\;\vcenter{\hbox{\includegraphics[scale=0.2]{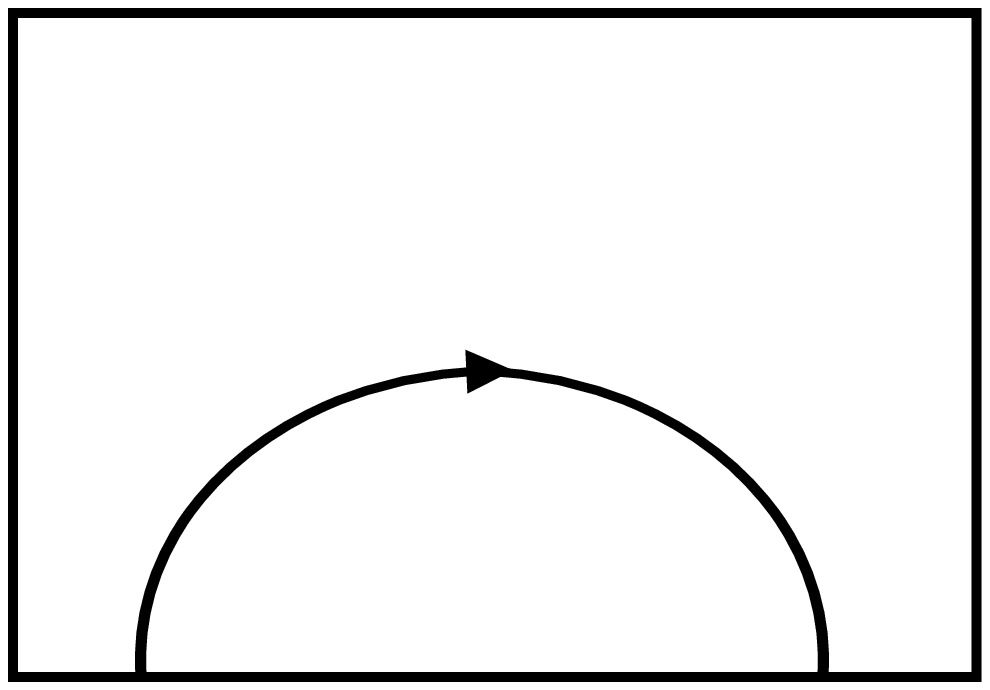}}}\qquad \widetilde{\text{coev}}\;=\;\vcenter{\hbox{\includegraphics[scale=0.2]{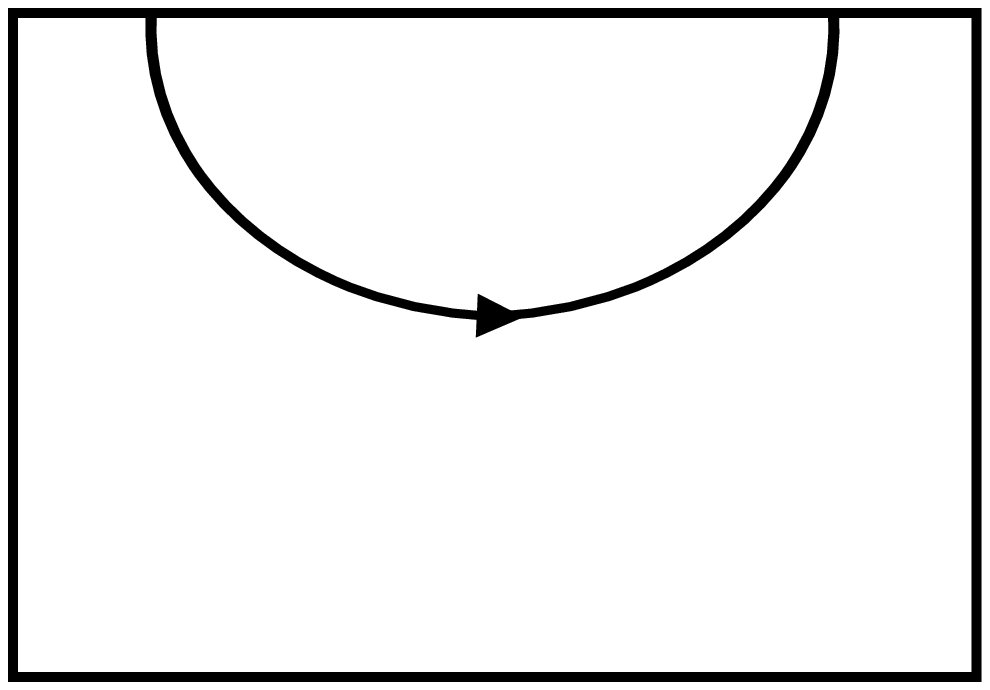}}}\nonumber
\end{align}
where the last 4 diagrams represent the duality maps, see more below and in Appendix~\ref{sec:quantumgroupconventions}.
By horizontal juxtaposition, we mean placing two diagrams next to each other horizontally. On the operator side, this means taking the tensor product of the associated linear maps. The vertical concatenation, or composition, means placing two diagrams on top of each other if their boundaries agree. On the operator side, it means taking the composition of the associated linear maps. For instance, consider the following open webs:
\begin{align}
    A=\vcenter{\hbox{\includegraphics[scale=0.2]{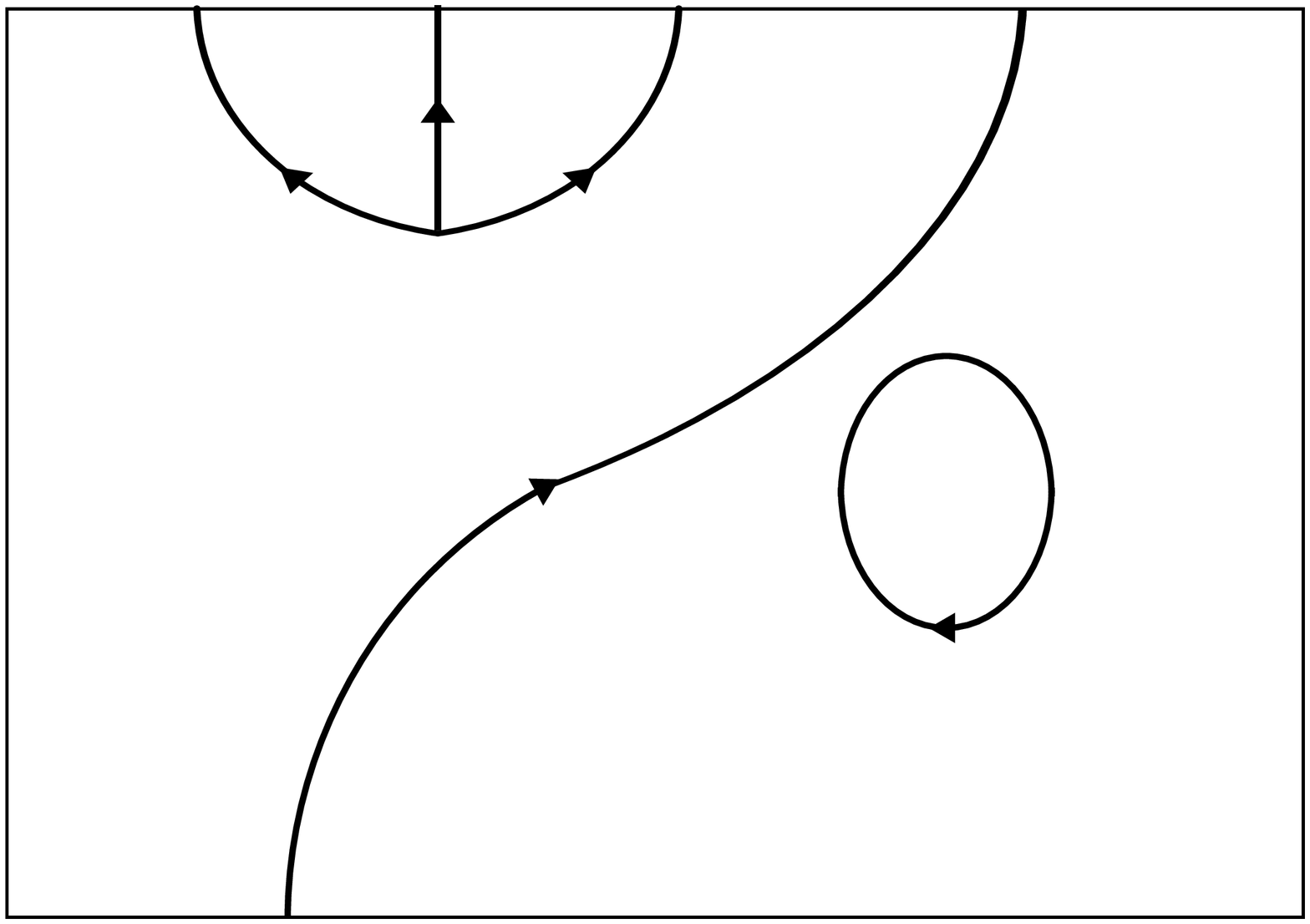}}}\quad \quad
    B=\vcenter{\hbox{\includegraphics[scale=0.2]{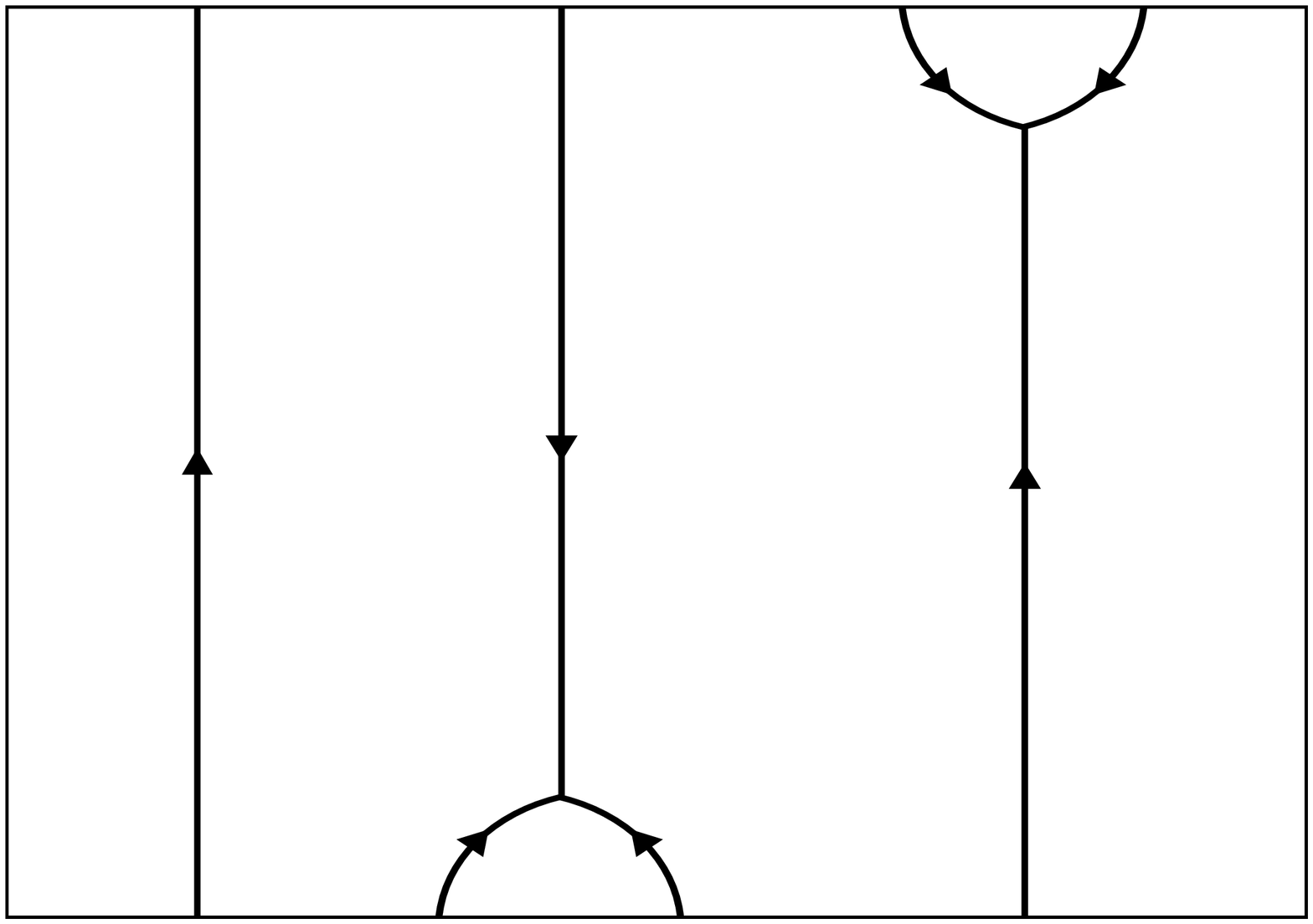}}}
\end{align}
Then their composition is
\begin{align}
    BA=\vcenter{\hbox{\includegraphics[scale=0.2]{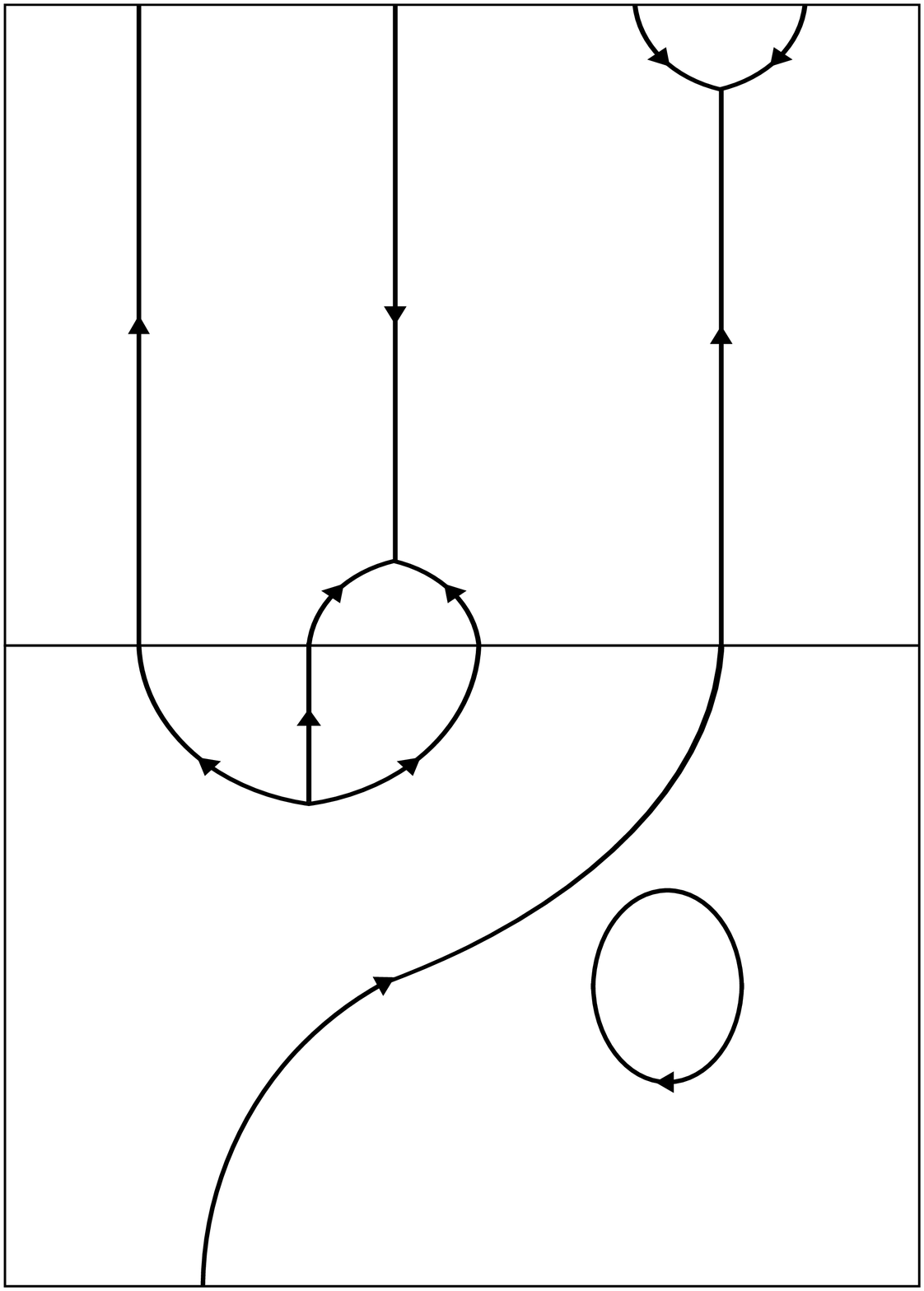}}}=[3]_q[2]_q\vcenter{\hbox{\includegraphics[scale=0.2]{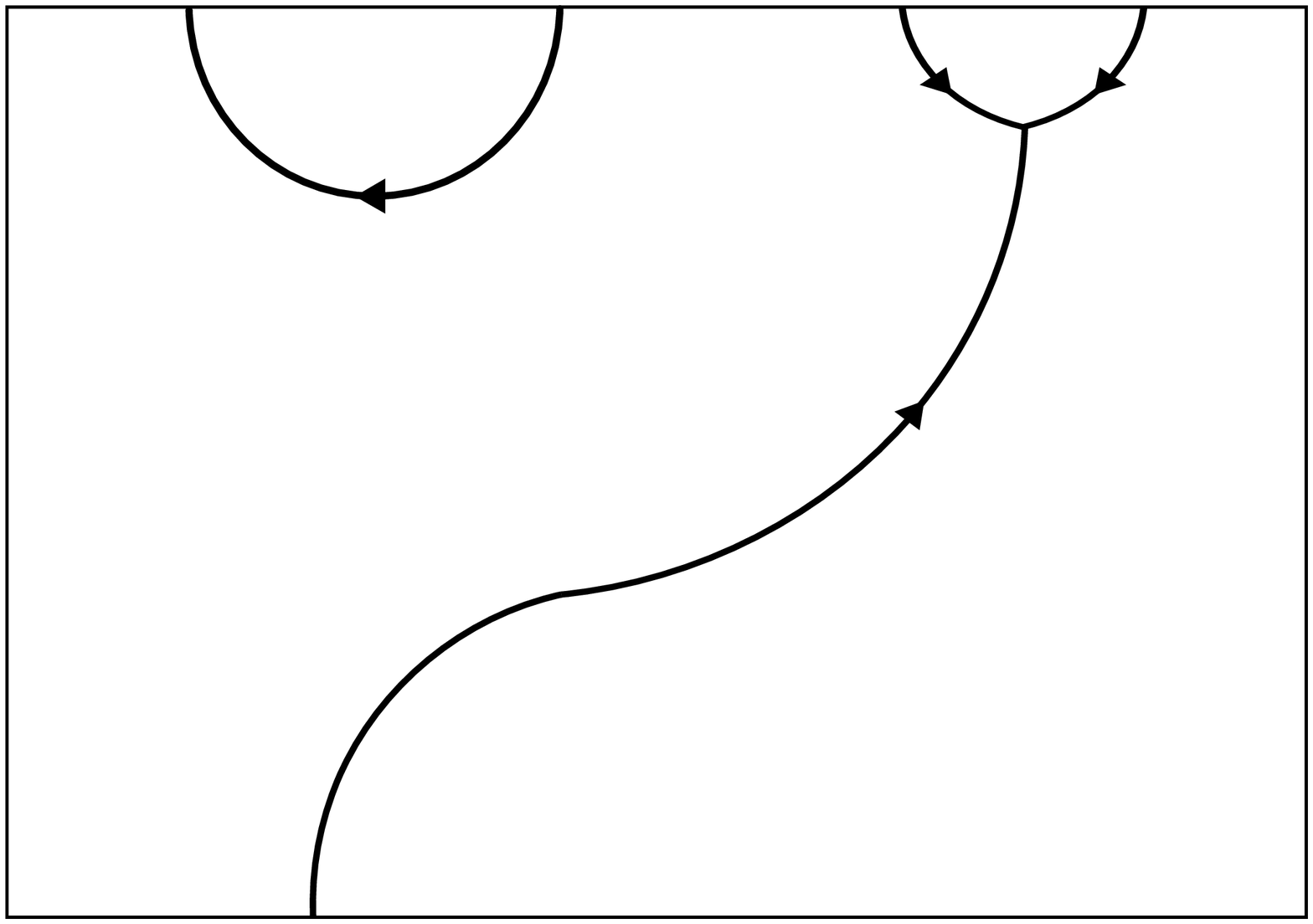}}}
\end{align}
As can be seen in the last equation, the open webs, like the closed ones, are subject to the Kuperberg rules~\eqref{3rules}.

These generating webs in \eqref{kupgendiagrams} represent the following intertwiners:
\begin{subequations}
\label{kupgen}
\begin{eqnarray}
    v&: \quad   1 \quad \mapsto&  q^{\frac{3}{2}}v_1\otimes v_2\otimes v_3+q^{\frac{1}{2}}v_2\otimes v_1\otimes v_3+q^{\frac{1}{2}}v_1\otimes v_3\otimes v_2\nonumber\\
    & &\;+\;q^{-\frac{1}{2}}v_2\otimes v_3\otimes v_1+q^{-\frac{1}{2}}v_3\otimes v_1\otimes v_2+q^{-\frac{3}{2}}v_3\otimes v_2\otimes v_1 \,,  \label{source}\\
    w& \!\!\! = \quad \qquad&q^{\frac{3}{2}}w_1\otimes w_2\otimes w_3+q^{\frac{1}{2}}w_2\otimes w_1\otimes w_3+q^{\frac{1}{2}}w_1\otimes w_3\otimes w_2\nonumber\\
    & &+q^{-\frac{1}{2}}w_2\otimes w_3\otimes w_1+q^{-\frac{1}{2}}w_3\otimes w_1\otimes w_2+q^{-\frac{3}{2}}w_3\otimes w_2\otimes w_1 \,, \label{sink}\\
    \text{coev}&:\quad  1 \quad \mapsto&    v_1\otimes w_1 + v_2\otimes w_2 + v_3\otimes w_3\,, \\
    \widetilde{\text{coev}}&:\quad 1 \quad \mapsto &   q^{-2}w_1\otimes v_1 + w_2\otimes v_2 + q^2w_3\otimes v_3 \,,\\
        \text{ev}&:\qquad  \qquad &w_i\otimes v_j \quad \mapsto \quad \delta_{ij} \,,\\
    \widetilde{\text{ev}}& :\qquad  \qquad &v_i\otimes w_j\quad \mapsto \quad q^{4-2i}\delta_{ij}\,,
\end{eqnarray}
\end{subequations}
where the element $w$ is considered as a linear form on $V_1^{\otimes 3}$. For the general definition of left duality maps, $\mathrm{ev}$ and $\mathrm{coev}$, and right ones  $\widetilde{\mathrm{ev}}$ and $\widetilde{\mathrm{coev}}$ that use the pivotal element, we refer to Appendix~\ref{sec:quantumgroupconventions}, see~\eqref{E:DualitiesC} and~\eqref{E:DualitiesC-right}.

Now we can understand which intertwiners are represented by the diagrams in \eqref{kuptransfermatrix}. For instance, the first diagram in \eqref{kuptransfermatrix1} represents the projection from $V_1\otimes V_1$ into the direct summand $V_1^*$. It is graphically obtained by a composition of, for instance, ${\rm coev}$ and $w$ as
\begin{align}
    \vcenter{\hbox{\includegraphics[scale=0.2]{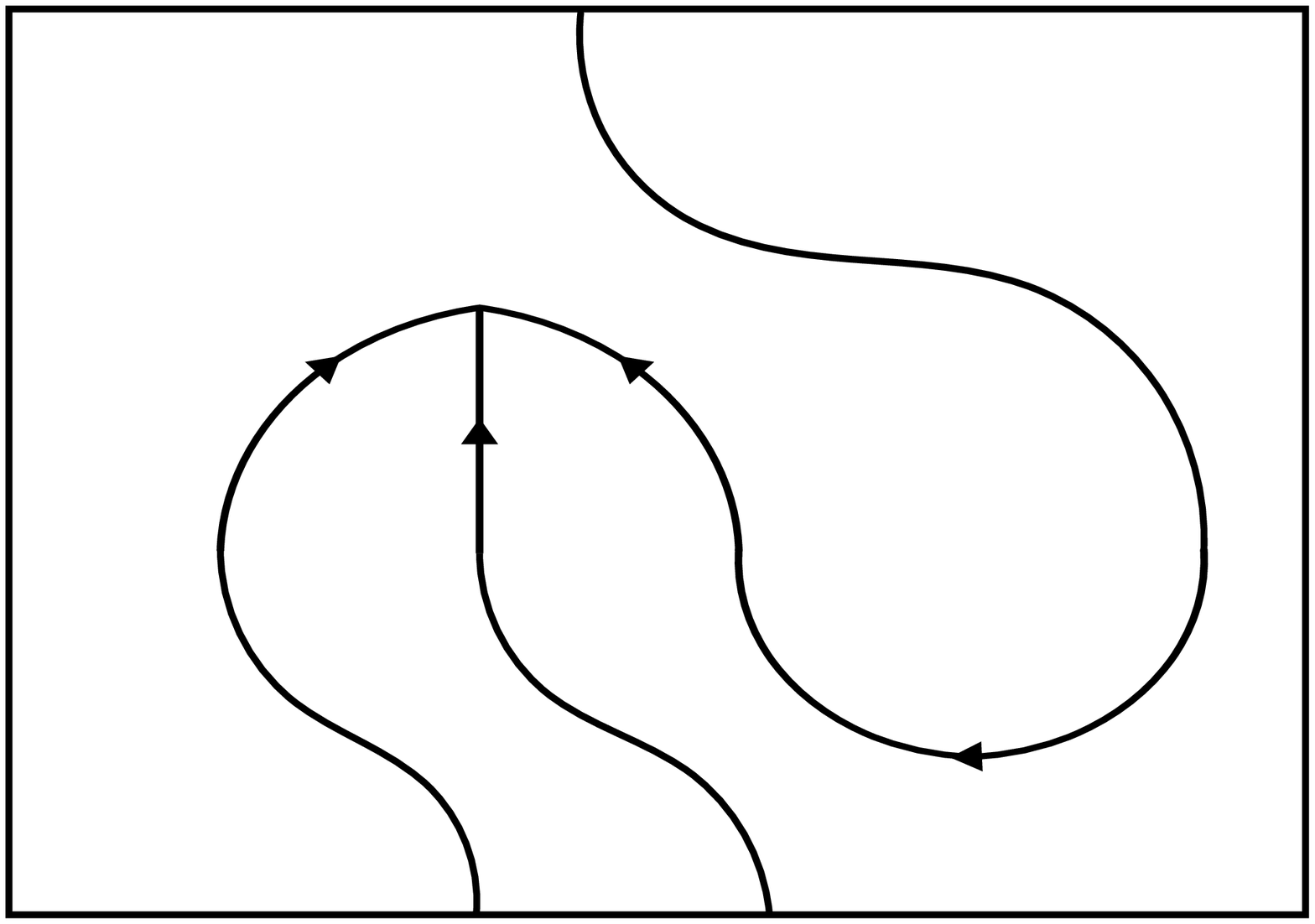}}}
\end{align}
As a $3\times9$ matrix in the bases $\{v_1\otimes v_1,v_1\otimes v_2,v_1\otimes v_3,v_2\otimes v_1,v_2\otimes v_2,v_2\otimes v_3,v_3\otimes v_1,v_3\otimes v_2,v_3\otimes v_3\}$ of $V_1\otimes V_1$ and $\{w_1,w_2,w_3\}$ of $V_1^*$, it reads
\begin{align}
    \begin{pmatrix} 0 & 0 & 0 & 0 & 0 & q^{-\frac{1}{2}} & 0 & q^{-\frac{3}{2}} & 0\\
    0 & 0 & q^{\frac{1}{2}} & 0 & 0 & 0 & q^{-\frac{1}{2}} & 0 & 0\\
    0 & q^{\frac{3}{2}} & 0 & q^{\frac{1}{2}} & 0 & 0 & 0 & 0 & 0
    \end{pmatrix}\,,
\end{align}
while in the bases $\{\ket{{\color{red} \uparrow}\color{red} \uparrow},\ket{{\color{red} \uparrow}\color{blue} \uparrow},\ket{{\color{red} \uparrow}\color{green} \uparrow},\ket{{\color{blue} \uparrow}\color{red} \uparrow},\ket{{\color{blue} \uparrow}\color{blue} \uparrow},\ket{{\color{blue} \uparrow}\color{green} \uparrow},\ket{{\color{green} \uparrow}\color{red} \uparrow},\ket{{\color{green} \uparrow}\color{blue} \uparrow},\ket{{\color{green} \uparrow}\color{green} \uparrow}\}$ of $V_1\otimes V_1$ and $\{\ket{\color{red} \downarrow},\ket{\color{blue} \downarrow},\ket{\color{green} \downarrow}\}$ of $V_1^*$, it becomes
\begin{align}
    \begin{pmatrix} 0 & 0 & 0 & 0 & 0 & q^{\frac{1}{6}} & 0 & q^{-\frac{1}{6}} & 0\\
    0 & 0 & q^{\frac{1}{6}} & 0 & 0 & 0 & q^{\frac{1}{6}} & 0 & 0\\
    0 & q^{\frac{1}{6}} & 0 & q^{-\frac{1}{6}} & 0 & 0 & 0 & 0 & 0
    \end{pmatrix} \,.
\end{align}
From the latter expression we thus see that we recover the correct vertex weigths \eqref{coloredkupvertex} for the corresponding states. One can show that this is true for all intertwiners.

In the strip geometry, the row-to-row transfer matrix is defined in a similar way as in the loop case,
\begin{align}
\label{kuptransfermatrixstrip}
    T_{\text{K}}=\left(\prod_{k=0}^{L-1}t^{\text{K}}_{2k+1}\right)\left(\prod_{k=1}^{L-1}t^{\text{K}}_{2k}\right) \,,
\end{align}
with $t^{\text{K}}_{i}=t^{\text{K}}_{(2)}t^{\text{K}}_{(1)}$ and the subscript denotes the position of the local transfer matrix. It is thus a $U_{-q}(\mathfrak{sl}_3)$ intertwiner. Define the vacuum by $\ket{\ }^{\otimes 2L}$. We see from \eqref{kuptransfermatrixstrip} that when we take the vacuum expectation value of a product of $M$ row-to-row transfer matrices, the result can be understood as the unique matrix element of a sum of intertwiners from the trivial representation to itself. These intertwiners are the ones represented by all possible closed webs embedded in $\mathbb{H}$ with some prefactors accounting for bond and vertex fugacities. We thus recover the partition function~\eqref{Z_K} on a lattice with $2M-2$ rows:\footnote{As in the loop model case, the original partition function is recovered for $2M-2$ rows instead of $2M$ rows because of our choice of vacuum.}
\begin{equation}
\label{KPF-TM}
    Z_{\text{K}}=\braket{\ T_{\text{K}}^M\ } \,.
\end{equation}

\medskip

In the cylinder geometry, the seam line operator is given by the pivotal element of $U_{-q}(\mathfrak{sl}_3)$, for the definition we refer to Appendix~\ref{sec:quantumgroupconventions},
\begin{align}
    S_{\rm K}=(-q)^{2H_{\bm{\rho}}}=q^{2H_{\bm{\rho}}} \,,
\label{kuppivot}
\end{align}
where $\bm{\rho} = \bm{\alpha}_1 + \bm{\alpha}_2$ is the Weyl vector of $\mathfrak{sl}_3$. The last equality follows because $H_{\bm{\rho}}=H_1+H_2$ is diagonalisable with integer eigenvalues on $\mathcal{H}_{\rm K}$. Since $S_{\rm K}$ belongs to the Cartan subalgebra and local transfer matrices are intertwiners, the seam line can be deformed through nodes of $\mathbb{H}$. The row-to-row transfer matrix is then defined as
\begin{align}
\label{kuptransfermatrixcyl}
    T_{\text{K}}=\left(\prod_{k=0}^{L-1}t^{\text{K}}_{2k+1}\right)\left(\prod_{k=1}^{L-1}t^{\text{K}}_{2k}\right)S_{\text{K}}t^{\text{K}}_{2L}S_{\text{K}}^{-1} \,.
\end{align}

The pivotal element is the one implementing the quantum trace $\mathrm{qtr}$, see~\eqref{eq-qtr}, and its role in the diagrammatic setting is to give to the closed webs embedded in the cylinder the weight given by \eqref{3rules}, as if they were unfolded on the plane. Hence we see that, again, taking the vacuum expectation of $M$-th power of \eqref{kuptransfermatrixcyl}  recovers \eqref{Z_K}.

However, in the cylinder geometry, the row-to-row transfer matrix will, in general, not posses the full quantum group symmetry of the local transfer matrices. Yet the invariance with respect to the action of the Cartan subalgebra remains.

\subsection{Relation with the FPL on $\mathbb{H}$}

We now tune $x_1=x_2=x$, $y=z=1$ in the Kuperberg web model and consider the $x\rightarrow +\infty$ limit. In this case, the configurations are webs that completely cover $\mathbb{H}$. There are two such webs that are related by a reflection of all of their arrows: in the first, each type 1 node is a source and each type 2 node is a sink, while in the second it is the other way around. Both of those webs have the same weight, hence the partition function reads
\begin{align}
 \label{FPLlim}
    \lim_{x\rightarrow +\infty} \frac{1}{x^{N_l}} Z_{\rm K} =2w_{\rm K}(\mathbb{H}) \,,
\end{align}
where $N_l$ is the total number of links of $\mathbb{H}$. This limit is thus described by the whole lattice $\mathbb{H}$ acting as a unique web, so it is interesting to regard this web in the refined model of coloured webs. The configurations are then all three-colourings of the hexagonal lattice. Such a three-colouring model was first studied by Baxter, who found the exact asymptotic equivalent of the partition function in the special case where each three-colouring has the same statistical weight \cite{Baxter3C}. If one further considers blue links as empty, one gets a collection of cycles made of alternating red and green links that are jointly covering each node of $\mathbb{H}$. We thus obtain the configuration space of the fully-packed loop (FPL) model on $\mathbb{H}$. The equal-weighted case would correspond to giving a fugacity $N=2$ to each of these loops (since each loop is invariant upon permuting red and green along the corresponding alternating cycle).

We now investigate closer which weighting of the FPL model is really obtained in the limit \eqref{FPLlim}. By reversing the orientation of red links, one gets oriented loops that cover every node of $\mathbb{H}$. According to \eqref{coloredkupvertex}, these loops pick a factor $q^{-\frac{1}{6}}$ when they turn left and a factor $q^{\frac{1}{6}}$ when they turn right. Hence, summing over both orientations, contractible unoriented loops are all weighted by $[2]_q$. We thus recover in this limit the more general FPL model on the hexagonal lattice with an adjustable loop fugacity, $N=[2]_q$:
\begin{align}
    Z_{\text{FPL}}=\lim_{x\rightarrow +\infty} \frac{1}{2x^{N_l}} Z_{\rm K} \,.
\end{align}
This mapping is originally due to Reshetikhin \cite{Reshetikhin_1991}. In our case, when $\mathbb{H}$ is embedded in the cylinder, non-contractible loops are given a different weight, $\widetilde{N} = q^2+q^{-2}=[2]_q^2-2$. The scaling limit of the FPL model has been studied by Coulomb Gas (CG) techniques in \cite{KdGN96}, and the particular choice of $\widetilde{N}$ was further shown in \cite{DEI16} to lead to a CFT with an extended $W_3$ symmetry.

The FPL model is in fact integrable. In order to make this apparent, consider the local transfer matrix $t^{\rm K}=t^{\rm K}_{(2)}t^{\rm K}_{(1)}$ in our limit. As we have seen that we can regard $\mathbb{H}$ as a unique web, one can write $t^{\rm K}$ as
\begin{align}
    t^{\rm K}=x^3\vcenter{\hbox{\includegraphics[scale=0.15]{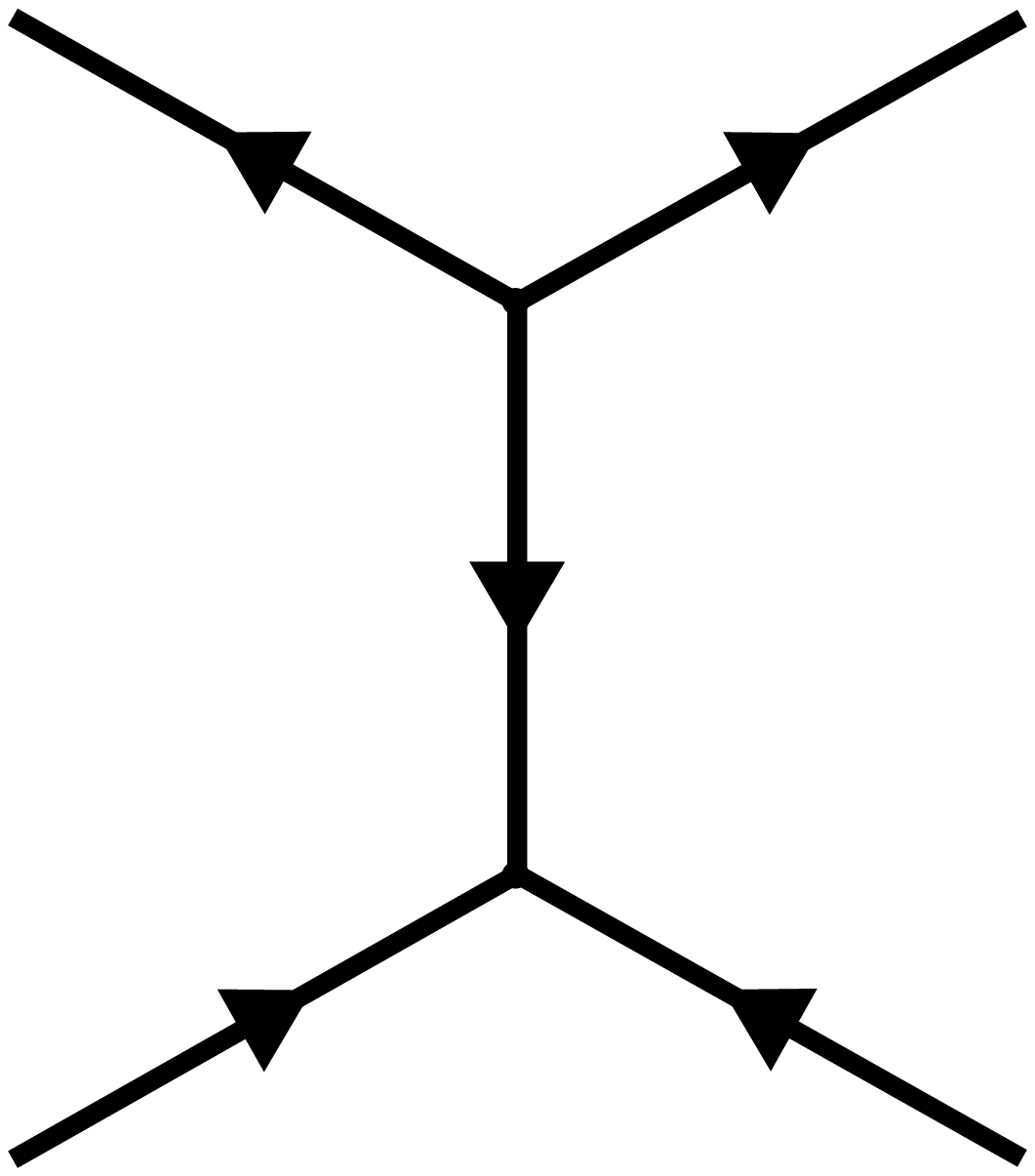}}} \,.
\end{align}
Now consider the integrable trigonometric R-matrix, $R_{15}$, of the fifteen-vertex model with $U_{-q}(\mathfrak{sl}_3)$ symmetry~\cite{Jimbo86}. It intertwines between $V_1\otimes V_1$ and itself, seen as $U_{-q}(\mathfrak{sl}_3)$ representations. In terms of Kuperberg web it reads
\begin{align}
    R_{15}(u)=\sin(\gamma -u)\vcenter{\hbox{\includegraphics[scale=0.15]{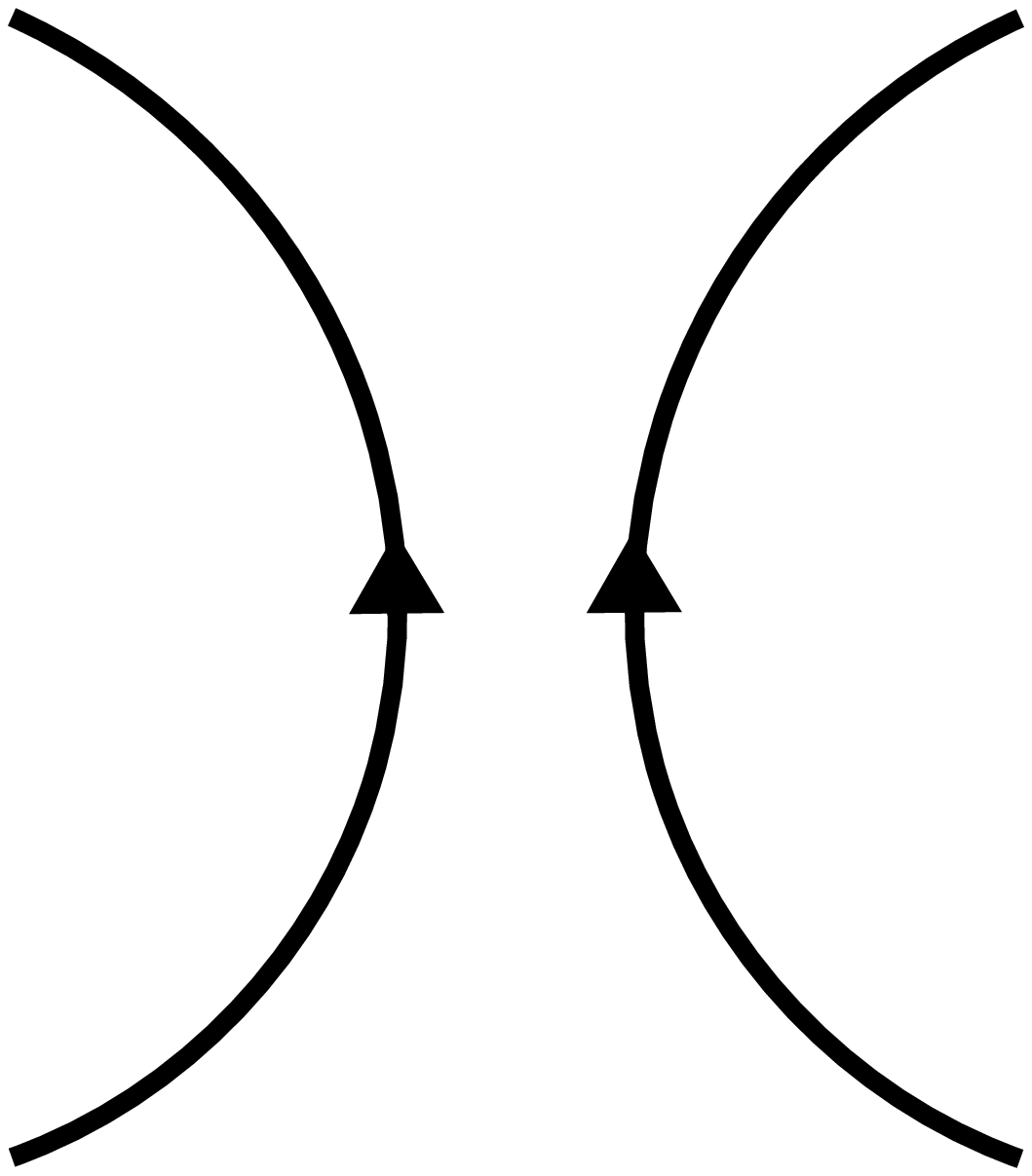}}} + \sin (u) \vcenter{\hbox{\includegraphics[scale=0.15]{diagrams/kupFPLtm2.eps}}} \,,
\end{align}
where $u$ denotes the additive spectral parameter and $q$ is parameterised by $q=e^{i\gamma}$. At $u=\gamma$, one recovers, up to a scalar, $t^{\rm K}$.

\medskip

We may also ask what kind of loop model one would obtain by the above procedure, for a general choice of local web fugacities $x$, $y$ and $z$. The loop configurations are given by sets $\mathcal{L}_c$ of cycle coverings of webs $c$ embedded in $\mathbb{H}$, i.e. $c\in \mathcal{K}$. Hence the partition function reads
\begin{align}
    Z =\sum_{l\in \mathcal{L}_c\, |\, c\in \mathcal{K}} x^{N}(yz)^{N_V}w(l)\, ,
\end{align}
where $w(l)$ denotes the weight of a loop configuration. In this general case, all contractible loops do not get the same weight, and hence $w(l)$ does not take a simple form. Indeed, a given oriented loop picks a factor $q^{\frac{1}{3}}$ when turning left at a node that is not a vertex of the underlying web $c$, but a factor $q^{-\frac{1}{6}}$ when the node is a vertex of $c$.

\section{Phase diagram of the Kuperberg web model}
\label{sec:pd}

\begin{figure}
\begin{center}
    \includegraphics[scale=0.4]{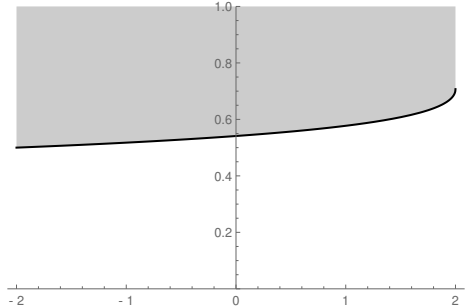}
\end{center}
    \caption{The phase diagram of the $O(N)$ loop model. $N$ is shown on the horizontal axis and $x$ is shown on the vertical axis. The black curve indicates the loci of dilute critical points. The grey area corresponds to the dense phase.}
    \label{fig:loopphasediagram}
\end{figure}

In this section, we give an exposition of the phase diagram of the web model. A fruitful comparison can be made with the phase diagram of the $O(N)$ loop model, with loop weight $N=[2]_q$ and bond fugacity $x \ge 0$. Recall that on the hexagonal lattice $\mathbb{H}$, one can identify three critical phases in the range $N\in[-2,2]$. The phase diagram is shown in Figure~\ref{fig:loopphasediagram}. The so-called dilute phase occurs at $x=x_{\rm c}$, with \cite{Nienhuis,DCS12}
\begin{align}
\label{xc_loop}
   x_{\rm c}=\frac{1}{\sqrt{2+\sqrt{2-N}}} \,,
\end{align}
corresponding to a critical continuum limit.
For $x<x_{\rm c}$, the model is not critical, and will in fact flow under the Renormalisation Group (RG) to the trivial fixed point $x = 0$. For $x_{\rm c} < x < \infty$, the model is critical and in the so-called dense phase, governed by the attractive fixed point $x^\star = (2 - \sqrt{2-N})^{-1/2}$ \cite{Nienhuis}. At $x=+\infty$, the model is also critical and in its fully-packed phase \cite{KdGN96}. The three phases---dilute, dense and fully-packed---are described by three distinct CFTs \cite{LoopReview}.

Summarising, we see that for each fixed value of $q$ satisfying \eqref{crit-reg-loop}, the model is not critical for small values of $x$. Then, increasing $x$, we cross the dilute critical point $x=x_{\rm c}$ and enter into an extended dense phase for $x_{\rm c}<x<\infty$. We shall see that the phase diagram of the Kuperberg web model exhibits very similar features.

\medskip

The phase diagrams of the web model presented below have been obtained thanks to the numerical diagonalisation of the row-to-row transfer matrix. To be precise, the transfer matrix we have used in our numerical work is slightly different from the one depicted in Figure \ref{fig:transfermatrix}. It is given by a product of local transfer matrices and the seam operator \eqref{kuppivot}, as depicted below for size $L=5$:
\begin{center}
    \includegraphics[scale=0.4]{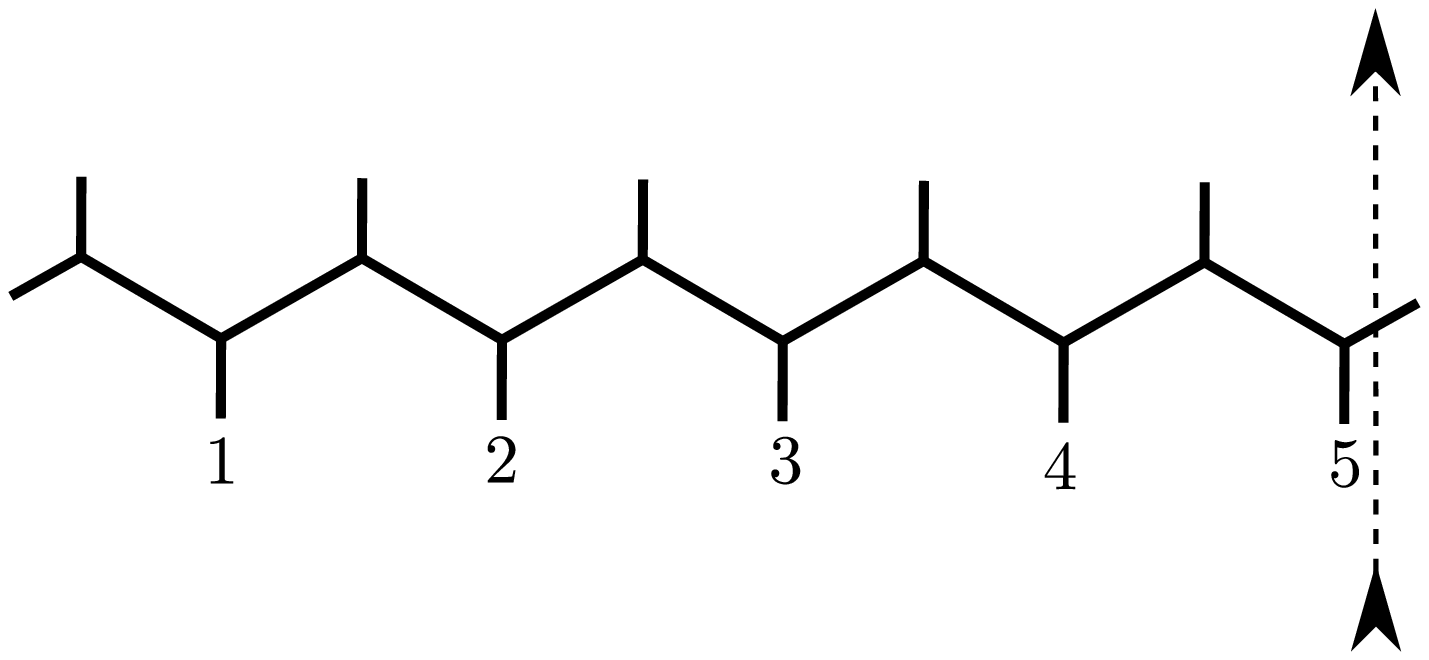}
\end{center}
It is an endomorphism of $\mathcal{H}_{\rm K}^{\otimes L}$. All numerical results are given thanks to this transfer matrix, or a modification thereof where the seam operator is changed (see Section \ref{sec:em}).

There are two differences between this transfer matrix and the one described in Section~\ref{sec:localtmKup} and depicted in
Figure~\ref{fig:transfermatrix}. First, we are now transfering between two rows of $L$ vertical edges, rather than between two rows of $2L$ edges
with alternating inclinations $\diagup$ and $\diagdown$. From a numerical perspective this has the advantage of considerably diminishing the dimension of
the matrix, thus enabling us to study larger $L$ than would be possible otherwise. It is  clear that the two conventions construct the same lattice
and hence are physically equivalent. 
It can be seen from the picture above that this transfer matrix will have the labels of the spaces drift towards the left (by one half  lattice spacing
per row), so its square is related to the product of the former transfer matrix with a shift operator. 
This will however not entail any modification for eigenvalues corresponding to the vanishing lattice momentum sector, the only one to be studied in this section.
In this sector, the spectrum of the squared transfer matrix  is included within that of the former matrix, and moreover it is
not hard to see that their dominant eigenvalue (the one of largest norm) coincide.
 So, summarising, the change of transfer
matrix makes the numerical work much more efficient, without modifying the physical quantities to be studied.

\medskip

The effective central charge $c_{\text{eff}}$ provides a convenient means of investigating properties of the phase diagram and the corresponding RG flows.
We first describe how $c_{\rm eff}$ can be approximated using finite-size scaling. The free energy density for the model defined on a cylinder with a circumference of $L$ hexagons is given by 
\begin{align}
    f_L&=-\frac{2}{\sqrt{3}L}\log(\Lambda_{\max}) \,,
\end{align}
where the numerical prefactor is related to the geometry of the hexagonal lattice, and $\Lambda_{\max}$ denotes the real part of the dominant eigenvalue of the transfer matrix in a subspace of the spectrum. Indeed, to gain in efficiency we have restricted the transfer matrix to a specific sector of vanishing magnetisation (see Section \ref{sec:em}), or more precisely, to the subspace of states having weight $\bm{0}$ with respect to the Cartan subalgebra symmetry. The free energy density has the finite-size scaling \cite{Cardy_c,Affleck_c}
\begin{align}
 \label{FSS_form}
    f_L&=f_{\infty}-\frac{\pi c_{\text{eff}}}{6L^2} + o\left(\frac{1}{L^2}\right) \,,
\end{align}
with $f_{\infty}$ being the free energy in the thermodynamical limit. Hence, by diagonalising the transfer matrices for two consecutive sizes, $L=5$ and $L=6$, we can extract the two constants, $f_{\infty}$ and $c_{\text{eff}}$. The sizes are chosen in a compromise between being sufficiently close to the thermodynamical limit for the scaling behaviour to be visible, and yet being able to perform the required number of diagonalisations in a reasonable time. The dimension of the vacuum sector (of vanishing magnetisation, see Section \ref{sec:em}) used here is $5881$ for $L=6$, and each of the phase diagrams presented in the following figures is based on computing $f_L$ for $22500$ different parameter values. For the diagonalisation itself we employ the Arnoldi method for non-symmetric complex matrices, in combination with standard sparse matrix and hashing techniques.

\medskip

In the following we set $x=x_1=x_2$ and $y=z$, so that the web model is isotropic and invariant under the global reversal of orientations. We moreover restrict
to non-negative parameters ($x,y \ge 0$). We shall depict the phase diagrams in the $(\sqrt{x},y)$ plane, with $\sqrt{x} \in [0,3]$ shown on the horizontal axis
and $y \in [0,3]$ on the vertical axis of the figures. To sample the critical region \eqref{crit-reg} we focus on three different values of $q$, viz.\ 
$q={\rm e}^{i \pi/5}$, $q={\rm e}^{i \pi/4}$ and $q={\rm e}^{i \pi/3}$. The corresponding weights of an oriented loop, $[3]_q$ from \eqref{3rulesa}, are
$(1+\sqrt{5})/2 \simeq 1.618$, $1$ and $0$.

We remark that on the horizontal axis, $y=0$, vertices are suppressed and the web model is equivalent, at the level of partition functions, to the $O(N)$ loop model with a loop weight given by $N=2[3]_q$, since loops come with two orientations in the web model. The three values of $q$ hence correspond to cases $N>2$,
$N=2$ and $N<2$, respectively.

\begin{figure}
\begin{center}
    \includegraphics[scale=0.5]{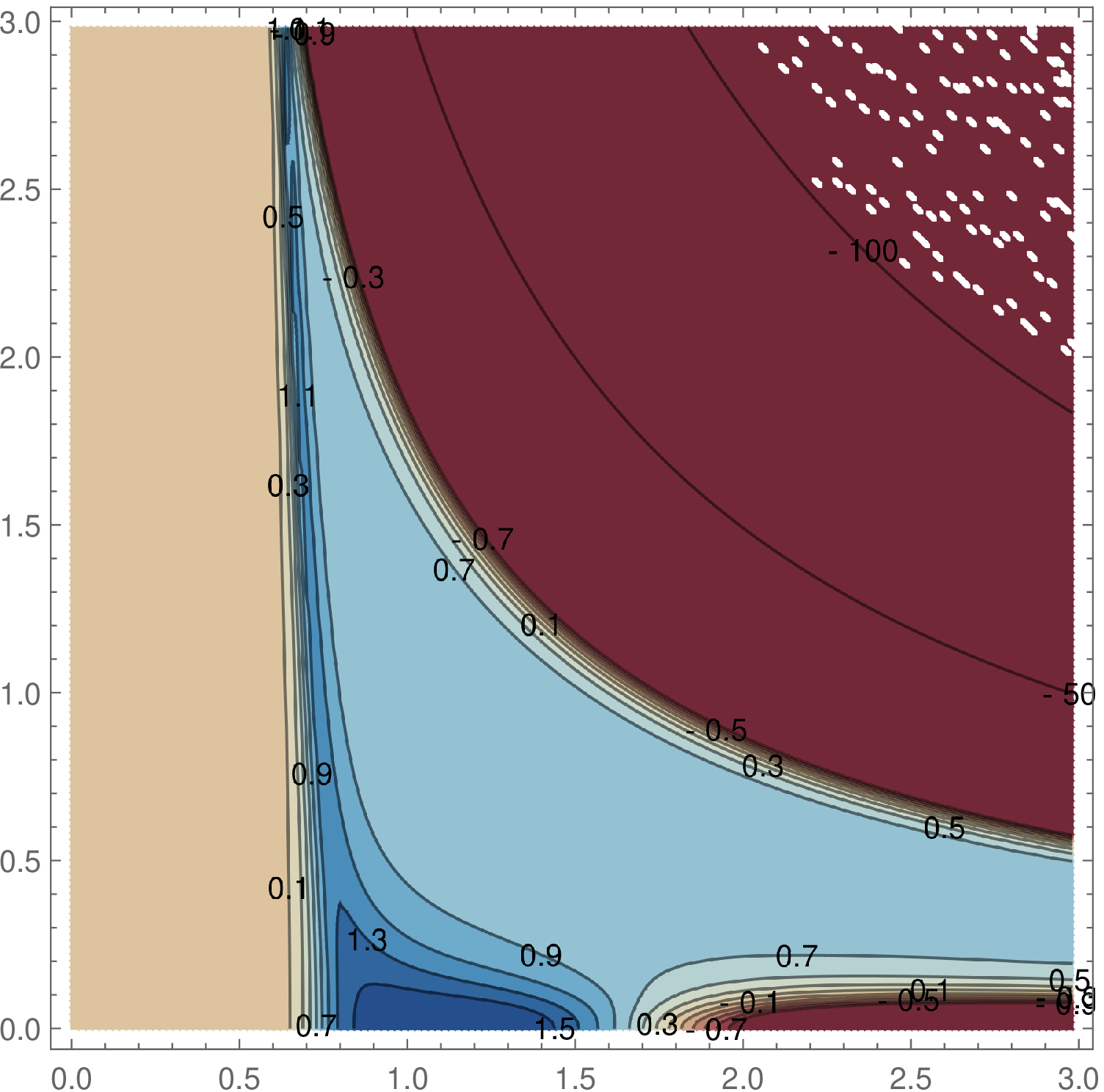} \qquad \includegraphics[scale=0.5]{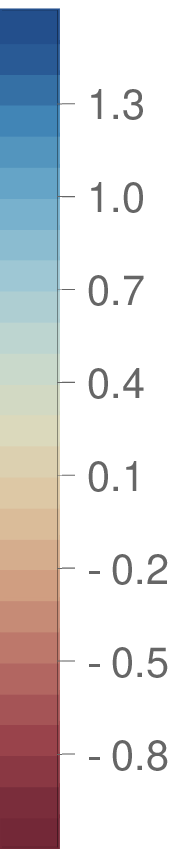}
\end{center}
    \caption{Phase diagram of the web model with $q = {\rm e}^{i \pi/5}$ in the $(\sqrt{x},y)$ plane, as given by the contour plot of the effective central charge for sizes $L=5$ and $L=6$. The interpretation is given in the main text.}
    \label{fig:pd-eipi5}
\end{figure}

With these conventions, the phase diagram obtained for the first case, $q={\rm e}^{i \pi / 5}$, can be inferred from the corresponding values of the effective central charge, shown as a contour plot in Figure~\ref{fig:pd-eipi5}. At first sight, three different regions can be distinguished:
\begin{enumerate}
 \item To the left of an almost vertical line, $\sqrt{x} \lesssim 0.6$, we have $c_{\rm eff} \approx 0$ (sand coloured region).
 \item To the right of a curve that resembles a hyperbola and extends from $(0.7,3.0)$ to $(3.0,0.5)$ approximately,
 $c_{\rm eff}$ takes large negative values (dark red region).
 \item In between those, $c_{\rm eff}$ takes predominantly values between 0 and 1.5 (region with shades of blue).
\end{enumerate}
To interpret these regions, we refer to results and experience gathered in the study of vertex models \cite{BaxterBook} and some related numerical investigations \cite{JC98,J00}.
Vertex models generally possess two types of non-critical regions. In the former, there is a finite correlation length, and using nevertheless the finite-size
scaling form~\eqref{FSS_form} one sees that $c_{\rm eff} \to 0$ exponentially fast in $L$. This agrees with the first region identified above.
In the latter, the system is frozen into long-range (``ferroelectric'', in the context of the six-vertex model) order, and the orientational degrees of freedom
are correlated throughout the system. In this case, the hypotheses leading to \eqref{FSS_form} are inapplicable and one observes large (positive or negative)
values of $c_{\rm eff}$. This behaviour agrees with the second region identified above. Finally, the third region is the most interesting one, inside which
the system exhibits critical behaviour characterised by an infinite bulk correlation length.

\begin{figure}
\begin{center}
    \includegraphics[scale=0.5]{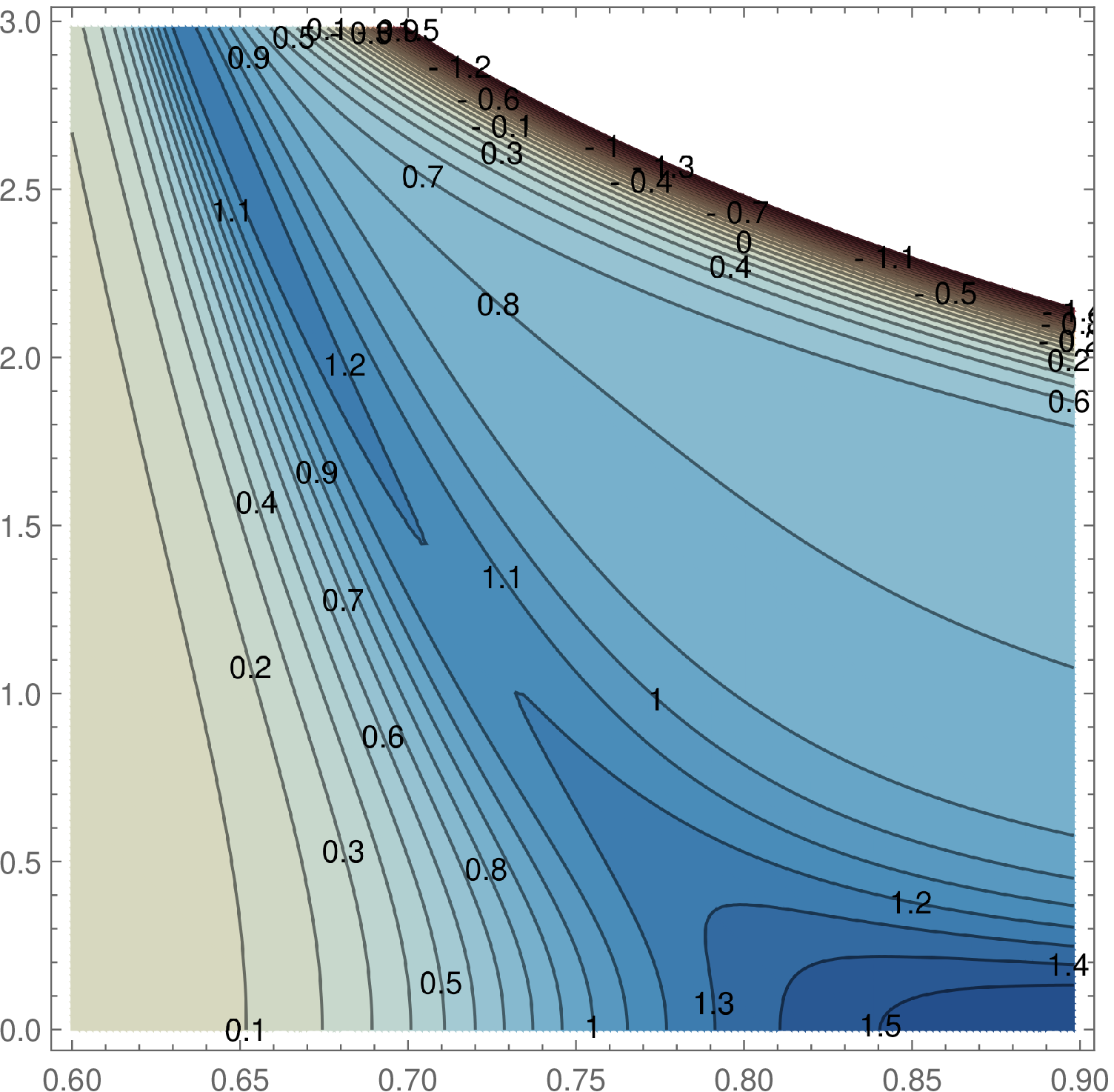} \qquad \includegraphics[scale=0.5]{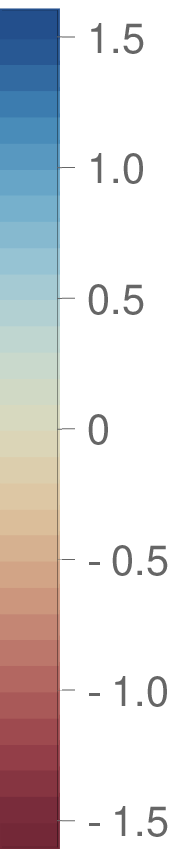}
\end{center}
    \caption{Close-up of the ridge region of the web model with $q = {\rm e}^{i \pi/5}$ in the $(\sqrt{x},y)$ plane, as given by the contour plot of the effective central charge for sizes $L=5$ and $L=6$. The interpretation is given in the main text.}
    \label{fig:pd-eipi5zoom}
\end{figure}

We therefore discard the non-critical regions and focus on the third, critical region. Consider first the part that is not too close to the horizontal axis.
We observe an almost vertical curve around $\sqrt{x}\approx 0.7$ with a central charge $c_{\text{eff}}\approx 1.2$. This curves takes the form of a
``mountain ridge'' in the landscape of $c_{\rm eff}$. A close-up of the ridge region, shown in Figure~\ref{fig:pd-eipi5zoom}, gives better evidence for our
estimate for the value of $c_{\rm eff}$ and the claim that it is almost constant along the ridge. We identify this ridge as the {\em dilute critical phase}.
Notice that in the loop model it was situated at $x = x_{\rm c}$ in \eqref{xc_loop}, that is, attained by adjusting one parameter. The situation in the web 
model is similar, except that we now have two parameters, $x$ and $y$, at our disposal. Hence adjusting one parameter will leave us with a critical
curve, instead of just a critical point. Moving along this curve corresponds to perturbing the fixed point theory by an irrelevant operator.

To the right of the dilute critical phase, and below the non-critical region 2, we observe a plateau with $c_{\text{eff}}\approx 0.8$.
We identify this as the {\em dense critical phase}. As in the loop model, it is obtained by adjusting no parameter (within a given range),
and therefore it here takes the form of a two-parameter critical surface. Displacements along this surface correspond to the perturbation
by irrelevant operators.

In a subsequent paper \cite{LGJ-CG} we shall propose a Coulomb Gas description of the web model. It will turn out that the exact value
of the central charge is $c = \frac{6}{5}$ in the dilute critical phase and $c = \frac{4}{5}$ in the dense critical phase, both in fine agreement
with the above numerical results.

To conclude the discussion of Figure~\ref{fig:pd-eipi5} we now focus on the horizontal axis, $y = 0$. As already mentioned, along this line the web model is equivalent to a loop model
with monomer fugacity $x$ and loop weight $N = 1 + \sqrt{5} \simeq 3.236 > 2$. Rather interestingly, the O($N$) model on $\mathbb{H}$ can exhibit
critical behaviour even though $N>2$ \cite{GBW00}. This comes about because $N$ can flow to infinity under the RG, from any starting value $N>2$, and provided $x$ is adjusted accordingly
the model hits the phase transition in the hard hexagon (HH) model \cite{BaxterHH}, which is known to be in the universality class of the critical three-state Potts
model with $c = \frac{4}{5}$. The table of \cite{GBW00} contains numerical estimates of the corresponding critical value, $x = x_{\rm HH}$, for selected values $N \ge 4$.
For $N=4$, finite-size effects are found to be severe, even using sizes as large as $L=15$ (thus far larger than $L=6$ attained in our study of the web model),
and the situation would be worse for the value $N \simeq 3.236$ of interest here.
Fitting the values for $x_{\rm HH}(N)$ given in the table to a polynomial in $1/N$, we can expect $x_{\rm HH} \sim 550$.
Despite the obvious difficulties of making numerical observations in this case, the conclusion is nevertheless clear: there should be point on the horizontal axis
which is in the universality class of the HH model.

In addition to the identification of critical points, the contour plot of $c_{\rm eff}$ also contains information about the RG flows. According to
Zamolodchikov's $c$-theorem \cite{C-theorem}, the RG fixed points correspond to saddle points of $c_{\rm eff}$, and away from those---under the
assumption of reflexion positivity, or unitarity---the RG flows will be in the direction of decreasing $c_{\rm eff}$. This result is applicable even
though $c_{\rm eff}$ is here a finite-size approximation to the true $c$-function. 

\medskip

\begin{figure}
\begin{center}
    \includegraphics[scale=0.5]{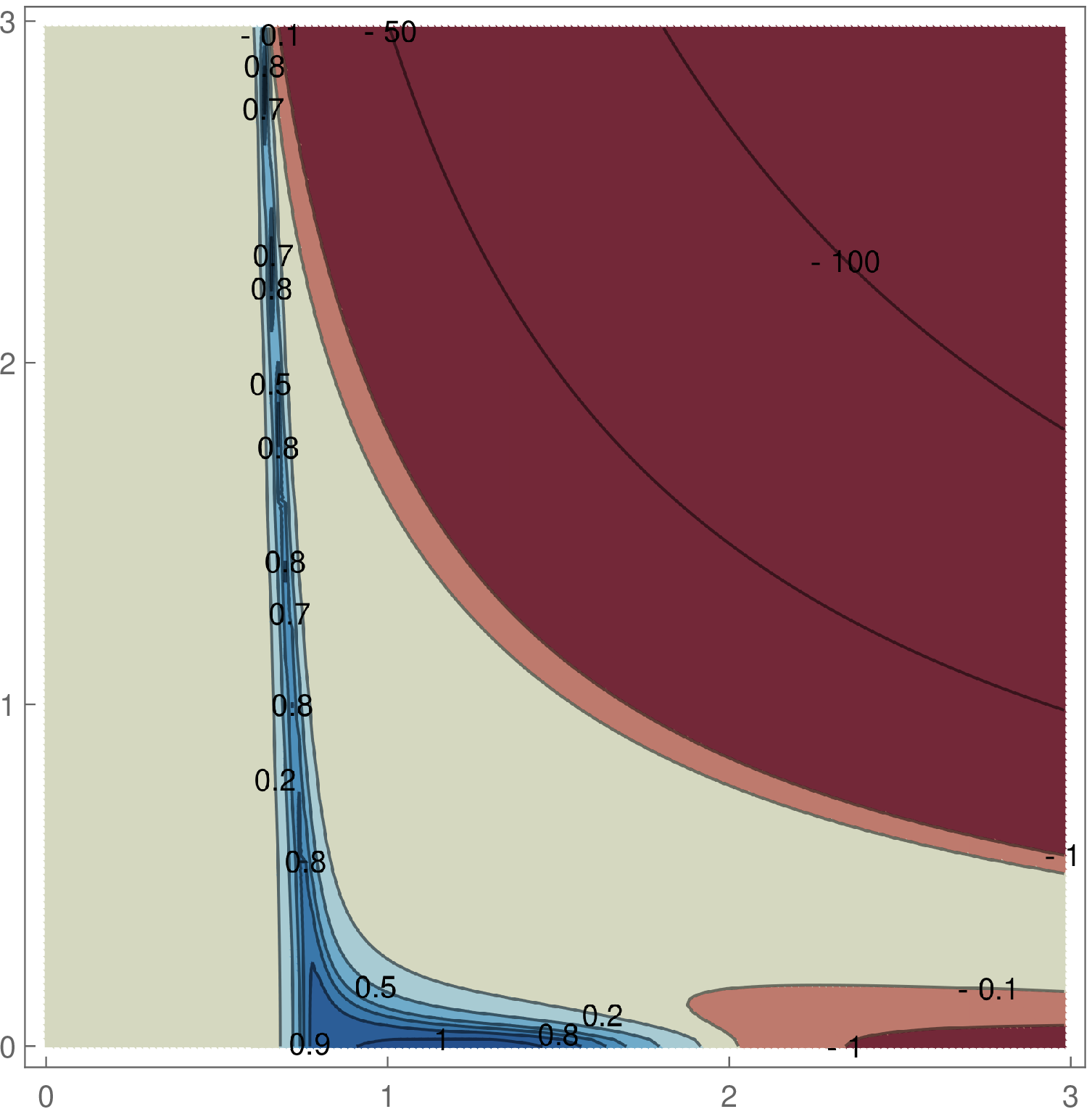} \qquad \includegraphics[scale=0.5]{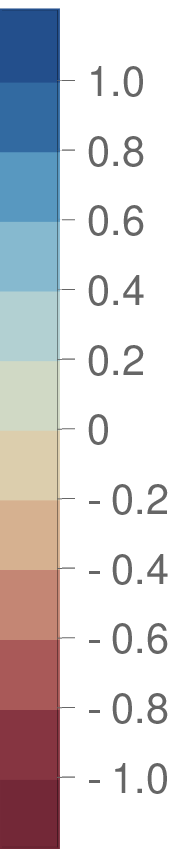}
\end{center}
    \caption{Phase diagram of the web model with $q = {\rm e}^{i \pi/4}$. Conventions are as in Figure~\ref{fig:pd-eipi5}.}
    \label{fig:pd-eipi4}
\end{figure}

We now turn to our second value of $q$, namely $q={\rm e}^{i \pi/4}$, which corresponds to $[3]_q=1$.
The phase diagram is shown in Figure~\ref{fig:pd-eipi4}. The structure is very similar to the preceeding case, with the two non-critical regions
having similar characteristics.  We again observe the presence of a dilute critical phase, this time with $c_{\text{eff}}\approx0.8$. The corresponding
fixed point can in fact be identified, in this particular case, with the ``special point'' given by  \cite[eq.\,(6)]{LGJ21}. It was shown in that reference
that fixing $y z = 2^{-1/2}$, corresponding here to the horizontal line $y = 2^{-1/4} \simeq 0.841$, makes the web model equivalent to the three-state Potts model
on the triangular lattice $\mathbb{T}$, dual to $\mathbb{H}$.
The position of the critical point along this line is known \cite{BTA78}.
In terms of the variable $v = {\rm e}^J - 1$, it is the (unique) positive solution of $v^3 + 3 v^2 = 3$, where $J$ denotes the usual
coupling between nearest-neighbour spins. The corresponding weight of a piece of domain wall on the lattice dual to~$\mathbb{T}$, hence
$\mathbb{H}$ again, is finally $x = {\rm e}^{-J} \simeq 0.532$. So the critical three-state Potts model with $c=\frac{4}{5}$ is situated in our phase diagram
at $(\sqrt{x},y) \simeq (0.729,0.841)$, in fine agreement with Figure~\ref{fig:pd-eipi4}.  Because of the high symmetry of this critical model,
it is tempting to conjecture that it may act as an attractive fixed point controlling the whole dilute
critical curve.

The other fixed point of interest along the three-state Potts line, $y = 2^{-1/4}$ is situated at infinite temperature, i.e., $J=0$ and $x = 1$.
At this fixed point all the lattice sites are coloured independently with uniform probability, using the three colours. Although this is a trivial fixed
point from the point of view of the spin degrees of freedom, it may cause the corresponding geometrical description in terms of domain walls
to exhibit critical fluctuations (the infinite-temperature Ising model and site percolation are similarly related). We can infer from this that
the point $(\sqrt{x},y) = (1,2^{-1/4})$ has central charge $c=0$, and it is conceivable that this may in fact be the attractive fixed point controlling the dense critical phase. Indeed, the latter phase is seen to have $c_{\text{eff}}\approx0$ from Figure~\ref{fig:pd-eipi4}.

Finally, on the horizontal axis $y=0$, we observe a set of critical points with $c_{\text{eff}}\approx1$. This can be explained by the corresponding
loop model having loop weight $N=2$. Indeed, for this loop model the dense and dilute fixed points coincide and are situated at 
$x_{\rm c} = 2^{-1/2}$ according to \eqref{xc_loop}, whence $\sqrt{x_{\rm c}} = 2^{-1/4} \simeq 0.841$. The corresponding central charge is $c=1$
indeed. For $y=0$ and $x > x_{\rm c}$ the loop model remains in the dense phase with $c=1$, and critical exponents that are independent of $x$ \cite{BN89pd}.
This $c=1$ line is visible in figure~\ref{fig:pd-eipi4}. However, the numerical data seem to indicate that the $c=1$ line terminates at a finite value of $\sqrt{x}$,
but since this is inconsistent with the analytical argument,  it must be a finite-size artifact.
By the $c$-theorem, the RG flows are orthogonal to the contour lines of constant $c_{\rm eff}$. This appears to be consistent with an RG flow
from the $c=1$ point at $(\sqrt{x_{\rm c}},0) \simeq (0.841,0)$ towards the dilute critical phase with $c=\frac{4}{5}$.

\medskip

For the third and last value of $q$, namely $q={\rm e}^{i \pi/3}$ corresponding to $[3]_q=0$,
the phase diagram is similar to the previous ones, with the presence of a dilute phase at $c_{\text{eff}}\approx0$ and a dense phase at $c_{\text{eff}}\approx -2$.
These values of the effective central charges that we have read from our numerical investigation are in fact exact.
This will be shown thanks to a Coulomb Gas description of these phases in our subsequent paper \cite{LGJ-CG}.

\section{Electromagnetic operators in the $U_{-q}(\mathfrak{sl}_3)$ vertex model}
\label{sec:em}

In this section we define modified partition functions of the loop model and the Kuperberg web model. We denote them by $Z_{\text{loop}}^{\bm{e},\bm{m}}$ (respectively $Z_{\rm K}^{\bm{e},\bm{m}}$) in the cylinder geometry and $Z_{\text{loop}}^{\bm{m}}$ (respectively $Z_{\rm K}^{\bm{m}}$) in the strip geometry. 

At the critical points of the $O(N)$ loop model, these objects are well known. In the cylinder geometry, when $\frac{M}{L}$ becomes large, one has the asymptotic equivalent \cite{Cardy_x}
\begin{align}
\label{scalingloopcylinder}
    \frac{Z_{\text{loop}}^{\bm{e},\bm{m}}}{Z_{\text{loop}}}\sim {\rm e}^{-\frac{\sqrt{3}\pi M}{4L}(h_{\bm{e},\bm{m}}+\Bar{h}_{\bm{e},\bm{m}})} \,,
\end{align}
where $(h_{\bm{e},\bm{m}},\Bar{h}_{\bm{e},\bm{m}})$ are the conformal weights of the so-called electromagnetic operators of electric and magnetic charges, $\bm{e}$ and $\bm{m}$ \cite{LoopReview}. In the usual field-theory normalisation the prefactor in the exponential would be $\frac{2 \pi  M}{L}$,
but the aspect ratio must here be modified in order to account for the specific choice of lattice $\mathbb{H}$. 
Recall that Figure~\ref{fig:transfermatrix} depicts two rows and $2L$ columns. Hence, in the presence of $2M$ rows, the aspect ratio is given by $\frac{\sqrt{3}M}{4L}$ because the height of an equilateral triangle of side $1$ is $\frac{\sqrt{3}}{2}$.

The electromagnetic operators are described in the Coulomb Gas (CG) formulation of the continuum limit of the loop model. In this picture, $Z_{\text{loop}}^{\bm{e},\bm{m}}$ is the lattice version of the CG partition  function with a pair of electromagnetic operators inserted at the bottom and top ends of the cylinder.

In the strip geometry, one has instead
\begin{align}
\label{scalingloopstrip}
    \frac{Z_{\text{loop}}^{\bm{m}}}{Z_{\text{loop}}}\sim {\rm e}^{-\frac{\sqrt{3}\pi M}{8L}h_{\bm{m}}} \,,
\end{align}
where $h_{\bm{m}}$ is the conformal weight of the boundary magnetic operator of charge $\bm{m}$. One can again look at $Z_{\text{loop}}^{\bm{m}}$ as the lattice version of a partition function modified by the insertion of magnetic operators at both ends of the strip. Thus, we will borrow the vocabulary of electromagnetic operators when discussing these lattice modified partition functions.

These scaling formulae are similar in the case of Kuperberg webs. Although a CG description of the web model will only be given in a subsequent paper \cite{LGJ-CG},
it is appropriate to discuss the precursors of field-theory operators in the context of the lattice model. In this section we therefore consider the modification of
the partition function due to the insertion of a pair of electromagnetic operators. The scaling formulae
\begin{subequations}
\begin{eqnarray}
    \frac{Z_{\rm K}^{\bm{e},\bm{m}}}{Z_{\rm K}} &\sim& {\rm e}^{-\frac{\sqrt{3}\pi M}{4L}(h_{\bm{e},\bm{m}}+\Bar{h}_{\bm{e},\bm{m}})} \,, \\
    \frac{Z_{\rm K}^{\bm{m}}}{Z_{\rm K}} &\sim& {\rm e}^{-\frac{\sqrt{3}\pi M}{8L}h_{\bm{m}}} \,,
\end{eqnarray}
\end{subequations}
valid for the cylinder and strip geometries respectively, then
define conformal weights of electromagnetic excitations at critical points of the web model.

The aim of this section is to elaborate on the definition of such electromagnetic partition functions and to provide their geometrical interpretation.
For this reason, we shall sometimes refer to the modifications of the partition functions as the insertions of {\em geometrical defects}.
We shall treat the loop and web models in parallel, discussing first the easiest case of the strip, before moving on to the cylinder geometry.

\subsection{The strip geometry}
\subsubsection{Magnetic defects}

As we have seen in Section \ref{sec:kuploc}, the row-to-row transfer matrix of the $O(N)$ loop model in the strip geometry possesses a symmetry under $U_{-q}(\mathfrak{sl}_2)$. The Hilbert space therefore decomposes in weight subspaces, i.e., eigenspaces of the Cartan element $H_1$. Let $\mathcal{R}_2^*$ be the weight lattice of $\mathfrak{sl}_2$, dual to the root lattice $\mathcal{R}_2=\mathbb{Z}\bm{\alpha}_1$ which is generated by $\bm{\alpha}_1$. The weight lattice is also generated by one vector, $\bm{w}_1$ satisfying $(\bm{w}_1,\bm{\alpha}_1)=1$,%
\footnote{See Appendix~\ref{sec:quantumgroupconventions} for our conventions on the scalar product $(\ ,\ )$.}
called the fundamental weight, that is, $\mathcal{R}_2^*=\mathbb{Z}\bm{w}_1$.
A weight vector  of weight $\bm{m}=n\bm{w}_1$, with $n$ integer, is an eigenstate of $H_1$ with eigenvalue $(\bm{m},\bm{\alpha}_1)=n$. 
In the $SU(2)$ spin projection notations it corresponds to the spin $n/2$.

The eigenspace of $H_1$ comprised of weight vectors of weight $\bm{m}$ will be called a \textit{sector}. It is stable under the action of the transfer matrix and contains excitations that are lattice precursors of the ones created by magnetic operators in the Coulomb Gas formalism. We call a \textit{magnetic defect} state $\ket{\bm{m}}$ (or simply magnetic defect) of \textit{magnetic charge} $\bm{m}$, a pure tensor state in the sector of weight $\bm{m}$, such that any two sites labelled by $2i$ and $2i+1$ cannot be occupied (i.e., non-empty) simultaneously and there are exactly $|n|$ occupied sites. Here are some examples with $\bm{m}=n\bm{w}_1$:
\begin{subequations}
\label{loopmagneticdefect}
\begin{eqnarray}
\label{loopmagneticdefectconv1}
    \ket{\bm{m}} &=&  (\ket{\uparrow}\otimes\ket{\ })^{\otimes n}\otimes \ket{\ }^{\otimes 2L-2n} \,, \quad \quad \mbox{if } n\geq 0 \,, \\
\label{loopmagneticdefectconv2}
    \ket{\bm{m}} &=& (\ket{\downarrow}\otimes\ket{\ })^{\otimes |n|}\otimes \ket{\ }^{\otimes 2L-2|n|} \,, \quad \mbox{if } n\leq 0 \,.
\end{eqnarray}
\end{subequations}
The dilution of the insertion sites is required in order to avoid a trivial propagation. For instance, the state $\ket{\uparrow}^{\otimes 3}\otimes \ket{\ }^{\otimes 2L-3}$ is not a magnetic defect, as it is mapped to $0$ by the transfer matrix. 

The partition function modified by the insertion of the magnetic defect is then
\begin{align}
\label{loopMPF}
    Z_{\text{loop}}^{\bm{m}}=\bra{\bm{m}}T_{\text{loop}}^M\ket{\bm{m}} \,,
\end{align}
with $T_{\text{loop}}$ being defined in \eqref{looptransfermatrixstrip}.
Because any magnetic defect of charge $\bm{m}$ becomes a magnetic defect of charge $-\bm{m}$ under the action of raising and lowering operators, $E_1$ and $F_1$, the sectors of opposite magnetic charges contain the same excitations. It is thus possible to focus only on positive magnetic charges.

The different choices for $\ket{\bm{m}}$ are physically equivalent. Every magnetic defect state having a non-zero overlap with the dominant eigenvector (that eigenvector of the transfer matrix whose eigenvalue is the largest in norm) will lead to the same scaling behaviour \eqref{scalingloopcylinder}--\eqref{scalingloopstrip}. We believe that, in the loop models, every magnetic defect state has a non-zero overlap with the dominant eigenvector. 

\begin{figure}
\begin{center}
    \includegraphics[scale=0.3]{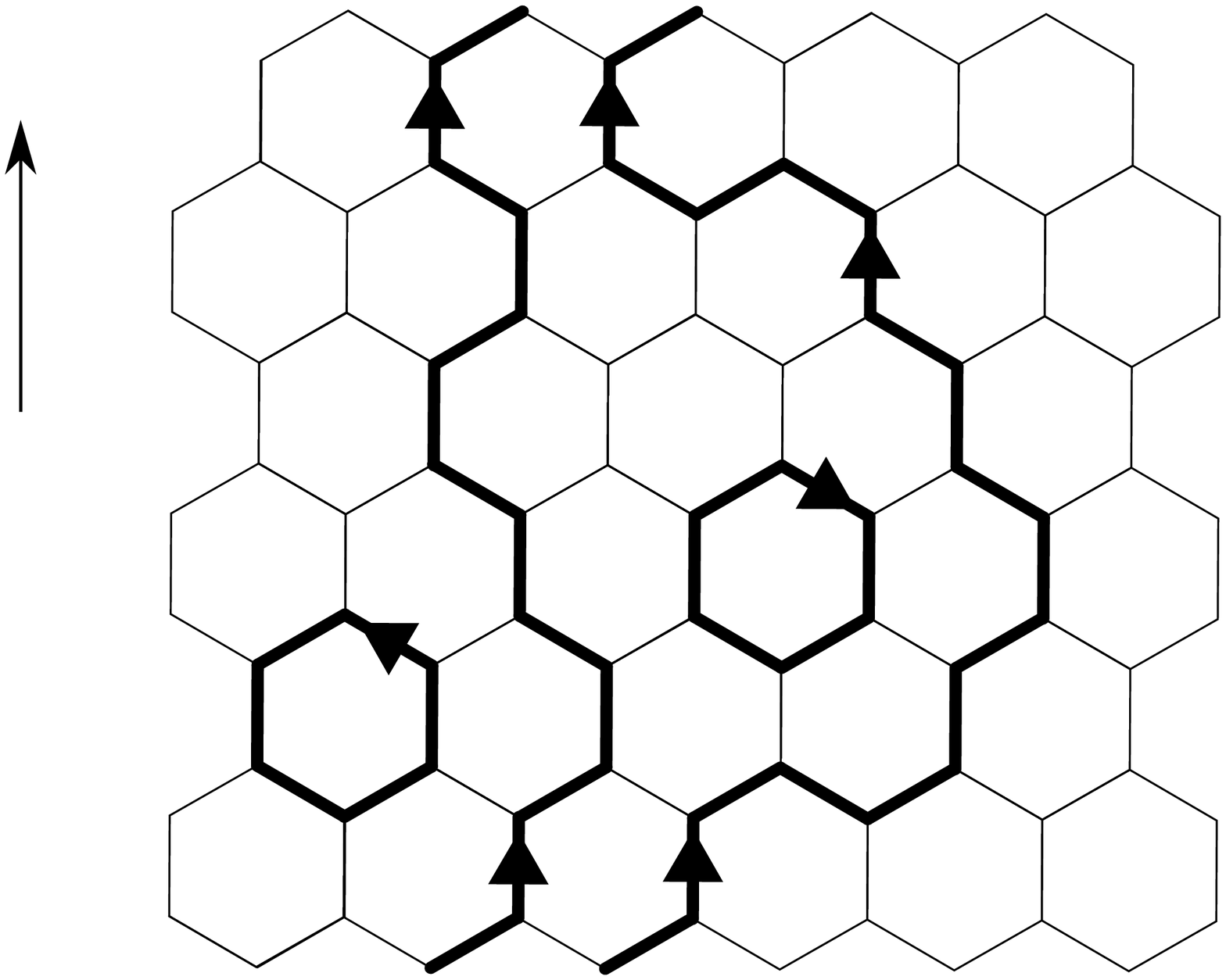}\qquad\qquad \includegraphics[scale=0.3]{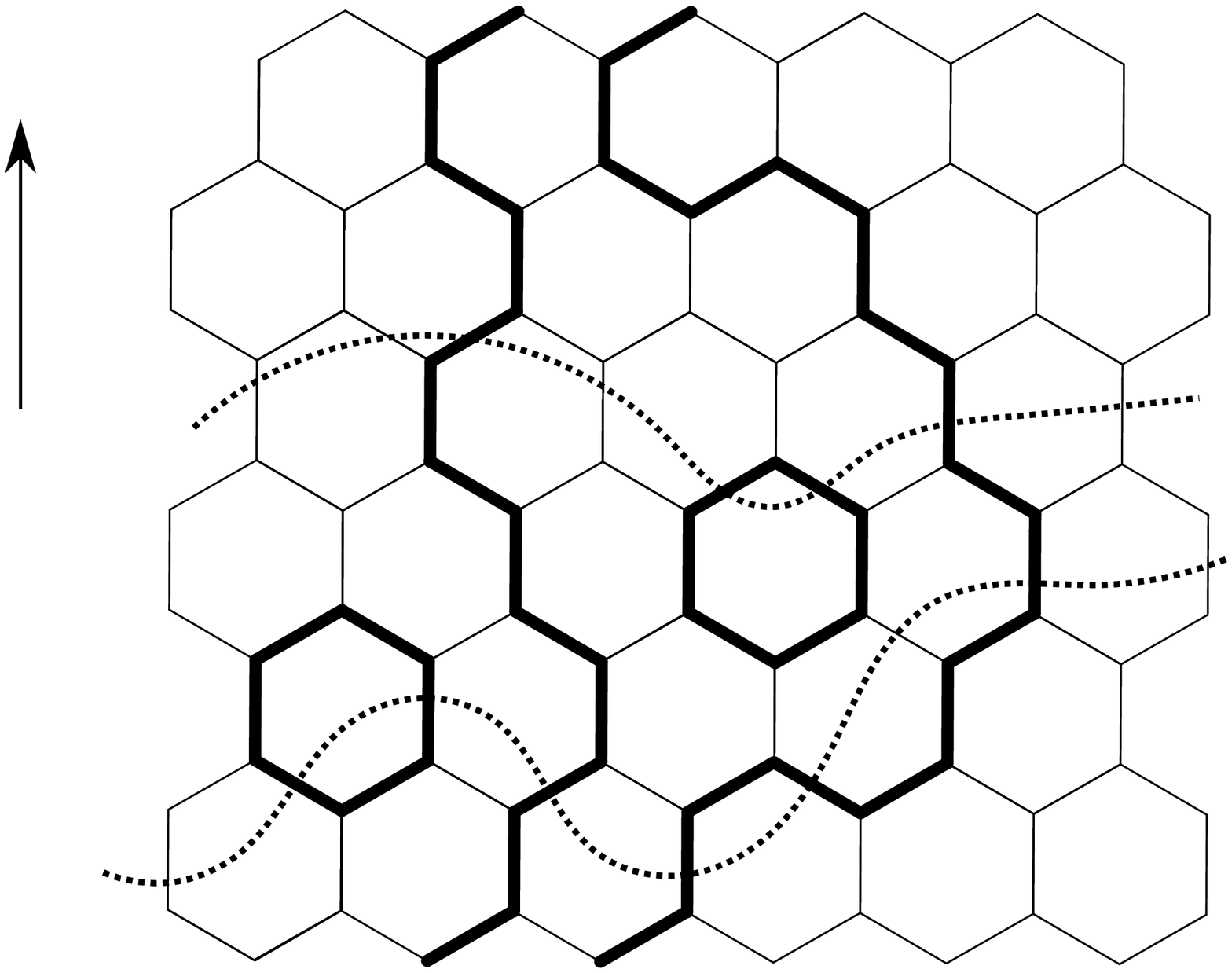}
\end{center}
    \caption{Left panel: A configuration in $\mathcal{L}_{\rm col}^{\bm{m}}$ for $\bm{m}=2\bm{w}_1$ in the strip geometry. Right panel: A configuration in $\mathcal{L}^{\bm{m}}$ obtained from the left one by forgetting orientations. We also show two examples of cuts.}
    \label{fig:loopmagdefectstrip}
\end{figure}

One can write \eqref{loopMPF} as a sum over trajectories of transition amplitudes. Denote by $\mathcal{L}_{\rm col}^{\bm{m}}$ the set of coloured (cf.\ Remark~\ref{rem:1}) subgraphs of $\mathbb{H}$ whose connected components are either coloured loops or coloured lines, such that loops cannot touch the bottom and top boundaries, and lines touch the bottom and top boundaries only at their end points corresponding to the occupied sites in $\ket{\bm{m}}$ and with the inherited orientations; see Figure \ref{fig:loopmagdefectstrip} for an example with $\ket{\bm{m}}=\ket{\ }^{\otimes 3} \otimes \ket{\uparrow} \otimes \ket{\ } \otimes \ket{\uparrow} \otimes \ket{\ }^{\otimes 6}$. We call these configurations  \textit{coloured}. We have then 
\begin{align}
    Z_{\text{loop}}^{\bm{m}}=\sum_{c\in \mathcal{L}_{\rm col}^{\bm{m}}} w_{\rm col}^{\bm{m}}(c) \,,
\end{align}
where the weight $w_{\rm col}^{\bm{m}}(c)$ is given by the local rules \eqref{loopvertexweights}. In any given row, the number of arrows pointing upward minus the number of arrows pointing downward is conserved, manifesting the magnetic charge conservation. Remark that, as in the case without defects \eqref{loopPF-TM}, the modified partition function can be interpreted as one for a loop model on a lattice with two rows less. This is because the degrees of freedom are completely constrained on the first and last rows, due to our choice of magnetic defect state. Yet, it appears more convenient to keep working with the model defined by \eqref{loopMPF} on a lattice with two more rows. We will do the same in the other settings of loop or Kuperberg web models in the strip or cylinder geometry.

\medskip

In the case of the Kuperberg web model, and still in the strip geometry, the row-to-row transfer matrix of the local formulation is symmetric under $U_{-q}(\mathfrak{sl}_3)$. Hence we define magnetic charges belonging to the weight lattice $\mathcal{R}_3^*$ of $\mathfrak{sl}_3$. It is generated by two fundamental weights, $\bm{w}_1$ and $\bm{w}_2$. 

Again, by using the action of raising and lowering operators, it is enough to focus on magnetic charges inside the fundamental Weyl chamber, i.e, dominant weights
\begin{align}
    \bm{m}=n_1\bm{w}_1+n_2\bm{w}_2 \,,
\end{align}
where $n_1$ and $n_2$ are non-negative integers. We call a  magnetic defect state of charge $\bm{m}$ a pure tensor state of weight $\bm{m}$, such that sites labelled by $2i$ and $2i+1$ are not both occupied simultaneously, and there are exactly $n_1+n_2$ occupied sites. Here is an example:
\begin{align}
\label{kupmodelmagdefect}
    \ket{\bm{m}}=(\ket{\color{red} \uparrow}\otimes\ket{\ })^{\otimes n_1} \otimes (\ket{\color{green} \downarrow}\otimes\ket{\ })^{\otimes n_2} \otimes  \ket{\ }^{\otimes 2L-2n_1-2n_2} \,.
\end{align} 
To understand the choice of colours, recall that whatever the inclination of the link ($\diagdown$, $\diagup$ or $|$), by the relations \eqref{reptocolor1}--\eqref{reptocolor3}, $\ket{\color{red} \uparrow}$ has weight $\bm{w}_1$ and $\ket{\color{green} \downarrow}$ has weight $\bm{w}_2$. In fact, for any magnetic defect, there are necessarily $n_1$ sites occupied by upward oriented red arrows, $n_2$ sites occupied by downward oriented green arrows, and no blue arrows.
As in the loop model case, the occupied sites have to respect some dilution in order to avoid being mapped to zero by the transfer matrix. Therefore a
state such as $\ket{\color{red}\uparrow}^{\otimes 3}\otimes \ket{\ }^{\otimes 2L-3}$ is not a magnetic defect, according to the above definition.

Different choices for $\ket{\bm{m}}$ having a non-zero overlap with the transfer matrix eigenvector whose eigenvalue is the largest in norm are physically equivalent. Based on experience with the loop models we expect such a non-zero overlap to hold for any magnetic defect state. The partition function modified by the presence of the magnetic defect is
\begin{equation}\label{ZKm-T}
    Z_{\rm K}^{\bm{m}}=\bra{\bm{m}}T_{\rm K}^M\ket{\bm{m}} \,,
\end{equation}
with $T_{\rm K}$ defined in \eqref{kuptransfermatrixstrip}.

As in the loop models case, the next step is to rewrite $Z_{\rm K}^{\bm{m}}$ in terms of coloured open web configurations. We begin with a definition:
\textit{an open Kuperberg web in a rectangle}\footnote{For brevity, we will call it just `open web'.} is a planar oriented bipartite graph with trivalent and univalent vertices such that the univalent vertices are only at the top and bottom boundaries of the rectangle. Assume $c$ is an open web with $2(n_1+n_2)$ univalent vertices such that the following holds for both bottom and top boundaries: $c$ has $n_1$ (respectively $n_2$) upward (respectively downward) oriented edges incident on the univalent vertices.
A {\em three-colouring} of such an open web $c$ is a map from the edges of $c$ into the set $\{\rm red, blue, green\}$, such that all three colours are present around any trivalent vertex, and such that at every boundary side the $n_1$ (respectively $n_2$) upward (respectively downward) oriented edges  incident on the univalent vertices are red (respectively green). 

\begin{figure}
\begin{center}
    \includegraphics[scale=0.3]{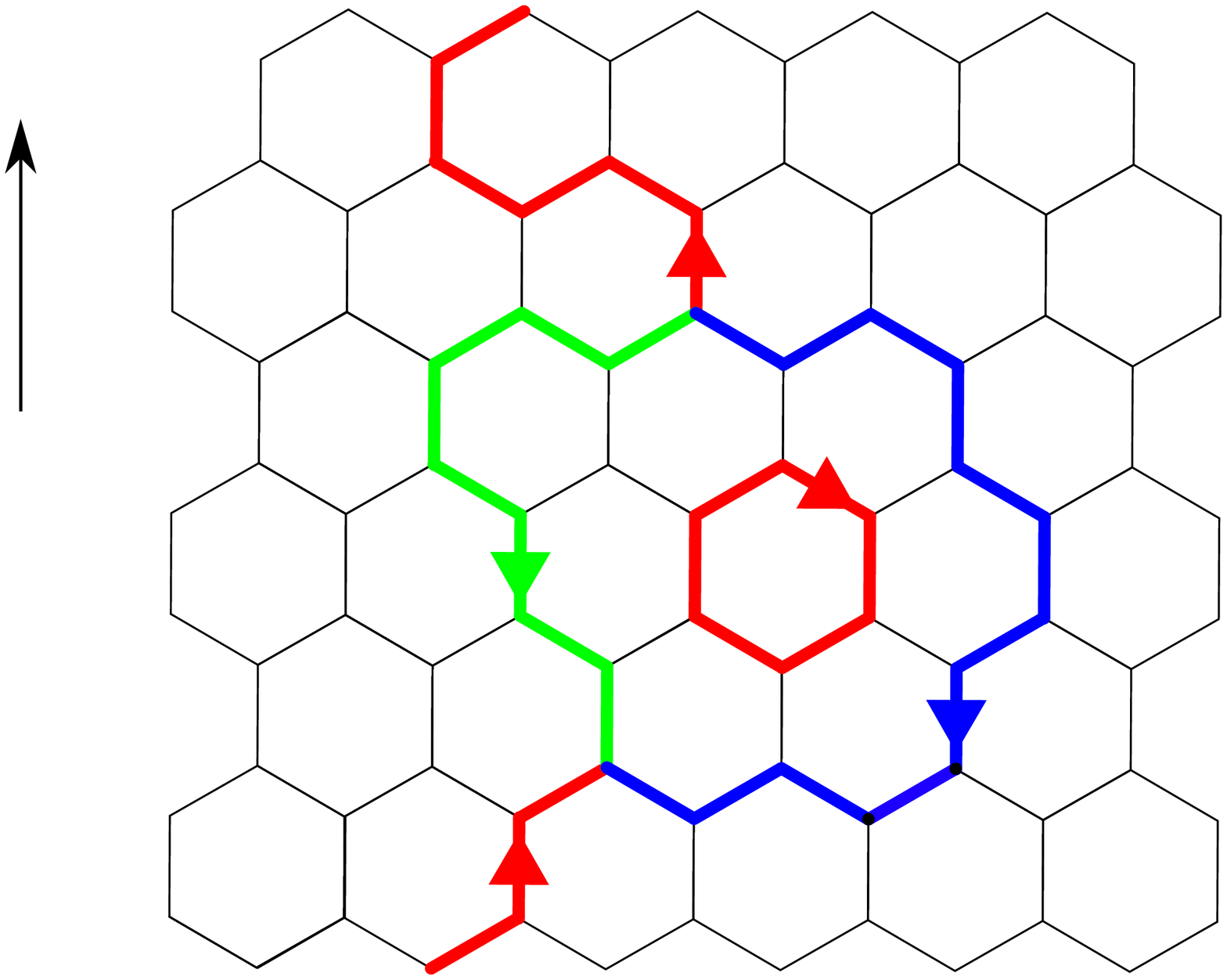}\qquad\qquad \includegraphics[scale=0.3]{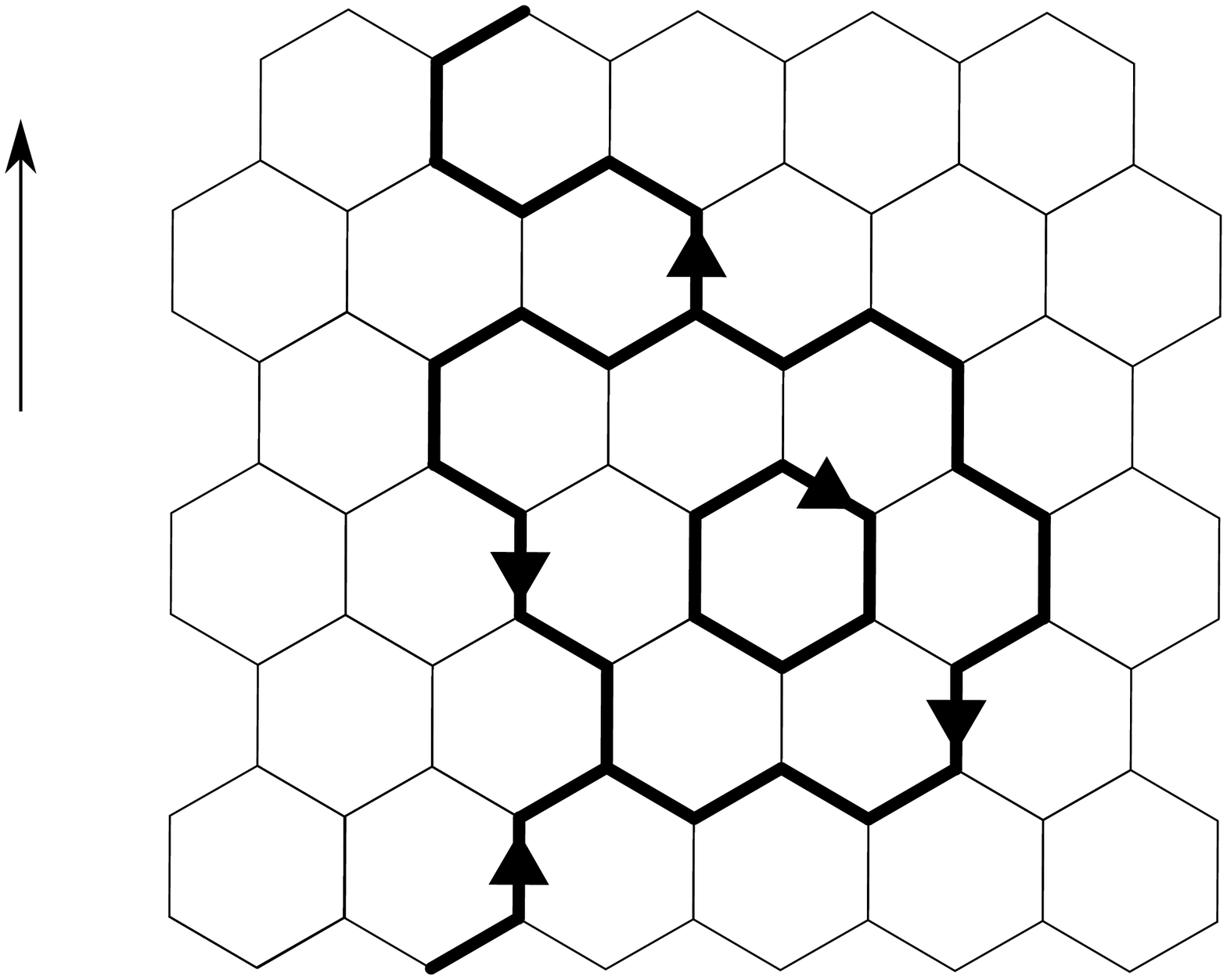}
\end{center}
    \caption{Left panel: A configuration of $\mathcal{K}_{\rm col}^{\bm{m}}$ for $\bm{m}= \bm{w}_1$ in the strip geometry. Right panel: A configuration of $\mathcal{K}^{\bm{m}}$ obtained from the left one by forgetting colours.}
    \label{fig:kupmagdefectstrip}
\end{figure}

Now, we can  rewrite the partition function~\eqref{ZKm-T} in terms of the coloured open webs (i.e., via amplitudes of trajectories):
\begin{align}\label{ZKm-T-col}
    Z_{\rm K}^{\bm{m}}=\sum_{c\in \mathcal{K}_{\rm col}^{\bm{m}}} w_{\rm col}^{\bm{m}}(c) \,,
\end{align}
where $\mathcal{K}_{\rm col}^{\bm{m}}$ denotes the set of subgraphs of $\mathbb{H}$ whose connected components are either open or closed three-coloured webs, such that webs cannot touch the bottom and top boundaries, except for open webs that 
have their univalent vertices  at  the occupied sites in $\ket{\bm{m}}$ with the  colours and orientations inherited from $\ket{\bm{m}}$.
The weight of a coloured configuration $w_{\rm col}^{\bm{m}}(c)$ is given by the local weights~\eqref{kupvertex}, in the case $x_1=x_2=x$ (or by~\eqref{kupvertex-gen} in the general case).
A sample configuration is shown in Figure \ref{fig:kupmagdefectstrip} for $\bm{m}=\bm{w}_1$.
We remark that the equality between~\eqref{ZKm-T} and~\eqref{ZKm-T-col} follows straightforwardly from the construction of the transfer matrix $T_{\rm K}$ which matrix elements are given by the expressions from~\eqref{kupvertex}-\eqref{kupvertex-gen}.

\subsubsection{Geometrical interpretation}
\label{sec:StripGeoInterpretation}

We now give a geometrical interpretation of the magnetic partition functions defined in the last subsection. More precisely, we show how to define and evaluate a non-coloured open loop or Kuperberg configuration (similar to the closed case, but with modified rules) such that we recover $Z_{\rm loop}^{\bm{m}}$ or $Z_{\rm K}^{\bm{m}}$, respectively. We also describe how such configurations are geometrically constrained. We begin with the loop model. 

The idea is to group coloured configurations in $\mathcal{L}_{\rm col}^{\bm{m}}$ that differ only by the colours (also called `orientations' prior to Remark~\ref{rem:1}) of loops. This is exactly what one does when going from the local vertex model to the non-local geometrical loop model. The difference with the usual argument for $Z_{\rm loop}$ is the presence of lines connected to the boundary, where their colour is fixed by the choice of magnetic defect $\ket{\bm{m}}$. In other words, the boundary-touching elements in $\mathcal{L}_{\rm col}^{\bm{m}}$ have constrained colours. Define by $\mathcal{L}^{\bm{m}}$ the set of subgraphs of $\mathbb{H}$ obtained by forgetting the colourings of
the non-constrained elements of $\mathcal{L}_{\rm col}^{\bm{m}}$. By summing the contributions coming from unconstrained colourings, we obtain 
\begin{align}
\label{loopStripGeoPF}
    Z_{\rm loop}^{\bm{m}}=\sum_{c\in \mathcal{L}^{\bm{m}}} w^{\bm{m}}(c) \,,
\end{align}
where the weight $w^{\bm{m}}(c)$ is the product of a non-local weight $q+q^{-1}$ for each loop and a fugacity $x$ for each monomer. Indeed, as we shall see, the open lines contribute to the weight only by the fugacities of the bonds they cover. An example of configuration in $\mathcal{L}^{\bm{m}}$ is given in Figure \ref{fig:loopmagdefectstrip}.

\medskip

In Section \ref{sec:localtmKup} we have seen that graphs in the $O(n)$ loop model can be understood as intertwiners of $U_{-q}(\mathfrak{sl}_2)$ representations. In this picture, we can think of a bond as the propagation of states inside the fundamental representation. It is then apparent that the insertion of a non-trivial magnetic defect will constrain the geometry of the configurations due to the condition of keeping unchanged the Cartan weight of a propagated state. 

More precisely, define a \textit{cut} as a smooth curve crossing the strip from left to right such that it avoids nodes and its projection onto the horizontal axis is injective (no overhangs). Some examples of cuts are depicted in Figure \ref{fig:loopmagdefectstrip}. A cut defines a Hilbert space that is the tensor product of the local Hilbert spaces of the links it crosses. The evolution operator between two disjoint cuts is a product of local transfer matrices. The row-to-row transfer matrix is a special case of such an evolution operator. A cut on a colored configuration defines a pure tensor state in the basis of up/down arrows. This pure tensor state is an eigenvector of the Cartan subalgebra, i.e.\ a weight vector. Moreover it has nonzero overlap with the evolution (by transfer matrices) of the magnetic defect state, which is of Cartan weight $\bm{m}$ by symmetry of the local transfer matrices. As two weight vectors of different Cartan weights must have zero overlap we conclude that the Cartan weight of the pure tensor state on any cut is equally $\bm{m}$.
The presence of $p$ bonds on a given cut indicates that the pure tensor state is a vector of the representation $V^{\otimes p}$. Hence, on any given cut, the magnetic charge $\bm{m}=n\bm{w}_1$ should satisfy
\begin{align}
    \bm{m}\preceq p\bm{w}_1 \,,
\end{align}
where $\preceq$ denotes the partial ordering on weights. Equivalently
\begin{align}
\label{loopconstraint}
    n \leq p \,, \quad \mbox{with }
    n \equiv p \text{ mod } 2 \,.
\end{align}

We note that it is insufficient to apply the constraint~\eqref{loopconstraint} on cuts intersecting only vertical links (i.e., on completed rows). 
This can be seen from the example
\begin{center}
    \includegraphics[scale=0.3]{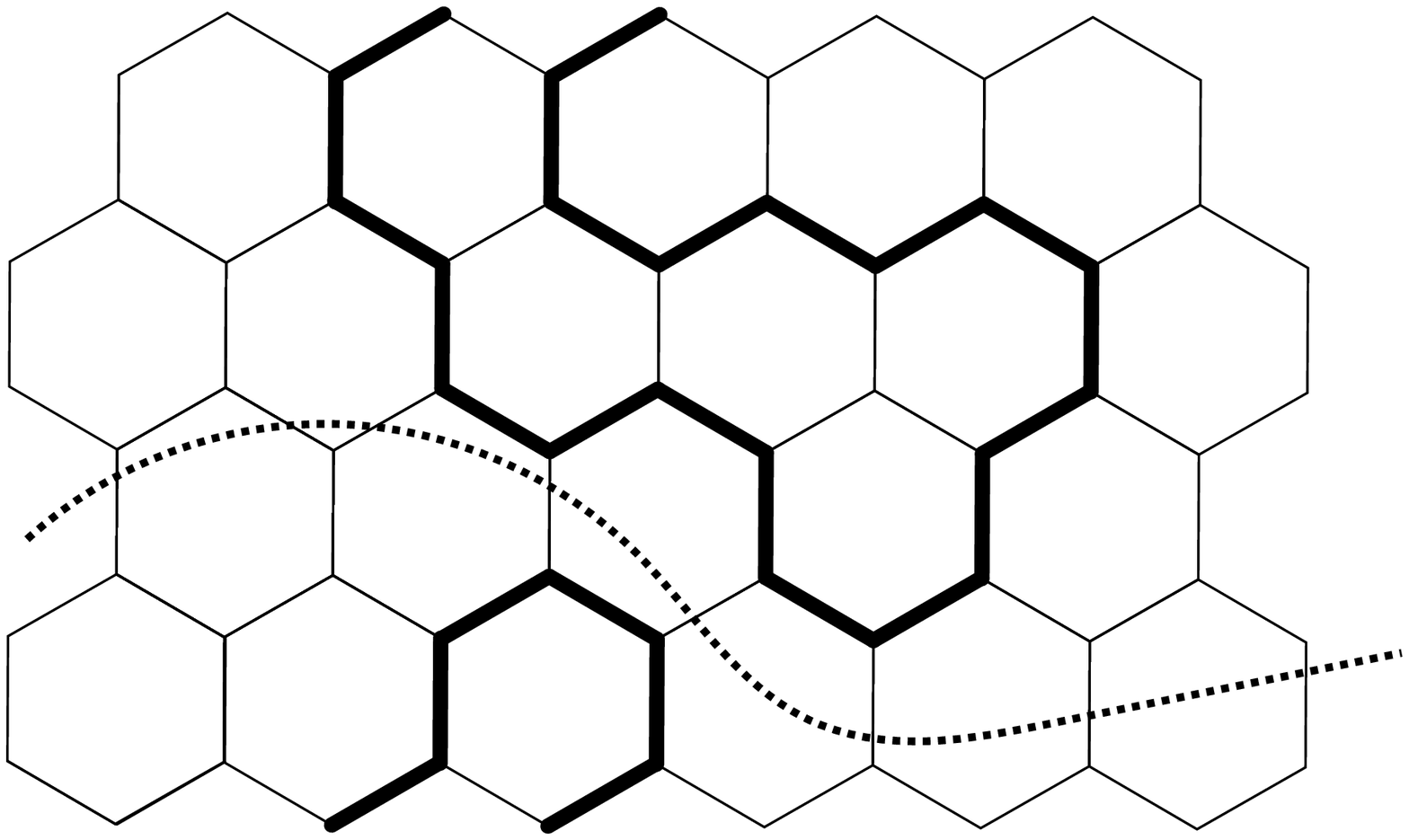}
\end{center}
where \eqref{loopconstraint} is satisfied (with $n=2$) on each completed row, but not on the cut depicted. This is why we impose \eqref{loopconstraint} on any cut. This stronger constraint imposes that each line connected to the bottom boundary is also connected to the top boundary; we call such lines \textit{through-lines}. Since the through-lines enter and leave the strip with the same inclination (due to our choice of using the same magnetic defect $\ket{\bm{m}}$ as initial and final state), they do not pick up any powers of $q$ from bending. Hence the through-lines contribute to the weight of a configuration in \eqref{loopStripGeoPF} only by the fugacities of the bonds that they cover.

\medskip

We are now ready to discuss the Kuperberg web model.
Denote by $\mathcal{K}^{\bm{m}}$ the set of webs embedded in $\mathbb{H}$ that are obtained from elements of $\mathcal{K}_{\rm col}^{\bm{m}}$ by forgetting their colourings. See Figure \ref{fig:kupmagdefectstrip} for an example. Remark that the colours of the edges connected to the boundary are constrained
by the choice of magnetic defect. Set $\bm{m}=n_1\bm{w}_1+n_2\bm{w}_2$, with $n_1, n_2\in \mathbb{Z}_{\geq 0}$. This boundary condition imposes constraints on the possible three-colourings of $c$. Different configurations $c_i$ in $\mathcal{K}_{\rm col}^{\bm{m}}$, that produce the open web~$c$ once one forgets their colourings, are exactly the three-colourings of $c$. By summing over their weight we obtain
\begin{align}
 \label{ZKstrip}
    Z_{\rm K}^{\bm{m}}=\sum_{c\in \mathcal{K}^{\bm{m}}} w^{\bm{m}}(c) \,,
\end{align}
with 
\begin{align}
\label{webweightstrip}
    w^{\bm{m}}(c)=\sum_i w^{\bm{m}}_{\rm col}(c_i) \,.
\end{align}
where the sum is over all three-colourings of $c$.

\smallskip

We now wish to understand \eqref{ZKstrip}--\eqref{webweightstrip} without making reference to colourings.
To this end, we first describe more closely what is the set $\mathcal{K}^{\bm{m}}$.
By analogy with the loop model case, we examine what the insertion of a defect of charge $\ket{\bm{m}}$ implies for the geometry of open webs. The Cartan weight is conserved between two cuts as evolution operators commute with the $U_{-q}(\mathfrak{sl}_3)$ action. 
Indeed, any given cut of a colored configuration gives a pure tensor state whose overlap with the evolution of the magnetic state is nonzero. States with different Cartan weights have no overlaps, hence the pure tensor state has Cartan weight $\bm{m}$. It therefore must belong to a direct summand (in the Hilbert space)  whose highest weight is higher or equal to $\bm{m}$. This means that, on a given cut, the numbers $p_1$ and $p_2$ of bonds pointing upward and downward, respectively, satisfy
\begin{align}\label{kupmodelconstraint-gen}
    \bm{m}\preceq p_1\bm{w}_1+p_2\bm{w}_2
\end{align}
with respect to the partial ordering on weights. Equivalently, this can be written
\begin{subequations}
\label{kupmodelconstraint}
\begin{align}
    2n_1+n_2&\leq 2p_1+p_2 \,, \qquad \mbox{with } 2n_1+n_2\equiv 2p_1+p_2 \text{ mod }3 \,, \\
    n_1+2n_2&\leq p_1+2p_2 \,, \qquad \mbox{with } n_1+2n_2\equiv p_1+2p_2 \text{ mod }3 \,.
\end{align}
\end{subequations}
We define a \textit{minimal cut} to be a cut such that the above constraints are satisfied as equalities.

Now, consider the vector space generated by open webs embedded in the rectangle, up to boundary preserving isotopy, such that there are $n_1$ upward oriented edges and $n_2$ downward oriented edges connected to the bottom (respectively top) boundary. The ordering of these oriented edges can be different on either boundary. We then quotient this space by the Kuperberg relations~\eqref{3rules} and by the rule that a web not satisfying the constraints \eqref{kupmodelconstraint} on any cut crossing the rectangle from left to right is set to zero.  We  call the quotient a space of {\em magnetised webs} of (magnetic) charge $\bm{m}=n_1\bm{w}_1+n_2\bm{w}_2$.\footnote{In the terminology introduced by Kuperberg, this quotient is called a {\em clasped web space} \cite{Kuperberg_1996}.} For instance, the following web is a magnetised web of charge $\bm{w}_1+\bm{w}_2$:
\begin{align}
     \vcenter{\hbox{\includegraphics[scale=0.3]{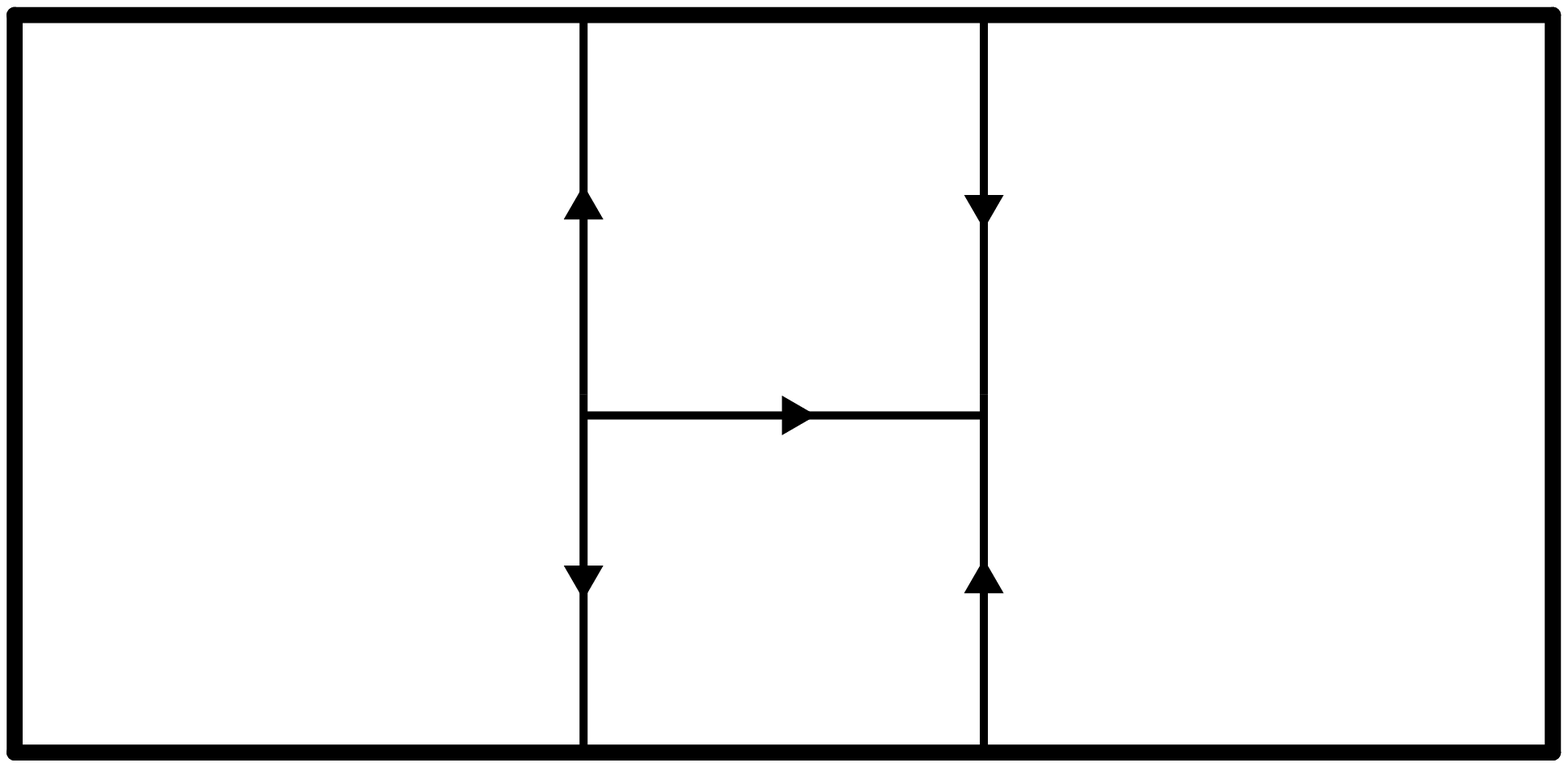}}}
    \label{H-web}
\end{align}
We conclude that the graphs appearing in $\mathcal{K}^{\bm{m}}$ are non-zero elements in the space of magnetised webs of charge $\bm{m}$ with the property that
their oriented edges are arranged in the same order at the top and bottom boundaries (since the same magnetic defect is used as initial and final state).

\smallskip

We next describe how to compute the weight $w^{\bm{m}}(c)$ for a given element $c \in \mathcal{K}^{\bm{m}}$,
without making reference to colourings. The bond and vertex fugacity part is obvious, so we omit this part of the weight from the following argument.
We can reduce a magnetised web to a linear combination of magnetised webs containing strictly less edges by applying the set of relations \eqref{3rules}. Remark that the use of the square rule \eqref{kupsquare} can result in a linear combination containing a web that is not magnetised, which is then set to zero. Hence, we can also think of this reduction as using an extension of the set of rules \eqref{3rules} by
\begin{align}\label{modsquarerule}
    \vcenter{\hbox{\includegraphics[scale=0.2]{diagrams/rel3kup1.eps}}}&\;=\;\vcenter{\hbox{\includegraphics[scale=0.2]{diagrams/rel3kup2.eps}}} \qquad \text{or} \qquad
    \vcenter{\hbox{\includegraphics[scale=0.2]{diagrams/rel3kup1.eps}}}\;=\;\vcenter{\hbox{\includegraphics[scale=0.2]{diagrams/rel3kup3.eps}}}
\end{align}
where the right-hand sides are assumed to form parts of magnetised webs.
A web resulting from the application of these rules that has a minimal number of edges is called {\em irreducible}.

By \cite[Thm.\,6.1]{Kuperberg_1996}, the space of magnetised webs of charge $\bm{m}$ for {\em any} choice of orderings of the bond orientations in the initial and final states has the same dimension as the space $\text{Inv}(V_{\bm{m}}\otimes V_{\bm{m}}^*)$ of $U_{-q}(\mathfrak{sl}_3)$ invariants, where $V_{\bm{m}}$ denotes the irreducible representation of highest weight $\bm{m}$. This latter space is one-dimensional.%
\footnote{The argument uses  duality and Schur's lemma: by definition we have $\text{Inv}(V_{\bm{m}}\otimes V_{\bm{m}}^*) = \mathrm{Hom}_U \bigl(\mathbb{C}, V_{\bm{m}}\otimes V_{\bm{m}}^*\bigr)$, where $U=U_{-q}(\mathfrak{sl}_3)$, and the latter space is isomorphic to $\mathrm{Hom}_U(V_{\bm{m}},V_{\bm{m}})$, by the duality $\mathrm{ev/coev}$ maps, and by  Schur's lemma it is one-dimensional.} 
Therefore, there exists an irreducible magnetised web $c_0$ corresponding to a given pair of initial and final states of charge $\bm{m}$ generating the whole space of magnetised webs.
In particular, in the case of $\mathcal{K}^{\bm{m}}$, one can choose $c_0$ as the web without vertices, i.e., the one
where a collection of $n_1+n_2$ disjoint edges (through-lines) connect the bottom boundary to the top boundary.

As we have seen, the space of  magnetised webs is one-dimensional, thus any magnetised web is proportional to the irreducible web $c_0$ with a given proportionality factor. We define the \textit{magnetised Kuperberg weight} $w_{\rm K}^{\bm{m}}(c)$ of a magnetised web $c$ to be this proportionality factor. Or, equivalently, the irreducible web $c_0$ is weighted by $1$, and this weighting extends linearly to any magnetised web.

\smallskip

We are now equipped to demonstrate
that the product of $w^{\bm{m}}_{\rm K}(c)$ and local fugacities for bonds and vertices is equal to $w^{\bm{m}}(c)$ from \eqref{webweightstrip}.
Consider a configuration $c \in \mathcal{K}^{\bm{m}}$ of charge $\bm{m}=n_1\bm{w}_1+n_2\bm{w}_2$. 
As was shown in Section \ref{sec:kupvertex}, if a loop, a digon or square is present, summing over the weights of the three-colourings is equivalent to applying the rules \eqref{3rules} as long as there is no constraint on the colouring. In fact, only the square rule is sensitive to the colouring constraint. Indeed, a loop is disconnected from the boundary and all its colourings are therefore admissible;
moreover, in the case \eqref{digon-gym} of the digon, even if the colours of the external edges are constrained, the admissible colourings of its internal edges always recover the original digon rule.
For the case \eqref{square-gym} of a square with external edges of the same colour, both diagrams on the right-hand side have zero weight on any cut,
so they do not break the constraint \eqref{kupmodelconstraint}.
On the contrary, for the case \eqref{square2colours} of a square with external edges of two different colours, 
only one diagram of the right-hand side of \eqref{kupsquare} contributes. This happens exactly when decomposing the square would produce a web that breaks the constraint \eqref{kupmodelconstraint}. We have thus shown that summing over all possible three-colourings is equivalent to the use of the additional rule \eqref{modsquarerule}. 

We conclude that by summing over three-colourings of an open web, the local weights \eqref{coloredkupvertex}--\eqref{coloredkupvertexgeneralized} recover the magnetised Kuperberg weight.

\begin{remark}
The use of the additional rule \eqref{modsquarerule} does not occur if $(n_1,n_2)=(1,0)$ or $(n_1,n_2)=(0,1)$ as in the former case (the latter being similar) we have
\begin{align}
\label{kupconstraintexample}
    2&\leq 2p_1+p_2 \,, \qquad \mbox{with } 2\equiv 2p_1+p_2 \text{ mod }3 \,, \nonumber\\
    1&\leq p_1+2p_2 \,, \qquad \mbox{with } 1\equiv p_1+2p_2 \text{ mod }3
\end{align}
on any cut. The use of the square rule on a given cut crossing $(p_1,p_2)$ edges implies $p_1>0$ and $p_2>0$. This implies
\begin{align}
    5&\leq 2p_1+p_2 \,, \qquad \mbox{with } 2\equiv 2p_1+p_2 \text{ mod }3 \,, \nonumber\\
    4&\leq p_1+2p_2 \,, \qquad \mbox{with } 1\equiv p_1+2p_2 \text{ mod }3 \,.
\end{align}
Hence, a web resulting from the use of the square rule having $(p_1-1,p_2-1)$ edges crossing the cut still satisfies \eqref{kupconstraintexample}. Indeed, we have
\begin{align}
    2&\leq 2(p_1-1)+(p_2-1) \,, \qquad \mbox{with } 2\equiv 2(p_1-1)+(p_2-1) \text{ mod }3 \,. \nonumber\\
    1&\leq (p_1-1)+2(p_2-1) \,, \qquad \mbox{with } 1\equiv (p_1-1)+2(p_2-1) \text{ mod }3 \,.
\end{align}
In other cases, there are in general webs such that the rule \eqref{modsquarerule} has to be used in order to weight them.
\end{remark}

We end this section by stressing that the partition functions defined above differ fundamentally from the defect partition functions defined in our first paper \cite{LGJ21} by their geometrical constraints \eqref{kupmodelconstraint}.

\subsection{The cylinder geometry}
\subsubsection{Magnetic defects}

The discussion on magnetic defects in the strip geometry mostly applies to the cylinder case as well. The main difference is that it is not sufficient anymore to consider only dominant weights as magnetic charges. Indeed the evolution operators, such as the row-to-row transfer matrix, are no longer symmetric under the full quantum group. Yet, the symmetry with respect to the Cartan subalgebra still holds. This means that we can again define sectors for a given magnetic charge but it can be any weight of $\mathcal{R}_2^*$ in the loop model case and any weight of $\mathcal{R}_3^*$ in the Kuperberg web model case. In general, two charges present in the same representation will describe inequivalent sectors. 
This can be seen by looking at the weight which a magnetic defect $\ket{\bm{m}}$ picks up when winding around the cylinder in the Kuperberg web model:
\begin{align}
    q^{2(\bm{\rho},\bm{m})} \,.
\end{align} 
Then consider, for example, the first fundamental representation $V_1$ of $U_{-q}(\mathfrak{sl}_3)$. It contains three weight vectors of weights 
\begin{subequations}
\label{threeweightvectors}
\begin{eqnarray}
    \bm{h}_1 &=&\bm{w}_1 \,, \\
    \bm{h}_2 &=&\bm{w}_2-\bm{w}_1 \,, \\
    \bm{h}_3 &=&-\bm{w}_2 \,.
\end{eqnarray}
\end{subequations}
These weights do not lead to the same winding phases: we have $(\bm{\rho},\bm{h}_1)=1$, whilst $(\bm{\rho},\bm{h}_2)=0$ and $(\bm{\rho},\bm{h}_3)=-1$.

We define a magnetic defect state in the loop model the same way as in the strip geometry, see \eqref{loopmagneticdefect}.
The partition function modified by the presence of this magnetic defect is then given by
\begin{align}
    Z_{\text{loop}}^{\bm{0},\bm{m}}=\bra{\bm{m}}T_{\text{loop}}^M\ket{\bm{m}} \,,
\end{align}
where $T_{\text{loop}}$ is defined in \eqref{looptransfermatrixcyl}.

Consider now the Kuperberg web model. Let $\bm{m}\in \mathcal{R}_3^*$ and denote by $d(\bm{m})=n_1\bm{w}_1+n_2\bm{w}_2$ the unique dominant weight in the Weyl orbit of $\bm{m}$. As in the strip geometry, we define a magnetic defect state $\ket{\bm{m}}$ of charge $\bm{m}$ to be a pure tensor state of weight $\bm{m}$, such that two sites labelled by $2i$ and $2i+1$ cannot be occupied simultaneously, and such that there are exactly $n_1+n_2$ occupied sites. There are necessarily $n_1$ sites occupied by equally coloured upward arrows, and $n_2$ sites occupied by equally coloured downward arrows, with the two colours being different. In fact, the only difference with the strip geometry resides in the possible colours of the occupied sites in $\ket{\bm{m}}$. For instance, the following is a magnetic defect:
\begin{align}
\label{kupmodelcylmagdeffect}
    \ket{\bm{m}}=(\ket{ \uparrow}\otimes\ket{\ })^{\otimes n_1} \otimes (\ket{\downarrow}\otimes\ket{\ })^{\otimes n_2} \otimes  \ket{\ }^{\otimes 2L-2n_1-2n_2} \,,
\end{align}
where arrows are coloured in some way that depends on the Weyl chamber to which $\bm{m}$ belongs. There are six different choices of pairs of colours corresponding to the six Weyl chambers of $\mathfrak{sl}_3$. As the state $\ket{d(\bm{m})}$ is highest-weight, using our standard convention we  write 
\begin{align}
    \ket{d(\bm{m})}=(\ket{\color{red} \uparrow}\otimes\ket{\ })^{\otimes n_1} \otimes (\ket{\color{green} \downarrow}\otimes\ket{\ })^{\otimes n_2} \otimes  \ket{\ } ^{\otimes 2L-2n_1-2n_2} \,.
\end{align}
Denote by $s_1$ (respectively $s_2$) the Weyl reflection with respect to the hyperplane orthogonal to $\bm{\alpha}_1$ (respectively $\bm{\alpha}_2$). Let $w$ be an element of the Weyl group mapping $d(\bm{m})$ to $\bm{m}$; it can be written as a product of the generators $s_1$ and $s_2$. Applying $s_1$ corresponds to swapping red and blue, whereas applying $s_2$ corresponds to swapping blue and green. This procedure determines the choice of colours in~\eqref{kupmodelcylmagdeffect}. For instance, we have that $d(\bm{h}_2)=\bm{w}_1$ and $\bm{h}_2=s_1(\bm{w}_1)$, so
\begin{align}
    \ket{\bm{w}_1}= (\ket{\color{red} \uparrow}\otimes\ket{\ }) \otimes \ket{\ }^{\otimes 2L-2}
\end{align} gives
\begin{align}
    \ket{\bm{h}_2}= (\ket{\color{blue} \uparrow}\otimes\ket{\ }) \otimes \ket{\ }^{\otimes 2L-2} \,.
\end{align}
The partition function modified by the insertion of the magnetic defect is then
\begin{align}
    Z_{K}^{\bm{0},\bm{m}}=\bra{\bm{m}}T_{\rm K}^M\ket{\bm{m}} \,,
\end{align}
where $T_{\rm K}$ is defined in \eqref{kuptransfermatrixcyl}.

\subsubsection{Electric charges}
We define \textit{electric charges} as vectors in the space $\mathbb{C}^k$ generated by the basis of fundamental weights, where $k=1$ in the loop case and $k=2$ in the Kuperberg web case. The seam line operators in \eqref{looppivot} are
\begin{align}
    S_{\text{loop}}=q^{2H_{\bm{\rho}}}={\rm e}^{2i\gamma(\bm{\rho},\bm{w}_1) H_1} \,,
\end{align}
with $q=e^{i\gamma}$. For any $\bm{e}\in \mathbb{C}$, define also the generalised seam line operators
\begin{align}
\label{loopseamopelectriccharge}
    S_{\text{loop}}^{\bm{e}}={\rm e}^{-2i\pi(\bm{e}-\bm{e}_0,\bm{w}_1) H_1} \,,
\end{align}
where $\bm{e}_0=\frac{\gamma}{\pi}\bm{\rho}$. Then $S_{\text{loop}}$ is recovered for $\bm{e}=\bm{0}$.

If we define the row-to-row transfer matrix with \eqref{loopseamopelectriccharge} instead of \eqref{looppivot}, we obtain 
\begin{align}
    T_{\text{loop}}^{\bm{e}}=\left(\prod_{k=0}^{L-1}t^{\text{loop}}_{2k+1}\right)\left(\prod_{k=1}^{L-1}t^{\text{loop}}_{2k}\right)S_{\text{loop}}^{\bm{e}}t^{\text{loop}}_{2L}(S_{\text{loop}}^{\bm{e}})^{-1} \,.
\end{align}
The modified partition function then reads
\begin{align}
    Z_{\text{loop}}^{\bm{e},\bm{0}}=\braket{\ (T_{\text{loop}}^{\bm{e}})^M\ } \,.
\end{align}
It is standard usage in the Coulomb Gas context to refer to this modified partition function by saying that a pair of opposite electric charges $\bm{e}-\bm{e}_0$ and $-\bm{e}+\bm{e}_0$ have been inserted, one at the top of the cylinder, the other at the bottom. When $\bm{e}=0$, we say that we are in presence of {\em background (electric) charges} $\bm{e}_0$ and $-\bm{e}_0$.

The case of the Kuperberg web model is analogous. We define a modified seam line operator 
\begin{align}
    S_{\rm K}^{\bm{e}}={\rm e}^{-2i\pi[(\bm{e}-\bm{e}_0,\bm{w}_1) H_1 + (\bm{e}-\bm{e}_0,\bm{w}_2) H_2]} \,.
\end{align}
The transfer matrix is then given by 
\begin{align}
    T_{\rm K}^{\bm{e}}=\left(\prod_{k=0}^{L-1}t^{\rm K}_{2k+1}\right)\left(\prod_{k=1}^{L-1}t^{\rm K}_{2k}\right)S_{\rm K}^{\bm{e}}t^{\rm K}_{2L}(S_{\rm K}^{\bm{e}})^{-1} \,,
\end{align}
leading to the partition function
\begin{align}
    Z_{\rm K}^{\bm{e},\bm{0}}=\braket{\ (T_{\rm K}^{\bm{e}})^M\ } \,.
\end{align}

Finally, in both the loop model and Kuperberg web model, one can combine magnetic defects and electric charges to define modified partition functions 
\begin{subequations}
\begin{eqnarray}
    Z_{\text{loop}}^{\bm{e},\bm{m}} &=& \bra{\bm{m}}(T_{\text{loop}}^{\bm{e}})^M\ket{\bm{m}} \,, \\
    Z_{\rm K}^{\bm{e},\bm{m}} &=& \bra{\bm{m}}(T_{\rm K}^{\bm{e}})^M\ket{\bm{m}} \,.
\end{eqnarray}
\end{subequations}

As in the strip geometry, we can rewrite these partition functions in terms of a sum over trajectories of transition amplitudes. For the loop case, denote again by $\mathcal{L}_{\rm col}^{\bm{m}}$ the set of oriented subgraphs of $\mathbb{H}$ whose connected components are either coloured loops (cf.\ Remark~\ref{rem:1}) or coloured lines, such that loops cannot touch the bottom and top boundaries, whereas lines touch the bottom and top boundaries only at their end points corresponding to the occupied sites in $\ket{\bm{m}}$. We stress that, although we have used the same notation as for the strip geometry, the elements of $\mathcal{L}_{\rm col}^{\bm{m}}$ are now embedded in the cylinder. We then have 
\begin{align}
\label{loopCylGeoPF}
    Z_{\text{loop}}^{\bm{e},\bm{m}}=\sum_{c\in \mathcal{L}_{\rm col}^{\bm{m}}} w_{\rm col}^{\bm{e},\bm{m}}(c) \,.
\end{align}
The weight $w_{\rm col}^{\bm{e},\bm{m}}(c)$ is given by the local weights \eqref{loopvertexweights} as well as modified weights for crossing the seam line:
\begin{align}
    &\vcenter{\hbox{\includegraphics[scale=0.2]{diagrams/crossseamleftrightloop1.eps}}}={\rm e}^{-2i\pi(\bm{e}-\bm{e}_0,\bm{w}_1)} \,, \qquad
    \vcenter{\hbox{\includegraphics[scale=0.2]{diagrams/crossseamleftrightloop2.eps}}}={\rm e}^{2i\pi(\bm{e}-\bm{e}_0,\bm{w}_1)} \,, \qquad
    \vcenter{\hbox{\includegraphics[scale=0.2]{diagrams/crossseamleftrightloop3.eps}}}=1 \,.
\end{align}

In the Kuperberg case, analogously to the strip geometry, define \textit{an open Kuperberg web on a cylinder} to be a planar oriented bipartite graph with trivalent and univalent vertices such that the univalent vertices are only at the top and bottom boundaries of the cylinder. Assume $c$ is an open web on a cylinder with $2(n_1+n_2)$ univalent vertices such that the following holds for both bottom and top boundaries: $c$ has $n_1$ (respectively $n_2$) upward (respectively downward) oriented edges incident on the univalent vertices.
A {\em three-colouring} of such an open web $c$ is a map from the edges of $c$ into the set $\{\rm red, blue, green\}$, such that all three colours are present around any trivalent vertex, and such that at every boundary side all the  upward oriented edges are of a colour $\mathfrak{c}_1$  while the downward oriented ones are of a colour $\mathfrak{c}_2$ such that $\mathfrak{c}_1\neq \mathfrak{c}_2$. 

\begin{figure}
\begin{center}
    \includegraphics[scale=0.3]{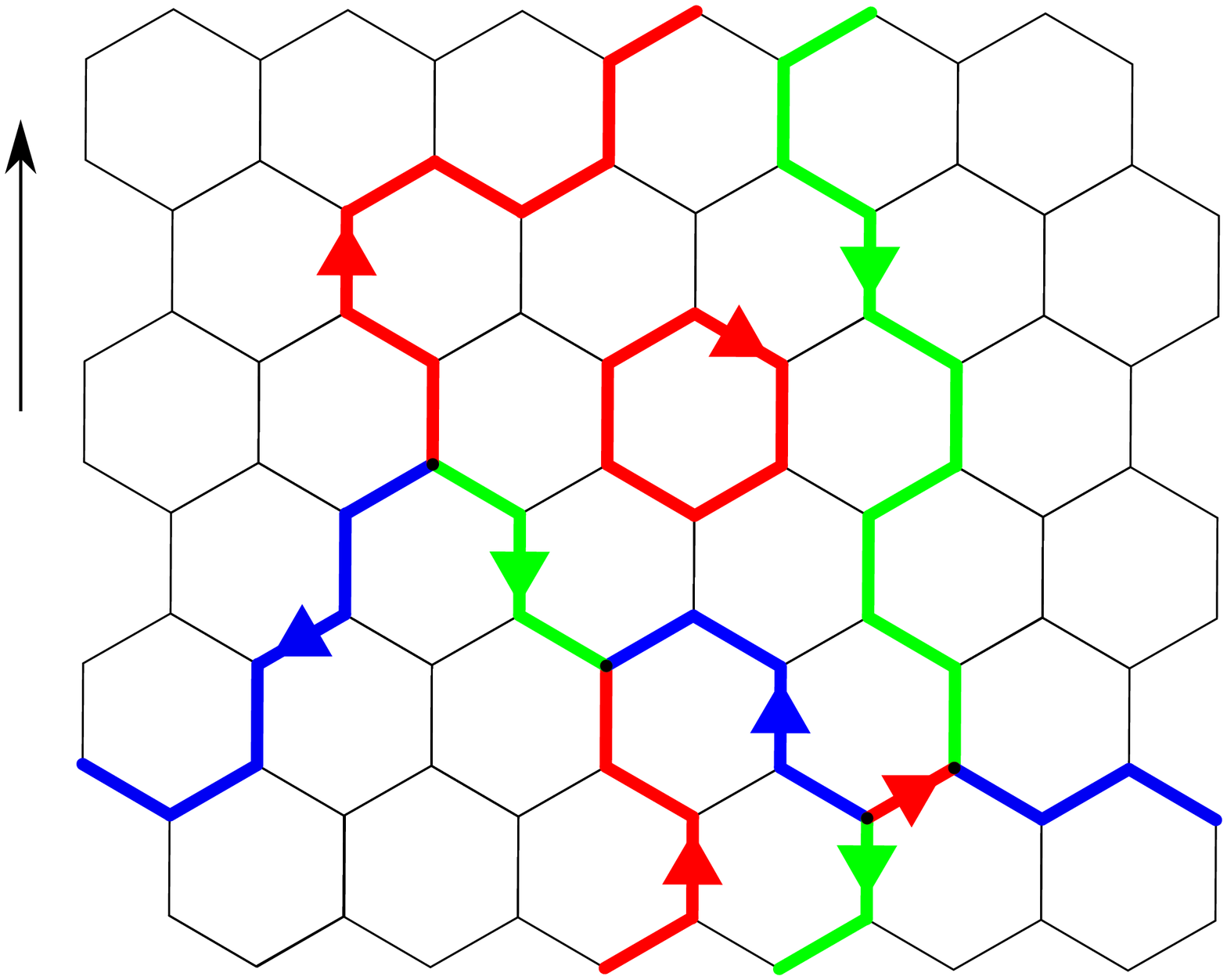}\qquad\qquad \includegraphics[scale=0.3]{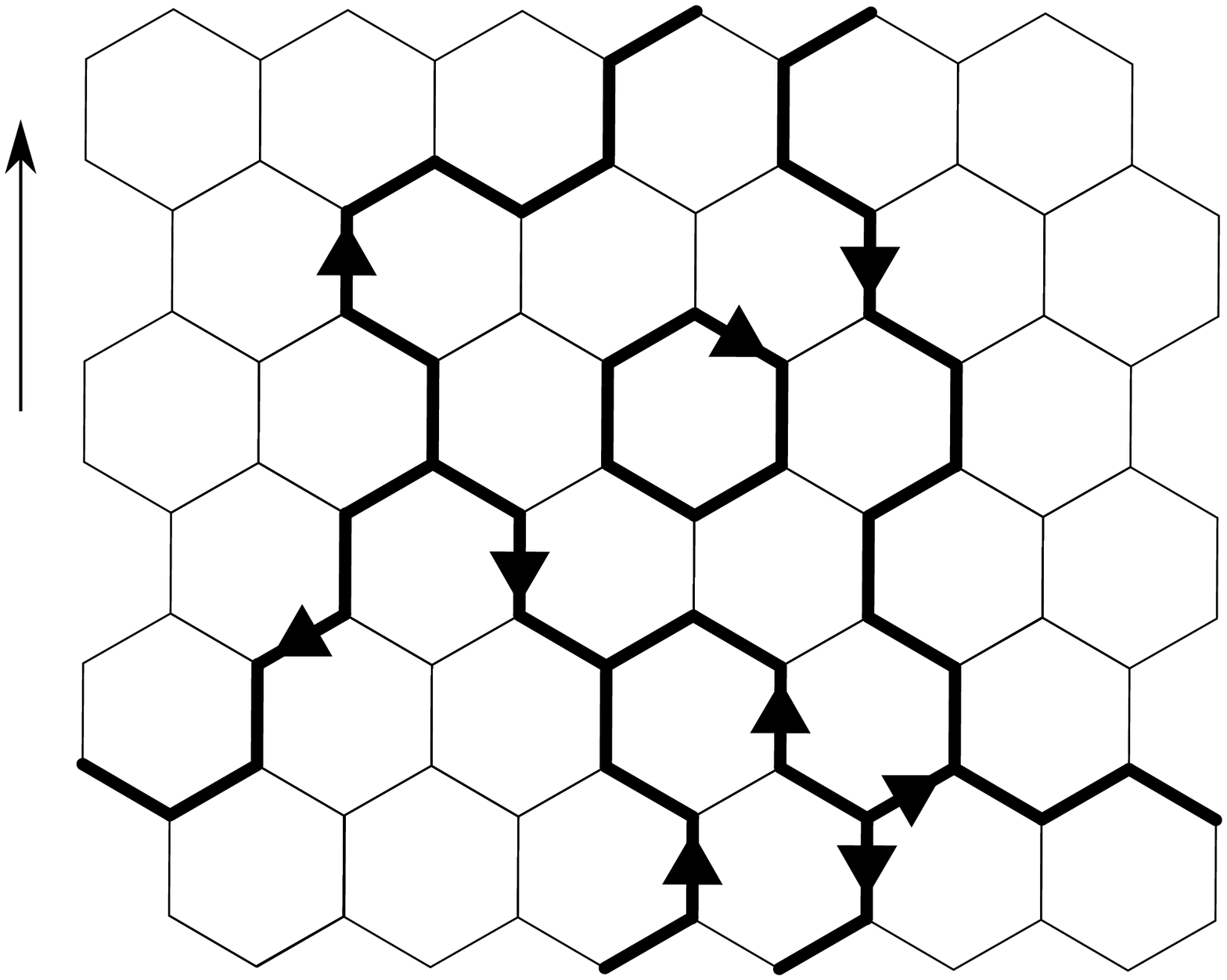}
\end{center}
    \caption{Left panel: A configuration of $\mathcal{K}_{\rm col}^{\bm{m}}$ for $\bm{m}= \bm{w}_1+\bm{w}_2$ in the cylinder geometry. Right panel: A configuration of $\mathcal{K}^{\bm{m}}$ obtained from the left one by forgetting colours.}
    \label{fig:kupmagdefectcylinder}
\end{figure}

Denote by $\mathcal{K}_{\rm col}^{\bm{m}}$, the set of subgraphs of~$\mathbb{H}$ whose connected components are either open or closed coloured webs on a cylinder, such that webs cannot touch the bottom and top boundaries, except for open webs that touch the bottom and top boundaries at their end points corresponding to the occupied sites in $\ket{\bm{m}}$, and with the inherited colours\footnote{See the discussion below \eqref{kupmodelcylmagdeffect}.}. We again use the same notation as in the strip geometry case, although it is clear that the two sets are different. We then have that
\begin{align}
    Z_{\rm K}^{\bm{e},\bm{m}}=\sum_{c\in \mathcal{K}_{\rm col}^{\bm{m}}} w_{\rm col}^{\bm{e},\bm{m}}(c) \,.
\end{align}
An example of configuration is given in Figure \ref{fig:kupmagdefectcylinder}. The weight of a coloured configuration $w_{\rm col}^{\bm{e},\bm{m}}(c)$ is given by the local weights \eqref{kupvertex} and the modified weights for crossing the seam line
\begin{align}
\label{electriccoloredseamweight}
    &\vcenter{\hbox{\includegraphics[scale=0.2]{diagrams/crossseamleftrightcolored1.eps}}}={\rm e}^{-2i\pi(\bm{e}-\bm{e}_0,\bm{h}_1)} \,, \qquad \vcenter{\hbox{\includegraphics[scale=0.2]{diagrams/crossseamleftrightcolored2.eps}}}={\rm e}^{-2i\pi(\bm{e}-\bm{e}_0,\bm{h}_2)} \,, \qquad \vcenter{\hbox{\includegraphics[scale=0.2]{diagrams/crossseamleftrightcolored3.eps}}}={\rm e}^{-2i\pi(\bm{e}-\bm{e}_0,\bm{h}_3)} \,, \nonumber\\
    &\vcenter{\hbox{\includegraphics[scale=0.2]{diagrams/crossseamleftrightcolored4.eps}}}={\rm e}^{2i\pi(\bm{e}-\bm{e}_0,\bm{h}_1)} \,, \ \ \qquad \vcenter{\hbox{\includegraphics[scale=0.2]{diagrams/crossseamleftrightcolored5.eps}}}={\rm e}^{2i\pi(\bm{e}-\bm{e}_0,\bm{h}_2)} \,, \ \ \qquad \vcenter{\hbox{\includegraphics[scale=0.2]{diagrams/crossseamleftrightcolored6.eps}}}={\rm e}^{2i\pi(\bm{e}-\bm{e}_0,\bm{h}_3)} \,, \\
    &\vcenter{\hbox{\includegraphics[scale=0.2]{diagrams/crossseamleftrightloop3.eps}}}=1 \,, \nonumber
\end{align}
where we have used the weight vectors \eqref{threeweightvectors}.

\subsubsection{Geometrical interpretation}

We now give a geometrical interpretation of electromagnetic partition functions. We begin with the known results for the $O(N)$ loop model.
Recall the expression \eqref{loopCylGeoPF} for $Z_{\rm loop}^{\bm{e},\bm{m}}$, obtained by summing the weight $w_{\rm col}^{\bm{e},\bm{m}}$ over
the configurations $\mathcal{L}_{\rm col}^{\bm{m}}$.
When no magnetic defect is present, $\mathcal{L}_{\rm col}^{\bm{0}}$ is the set of all possible unoriented loop configurations in $\mathbb{H}$. The weight of a configuration is given by the product of bond fugacities and loop weights. The weight of a contractible loop is clearly $q+q^{-1}$. On the other hand, the insertion of a pair of opposite electric charges $\pm(\bm{e}-\bm{e}_0)$ assigns a different weight
\begin{align}
    {\rm e}^{2i\pi(\bm{e}_0-\bm{e},\bm{w}_1)}+{\rm e}^{-2i\pi(\bm{e}_0-\bm{e},\bm{w}_1)}=[2]_t
\end{align}
to non-contractible loops. We have here parameterised $\bm{e}_0-\bm{e}=\frac{\mu}{\pi}\bm{\rho}$ and $t={\rm e}^{i\mu}$.

Consider now the insertion of a magnetic defect of charge $\bm{m}=n\bm{w}_1$, with $n\neq 0$. In the cylinder geometry, define a \textit{cut} to be a circle embedded in the cylinder and avoiding nodes of $\mathbb{H}$ such that its projection onto the horizontal circle generating the cylinder is bijective (no overhangs). As in the strip geometry, the insertion of magnetic defects implies a constraint on the number $p$ of bonds present on any given cut. Any representation containing a weight $\bm{m}$ contains the unique dominant weight in the Weyl orbit of the latter, $|n|\bm{w}_1$. Hence we must have that $|n|\bm{w}_1$ is lower than $p\bm{w}_1$. In other words,
\begin{align}
\label{loopgeodeffectconstraint}
    |n|&\leq p \,, \qquad \mbox{with } n\equiv p \text{ mod } 2 \,.
\end{align}
It follows that a line connected to the cylinder boundaries must connect the bottom and top boundaries. We can then say that $|n|$ through-lines propagate along the cylinder.

When $n\neq 0$, the presence of through-lines forbid non-contractible loops. Contractible loops are still weighted by $q+q^{-1}$. Moreover, 
each through-line acquires an additional weight
\begin{subequations}
\begin{align}
    {\rm e}^{ 2i\pi(\bm{e}_0-\bm{e},\bm{w}_1)}
\end{align}
when it winds around the periodic boundary condition from left to right (respectively right to left) if $n>0$ (respectively $n<0$); and a weight
\begin{align}
    {\rm e}^{- 2i\pi(\bm{e}_0-\bm{e},\bm{w}_1)}
\end{align}
\end{subequations}
when it winds from right to left (respectively left to right) if $n>0$ (respectively $n<0$).

A special class of operators, known as {\em watermelon operators}, see e.g.\ \cite{LoopReview}, are obtained if one suppresses the background charge by the additional electric charge, i.e., setting $\bm{e}=\bm{e}_0$. We then get a geometrical defect where through-lines do not get any additional weight when they wind around the cylinder. These operators have been useful in predicting, for instance, the fractal dimension of percolation hulls\cite{DS87}.

\medskip

Let us now discuss the Kuperberg web model. Denote by $\mathcal{K}^{\bm{m}}$ the set of webs embedded in $\mathbb{H}$ obtained from elements of $\mathcal{K}_{\rm col}^{\bm{m}}$ by forgetting their colours. See Figure \ref{fig:kupmagdefectcylinder} for an example. Let $c$ be a configuration in~$\mathcal{K}^{\bm{m}}$. The different configurations $c_i$ in $\mathcal{K}_{\rm col}^{\bm{m}}$ that give the open web $c$, once one forgets their colourings, are exactly the three-colourings of $c$ whose edges touching the boundary are coloured according to $\bm{m}$. By summing their weight, we obtain
\begin{align}\label{eq:Z-K-geom}
    Z_{\rm K}^{\bm{e},\bm{m}}=\sum_{c\in \mathcal{K}^{\bm{m}}} w^{\bm{e},\bm{m}}(c) \,,
\end{align}
where we defined
\begin{align}
\label{webweightstrip_em}
    w^{\bm{e},\bm{m}}(c)=\sum_i w^{\bm{e},\bm{m}}_{\rm col}(c_i) \,.
\end{align}

Remark that, as in the strip geometry, a magnetic defect of charge $\bm{m}$ implies a constraint on the geometry of webs. We write the corresponding dominant weight as $d(\bm{m})=n_1\bm{w}_1+n_2\bm{w}_2$. If we denote again by $p_1$ (respectively $p_2$), the number of bonds pointing upward (respectively downward) on a given cut, we must have the constraint~\eqref{kupmodelconstraint-gen} applied to the dominant weight:
\begin{align}
    d(\bm{m})\preceq p_1\bm{w}_1+p_2\bm{w}_2
\end{align}
This is again due to the Cartan subalgebra symmetry which implies that $\bm{m}$, hence $d(\bm{m})$, must be among the weights of the tensor product representation with $p_1$ factors $V_1$ and $p_2$ factors $V_2$ that are present on the given cut. Any magnetic defect configuration $c\in\mathcal{K}^{\bm{m}}$ of charge $\bm{m}$ satisfies the above constraint on any cut.
\medskip

The weight $w^{\bm{e},\bm{m}}(c)$ is a product of local fugacities as well as a part given by \eqref{coloredkupvertex}-\eqref{coloredkupvertexgeneralized} and \eqref{electriccoloredseamweight}. We now discuss this part of the weight that we name the \textit{electromagnetic Kuperberg weight} (or Kuperberg weight for short) $w^{\bm{e},\bm{m}}_{\rm K}(c)$ of a web $c$. 

Firstly, we can ask what is the Kuperberg weight of non-contractible webs when no magnetic defect is present. For simplicity, consider the case of a single non-contractible loop separating the pair of charges $\bm{e}-\bm{e}_0$ and $-\bm{e}+\bm{e}_0$. It will be weighted by 
\begin{align}
\label{noncontractibleloop}
    {\rm e}^{2i\pi(\bm{e}_0-\bm{e},\bm{h}_1)}+{\rm e}^{2i\pi(\bm{e}_0-\bm{e},\bm{h}_2)}+{\rm e}^{2i\pi(\bm{e}_0-\bm{e},\bm{h}_3)} \,,
\end{align}
as it gets a contribution from a red, a blue and a green edge, all oriented the same way crossing the seam line, corresponding to the three weights~\eqref{threeweightvectors} of the fundamental representation. For a charge parallel to the Weyl vector, $\bm{e}_0-\bm{e}=\frac{\mu}{\pi}\bm{\rho}$, this gives $[3]_t$ with the parametrisation $t={\rm e}^{i\mu}$.

One can show \cite{Kuperberg_1996} that any connected component of a planar web that is not simply a loop contains a face that is either a digon or a square.
In the absence of any electric charges, the strategy to obtain its weight is then to apply the second and third rules of \eqref{3rules} to reduce the
connected component, until a loop is obtained, which can finally be replaced by its respective weight from \eqref{3rulesa}. 
But for the system modified by a pair of electric charges, we must be more careful. The cylinder geometry can be represented as an annulus,
so the web is still a planar graph. However, we cannot immediately apply the reductions of the second and third rules of \eqref{3rules} in case the corresponding
face is the internal or external face of the annulus, the ones where the electric charges are situated. Fortunately one can show that if a connected closed web
is not a loop, it contains a face {\em different} from the internal or external ones that is either a digon or a square.%
\footnote{Proof: A connected web that is not a loop always contains at least $3$ faces surrounded by $4$ or less vertices. Indeed, suppose this is not the case for a given web $c$ satisfying the precedent conditions. Denote by $F$, $E$ and $V$, the number of faces, edges and vertices of $c$. By the hand-shake lemma, one has $2E=3V$. The graph being also planar, the Euler relation gives $F-E+V=F-\frac{1}{2}V=2$. Because at least $F-2$ faces are surrounded by $6$ or more vertices, one has $3V\geq 6(F-2)+R$, where $R$ is the number of vertices surrounding the two other faces. One has $R\geq 4$ implying $3V\geq 6(F-2)+4=6F-8$. This gives $8\geq 6(F-\frac{1}{2}V)=12$, a contradiction.}
So the reduction to a loop is still possible. This loop is finally replaced by the weight $[3]_q$ if it is contractible (i.e., homotopic to a point),
or by the weight \eqref{noncontractibleloop} if it is non-contractible (i.e., it wraps around the cylinder).

\medskip

We next discuss the electromagnetic Kuperberg weight of open webs in the presence of a defect with charges $\bm{e}$ and $\bm{m}\neq \bm{0}$. As in the strip geometry, we can obtain the Kuperberg weight of a given web by reducing it thanks to the rules \eqref{3rules} and \eqref{modsquarerule}. We then obtain a linear combination of webs that are irreducible, i.e., that do not contain loops, digons or squares. The Kuperberg weight of the original web is then obtained by weighting the irreducible webs. We will thus focus on such webs by characterising them in terms of elementary blocks and giving the Kuperberg weight of each block. Let $d(\bm{m})=n_1\bm{w}_1+n_2\bm{w}_2$, and consider an irreducible open web $c$. It is clear that $c$ is connected and connects the bottom and top boundary. On the cylinder we can always decompose $c$ into a number $j$ of cylindrical blocks, shown here as grey ribbons
\begin{center}
    \includegraphics[scale=0.3]{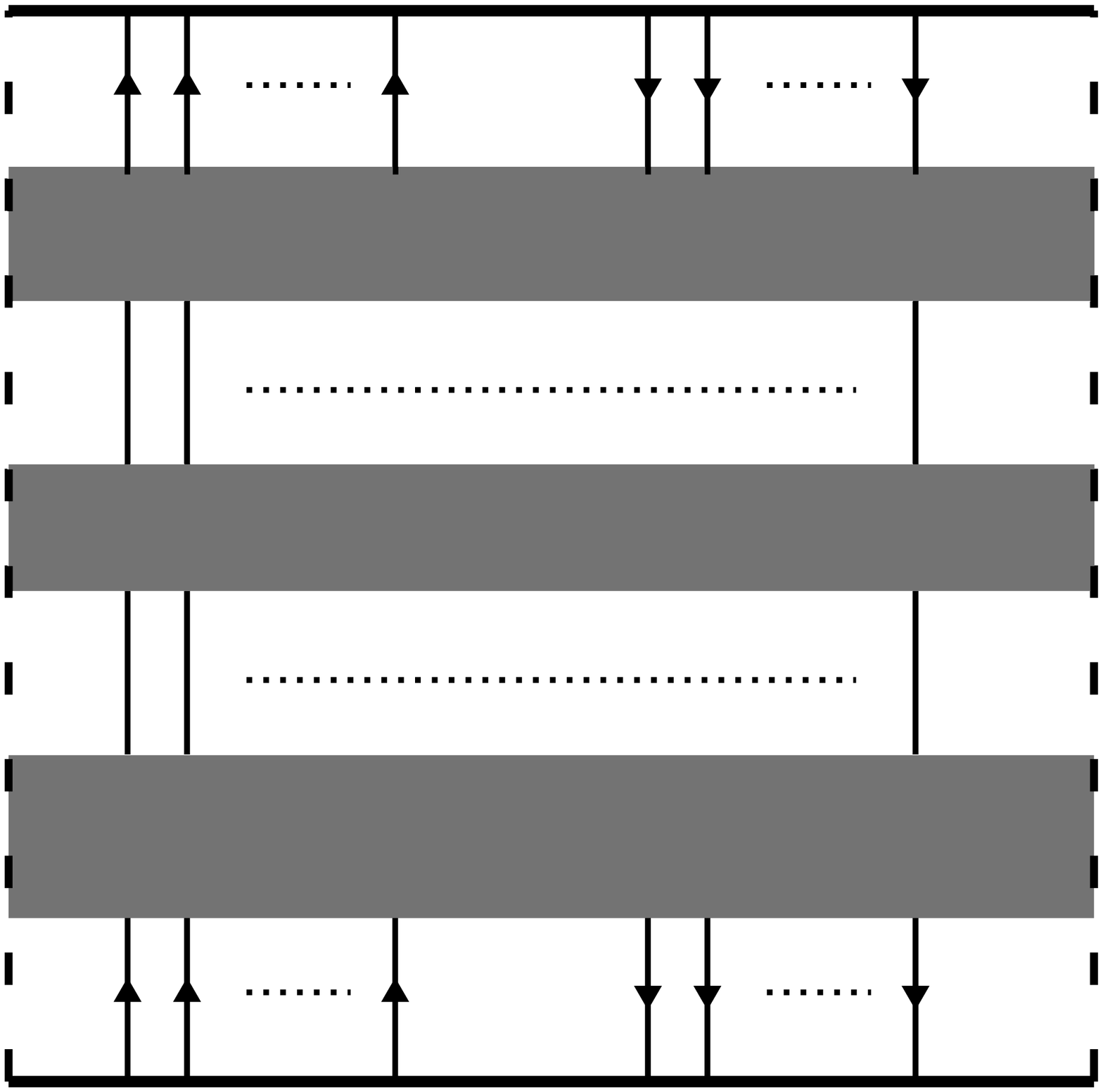}
\end{center}
for the case $j=3$. Blocks are connected to each other and to the top and bottom boundary by $n_1$ upward oriented edges and $n_2$ downward oriented edges, ordered arbitrarily, except for the top-most and bottom-most edges which must follow the order imposed by $\ket{\bm{m}}$. Hence, each block is separated from the others by a minimal cut. Of course, such a block decomposition is not unique, however, the overall weight (as we describe below) does not depend on a choice of decomposition. 

We now define three classes of blocks and give their contribution to the Kuperberg weight of the whole web. Firstly, a block of type A is constituted by an edge winding once around the cylinder. This gives a weight that depends on the electric charge and the colour and orientation of the edge\footnote{Note that, on any minimal cut, the colour of an edge is fixed by the choice of magnetic charge in the same way it is fixed for edges connected to the top and bottom boundaries, however the ordering of orientations might be different for the middle blocks.}. This weight is
\begin{subequations}
\begin{align}
    {\rm e}^{ 2i\pi(\bm{e}_0-\bm{e},\bm{w})}
\end{align}
for a coloured edge carrying a weight $\bm{w}$ winding from left to right (this weight is one of the weights from the fundamental or its dual representation of $U_{-q}(\mathfrak{sl}_3)$, depending on the orientation of the edge) and
\begin{align}
   {\rm e}^{-2i\pi(\bm{e}_0-\bm{e},\bm{w})}
\end{align}
\end{subequations}
for an edge winding from right to left.

The second class of blocks, called type B, is constituted by webs that do not wrap or wind around the cylinder. That is, webs that can be bounded by a rectangle. In Section \ref{sec:StripGeoInterpretation}, we have seen that the space of such magnetised webs of charge $\bm{m}$ bounded by a rectangle is of dimension $1$.\footnote{In the cylindrical case the colours of arrows of the magnetic defect state might be different from those of the strip geometry case but only the orientations matter for the argument.}. We will call \textit{H-web} any web comprised of a number of vertical strands oriented in some way, to the left and to the right of the H-shaped web~\eqref{H-web}, or this H-shaped web with its arrows reversed. It is clear that an H-web is irreducible. In fact, it is shown in Appendix \ref{App:IrrWebStrip} that for any choice of orientations of edges incident on univalent vertices of a magnetised web in a rectangle,  there exists an irreducible one that is a concatenation of H-webs\footnote{Or, of course, the trivial web consisting of only vertical edges if the two ordered sequences of orientations in the top and bottom boundaries are the same.}. One can see that such a web admits a unique colouring (recall our definition of three-colourings). Indeed, the colours of the top-most and bottom-most edges of any H-shaped web are fixed by the Weyl chamber of the charge $\bm{m}$. For instance, for a charge in the fundamental Weyl chamber, one obtains H-shaped webs coloured as
\begin{center}
\includegraphics[scale=0.3]{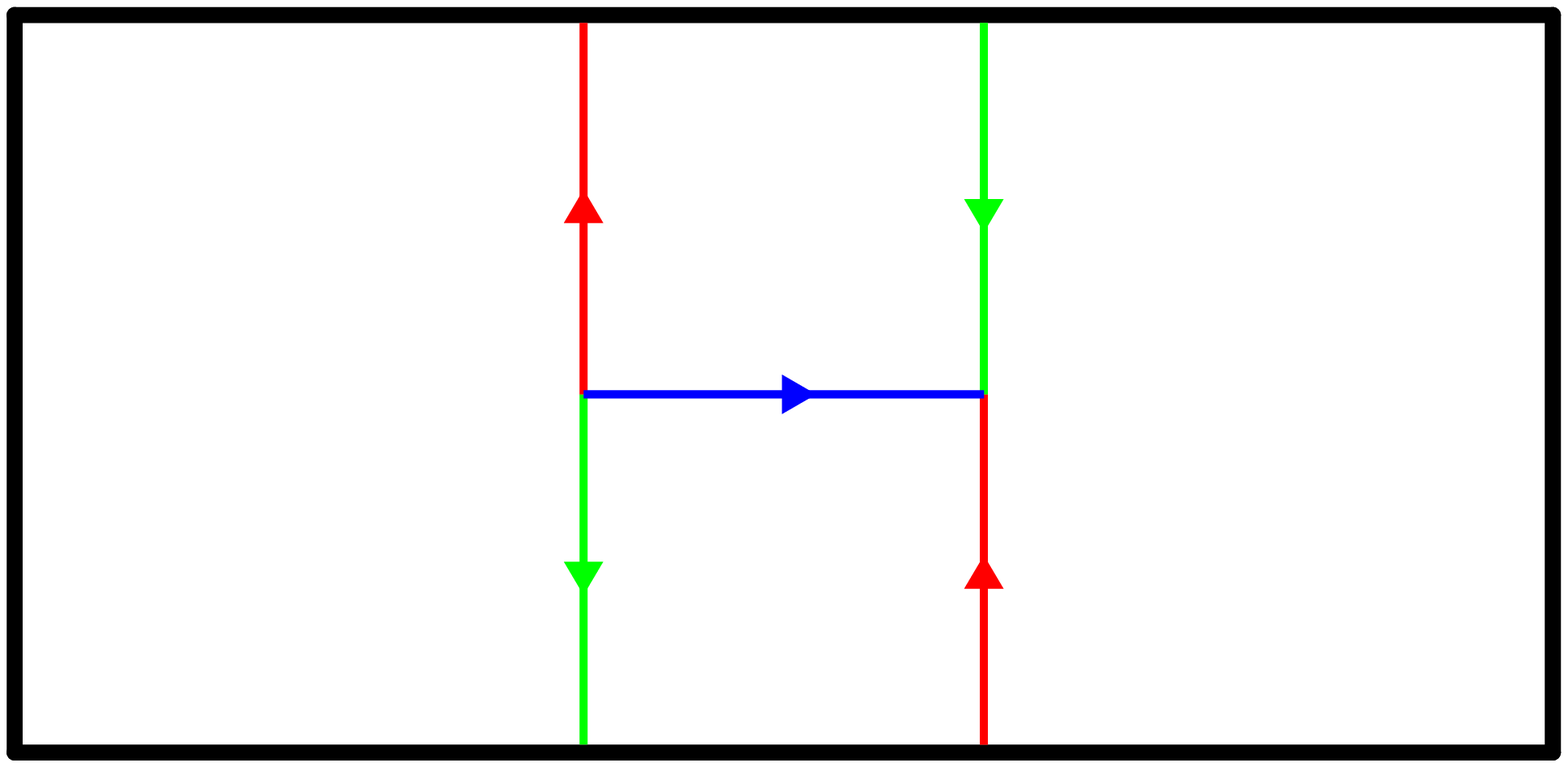}
\end{center}
or a vertical reflection thereof.
By \eqref{coloredkupvertexgeneralized}, the contributions of the two vertices of a given H-shaped web compensate to give a weight $1$. Hence, the weight of a block of type B is $1$.

The third class of blocks, of type C, is constituted by webs that wrap around the cylinder the following way. A block of type C is made of $n_1+n_2$ edges connected to the bottom boundary of the block and $n_1+n_2$ edges connected to the top boundary of the block such that all of these edges are connected to a wrapping cycle of edges that we denote by $\mathfrak{L}$. Here is an example with $n_1+n_2=2$:
\begin{align}
    \vcenter{\hbox{\includegraphics[scale=0.3]{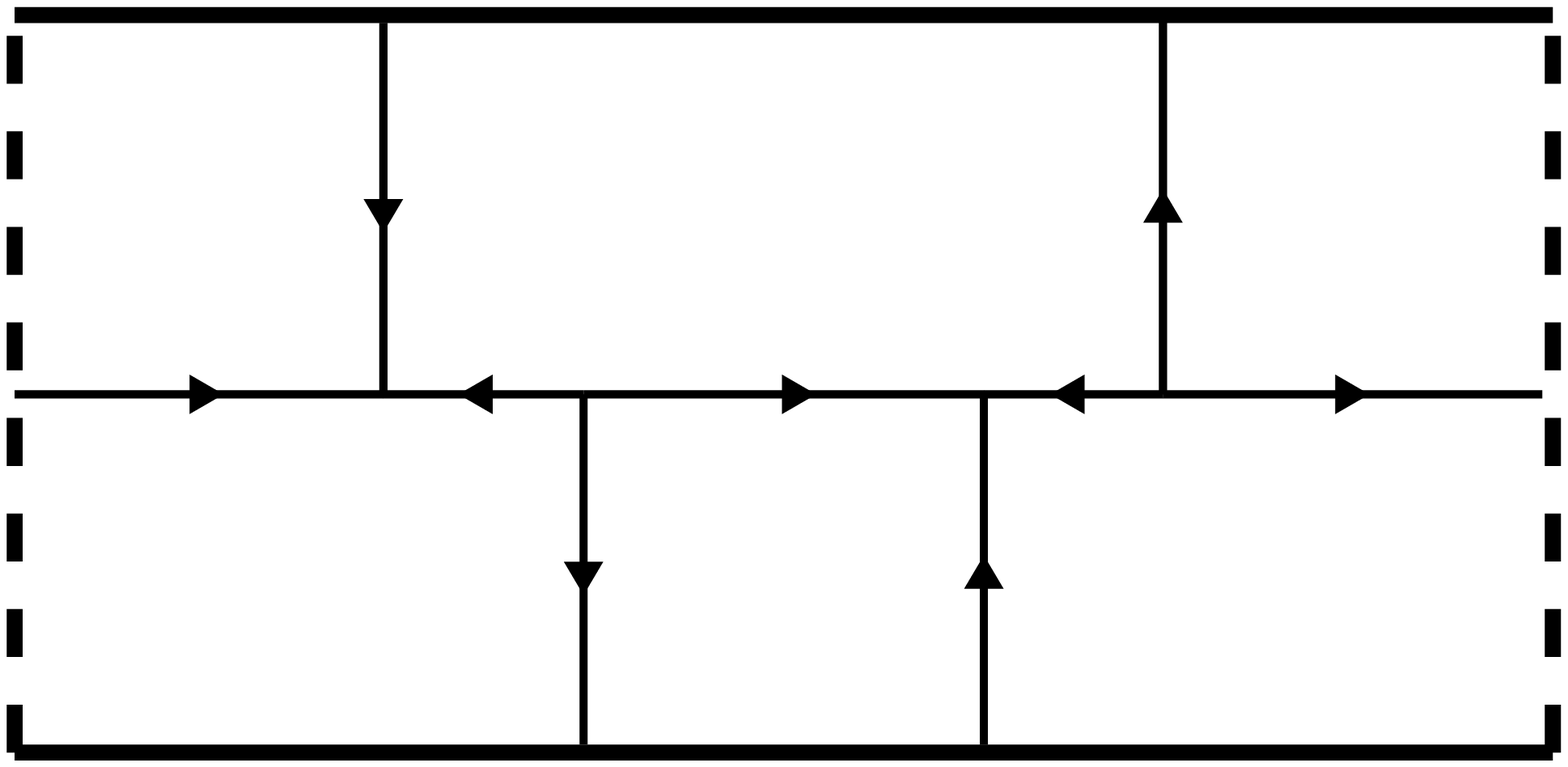}}}
\end{align}
In this example, the cycle $\mathfrak{L}$ is made of $4$ edges. We note that the situation here has no analogue in the loop models where the non-zero defect sectors have no wrapping structures.

In Appendix \ref{App:ColoringWrapWebs}, we show that given a colour for the upward oriented edges connected to the boundaries and a different one for the downward oriented ones, there are $2$ possible colourings for the cycle $\mathfrak{L}$ when $n_1=0$ or $n_2=0$, and only $1$ possible colouring for $\mathfrak{L}$ otherwise. The sum over the possible colourings of the weights of the coloured webs gives the contribution of the type C block to the Kuperberg weight of the whole open web. 

In more details, consider the case when both $n_1$ and $n_2$ are non-zero and there is only one colouring. For instance:
\begin{align}
\label{wrappingweb1}
    \vcenter{\hbox{\includegraphics[scale=0.3]{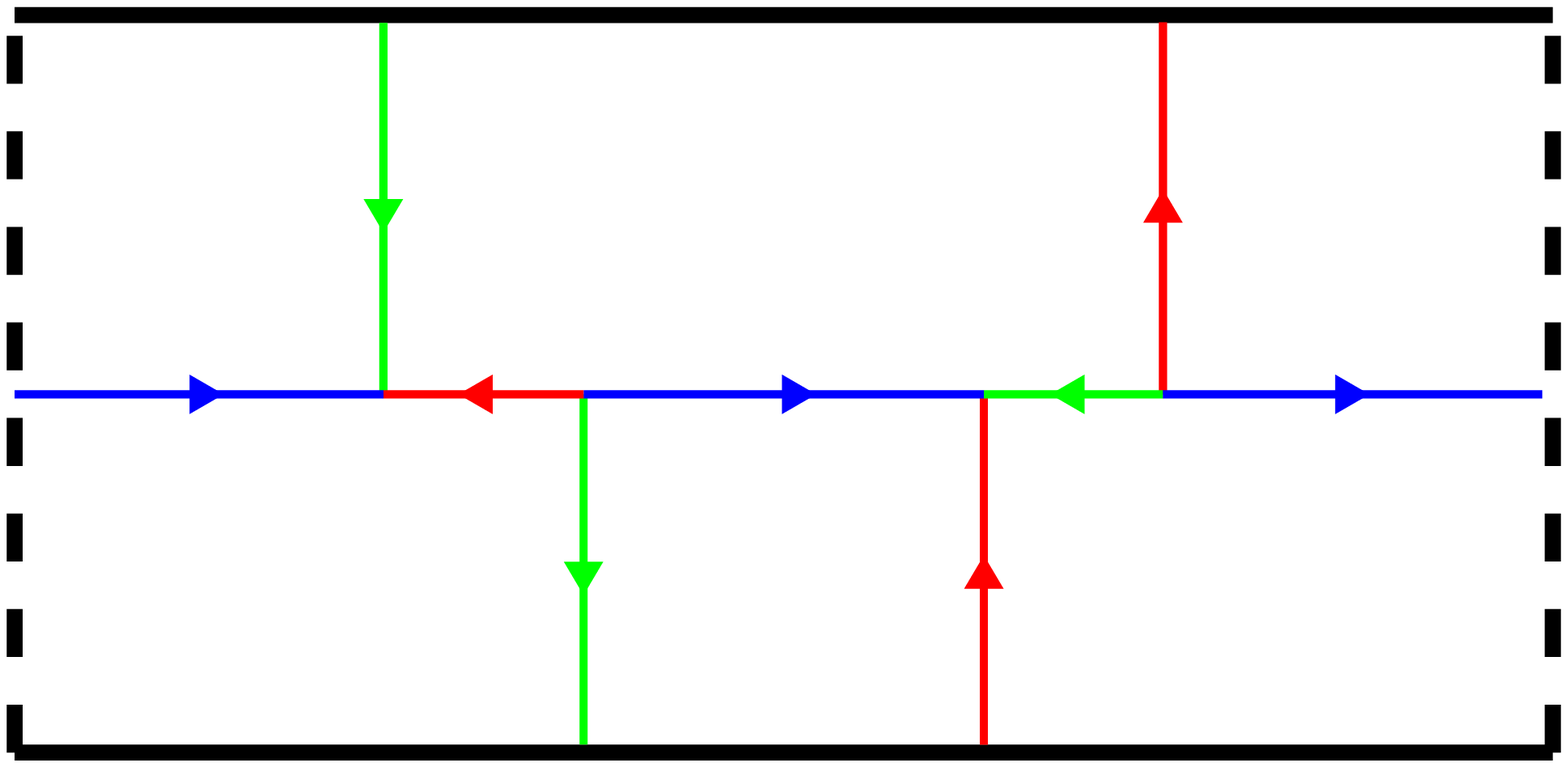}}}
\end{align}
One can see that the set of vertices can be partitioned into pairs, such that the weights of the two vertices of a given pair compensate each other (using \eqref{coloredkupvertexgeneralized}). Thus, the weight of the web solely depends on the colour and orientation of the edge crossing the seam line. This weight is
\begin{subequations}
\begin{align}
    {\rm e}^{ 2i\pi(\bm{e}_0-\bm{e},\bm{w})}
\end{align}
for a coloured edge carrying a weight $\bm{w}$ going from left to right and
\begin{align}
    {\rm e}^{-2i\pi(\bm{e}_0-\bm{e},\bm{w})}
\end{align}
\end{subequations}
for an edge going from right to left. In our example, it is given by 
\begin{align}
\label{weightwrappingsquare}
    {\rm e}^{2i\pi(\bm{e}_0-\bm{e},\bm{h}_2)} \,.
\end{align}

When $n_2=0$, the case of $n_1=0$ being similar, there are two possible colourings for the edges constituting $\mathfrak{L}$ and the weight is the sum of these two contributions. For instance:
\begin{align*}
    \vcenter{\hbox{\includegraphics[scale=0.3]{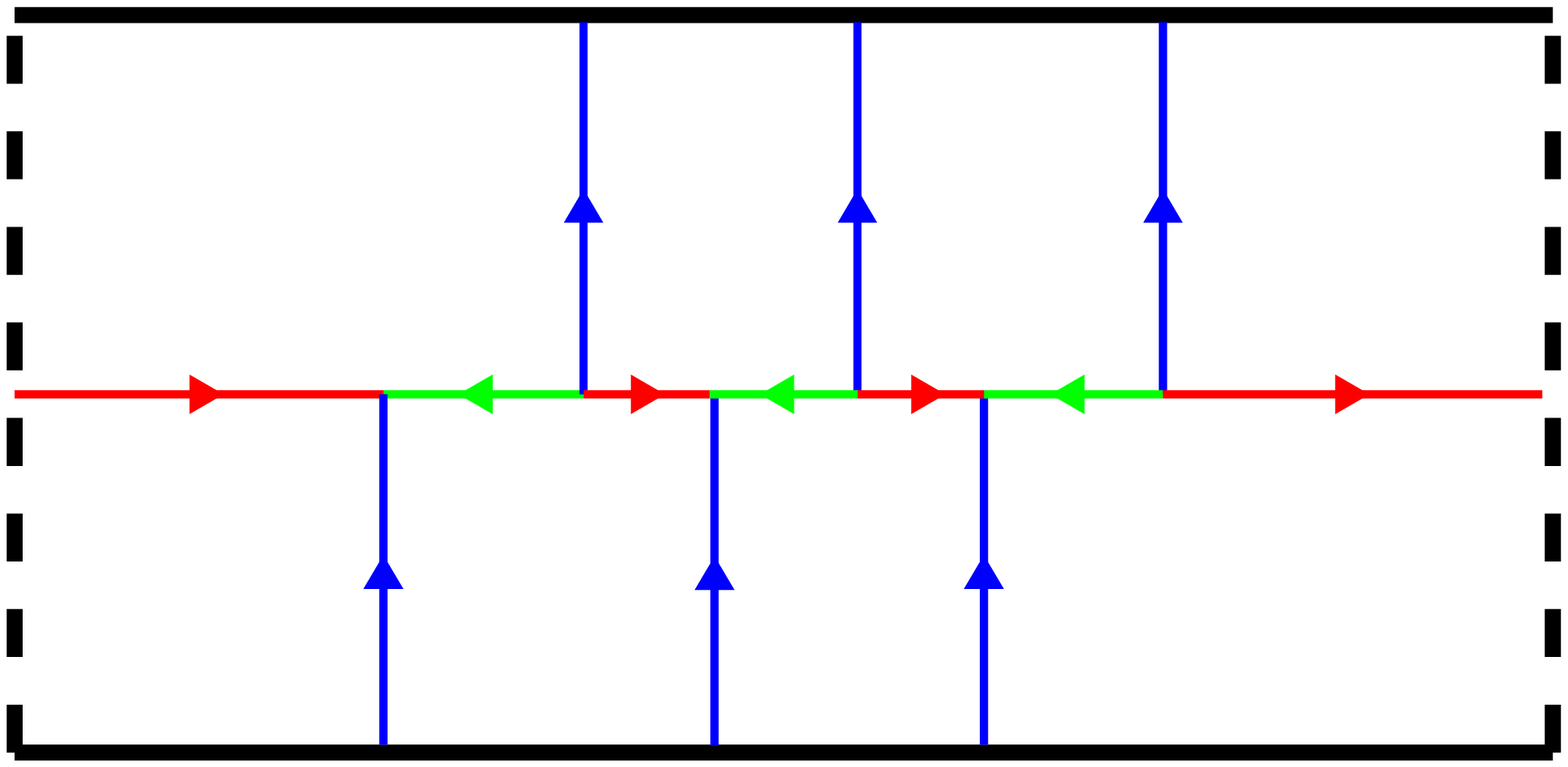}}}\quad+\quad \vcenter{\hbox{\includegraphics[scale=0.3]{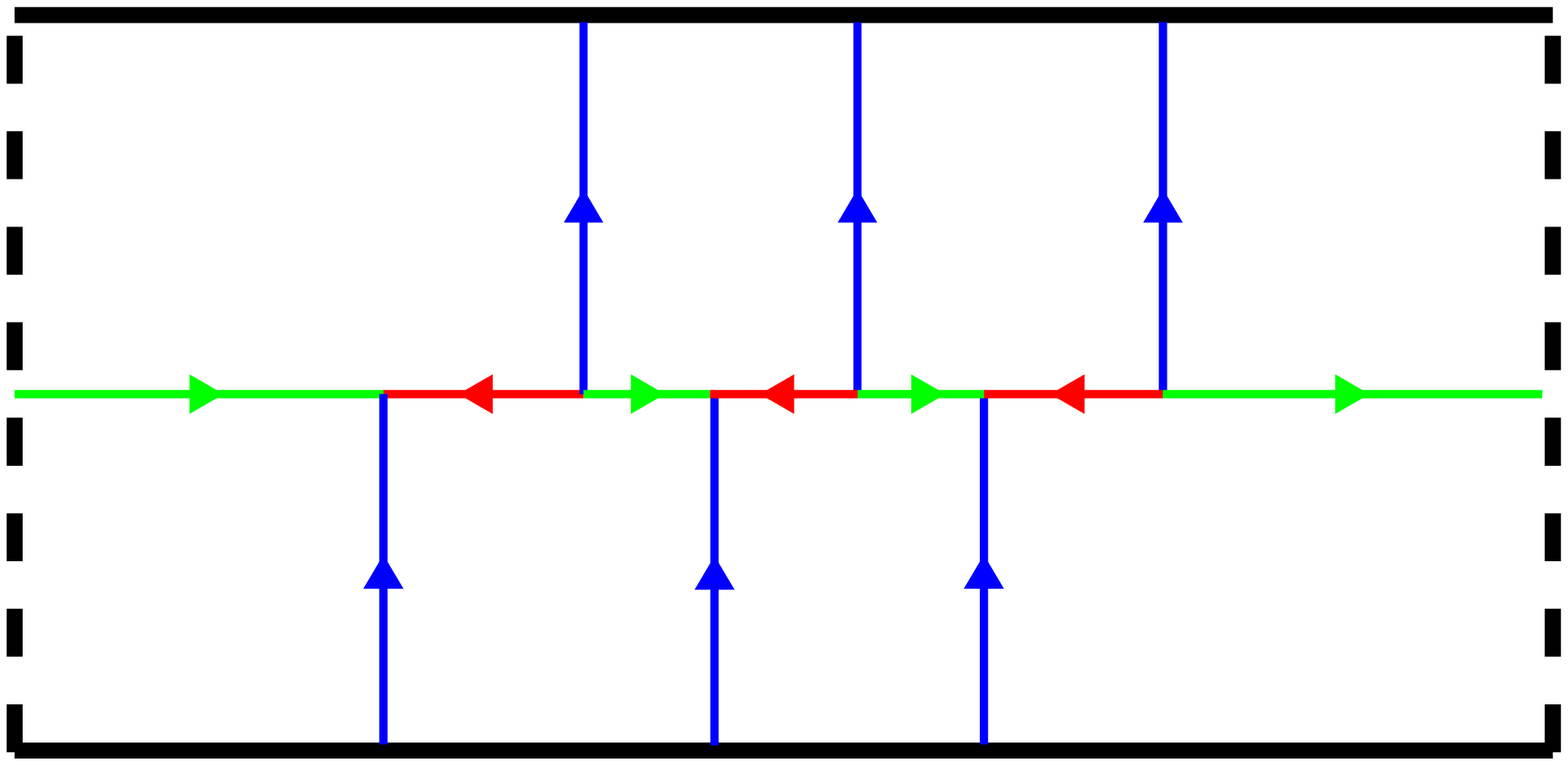}}}
\end{align*}
Again, one can see that vertices come in pairs that compensate. Hence the weight is again given by the edge crossing the seam line. This edge crosses the seam line in the same direction in both cases and it is coloured with the two colours that are different from the unique colour of the top-most and bottom-most edges. When these are blue, as in our example, the non-contractible cycle of edges contributes to the Kuperberg weight by a factor
\begin{align}
\label{cycleweight}
    {\rm e}^{2i\pi(\bm{e}_0-\bm{e},\bm{h}_1)}+{\rm e}^{2i\pi(\bm{e}_0-\bm{e},\bm{h}_3)} \,.
\end{align}
It is not hard to see what is the weight for other colours of bottom-most and top-most edges.

In Appendix \ref{App:IrrWebCyl}, we show that any irreducible web is a concatenation of blocks of type A, B and C.

\medskip

We now examine the set of open webs which do not get any non-trivial weight when one of the bottom-most or top-most edges are winding around the cylinder. This constitutes the analogue of watermelon operators present in the loop model. There are two ways to get a trivial weight when an edge winds around the cylinder. Either, we tune the electric charge to get a vanishing total charge as in the loop model, i.e., we set $\bm{e}=\bm{e}_0$. In this case, we see that wrapping webs get a weight $1$ (respectively $2$) for each type C block when both $n_1$ and $n_2$ are non-zero (respectively $n_1=0$ or $n_2=0$). Or we can consider defects coloured by a weight orthogonal to the electric charge. For instance, consider a total electric charge $\bm{e}_0-\bm{e}=\frac{\mu}{2\pi}\bm{\rho}$ parallel to the Weyl vector. Then, magnetic defects of charge $n\bm{h}_2$, with $n\neq 0$, suit the condition because $(\bm{\rho},\bm{h}_2)=0$. In this case, by~\eqref{cycleweight} a type~C block then has a weight
\begin{align}
    n_{\bm{e}}={\rm e}^{2i\pi(\bm{e}_0-\bm{e},\bm{h}_1)}+{\rm e}^{2i\pi(\bm{e}_0-\bm{e},\bm{h}_3)}=[2]_{t} \,,
\end{align}
where $t={\rm e}^{i\mu}$.

\section{Geometrical defects in $\mathbb{Z}_3$ spin models}
\label{sec:defect1}
In \cite{LGJ21}, we formulated an equivalence between a $\mathbb{Z}_3$ symmetric spin model on the triangular lattice dual to~$\mathbb{H}$ and the Kuperberg web model at the special point
\begin{subequations}
\begin{eqnarray}
    q&=& {\rm e}^{i\frac{\pi}{4}} \,, \\
    y&=&z=2^{-\frac{1}{4}} \,.
\end{eqnarray}
\end{subequations}
At this point, the topological weight of any configuration was shown to be $1$ and the Kuperberg web model partition function is then simply
\begin{align}
\label{KupPFSpecialPoint}
    Z_{\rm K} =\sum_{c\in \mathcal{K}} x_1^{N_1}x_2^{N_2} \,.
\end{align}
This equals, up to an overall factor, the partition function of the $\mathbb{Z}_3$ spin model written thanks to its low-temperature expansion.

We now exhibit another mapping between \eqref{KupPFSpecialPoint} and the partition function of a three-state Potts model, this time defined on $\mathbb{H}$ itself. This mapping results from a high-temperature expansion of the latter and is detailed in Section \ref{sec:HighT}. The equality of two partition functions of the model defined on dual lattices constitutes an example of duality \cite{MS71}.

In Section \ref{sec:3PottsDefects}, we consider the possible connections between geometrical defects of the web models and observables of the three-state Potts model in either its low- or high-temperature expansion. 

\subsection{Relation with a $\mathbb{Z}_3$ spin model via high-temperature expansion}\label{sec:HighT}

We now formulate an equivalence with a $\mathbb{Z}_3$ spin model on the same lattice $\mathbb{H}$ using a high temperature expansion. Consider spins in the set of third roots of unity $\{1,e^{2 \pi i/3},e^{4 \pi i/3}\}$. The group $\mathbb{Z}_3\cong\{1,e^{2 \pi i/3},e^{4 \pi i/3}\}$ acts on a given spin by left multiplication. The global symmetry is given by acting the same way on every spin. Denote by $\langle ij \rangle$ a pair of adjacent nodes in $\mathbb{H}$, such that $j$ is situated higher than $i$. This is well defined given our convention on the orientation of $\mathbb{H}$. The model is defined by its nearest-neighbour Boltzmann weights $W(\sigma_i,\sigma_j)$ associated to a pair of spins $\sigma_i$ and $\sigma_j$ linked by $(ij) \in \langle ij \rangle$. The most general $\mathbb{Z}_3$-symmetric local Boltzmann weight can be written as (up to an overall scalar)
\begin{align}
    W(\sigma_i,\sigma_j)= 1 + x_1\sigma_i\Bar{\sigma}_j+ x_2\Bar{\sigma}_i\sigma_j \,,
\end{align}
where the bar denotes complex conjugation.
The partition function of the spin model then reads
\begin{align}
Z_{\text{spin}}=\sum_{\sigma\in \mathcal{S}} \prod_{<ij>}W(\sigma_i,\sigma_j) = \sum_{\sigma\in \mathcal{S}} \prod_{<ij>} \left(1 + x_1\sigma_i\Bar{\sigma}_j+ x_2\Bar{\sigma}_i\sigma_j\right) \,,
\end{align}
where $\mathcal{S}$ denotes the set of spin configurations.

The high-temperature expansion consists in developing the product $\prod_{\langle ij \rangle} \left(1 + x_1\sigma_i\Bar{\sigma}_j+ x_2\Bar{\sigma}_i\sigma_j\right)$ and pictorially represent the term associated to a link $(ij)$. The product of these terms will then be represented by a subgraph $G$ of $\mathbb{H}$. If the summand $1$ is chosen, then the link $(ij)$ is empty, i.e., it is not part of $G$. If the summand $x_1\sigma_i\Bar{\sigma}_j$ is chosen, then $(ij)$ is part of $G$ and oriented upward. If the summand $x_2\Bar{\sigma}_i\sigma_j$ is chosen, then $(ij)$ is part of $G$ and oriented downward.

When the sum over all spin configurations is done, some subgraphs $G$ give no contribution. Indeed any term associated to a graph $G$ that contains a factor $\sigma_i$ or $\Bar{\sigma}_i$ for a given node $i$ will give $0$ when the the sum over $\sigma_i$ is applied. Hence, one can see that the only surviving graphs are closed Kuperberg webs. We can thus rewrite the partition function as 
\begin{align}
    Z_{\text{spin}} =\sum_{G\in \mathcal{K}} x_1^{N_1}x_2^{N_2} \,,
\end{align}
which is exactly \eqref{KupPFSpecialPoint}.

\medskip

Consider now a general correlation functions in the spin model,
\begin{align}
    \left \langle \sigma_{i_1}^{k_1}\sigma_{i_2}^{k_2}\cdots\sigma_{i_p}^{k_p} \right \rangle=\frac{1}{Z_{\rm spin}} \sum_{\sigma\in \mathcal{S}}\left(\sigma_{i_1}^{k_1}\sigma_{i_2}^{k_2}\cdots\sigma_{i_p}^{k_p}\right) \prod_{<ij>} W(\sigma_i,\sigma_j) \,,
\end{align}
where $k_j \in \{1,2\}$ and $i_j$ denotes a node of $\mathbb{H}$. Denote the data of spin insertions by $\Gamma=\{(i_j,k_j), j \in \llbracket1,p\rrbracket\}$. The global $\mathbb{Z}_3$ symmetry ensures that non-zero correlators satisfy 
\begin{align}
    \sum_j k_j \equiv 0 \text{ mod } 3 \,.
\end{align}

Doing again the high-temperature expansion, a graph $G$ surviving is an $\textit{open}$ Kuperberg web embedded in $\mathbb{H}$. Denote by $\mathcal{K}_{\Gamma}$ the set of such open webs. The correlation function can then be written as
\begin{align}
    \left \langle \sigma_{i_1}^{k_1}\sigma_{i_2}^{k_2}\cdots\sigma_{i_p}^{k_p} \right \rangle =\frac{1}{Z_{\rm K}} \sum_{G\in \mathcal{K}_{\Gamma}} x_1^{N_1}x_2^{N_2} \,.
\end{align}

Consider in particular the spin two-point function  $\langle \sigma_i\Bar{\sigma}_j \rangle$, with one operator inserted at a node $i$ situated at the bottom boundary of the cylinder (or the strip), and the other one inserted at a node $j$ situated at the top boundary of the cylinder (or the strip). In terms of webs, the correlation reads
\begin{align}
\label{spincorr}
    \left \langle \sigma_i\Bar{\sigma}_j \right \rangle =\frac{1}{Z_{\rm K}} \sum_{G\in \mathcal{K}_{\sigma}} x_1^{N_1}x_2^{N_2} \,,
\end{align}
where $\mathcal{K}_{\sigma}$ denotes the webs present in the high-temperature expansion. Here $\mathcal{K}_{\sigma}$ can be partitioned into $2$ sets,
\begin{align}
    \mathcal{K}_{\sigma}=\mathcal{K}_{1}\cup \mathcal{K}_{2} \,,
\end{align}
where $\mathcal{K}_{1}$ is the set of open webs with one edge incident on the node $i$ and one edge incident on the node~$j$. $\mathcal{K}_{2}$ is the set of open webs with more edges incident on the nodes $i$ or $j$. Nevertheless, remark that in all cases, on any cut, the edges satisfy the constraint \eqref{kupmodelconstraint} with $n_1=1$ and $n_2=0$. The correlation functions have the following scaling form 
\begin{subequations}
\begin{equation}
    \langle \sigma_i\Bar{\sigma}_j \rangle \sim {\rm e}^{-\frac{2\pi M}{\sqrt{3}L}(h_{\sigma}+\Bar{h}_{\sigma})}
\end{equation}
in the cylinder geometry, and 
\begin{equation}
    \langle \sigma_i\Bar{\sigma}_j \rangle \sim {\rm e}^{-\frac{\pi M}{2\sqrt{3}L}h_{\sigma}}
\end{equation}
\end{subequations}
in the strip geometry.

\subsection{Relations with geometrical defects of the Kuperberg web model}\label{sec:3PottsDefects}

It is apparent from the last subsection that spin-spin correlators in both the strip and cylinder geometry are related to geometrical defects in the vertex formulation. Indeed, the set $\mathcal{K}_{1}$ denotes precisely the set of webs given by the insertion of a geometrical defect of charge $\bm{m}$ with $d(\bm{m})=\bm{w}_1$. The other set $\mathcal{K}_{2}$ is not directly related to a geometrical defect. However since they all satisfy the same geometrical constraint, we expect that they contribute to the same sector in the continuum limit.

In order to relate the spin-spin correlation to a geometrical defect, we must ensure that all the open webs involved in the partition function have a topological weight $1$. Recall that the topological weight is given by the product of the Kuperberg weight \eqref{coloredkupvertexgeneralized} and the vertices fugacities, $y$ and $z$.

In the strip geometry, we have seen that when a geometrical defect of charge $\bm{w}_1$ is present, one does not need the additional rule \eqref{modsquarerule}. Hence, the trick of~\cite[Sec\, 2.2]{LGJ21}  can be applied to see that all open webs have topological weight $1$. Thus, in the strip geometry
\begin{align}
    h_{\sigma}=h_{\bm{w}_1} \,.
\end{align}
Remark also that, in the low-temperature expansion point of view, the same operator can be viewed as a boundary condition changing operator that takes a fixed boundary condition to another.

\medskip

In the cylinder geometry, consider a geometrical defect of charge $\bm{h}_2$ and an electric charge such that
\begin{align}
    n_{\bm{e}}=\sqrt{2} \,.
\end{align}
Again, one does not need the additional rule \eqref{modsquarerule} in this case. Moreover, because on any non-contractible cycle, there are exactly two vertices with fugacities $y$ and $z$, the weight of such a cycle is $yzn_{\bm{e}}=1$. Then, again by a trick analogous to the one introduced in \cite{LGJ21}, one can see that all webs are given a topological weight $1$. This means that 
\begin{align}
    (h_{\sigma}, \Bar{h}_{\sigma}) = (h_{\bm{e},\bm{h}_2},\Bar{h}_{\bm{e},\bm{h}_2}) \,.
\end{align}
The following table gives the numerical estimation of $h_{\bm{e},\bm{h}_2}+\Bar{h}_{\bm{e},\bm{h}_2}$ :
\begin{center}
\begin{tabular}{ |c|c| } 
 \hline
 Size $L$ & $h_{\bm{e},\bm{h}_2}+\Bar{h}_{\bm{e},\bm{h}_2}$  \\ 
 \hline
  5 & 0.134425217550764  \\ 
 \hline
  6 & 0.134307263093286   \\ 
 \hline
  7 & 0.134209499407688  \\ 
 \hline
  $\infty$ & 0.13338\phantom{3333333333} \\
 \hline
\end{tabular}
\end{center}
The values are obtained thanks to a numerical diagonalisation of the row-to-row transfer matrix for two consecutive sizes, $L$ and $L+1$. The extrapolation to the thermodynamical limit is obtained by fitting the finite-$L$ values to a second-order polynomial in $1/L$. It matches the exact value $h_{\sigma}+\Bar{h}_{\sigma}=2/15 \simeq 0.13333$ rather precisely.

\medskip

Remark finally that the presence of the modified rule \eqref{modsquarerule} impedes the use of the argument of \cite{LGJ21} to give a weight $1$ to all open webs for a defect with $(n_1,n_2)\notin \{(1,0),(0,1)\}$. Hence the insertion of such a defect will not only constrain the webs present in the configuration space of the modified partition function but also give them a weight that is different from the one they would get in the unconstrained partition function. This seems to prevent us from finding simple geometrical observables in the three-state Potts model related to some connected subsets of interfaces as was studied, for instance, in \cite{GC07}.

\section{Geometrical defect in the $q={\rm e}^{i\frac{\pi}{3}}$ model}
\label{sec:defect2}

We now discuss another application of the web models, concerning the points of parameter space satisfying
\begin{align}
\label{eipi3}
    q ={\rm e}^{i\frac{\pi}{3}} \,.
\end{align}
These points are the higher-rank analogues of dense and dilute polymers in the $O(N)$ loop model case. In the case, we
have $[2]_q=1$ and $[3]_q=0$. This implies that any non-empty web $c$ gets a vanishing Kuperberg weight $w_{\rm K}(c)=0$, since, at the very least,
one of its components picks up one factor of $[3]_q$ when it has been reduced to a loop by application of \eqref{3rules}. The partition function is then equal to $1$ as only the empty web configuration contributes. However, a
non-trivial model can be obtained by defining the following renormalised partition function
\begin{align}
\label{modpartfunction}
    Z=\lim_{[3]_q\rightarrow 0} \frac{1}{[3]_q}(Z_{\rm K}-1) \,.
\end{align}
In this model, only connected  webs get a non-trivial weight (we call a web {\em connected} when it consists of only one connected component). Indeed, the Kuperberg weight of a general web can be computed by using the rules 
\begin{subequations}
\begin{align}
    \vcenter{\hbox{\includegraphics[scale=0.2]{diagrams/rel2kup1.eps}}}&\;=\;\vcenter{\hbox{\includegraphics[scale=0.2]{diagrams/rel2kup2.eps}}}\\
    \vcenter{\hbox{\includegraphics[scale=0.2]{diagrams/rel3kup1.eps}}}&\;=\;\vcenter{\hbox{\includegraphics[scale=0.2]{diagrams/rel3kup2.eps}}}\;+\;\vcenter{\hbox{\includegraphics[scale=0.2]{diagrams/rel3kup3.eps}}}
\end{align}
\end{subequations}
to reduce the web to a collection of loops, giving weight $1$ if there is only one loop or $0$ if there are more. 
We can look at the weighting procedure differently. First, digons can be removed whenever they appear without introducing any weight.
Secondly, when applying the square rule, if a resulting web has more than one connected component, set it to $0$. In this case, we can rephrase it by invoking modified rules 
\begin{align}\label{modsquarerules2}
    \vcenter{\hbox{\includegraphics[scale=0.2]{diagrams/rel3kup1.eps}}}&=\vcenter{\hbox{\includegraphics[scale=0.2]{diagrams/rel3kup2.eps}}} \qquad \text{or} \qquad
    \vcenter{\hbox{\includegraphics[scale=0.2]{diagrams/rel3kup1.eps}}}=\vcenter{\hbox{\includegraphics[scale=0.2]{diagrams/rel3kup3.eps}}}
\end{align}
where the right-hand sides are assumed to form parts of a one-component web.
Observe the similarity with~\eqref{modsquarerule}, however, the difference here is that we consider only closed webs.
From this procedure it is clear that the weight of a connected web is always a positive integer. For instance, the ``cube'' gets weight~$2$:
\begin{center}
\includegraphics[scale=0.3]{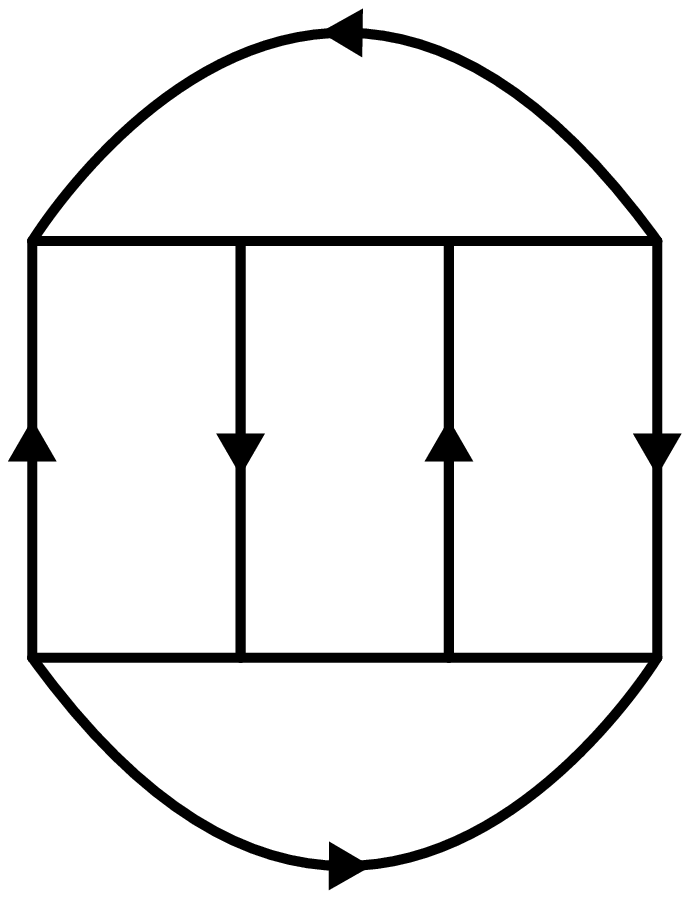}
\end{center}
Moreover any positive integer $k$ is the weight of some web. For instance, it can be showed by induction that the following web has weight $k$:
\begin{center}
\includegraphics[scale=0.3]{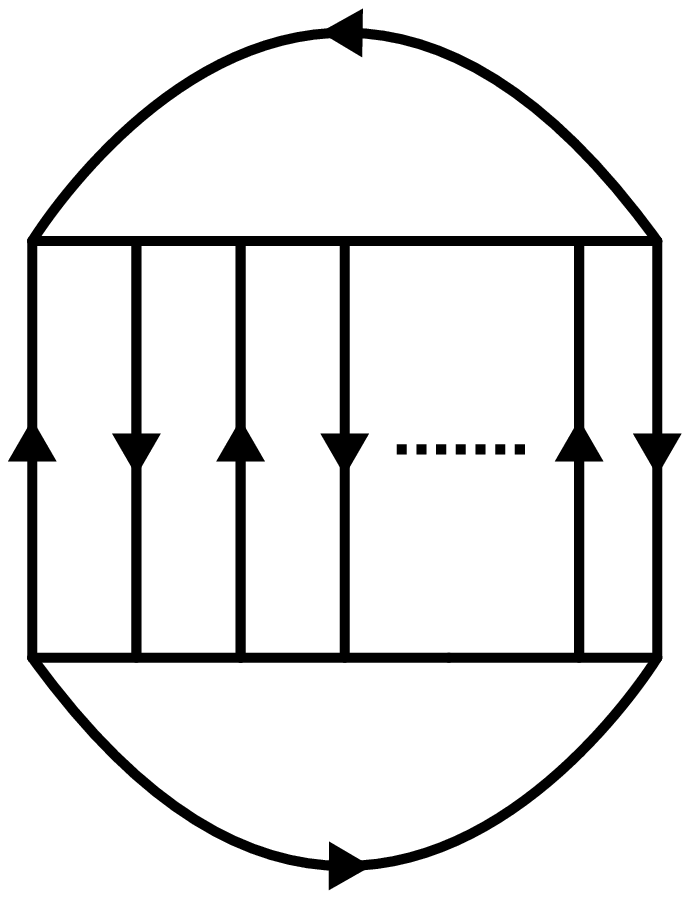}
\end{center}
where the dots represent $k-3$ pairs of vertical edges with opposite orientation (in addition to the three pairs explicitly drawn).

Consider now the insertion of a defect of magnetic charge $\bm{\rho}=\bm{w}_1+\bm{w}_2$ (i.e., $n_1=n_2=1$) and a vanishing total electric charge,
$\bm{e}_0 - \bm{e} = \bm{0}$.  Here, we define the electromagnetic partition functions as before via~\eqref{eq:Z-K-geom}. This means that if there is a web component disconnected from the defect, the weight of the configuration is~$0$. In addition, one can see from \eqref{kupmodelconstraint} with $n_1=n_2=1$ that the modified square rules \eqref{modsquarerule} are used exactly when two opposite sides of a square are cut by a minimal cut.  This means that they are used whenever the original square rule would produce an open web having two connected components, one connected to the bottom boundary and one to the top boundary. Moreover, there are no non-trivial winding weights when the total electric charge vanishes, and the weight of a wrapping web \eqref{weightwrappingsquare} is $1$ as well. Finally, wrapping webs as in \eqref{wrappingweb1} are weighted by $1$.

Hence, we can summarise the procedure for computing the Kuperberg weight of an open web in our setting. One can remove any digon or wrapping web, and use the original square rule whenever the resulting webs are connected. Otherwise, one uses the modified square rules \eqref{modsquarerule} and \eqref{modsquarerules2}. When a web cannot be reduced further by these rules, its weight is $1$.
Since \eqref{modsquarerule} and \eqref{modsquarerules2} apply in the same case, that is, when a web or a defect could produce two connected components by application of the original square rule, we expect the defect considered above to be related to a two-point function indicating whether two points are both on the single connected closed web of the  configuration.

More precisely, consider the Kuperberg web model for $q=e^{i\frac{\pi}{3}}$ in the plane geometry. Define the two-point correlation function $\mathcal{O}(a,b)$, with respect to $Z$ in~\eqref{modpartfunction}, of the indicator function $I(a,b)$ that takes the value~$1$ whenever the two points $a$ and $b$ are on the same connected web, and~$0$ otherwise. That is,
\begin{equation}
\label{2ptfunction}
    \mathcal{O}(a,b)=\frac{\lim_{[3]_q\rightarrow 0} \frac{1}{[3]_q}\sum_{c\in \mathcal{K}}\left[I(a,b)\, x_1^{N_1}x_2^{N_2}(yz)^{N_V}w_{\rm K}(c)\right]}{Z}\ ,
\end{equation}
where $w_{\rm K}(c)$ is defined in Section \ref{sec:kuploc}.
At criticality, denote by $(h,\Bar{h})$ the lowest conformal weights of the operators involved in the correlation function $\mathcal{O}(a,b)$. By analogy with the loop model case, we expect the transfer matrix \eqref{kuptransfermatrixcyl} to contain information on the continuum limit of the model defined by \eqref{modpartfunction} through the finite size scaling of its eigenvalues. Indeed, observe that the numerically estimated effective central charge in the dense phase is \textit{not} $0$ (see  the dense phase discussion at $q=\mathrm{e}^{i\pi/3}$ in Section~\ref{sec:pd}) as would have been the case for the non-renormalised partition function~\eqref{Z_K} which is equal to~1.

In the continuum, by mapping conformally the plane to the cylinder, with $a$ and $b$ mapped to the bottom and top of the cylinder, we can look at configurations contributing to the numerator of \eqref{2ptfunction} as connected open webs running along the cylinder. These are exactly the ones considered above and we can conjecture that the conformal weights $(h,\Bar{h})$ are determined by the asymptotic equivalent (in the limit when $M/L \gg 1$) of 
\begin{align}
    \frac{Z_{\rm K}^{\bm{e},\bm{\rho}}}{Z} \,, \qquad \text{with}\quad \bm{e}=\bm{e}_0=\frac{1}{3}\bm{\rho}\ .
\end{align}
Hence 
\begin{align}
    (h,\Bar{h})=(h_{\bm{e}_0,\bm{\rho}},\Bar{h}_{\bm{e}_0,\bm{\rho}}) \,.
\end{align}
We then obtain the fractal dimension $D_{\rm f}$ of critical webs at the point \eqref{eipi3} as the corresponding co-dimension,
\begin{align}
    D_{\rm f}=2-(h_{\bm{e}_0,\bm{\rho}}+\Bar{h}_{\bm{e}_0,\bm{\rho}}) \,.
\end{align}

\section{Conclusion}
\label{sec:disc}

We have continued our investigation of lattice models of webs, initiated in \cite{LGJ21}.
In this paper we have given a local formulation of the Kuperberg $A_2$ web models in terms of a vertex model. We have shown that the local transfer matrices possess a $U_{-q}(\mathfrak{sl}_3)$ symmetry. Thanks to a numerical diagonalisation of the row-to-row transfer matrices for various sizes, we have been able to explore the phase diagram of the model. For $q={\rm e}^{i \gamma}$ with $\gamma \in [0,\pi]$, we have identified two phases of interest where the model exhibits critical behaviour. These two phases are the higher-rank analogues of the so-called dense and dilute phases of the $O(N)$ loop model. In particular, it appears that a straightforward generalisation of the percolation model on the triangular lattice, obtained by considering the domain walls in the infinite-temperature three-state Potts model, is critical and forms part of the dense phase (with $q = {\rm e}^{i \pi / 4}$).

We have then defined electromagnetic partition functions and studied in detail their interpretation in terms of geometrical objects. In particular, we have studied potential applications of electromagnetic partition functions to random geometry. Contrarily to the loop-model case, the insertion of a magnetic defect does not only constrain the geometry of configurations but also modifies the weight of the open web containing the defect, due to the  rule~\eqref{modsquarerule} additional to the standard Kuperberg rules. This, in general, prevents us from relating electromagnetic partition function to indicator two-point functions of the web model in the plane (or half-plane). We have however identified some exceptions to this obstacle. In particular, we have been able to conjecture the fractal dimension of critical webs at $q={\rm e}^{i\pi/3}$. Unfortunately, we have not been able to reach interesting indicator two-point functions in the context of interfaces of the three-states Potts model, whose interfaces are known to exhibit interesting features that are yet only partially understood analytically \cite{GC07,Dubail10,Dubail10b}. It is also not clear whether the higher-rank $A_n$ web models \cite{LGJ-higherrank} can provide such results for $\mathbb{Z}_n$ spin models, with potential applications to interfaces in the four-state Potts and Ashkin-Teller models \cite{Dubail10,Dubail10b,PiccoSantachiara11}.

Our immediate continuation of the programme initiated in \cite{LGJ21} will be to relate
the dominant excitations of the electromagnetic partition functions, defined in the present paper,
to electromagnetic operators within a Coulomb Gas formulation of the continuum limit of web models \cite{LGJ-CG}. 
This will provide exact formulae for the central charges of dense and dilute critical points, as well as exact expressions of the conformal weights $h_{\bm{e},\bm{m}}$ and $\Bar{h}_{\bm{e},\bm{m}}$. The dense phase exhibits central charges equal to those of an integrable fifteen-vertex model \cite{DEI16}, closely related to a spin chain with the first fundamental representation of $U_{-q}(\mathfrak{sl}_3)$ living on every site. It thus seems that the dilution and the presence of the second fundamental representation at the dense critical point do not affect the continuum limit. Interestingly, as far as we know, no integrable representatives of the dilute phase exist. We remind that in the loop case, the dense phase is in the same universality class as the XXZ spin chain, whereas the dilute phase is obtained from a spin chain built from the trigonometric $A_2^{(2)}$ $R$-matrix, whose continuum limit has recently been shown to conceal a few surprises \cite{Vernier1,Vernier2}. In the $A_2$ web model case, it is not obvious what would be a candidate, if it exists, for an integrable representative of the dilute phase. 

It is also possible to define a local vertex-model formulation of the higher-rank $A_n$ web models\cite{LGJ-higherrank}. We then expect an analogous but more complicated phase diagram. Indeed, the $A_3$ web model has been shown to be equivalent to the Ashkin-Teller model for a certain tuning of the parameter $q$ and vertex fugacities\cite{LGJ21}. We expect this to constitute the ``dilute'' phase of the $A_3$ web model. In this correspondance, the critical line of the Ashkin-Teller model is swept when tuning a ratio of the bond fugacities corresponding to the different fundamental representation of $U_{-q}(\mathfrak{sl}_4)$. We can then ask whether there exists a ``dense'' phase containing a whole line of inequivalent critical points as well. The same questions appear in the general case of $A_n$ web models.

We can also wonder whether there exist integrable representatives of these would-be dense and dilute phases. Indeed, it is possible that, even in the dense phase, the presence of all the fundamental representations on each site of an integrable spin chain must be required. To our knowledge, no such case has been studied before.

\appendix

\newcommand{\balance}{\boldsymbol{g}}
\newcommand{\kk}{\mathbb{C}}
\newcommand{\coev}{\mathrm{coev}}

\section{Conventions for quantum groups}
\label{sec:quantumgroupconventions}
We recall here a definition of the Hopf algebra \UU\ and its pivotal structure. 
First, let $\bm{\alpha}_i$, $i\in\llbracket1,n-1\rrbracket$ be the simple roots of $\mathfrak{sl}_n$. We normalise them such that $(\bm{\alpha}_i,\bm{\alpha}_j)=2\delta_{ij}-\delta_{i,j+1}-\delta_{i,j-1}$. Denote fundamental weights by $\bm{w}_j$, $j\in\llbracket1,n-1\rrbracket$. They satisfy $(\bm{\alpha}_i,\bm{w}_j)=\delta_{ij}$.
Then, the $\mathbb{C}(q)$-algebra \UU\ is  generated by $E_i$, $F_i$, $q^{H_i}$ for $i\in\llbracket1,n-1\rrbracket$ satisfying the following relations:
\begin{subequations}
\begin{align}
    q^{H_i}q^{H_j} &= q^{H_j}q^{H_i} \,, \\
    q^{H_j}E_iq^{-H_j} = q^{(\bm{\alpha}_i,\bm{\alpha}_j)}E_i& \,, \qquad
    q^{H_j}F_iq^{-H_j} = q^{-(\bm{\alpha}_i,\bm{\alpha}_j)}F_i \,, \\
    [E_i,F_j] &= \delta_{ij}\frac{q^{H_i}-q^{-H_i}}{q-q^{-1}} \,, \\
    [2]_q E_iE_jE_i &= E_i^2E_j+E_jE_i^2 \,, \quad \text{if } |i-j|=1 \,, \\
    [E_i,E_j] &= 0 \,, \ \ \qquad \qquad \qquad \text{if } |i-j|>1 \,, \\
    [2]_qF_iF_jF_i &= F_i^2F_j+F_jF_i^2 \,, \ \quad \text{if } |i-j|=1 \,, \\
   [F_i,F_j] &= 0 \,, \ \ \qquad \qquad \qquad \mbox{if } |i-j|>1 \,.
\end{align}
\end{subequations}
It is a Hopf algebra with the coproduct
\begin{align}
    \Delta(E_i)=E_i\otimes q^{H_i} + 1\otimes E_i \,, \qquad \Delta(F_i)=F_i\otimes 1 + q^{-H_i}\otimes F_i \,, \qquad \Delta(q^{H_i})=q^{H_i}\otimes q^{H_i} \,,
\end{align}
the antipode
\begin{align}
    S(E_i)=-E_iq^{-H_i}\,, \qquad S(F_i)=-q^{H_i}F_i \,, \qquad S(q^{H_i})=q^{-H_i} \,,
\end{align}
and the counit
\begin{align}
    \epsilon(E_i)=0 \,, \qquad \epsilon(F_i)=0 \,, \qquad \epsilon(q^{H_i})=1 \,.
\end{align}

In what follows we use the notation $H_{\bm{\alpha}_i} := H_i$. In particular, for the Weyl vector $\bm{\rho} = \frac{1}{2}\sum_{i=1}^{n-1} i(n-i) \bm{\alpha}_i$, which is the half sum over all positive roots, we have $2H_{\bm{\rho}} = \sum_{i=1}^{n-1} i(n-i) H_i$.

A group-like\footnote{\textit{Group-like} means that
$\Delta(g)=g\otimes g$. It then follows that $g$ is invertible, $S(g)=g^{-1}$ and $\epsilon(g)=1$.} 
element $\balance$ of a Hopf algebra $U$ is called  \textit{pivotal} if its conjugation automorphism expresses the square of the antipode:
\begin{equation}
S^2(x) = \balance x \balance^{-1},
\end{equation}
for all $x\in U$.
The pivotal element of \UU\ is given by 
\begin{align}
 \balance =   q^{2H_{\bm{\rho}}}=q^{\sum_{1\leq i \leq n-1} {i(n-i)H_i}} \,.
\end{align}

For a  Hopf algebra $U$ with a pivotal element $\balance$,
each finite-dimensional $U$-module $V$ has a (left) dual
$V^*=\mathrm{Hom}(V,\mathbb{C})$ with the $U$ action defined by $(hf)(x)=f(S(h)x)$, for any $f\in V^*$, and $h,x\in U$.
With this, we define  the standard left duality maps:
\begin{align}\label{E:DualitiesC}
\mathrm{ev}_{V}&: \:\: V^*\otimes V\rightarrow \kk, &\text{ given by } \quad &f\otimes v \mapsto f(v), \\ 
\coev_{V} &: \:\: \kk \rightarrow V\otimes V^{*}, & \text{ given by }\quad &1 \mapsto \sum_{j\in J} v_j\otimes v_j^*, \notag
\end{align}
where $\{v_j \,|\, j\in J\}$ is a basis of~$V$ and $\{v_j^*\,|\, j\in J\}$ is the dual basis of $V^*$,
while the pivotal element $\balance$ of $U$ allows to define the right  duality maps as follows
\begin{align}\label{E:DualitiesC-right}
 \widetilde{\mathrm{ev}}_V&: \:\: V\otimes V^*\rightarrow \kk, &\text{ given by }\quad & v\otimes f \mapsto f
(\balance v)\\
\widetilde{\mathrm{coev}}_V&: \:\: \kk \rightarrow V^{*}\otimes V, &\text{ given by }\quad & 
1 \mapsto \sum_i v_i^*\otimes 
\balance^{-1}v_i \ . \notag
\end{align}
Note that the axioms on the pivotal element ensure that the last two maps are $U$-intertwiners.

Finally, the quantum trace on any endomorphism $f$ of a $U$-module $V$ is defined as follows:
\begin{equation}\label{eq-qtr}
\mathrm{qtr}_V(f):= \widetilde{\mathrm{ev}}_V\circ(f\otimes\mathrm{id})\circ\coev_V(1) = \mathrm{tr}_V(l_{\balance} \circ f)
\end{equation}
where $\mathrm{tr}_V(f)$ is the usual  trace and $l_{\balance}$ is the left action by $\balance$.

\section{Some results on irreducible magnetised webs}
\label{App:MagWebs}
\subsection{Irreducible magnetised webs in a rectangle}
\label{App:IrrWebStrip}

We show here that for any choice of orientations of edges incident on univalent vertices of a magnetised web in a rectangle,  there exists an irreducible one that is a concatenation of H-webs.
Consider finite sequences $s=(s_1,s_2,\cdots,s_n)$ and $s'=(s'_1,s'_2,\cdots,s'_n)$  of orientations of the edges at the bottom and top boundary of a magnetised web. In this case, we say that the web connects $s$ to $s'$.
For instance, the following web connects $s=(\downarrow,\uparrow)$ to $s'=(\uparrow,\downarrow)$.
\begin{equation}\label{eq:H-web}
      \vcenter{\hbox{\includegraphics[scale=0.3]{diagrams/kupmodelstripmagdefect1.eps}}}
\end{equation}

We will show by induction in $n$ that there exists an irreducible magnetised web connecting $s$ to $s'$ and that is a concatenation of H-webs. The statement is obvious for $n=1$. Suppose that the result is true for $n-1$ and consider $s$ and $s'$ of length $n$. If $s_1=s'_1$, then we can draw a vertical edge that connects $\{s_1\}$ to $\{s'_1\}$. Then, by the induction hypothesis, there is an irreducible magnetised web connecting $s\setminus\{s_1\}$ to $s'\setminus\{s'_1\}$ that is a concatenation of H-webs. By juxtaposing this web to the right of the vertical edge considered above, we obtain an irreducible web connecting $s$ to $s'$ that is a concatenation of H-webs. 

If $s_1\neq s'_1$, consider the minimum index $k\geq 2$ such that $s_k=s'_1$. Denote by $\tau_i(x)$, $i\in \llbracket 1,n-1\rrbracket$, the permutation of the $i$th and $(i+1)$th elements of the finite sequence $x$ of length $n$. Clearly, there is an irreducible web connecting $x$ to $\tau_i(x)$. Either $x_i$ and $x_{i+1}$ are the same, and the irreducible web is simply a bunch of vertical edges. Or $x_i$ and $x_{i+1}$ are different, and we can use an H-shaped web as in~\eqref{eq:H-web}, or with all arrows reversed depending on the initial orientations of edges, connecting the bottom $i$th and $(i+1)$th edges to the top ones, the others being connected by vertical edges. Consider then
\begin{align}
    r=\tau_1\circ \tau_2 \circ \cdots \circ \tau_{k-1} (s) \,.
\end{align}
By the above discussion, there is a concatenation of H-webs that we call $W_1$, connecting $s$ to $r$, one for each transposition. Moreover, as $s_k\neq s_i$ for $i\in \llbracket1,k-1\rrbracket$, it is clear that each H-web corresponding to each of the transpositions is not merely a set of vertical edges. Here is an example
\begin{center}
    \includegraphics[scale=0.3]{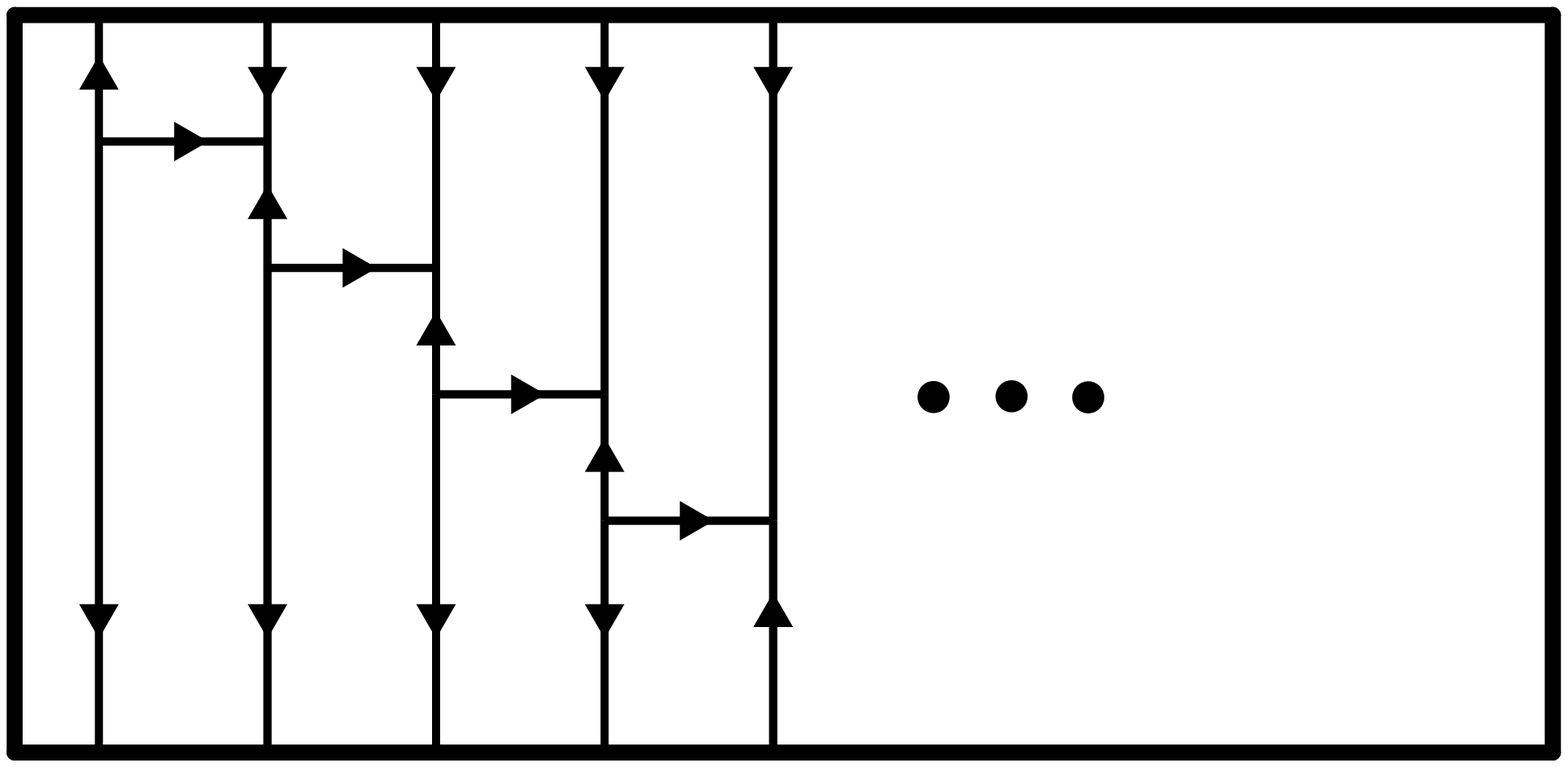}
\end{center}
Because $r_1=s'_1$, by the argument above, we know that there exists an irreducible web $W_2$ connecting $r$ to $s'$ that is a concatenation of H-webs. Moreover, we know that it connects $r_1$ to $s'_1$ by a vertical edge. Consider the concatenation of H-webs $W_2W_1$. It connects $s$ to $s'$ and it is irreducible. Indeed, no digons or loops appear in a concatenation of H-webs. For a square to appear in the concatenation $W_2W_1$, it is clear, due to the orientation of the edges at the top boundary of $W_1$ (the one linked to $W_2$), that it can only be the square containing the two left-most edges at the top  of $W_1$. However, the left-most edge of $W_2$ is a through-line (by the induction assumption) and so the two left-most edges at the bottom of $W_2$ cannot be a part of an $H$-shaped web, and therefore cannot produce a square while concatenating with $W_1$.

\subsection{Colourings of type C blocks}
\label{App:ColoringWrapWebs}

Consider a type C block with $n_1$ upward oriented and $n_2$ downward oriented edges connecting the bottom boundary to the wrapping cycle of edges $\mathfrak{L}$. We call these $n_1+n_2$ edges the bottom edges. Similarly there are $n_1$ upward oriented and $n_2$ downward oriented edges connecting the top boundary to the wrapping cycle of edges $\mathfrak{L}$, and we call them the top edges. Consider a segment of $\mathfrak{L}$ delimited by two consecutive points where two bottom edges, $e_1$ and $e_2$, meet $\mathfrak{L}$. We are interested in how many top edges are connected to this segment that we denote by $[e,e']$. If $e$ and $e'$ have the same orientation (both upward or both downward), then there must be an odd number of top edges connected to $[e,e']$. Whereas if $e$ and $e'$ have different orientations (one upward and one downward), then there must be an even number of top edges connected to $[e,e']$.

If $n_1=0$ or $n_2=0$, then all bottom edges have the same orientations and there must be an odd number of top edges connected to a segment delimited by two consecutive bottom edges. Hence this number must be equal to $1$. As all bottom edges and all top edges are coloured the same, it is clear that there are two possible colourings for edges in $\mathfrak{L}$. These edges are coloured in an alternate way, with the two colours that are different from the one of the bottom and top edges, and two such alternating colourings are possible.

If $n_1\neq 0$ and $n_2\neq 0$, then there exist two consecutive bottom edges $e$ and $e'$ such that the segment $[e,e']$ is not connected to any top edge. Indeed suppose the contrary. Denote by $e_i$, $i\in\llbracket1, n_1+n_2\rrbracket$, the bottom edges such that $e_i$ and $e_{i+1}$ are consecutive, where indices are taken modulo $n_1+n_2$. Then for all $i\in\llbracket1, n_1+n_2\rrbracket$, $[e_i,e_{i+1}]$ is connected to a number $m_i\geq 1$ of top edges. Moreover, every pair of consecutive edges with different orientations has $m\geq2$, therefore we get after summing over all $m$'s that the number of top edges is strictly larger than $n_1+n_2$, a contradiction.

 We thus have a pair of consecutive bottom edges $e$ and $e'$ such that the segment $[e,e']$ is not connected to any top edge. Necessarily $e$ and $e'$ have opposite orientations and thus different colours. The segment $[e,e']$ is constituted of one edge whose colour is fixed to be different to those of $e$ and $e'$. It is not hard to see that this fixes the colours of all edges in the cycle $\mathfrak{L}$, hence there is a unique possible colouring.

\subsection{Irreducible webs in the cylinder geometry}
\label{App:IrrWebCyl}
We show here that an irreducible web $c$ with nontrivial magnetic charge embedded in the cylinder is a concatenation of blocks of types A, B and C.

If there exists a path avoiding all edges and vertices of $c$ that goes from the bottom boundary to the top boundary of the cylinder, then, up to some winding, $c$ is contained inside a full rectangle. Hence it is a concatenation of blocks of type A and B.

Otherwise, the web, up to some winding, looks like
\begin{center}
    \includegraphics[scale=0.3]{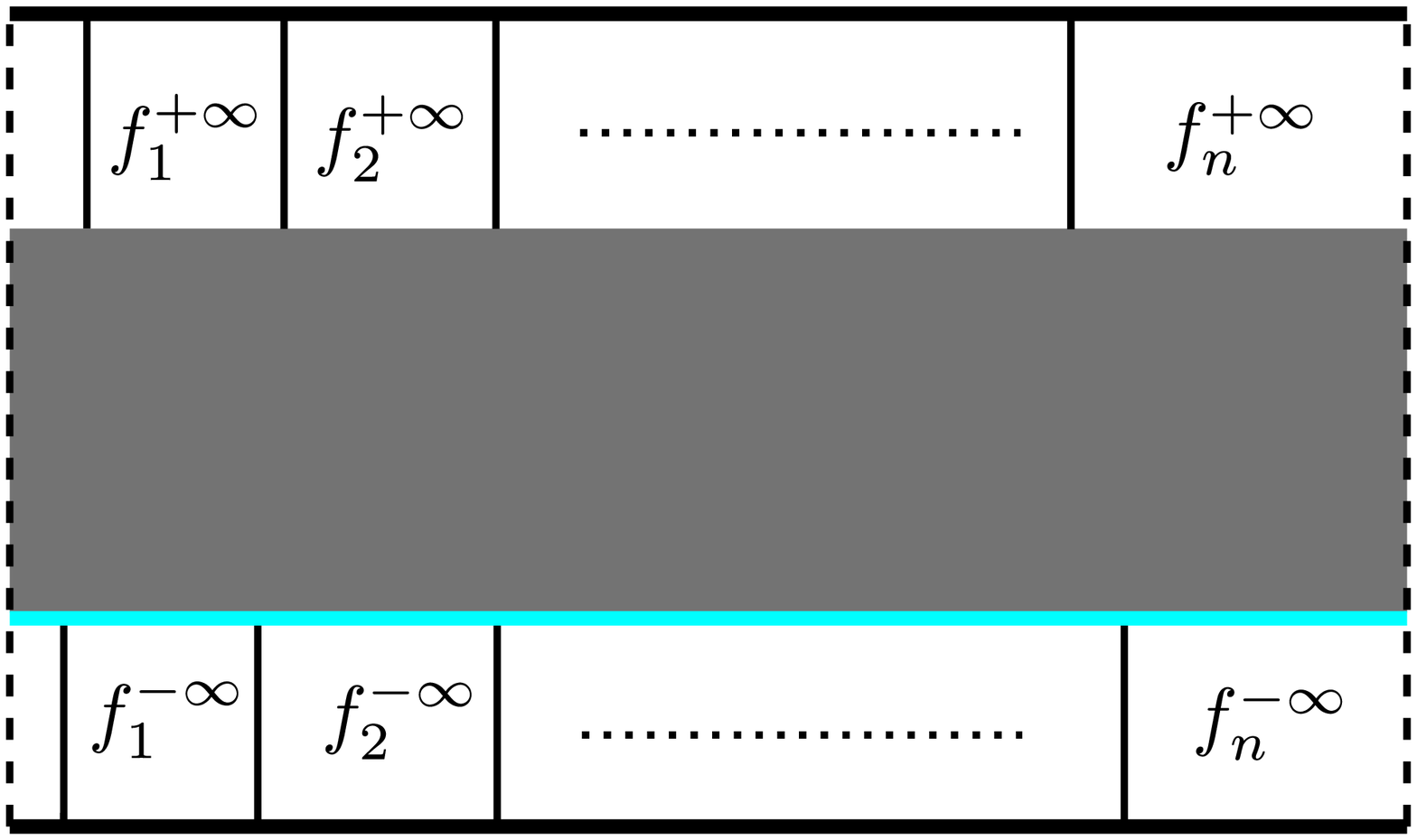}
\end{center}
where $n=n_1+n_2$ and we have labelled the bottom-most faces and top-most faces surrounded by the web. There are $n_1$ upward oriented edges and $n_2$ downward oriented edges connected to the bottom and to the top boundaries. Orientations are in any order. All the faces $f^{+\infty}_i$ and $f^{-\infty}_k$ are different. The cycle coloured in cyan at the bottom of the web must then be a union of edges. We call this cycle $\mathfrak{L}_1$. The web then looks like
\begin{center}
    \includegraphics[scale=0.3]{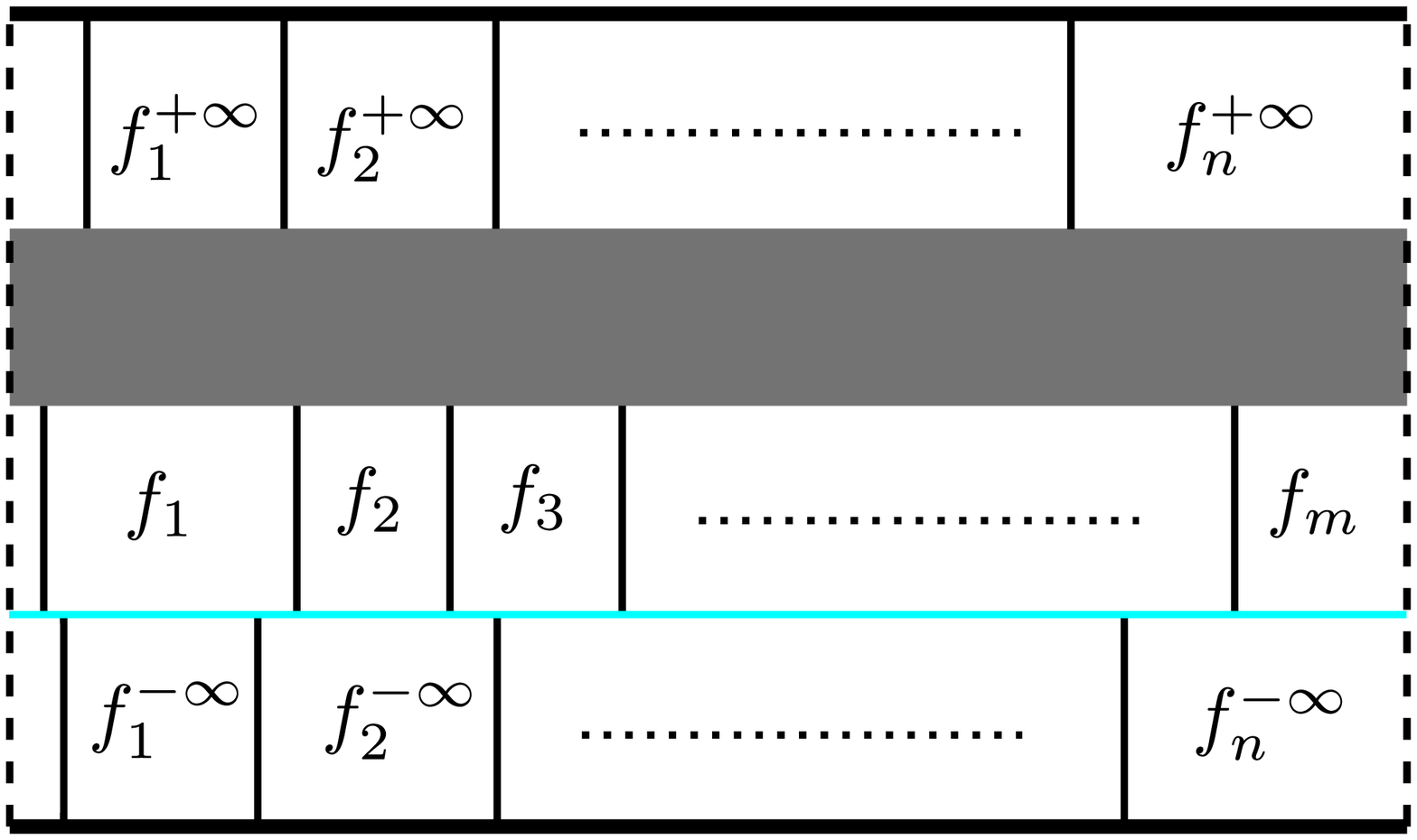}
\end{center}
where we have numbered the faces above $\mathfrak{L}_1$ connected to it. We have that $m\geq n$ because of the minimality of the cuts crossing the bottom-most and top-most edges. We claim that $m=n$. Indeed, suppose $m>n$.
Denote by $V^{\rm top}_1=m$ (respectively $V^{\rm bot}_1=n$) the number of edges connected to $\mathfrak{L}_1$ that are situated on top (respectively at the bottom) of $\mathfrak{L}_1$ but not in $\mathfrak{L}_1$. We then have that $V^{\rm top}_1-V^{\rm bot}_1>0$. There must be a face $f_k$ that is not in the set $\{f^{+\infty}_i,\ i \in \llbracket1,n\rrbracket\}$. This face must be closed, i.e., be surrounded by edges. Denote its surrounding edges present in $\mathfrak{L}_1$ by $e_i,\ i\in I$, and those absent by $e_j,\ j\in J$. Then consider the cycle $\mathfrak{L}_2=\mathfrak{L}_1\cup \{e_j,\ j\in J\} \setminus \{e_i,\ i\in I\}$. Here is an example with $k=1$:
\begin{center}
    \includegraphics[scale=0.3]{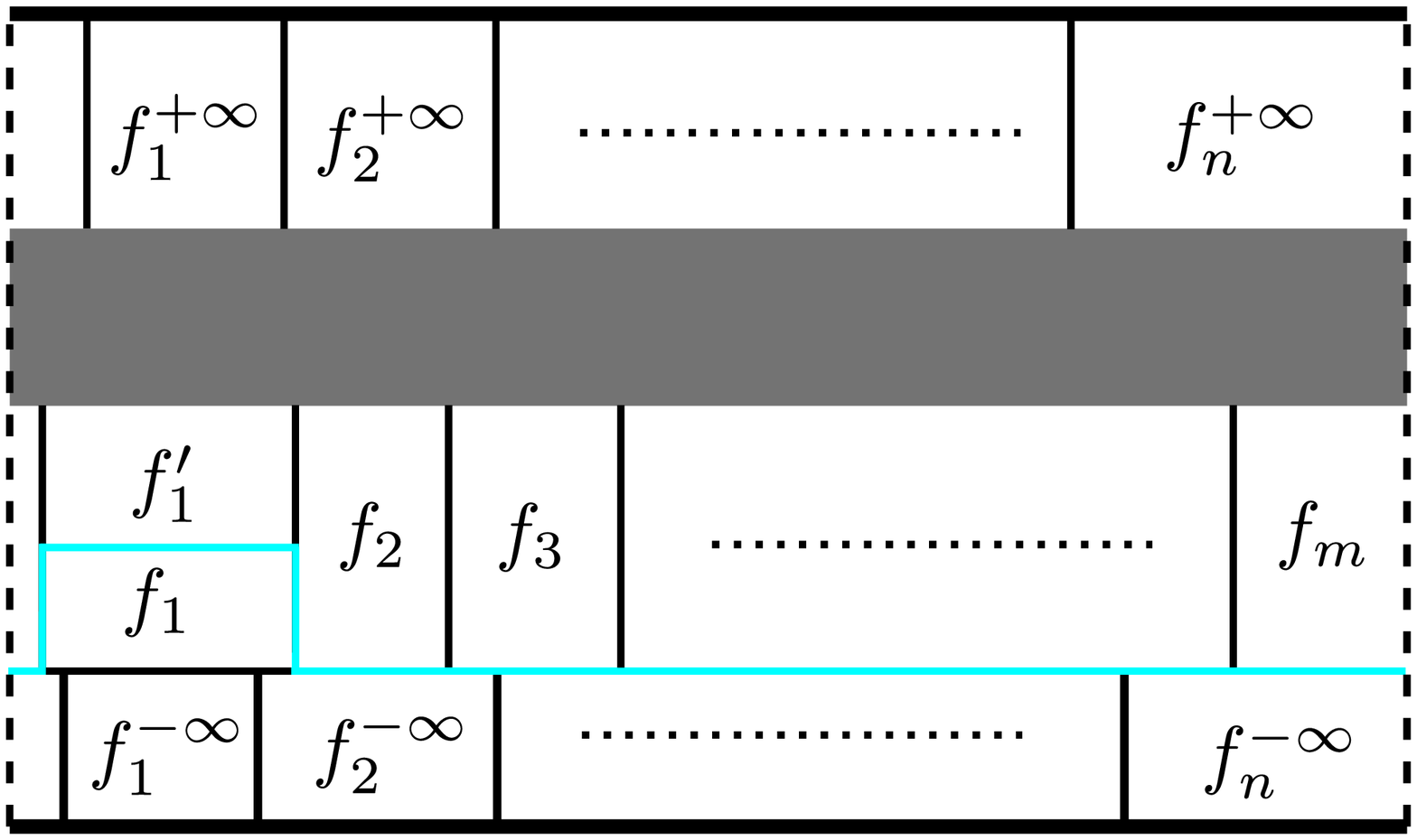}
\end{center}
where we depicted $\mathfrak{L}_2$ in cyan again.

Denote by $E^{\rm top}$ the number of edges that are connected to $f_k$ at the top of $\mathfrak{L}_2$ but not in $\mathfrak{L}_2$. Denote by $E^{\rm bot}$ the number of edges that are connected to $f_k$ at the bottom of $\mathfrak{L}_1$ but not in $\mathfrak{L}_1$. Because all digons and squares have been reduced, the face $f_k$ is surrounded by at least $6$ vertices, so we must have $E^{\rm top}+E^{\rm bot}\geq 4$. Denote by $V^{\rm top}_2$ (respectively $V^{\rm bot}_2$) the number of edges connected to $\mathfrak{L}_2$ that are situated on top (respectively at the bottom) of $\mathfrak{L}_2$ but not in $\mathfrak{L}_2$. We have that 
\begin{subequations}
\begin{align}
    V^{\rm top}_2&=V^{\rm top}_1+E^{\rm top}-2 \,, \\
    V^{\rm bot}_2&=V^{\rm bot}_1-E^{\rm bot}+2 \,,
\end{align}
\end{subequations}
which implies  
\begin{align}
    V^{\rm top}_2-V^{\rm bot}_2&=V^{\rm top}_1-V^{\rm bot}_1+E^{\rm top}+E^{\rm bot}-4>0 \,.
\end{align}
We can repeat the process and define cycles of edges $\mathfrak{L}_i$ such that $V^{\rm top}_i-V^{\rm bot}_i>0$. After $i_0$ iterations, the process terminates, and we have that $V^{\rm top}_{i_0}=n>V^{\rm bot}_{i_0}$. But this contradicts the minimal cut assumption. Hence $m=n$ and thus the web necessarily contains a block of type C. The procedure of constructing the wrapping cycle can be repeated until we are able to draw a path (from the top of the constructed concatenation of C type blocks) avoiding all edges and vertices that goes from  bottom to top, up to some winding: this means that the given web $c$ is a concatenation of blocks of type A, B and C.

\end{document}